\documentclass[prx,amsmath,amssymb,twocolumn,superscriptaddress]{revtex4-2}
\usepackage{bm,amsthm,amsmath,amsfonts,graphicx,caption,subcaption,xcolor,appendix,multirow}
\usepackage[colorlinks,bookmarks=true,citecolor=blue,linkcolor=blue,urlcolor=blue]{hyperref}

\newtheorem{theorem}{Theorem}[]
\newcommand{\mC}{\mathcal{C}}
\renewcommand{\bf}[1]{{\bm #1}}
\newcommand{\BZ}{\mathrm{BZ}}

\newcommand{\Gr}{\mathrm{Gr}}
\newcommand{\Ric}[1]{\mathrm{Ric}\; (#1)}
\newcommand{\ket}[1]{| #1 \rangle}
\newcommand{\bra}[1]{\langle #1|}
\DeclareMathOperator{\Tr}{tr}

\begin{document}

\title{Theory of Generalized Landau Levels and Implication for non-Abelian States}

\author{Zhao Liu}
\thanks{B.~M. and Z.~L. contributed equally to this work.}
\affiliation{Zhejiang Institute of Modern Physics, Zhejiang University, Hangzhou 310058, China}

\author{Bruno Mera}
\thanks{B.~M. and Z.~L. contributed equally to this work.}
\affiliation{Instituto de Telecomunica\c{c}\~oes and Departmento de Matem\'{a}tica, Instituto Superior T\'ecnico, Universidade de Lisboa, Avenida Rovisco Pais 1, 1049-001 Lisboa, Portugal}
\affiliation{Advanced Institute for Materials Research (WPI-AIMR), Tohoku University, Sendai 980-8577, Japan}

\author{Manato Fujimoto}
\affiliation{Department of Physics, Harvard University, Cambridge, MA 02138, USA}
\affiliation{Department of Applied Physics, The University of Tokyo, Hongo, Tokyo, 113-8656, Japan}

\author{Tomoki Ozawa}
\affiliation{Advanced Institute for Materials Research (WPI-AIMR), Tohoku University, Sendai 980-8577, Japan}

\author{Jie Wang}
\thanks{Author to whom correspondence should be addressed to: J.~W. (jie.wang0005@temple.edu).}
\affiliation{Department of Physics, Temple University, Philadelphia, Pennsylvania, 19122, USA}

\begin{abstract}
    Quantum geometry is a fundamental concept to characterize the local properties of quantum states. It is recently demonstrated that saturating certain quantum geometric bounds allows a topological Chern band to share many essential features with the lowest Landau level, facilitating fractionalized phases in moir\'e flat bands. In this work, we systematically extend the consequence and universality of saturated geometric bounds to arbitrary Landau levels by introducing a set of single-particle states, which we term as ``generalized Landau levels''. These generalized Landau levels exhibit exactly quantized values of integrated trace of quantum metric determined by their corresponding Landau level indices, regardless of the nonuniformity of their quantum geometric quantities. We derive all geometric quantities for individual and multiple generalized Landau levels, discuss their relations, and understand them in light of the theory of holomorphic curves and moving frames. We further propose a model by superposing few generalized Landau levels which is supposed to capture a large portion of the single-particle Hilbert space of a generic Chern band analogous to the first Landau level. Using this model, we employ exact diagonalization to identify a single-particle geometric criterion for permitting the non-Abelian Moore-Read phase, which is potentially useful for future engineering of moir\'e materials and beyond. We use a double twisted bilayer graphene model with only adjacent layer hopping term to show the existence of first generalized Landau level type narrow band and zero-field Moore-Read state at the second magic angle which serves as a promising starting point for more detailed future studies. We expect that generalized Landau levels will serve as a systematic tool for analyzing topological Chern bands and fractionalized phases therein.
\end{abstract}

\maketitle

\tableofcontents

\section{INTRODUCTION}
\label{sec: introduction}

Quantum geometry, an intrinsic and fundamental feature of quantum states, plays a significant role in various applications within both topological and non-topological quantum matter~\cite{HaldaneNobelLecture,Di_Review,Torma_Essay_23}. The influence of quantum geometry extends across a diverse range of phenomena, spanning from linear and nonlinear responses~\cite{NiuThoulessWu,Sundaram_Niu,Di_BerryDOS,haldaneanomaloushall,LiangFu_Sodemann_BerryDipole15} to non-equilibrium physics~\cite{GoldmanCrhistofNaturePhysics18,RepellinGoldman19,GoldmanOzawa24}, encompassing domains such as topological photonics and light-matter interactions~\cite{Nagaosa_ShiftCurrent16,RMP_TopologicalPhotonics19,Ahn:2022aa}, coherence bounds~\cite{Peotta:2015aa,Randeria_19PRX,Randeria_PNAS_21,BAB_Torma_NatureReview22,Debanjan_23_PNAS,LiangFu_bounds_24_PRX}, emergent degrees of freedom~\cite{Haldanegeometry,Son_PRR_21,LingjieDu_Graviton_Nature_24}, the stability of quantum matter~\cite{RahulRoy14,Jackson:2015aa,Martin_PositionMomentumDuality,ChinaHuaLeePRX15}, and beyond.

Important quantum geometric concepts include the Berry curvature and the Fubini-Study metric --- also known as the quantum metric. The quantum metric and the Berry curvature are related by a geometric bound~\cite{RahulRoy14}. Recent theoretical progress has clarified the mathematical meaning of saturating this bound, namely, the existence of a notion of momentum-space holomorphicity --- a complex structure --- induced from the geometry of the space of quantum states~\cite{kahlerband1,kahlerband2,kahlerband3}. The saturation of the trace bound is linked to the formation of lowest Landau level (LL) type wavefunctions, enhancing the stability of fractional Chern insulators (FCIs)~\cite{Regnault_Bernevig_CFI_PRX11,Titus_FCI_11,Sheng_FCI_11,zhao_review} under short-ranged interactions~\cite{Grisha_TBG2,JieWang_exactlldescription,LedwithVishwanathParker22,Jie_hierarchyidealband}. Recently, FCIs and composite Fermi liquid states~\cite{HalperinLeeRead,Son} have been theoretically proposed and experimentally realized in real materials at zero field, including the moir\'e transition metal dichalcogenides (TMD)~\cite{XiaodongXu_Signature23,KinfaiMak_Thermodynamic23,XiaoDongXu_Observation23,TingxinLiPRX23} and multilayer graphene~\cite{LongJuFCI23}. Engineering materials that saturate the geometric bound is an essential ingredient towards realising the lowest LL physics.

To date, much of the research has focused on the lowest LL physics and Abelian FCIs. More exotic topological orders, such as the non-Abelian ones described by the Moore-Read and Read-Rezayi states, are naturally present at higher LLs (specifically the first LL) in quantum Hall systems~\cite{MoreReadState,Read_Rezayi}, although none of them have yet been experimentally realized in topological flatbands at zero field. A critical challenge is ensuring stability against geometric fluctuations (non-uniformity of quantum geometries), a factor absent in standard LLs. Recently, there have been theoretical proposals for realizing Moore-Read non-Abelian phase in flat Chern bands in twisted MoTe$_2$ materials at small twist angles and multilayer moir\'e graphene~\cite{LiangFuNonabelian24,YangZhangNonabelian24,ChoNonabelian24,DiXiaoNonabelian24,Fujimoto24,Emil_2405}. However, to date, general discussions of the geometric stability for many-body phases especially non-Abelian fractionalized orders are still missing.

In this work, we systematically generalize the standard quantum geometric condition for lowest LL physics to arbitrarily higher LLs. We introduce an orthonormal basis, termed ``generalized Landau levels'', whose quantum geometry is allowed to fluctuate but, nevertheless, maintain a quantized value of their integrated trace of quantum metric as in the case of the standard LLs (see Theorem.~\ref{theorem:quantizedtrace}). The quantized integrated trace value determines the LL index of the generalized LLs. The quantum geometries of individual, as well as that of filling several generalized LLs are explicitly derived. The results are naturally understood in terms of the theory of holomorphic curves and Cartan's method of moving frames. The generalized LLs form a complete basis for generic topological bands~\cite{griffiths:74,griffiths:harris:14}. We propose a toy model constructed from generalized LLs and designed to resemble key features of the standard first LL. Within such model, we employed exact-diagonalization and identified a region of single-particle geometric quantities, in terms of the integrated trace of quantum metric and standard deviation of Berry curvature, permitting the Moore-Read non-Abelian fractionalized phase (see Fig.~\ref{fig:trgwindow}). Our geometric criteria for the non-Abelian phases is supposed to be general and may serve as a guiding principle for future material design.

The structure of this manuscript is as follows. We begin by reviewing the fundamental concepts of quantum geometry and band theory in Sec.~\ref{sec: quantum geometry, band theory and fractional Chern insulators}. Sec.~\ref{sec: QUANTIZED TRACE FORMULA AND CANONICAL BASIS} is pivotal, establishing the theoretical foundations of this work. It presents the proof of the integrated trace formula satisfied by the generalized LL basis states and explores various geometric quantities and their relationships. Holomorphic curves and moving frames are presented in that section to provide a unifying gauge invariant geometric formulation of the results. In Sec.~\ref{sec:suposed_state_geometry_and_decomposition}, we progress from generalized LL basis states to general states constructed by superposing two or multiple basis states. We present an algorithm to extract the weights of a Chern band on generalized LL basis states if it can be decomposed into a finite number of generalized LLs. Sec.~\ref{sec:nonabelian_fractionalization} discusses the implications for non-Abelian fractionalized states and proposes a quantum geometric criterion for the Moore-Read phase, which represents another focal point of this work. We summarize key results, discuss open problems and suggest future directions in Sec.~\ref{sec: conclusion}.
\section{REVIEW OF QUANTUM GEOMETRY AND RELATED BAND THEORIES}
\label{sec: quantum geometry, band theory and fractional Chern insulators}
In this section, we review the fundamental concept of quantum geometry in the context of band theory for both single-band and multiple-band settings, recall its interpretation in terms of the underlying geometry of complex projective spaces and Grassmannians, and recall the quantum geometric bounds involving the Berry curvature and quantum metric. We will also review various recently introduced notions, including K\"ahler bands~\cite{kahlerband1,kahlerband2,kahlerband3}, ideal bands~\cite{JieWang_exactlldescription,Jie_Origin22}, and vortexable bands~\cite{LedwithVishwanathParker22,Fujimoto24}. We will discuss their relationships and implications for the stability of FCIs.
\subsection{Quantum Geometry and Bounds}
\subsubsection{Quantum geometry of Bloch states}
We consider Bloch states $|\psi_{n\bm k}\rangle$ of two-dimensional (2D) electrons, where $n$ is the band index and momentum $\bm k$ is the lattice translation quantum number. The state $\ket{\psi_{n\bm{k}}}$ can carry internal components describing degrees of freedom such as spin, valley and sublattice. Bloch wavefunctions are orthogonal in the sense of $\langle \psi_{n\bm{k}}|\psi_{n^\prime \bm{k}^\prime}\rangle \propto \delta_{n,n^\prime}\delta_{\bm{k},\bm{k}^\prime}$. LLs can be incorporated into this description by regarding $\bm{k}$ as magnetic translation quantum numbers. We denote the first quantized wavefunction by $\psi_{n\bm k}(\bm r) = \langle\bm r|\psi_{n\bm k}\rangle$.

The cell-periodic functions $|u_{n\bm k}\rangle$ are defined by
\begin{equation}
    u_{n\bm k}(\bm r) \equiv e^{-i\bm k\cdot\bm r}\psi_{n\bm k}(\bm r),\quad u_{n\bm k}(\bm r) = \langle\bm r|u_{n\bm k}\rangle,\label{def_u}
\end{equation}
which obey, independently of $\bm{k}$, the same real-space translation properties and, hence, can be understood as vectors in the same Hilbert space. The overlaps $\langle u_{m\bm k}|u_{n\bm k'}\rangle$ are in general nonzero, and they give rise to gauge covariant quantum geometric quantities: the (non-Abelian) Berry curvature $\Omega_{mn}(\bm k)$ which characterizes the non-Abelian phase factor accumulated in an infinitesimal rectangle in momentum space, and the non-Abelian quantum metric $g^{ab}_{mn}(\bm k)$, with $a,b$ labelling momentum-space indices and $m,n$ band indices, which describes the decaying properties of quantum states via
\begin{equation}
    |\langle u_{m\bm k}|u_{n,\bm k+\delta\bm k}\rangle|^2 = 1 - \sum_{a,b=x,y}g^{ab}_{mn}(\bm k)\delta\bm k_a\delta\bm k_b + \mathcal{O}(|\delta\bm k|^3).
\end{equation}

The Berry curvature and quantum metric are respectively the imaginary anti-symmetric and real symmetric parts of the gauge-covariant, Hermitian, positive semi-definite quantum geometric tensor,
\begin{eqnarray}
    \chi^{ab}_{mn}(\bm k) &\equiv& \langle D^a u_{m\bm k}|D^b u_{n\bm k}\rangle,\nonumber\\
    &=& g^{ab}_{mn}(\bm k) + \frac{i}{2}\varepsilon^{ab}\Omega_{mn}(\bm k), \label{def_nonabelian_geometries}
\end{eqnarray}
where $A_{mn}(\bm k) = -i\langle u_{m\bm k}|\partial_{\bm k}^au_{n\bm k}\rangle$ is the non-Abelian Berry connection and the covariant derivative $|D^au_{n\bm k}\rangle$ is denoted as
\begin{equation}
    |D^au_{n\bm k}\rangle = |\partial_{\bm k}^au_{n\bm k}\rangle - i\sum_{m=1}^{r}|u_{m\bm k}\rangle A_{mn}(\bm k).
\end{equation}
Here the sum is restricted to the $r$ bands under consideration. 

The non-Abelian quantum geometric tensor $\chi^{ab}_{mn}$ becomes an Abelian quantum geometric tensor when either a single band or the fermionic many-body Slater determinant state constructed from the band complex is considered. For the latter, we have 
\begin{equation}
     \chi^{ab}_{\rm det}(\bm k) = \sum_{m=1}^{r}\chi^{ab}_{mm}(\bm k) \equiv \mathrm{Tr}\left[\chi^{ab}(\bm k)\right], \label{trace_nonabelian_det_abelian}
\end{equation}
where the trace $\mathrm{Tr}$ is over the band index. Mathematically, the determinant state describes the determinant line bundle (the highest exterior power bundle) associated to the vector bundle of the occupied Bloch states. Physically, it is the many-body state obtained by fully filling the band with fermions. A direct proof of Eq.~(\ref{trace_nonabelian_det_abelian}) and its relation to Pl\"ucker embedding can be found in Appendix~\ref{sec:quantummetricfilledbands}.

The fact that the Abelian quantum geometric tensor is a positive semi-definite Hermitian tensor implies a bound relating its real symmetric part (quantum metric) and imaginary anti-symmetric part (Berry curvature):
\begin{equation}
    \Tr_{\widetilde g} g(\bm k) \geq 2\sqrt{\det g(\bm k)} \geq |\Omega(\bm k)|. \label{eq: quantum geometric bounds}
\end{equation}
These bounds are valid for any ``general trace'' $\Tr_{\widetilde g}(g) \equiv \sum_{a,b} \widetilde g_{ab}g^{ab}$ defined with respect to a unimodular matrix satisfying $\sqrt{\det(\widetilde g)} = 1$, which is not necessarily the identity matrix. In later sections, we will review the physical implications from saturating the quantum geometric bounds. Notice that we have introduced distinct notations for two different traces: the notation $\Tr_{\widetilde g}$ denotes the generalized trace over spatial indices, and the notation {\rm Tr} denotes the trace over band indices, as shown in Eq.~(\ref{trace_nonabelian_det_abelian}).
\subsubsection{Geometry of Bloch bands}
\label{subsec: geometry of Bloch bands}
In this section, we provide a review of the geometry of Bloch bands. From a differential geometric perspective, Bloch bands can be interpreted as smooth maps from the Brillouin zone, a genus-1 smooth compact 2D manifold, to K\"ahler manifolds parameterizing certain spaces of quantum states. The quantum geometric quantities defined above are then naturally understood as the pullback of geometric quantities on the target K\"ahler manifold to the Brillouin zone: the Abelian quantum metric and Berry curvature are, respectively, the pullback of the Fubini-Study (FS) metric and symplectic form. In the single-band case, the corresponding K\"ahler manifold is the projective space, and, more generally, in the multiple-band case it is the Grassmannian of the rank equal to the number of bands. Such perspective plays a crucial role for the theory of K\"ahler bands~\cite{kahlerband1,kahlerband2,kahlerband3} which we will review in Sec.~\ref{subsubsec: Kaehler bands}.

We begin by reviewing the quantum geometry of a single band in this language. If the Hilbert space under consideration is $N$-dimensional, the unnormalized single-band state $|u_{\bm k}\rangle$ can be represented by a nonvanishing $N-$dimensional column vector
\begin{equation}
    |u_{\bm k}\rangle =Z(\bm k)\equiv \begin{bmatrix}
    Z_1(\bm k)\\ Z_2(\bm k)\\ \vdots\\ Z_N(\bm k)    
    \end{bmatrix}.
\end{equation}

There is a $\mathrm{GL}(1;\mathbb{C})=\mathbb{C}^*$ ambiguity in the state such that $Z(\bm k)$ and $G(\bm{k}) Z(\bm k)$, for $G(\bm{k})\in\mathbb{C}^*$ a nonvanishing complex number, represent the same physical state. $Z(\bm k)$ and all its equivalent vectors form a ray, which is a one-dimensional subspace of $\mathbb{C}^N$ to which $Z(\bm k)$ belongs to. The set of all rays forms the complex projective space consisting of all pure quantum states, denoted by $\mathbb{C}P^{N-1}$.

The space $\mathbb{C}P^{N-1}$ is a K\"ahler manifold. This means that it is a complex manifold and also that the FS metric and symplectic forms are compatible through the complex structure. Furthermore, they are both locally determined by derivatives with respect to the complex coordinates and its complex conjugates of the K\"ahler potential. In terms of the homogeneous complex coordinates $W_i \equiv Z_i/Z_1 \quad (2\leq i \leq N)$, where $Z_{1} \neq 0$, the local K\"ahler potential $\varphi$ is given by
\begin{equation}
    \varphi = \log  \left(1+\sum_i |W_i|^2\right), \label{eq: Kaehler potential Projective Space}
\end{equation}
and the FS symplectic form and metric are determined by
\begin{eqnarray}
    \omega_{FS} &=& \frac{i}{2} \sum_{ij} \frac{\partial^2 \varphi}{\partial W_i\partial \bar{W}_{j}} dW_{i}\wedge d\bar{W}_{j}, \nonumber\\
    g_{FS} &=& \sum_{ij} \frac{\partial^2 \varphi}{\partial W_{i}\partial \bar{W}_{j}} dW_{i}d\bar{W}_{j}. \label{eq: FS Kaehler stucture projective space}
\end{eqnarray}

The quantum geometry of electronic states as described by the Berry curvature and the quantum metric, is inherited from the geometry of the complex projective space through the map determined by the Bloch wavefunction. Regarding the state $|u_{\bm k}\rangle$ as a map $P: \mathrm{BZ}^2\cong \mathbb{T}^2 \rightarrow \mathbb{C}P^{N-1}$ from the Brillouin zone $\BZ^2$ to projective space, the associated Berry curvature and quantum metric are given, respectively, by the pullback of the FS symplectic form and metric. We use the notation $P$, because the state determined by the Bloch wavefunction is uniquely described by the orthogonal projector $P(\bm{k})=\ket{u_{\bm{k}}}\bra{u_{\bm{k}}}/\langle u_{\bm{k}}|u_{\bm{k}}\rangle$.

The generalization to multiple bands requires the notion of Grassmannian manifold. The periodic state of a rank-$r$ band complex --- {\it i.e.} a band having $r$ linearly independent states per momentum --- can be locally described by
\begin{equation}
    \begin{bmatrix}\ket{u_{1\bm{k}}},\dots, \ket{u_{r\bm{k}}}\end{bmatrix} = \begin{bmatrix}
        Z_{11}(\bm{k}) & \cdots & Z_{1r}(\bm{k}) \\
        \vdots & \ddots & \vdots \\
        Z_{N1}(\bm{k}) & \cdots & Z_{Nr}(\bm{k}) 
    \end{bmatrix}.\label{eq: local frame field for a rank r band}
\end{equation}
In this case, arbitrary linear combination amongst the $\ket{u_{n\bm{k}}}$'s does not change the band and hence the $N\times r$ matrix $Z(\bf{k})=\left[Z_{mn}(\bm{k})\right]_{1\leq m\leq N, 1\leq n\leq r}$ determines the bands up to multiplication on the right by an $r\times r$ invertible matrix with entries smooth functions in the Brillouin zone. In particular, we can locally  choose $Z(\bf{k})$ such that $Z(\bm{k})^{\dagger}Z(\bm{k})=I_{r}$, which is to say that the $\ket{u_{n\bm{k}}}$'s form an orthonormal basis for the band at $\bm{k}$. The projector $P(\bm{k}) = \sum_{n=1}^{r}|u_{n\bm k}\rangle\langle u_{n\bm k}|$, where the right-hand side assumes an orthonormal basis choice, uniquely determines the band. Furthermore, $P(\bm{k})$ determines a map from the Brillouin zone torus to the Grassmannian $\Gr_r(\mathbb{C}^N)$, a manifold which, as a set, consists of all $r$-dimensional subspaces in $\mathbb{C}^N$ which is in bijection with the $N\times N$ orthogonal projection matrices of rank $r$. When $r=1$, the Grassmannian $\Gr_r(\mathbb{C}^N)$ reduces to $\mathbb{C}P^{N-1}$. Similar to what happens in the $r=1$ case, here the Abelian Berry curvature and the (Abelian) quantum metric, given by the traces over band indices of their non-Abelian counterparts, are respectively, the pullback of (twice) the FS symplectic form and metric by the map $P:\BZ^2\to\Gr_r(\mathbb{C}^N)$. Just like $\mathbb{C}P^{N-1}=\Gr_{1}(\mathbb{C}^N)$, the Grassmannian $\Gr_r(\mathbb{C}^N)$, for general $r$, is also a K\"ahler manifold. In Appendix.~\ref{sec:review:grassmannian}, we review local complex coordinates and the local K\"ahler potential for Grassmannian.

The notion of K\"ahler band, as introduced in Ref.~\cite{kahlerband1,kahlerband2,kahlerband3}, corresponds to the special case where the locally defined matrix $Z(\bm{k})$ can always be taken to be holomorphic with respect to local complex coordinates in the Brillouin zone, coming from a global complex structure. See more about K\"ahler bands in Sec.~\ref{subsubsec: Kaehler bands} below.

\subsection{Standard Landau Levels and Ladder Operators} \label{subsec: flat Landau level states}
Now we review the quantum geometries of standard LLs which are formed in a 2D electron gas pierced by a uniform magnetic field. These LLs can be regarded as Chern bands exhibiting the simplest quantum geometric properties. Due to the magnetic translation invariance, their quantum geometric quantities are uniform. For comparison with lattice systems such as moir\'e materials and tight-binding models, we find it convenient to formulate the LL problem in the symmetric gauge and use 2D wave vectors inside a 2D Brillouin zone to label the Bloch-like LL wavefunctions. In this formulation, the lowest LL wavefunction on the torus can be written in terms of the Weierstrass sigma function~\cite{haldanemodularinv,haldaneholomorphic,Jie_MonteCarlo,JieWang_NodalStructure,Jie_Dirac},
\begin{equation}
    \Phi^{\rm LL}_{0\bm k}(\bm r) = \sigma(\zeta/\sqrt{\mC} + i\sqrt{\mC}k)e^{i\bar k \zeta - \frac{\mC}{2}|k|^2 - \frac{1}{2\mC}|\zeta|^2}, \label{def_LLL_wf}
\end{equation}
where $\bm k = (k_x, k_y)$ labels the magnetic translation quantum number and one takes $\mC = 1$ for the single-component lowest LL. The wavefunction Eq.~(\ref{def_LLL_wf}) can also be utilized (by summing over translation partners) to represent the color-entangled lowest LL wavefunction with Chern number $\mC > 1$~\cite{YangleWu_ColorEntanglement13,Yangle_haldanestatistics}, the unique wavefunction of an ideal K\"ahler band with constant Berry curvature~\cite{Mera_uniqueness_23}. Without loss of generality we only consider $\mC>0$ as those of negative Chern numbers are easily obtained by complex conjugation. Complex vectors $w_a$ and $w^a$ determine the complex coordinate $\zeta$ and complex momentum $k$ in Eq.~(\ref{def_LLL_wf}), respectively, through,
\begin{equation}
    \zeta = \sum_{a = x,y}w_ar^a, \quad k = \sum_{a = x,y}w^ak_a,
\end{equation}
where $i w_a = \sum_b\varepsilon_{ab}w^b$ and $\varepsilon_{xy} = -\varepsilon_{yx} = 1$ is the anti-symmetric tensor. The anisotropy of LLs is determined by the effective mass tensor $m_{ab}=\det(m)\widetilde g_{ab}$ of electrons in Galilean invariant LL models, where $\widetilde{g}_{ab} = \overline{w}_{a}w_b + w_a\overline{w}_{b}$ is its unimodular part. The mass tensor depends on the semiconductor material and is also tunable by altering the orientation of the magnetic field~\cite{Bo_BandMassAnisotropy_12,Shayegen_PRB_15,Shayegen_PRL_18,Shayegen_PRL_22}. In the isotropic case with $w_{x,y}=\left(1,i\right)/\sqrt{2}$, the complex coordinate and the complex momentum reduce to the standard ones $\zeta=(x+iy)/\sqrt{2}$ and $k=(k_x + ik_y)/\sqrt{2}$. 

It is important to notice that although the lowest LL wavefunction is a holomorphic function of real-space coordinates $\zeta$ (up to Gaussian factors), its ``periodic'' part $u^{\rm LL}_{0\bm k}(\bm r) \equiv e^{-i\bm k\cdot\bm r}\Phi^{\rm LL}_{0\bm k}(\bm r)$ is a holomorphic function of complex momentum up to Gaussian factors. This resembles the explicit position-momentum duality of LLs~\cite{Martin_PositionMomentumDuality,JieWang_NodalStructure}. The $u^{\rm LL}_{0\bm k}(\bm r)$ is
\begin{equation}
    u^{\rm LL}_{0\bm k}(\bm r) = \sigma(i\sqrt{\mC}k + \zeta/\sqrt{\mC})e^{-ik\bar\zeta - \frac{\mC}{2}|k|^2 - \frac{1}{2\mC}|\zeta|^2}.\label{def_uLL}
\end{equation}
Notably, the LL index can be raised and lowered by a momentum-space ladder operator via
\begin{equation}
    u^{\rm LL}_{n\bm k}(\bm r) = \frac{1}{\sqrt{n!}}(\hat a^\dag)^nu^{\rm LL}_{0\bm k}(\bm r),
\end{equation}
where $u^{\rm LL}_{n{\bm k}}(\bm r)$ is the ``periodic'' part of the $n$th LL wavefunction $\Phi^{\rm LL}_{n\bm k}(\bm r)$ and the ladder operators $\hat a$ and $\hat a^\dag$ are differential operators given in terms of holomorphic and anti-holomorphic derivatives with respect to momentum:
\begin{equation}
    \hat a^\dag = \frac{i}{\sqrt{\mC}}\frac{\partial}{\partial k} - \frac{i\sqrt{\mC}}{2}\bar k; \quad \hat a = \frac{i}{\sqrt{\mC}}\frac{\partial}{\partial\bar k} + \frac{i\sqrt{\mC}}{2}k.
\end{equation}

For LLs, the real-space formulation with $\Phi^{\rm LL}_{n\bm k}$ and the momentum-space formulation with $u^{\rm LL}_{n\bm k}$ are equally good. However, the momentum-space formulation is more general in the sense that it can be generalized to other systems including moir\'e models and tight-binding models: no matter whether the real space is a continuum or lattice, their momentum space in either case is a continuum in the thermodynamic limit~\cite{Martin_PositionMomentumDuality,ChingHuaLeePRB13,Lee_engineering_PRB17,ChinaHuaLeePRX15}. Therefore in what follows in this work, we will primarily focus on the ``periodic part'' of the Bloch wavefunction $u_{\bm k}(\bm r)$ and momentum-space ladder operators~\footnote{We added a quotation mark to ``periodic part'' because for Landau level type states, $u_{\bm k}(\bm r)$ rigorously speaking is not a periodic function upon translations; only periodic wavefunctions of Bloch states are. Nevertheless, the translation phase are independent on $\bm k$ hence $u_{\bm k}(\bm r)$ of different $\bm k$ are still within the same Hilbert space.}.

Since this work will be primarily focusing on momentum-space holomorphicity properties, we will use the notation $z$, which is often used for real-space complex coordinate, to denote the complex momentum-space coordinate $-i\sqrt{\mC}k\equiv z$. Note that $z$ and $k$ differ by scale, which is an invertible holomorphic transformation, so they determine the same complex structure. We will also denote by $\int$ the integration over the Brillouin zone acting on periodic functions. The convention is chosen so that the integration of Berry curvature gives the Chern number: $\int\Omega(\bm k) = \mC \in \mathbb{Z}$. The notations used in this work are summarized in below,
\begin{eqnarray}
    \int f(\bm{k}) &\equiv & \frac{1}{2\pi}\int f(\bm{k})\; d^2\bm k,\nonumber\\
    &=&\frac{1}{2\pi} \int_{\mathrm{BZ}^2}f(\bm{k})\;dk_x\wedge dk_y,\nonumber \\
    &=& \frac{i}{2\pi} \int_{\mathrm{BZ}^2} f(\bm{k})\; dk\wedge d\bar k,\nonumber\\
    k &\equiv& \sum_{a=x,y} w^a k_a,\nonumber\\
    z &\equiv& -i\sqrt{\mC}k,\quad \partial = \partial/\partial z, \quad \bar\partial = \partial/\partial {\bar z},\label{def_notation}
\end{eqnarray}
where $f(\bm{k})$ is any periodic function and $f(\bm{k})dk_x\wedge dk_y$ a $2$-form.
With the notation above, the ladder operators are rewritten as
\begin{equation}
    \hat a^\dag = \partial - \bar z/2,\quad\hat a = -\bar\partial - z/2. \label{def_ladder_k}
\end{equation}

Last but not least, although all standard LLs have the same Berry curvature, the quantum metric depends on the LL index. For the $n$th LL, the trace of quantum metric equals to $2n+1$ times of the Berry curvature: $\Tr_{\widetilde g}g_n(\bm k) = (2n+1)\Omega_n(\bm k)$ where both $g_n(\bm k)$ and $\Omega_n(\bm k)$ are independent of ${\bm k}$~\cite{LEDWITH2021168646}. The trace is defined with respect to $\widetilde{g}$ defining the anisotropy of LLs. Physically, this means that different LL states have distinct spatial spreading and wavefunction overlap. Such feature is crucial for the occurrence of non-Abelian and symmetry breaking orders in higher LLs~\cite{FoglerPRL96,FoglerPRB96,RezayiHaldanePRL00}.

Integrating both sides of the local trace condition $\Tr_{\widetilde g}g_n(\bm k) = (2n+1)\Omega_n(\bm k)$ over the Brillouin zone leads to
\begin{eqnarray}
    \int\Tr_{\widetilde g} g_{n}(\bm k) = (2n+1)\mC \label{def_quantized_LL}
\end{eqnarray}
for the $n$th standard LL. While this step looks redundant for standard LLs, we emphasize that the integration form Eq.~(\ref{def_quantized_LL}) of the trace condition is actually more general than the local form. One of the central achievements in our work is to systematically construct generalizations of standard LLs in the presence of non-uniform quantum geometries. As shown later, once the quantum geometric fluctuations are switched on, the local form of trace condition is generally violated for $n \geq 1$ and only holds for the $n = 0$ state, due to holomorphicity.
By contrast, the integration form Eq.~(\ref{def_quantized_LL}) is preserved for arbitrary LL index $n$ for the generalized LL states introduced in this work.

In the next sections, we will review different classes of bands which are closely tied to the geometric bounds discussed in Eq.~(\ref{eq: quantum geometric bounds}).

\subsection{Band Theory in the Holomorphic Setting}
\label{secsec: theory of bands}
In this section, we review different band theories where a notion of holomorphicity --- either momentum-space or real-space holomorphicity, or both --- is present. In particular, we will discuss the notions of K\"ahler bands~\cite{kahlerband1,kahlerband2,kahlerband3}, ideal bands~\cite{JieWang_exactlldescription,Jie_Origin22} and vortexable bands~\cite{LedwithVishwanathParker22,Fujimoto24}. We review their defining properties and discuss their interrelationships.
\subsubsection{K\"ahler bands} \label{subsubsec: Kaehler bands}
For a single band, saturation of the determinant bound $\sqrt{\det g(\bm k)} = |\Omega(\bm k)|/2$ at a particular $\bm k$ point, for non-vanishing $\Omega(\bm{k})$, implies compatibility of the quantum metric and Berry curvature, seen, respectively, as a metric and symplectic form in the tangent space at $\bm{k}$. More precisely, it means the compatibility of three objects in linear algebra: a symplectic form $\omega=\Omega/2$ (where we see $\Omega$ as a $2$-form), a scalar product $g$ and a complex structure $J$. Compatibility means that if one knows two of them, one is able to recover the third through $\omega J=g$ in matrix form. If we write
\begin{align}
    g=\begin{bmatrix}
    g_{xx} & g_{xy}\\
    g_{xy} & g_{yy}
    \end{bmatrix}   \text{ and } \omega=\begin{bmatrix}
    0 & \Omega_{xy}/2\\
    -\Omega_{xy}/2 & 0
    \end{bmatrix},
\end{align}
we then have
\begin{align}
    J=\frac{1}{\sqrt{\det(g)}}\begin{bmatrix}
    -g_{xy} & -g_{yy}\\
    g_{xx} & g_{xy}
    \end{bmatrix}. \label{eq: complex structure}
\end{align}
The matrix $J$ squares to minus the identity and provides the tangent space of the Brillouin zone at $\bm{k}$ with the structure of a complex vector space by declaring $J\cdot v:= i v$ for any tangent vector $v$ at $\bm{k}$.

If compatibility holds in a chart, we then have a local K\"ahler structure induced from the one in projective space. This comes from the fact that in that case, the wavefunction can, in that neighbourhood, be represented by a holomorphic function of a complex momentum coordinate $\xi$ (in general nonlinear in $k_x,k_y$), determined by solving the Beltrami differential equation for the quantum metric. The latter yields isothermal coordinates $\left[\xi_1(\bm{k}),\xi_2(\bm{k})\right]$ for which $g=e^{\psi(\xi_1,\xi_2)}(d\xi_1^2+d\xi_2^2)$ with some local smooth function $\psi$, and then one can set $\xi=\xi_1+i\xi_2$. The complex structure $J$ can here be thought as being determined by $w^a_{\bm k}$'s which now depend on $\bm{k}$, unlike the case of the lowest LL. 

A stronger condition is the saturation of the determinant bound and non-vanishing of the Berry curvature, both everywhere in the Brillouin zone. In this case, the matrix equation $\omega J= g$ holds everywhere, meaning that the quantum metric and the Berry curvature give the Brillouin zone the structure of a K\"ahler manifold, as induced by the K\"ahler manifold structure of the space of quantum states. This is the definition of a K\"ahler band~\cite{kahlerband1,kahlerband2,kahlerband3}. In 2D, the $J$ in Eq.~\eqref{eq: complex structure} is automatically ``integrable''. Integrability means there exist local complex coordinates with holomorphic coordinate changes in the overlaps making the Brillouin zone into a complex manifold. The saturation of the bound, for non-vanishing Berry curvature everywhere, implies that the map induced to projective space, $\bm{k} \mapsto P(\bm{k})$, is holomorphic with respect to this complex structure $J$. In contrast, this is no longer true in higher dimensions, where further integrability conditions are required. Therefore a K\"ahler band can be viewed as a holomorphic immersion (this property is equivalent to the statement of the Berry curvature being everywhere non-vanishing) from the Brillouin zone --- equipped with the structure of a Riemann surface {\it i.e.} a complex manifold of dimension one --- to a complex projective space (or, more generally, to a  Grassmannian for multiple bands), where the complex structure, determined by $w^a_{\bm k}$'s, for a general K\"ahler band is allowed to be varying as a function of $\bm k$. The gauge invariant condition for a band to be K\"ahler, with complex structure described locally by complex coordinates $\xi = \xi(\bm{k})$ (which can be nonlinear in $k_x,k_y$), is
\begin{equation}
    Q(\bm k)\frac{\partial P(\bm k)}{\partial \bar{\xi}}P(\bm k) = 0, \label{eq: Kaehler band condition}
\end{equation}
where $Q(\bm{k}) = 1-P(\bm{k})$ is the orthogonal complex projector of the band, supplemented with requiring positivity of the Berry curvature.

For a K\"ahler band, the pullback of FS symplectic form and metric can be described very simply in terms of a \emph{holomorphic gauge}. The condition Eq.~(\ref{eq: Kaehler band condition})
allows one to locally choose $Z(\bf{k})$ in Eq.~\eqref{eq: local frame field for a rank r band} to be holomorphic~\cite{kahlerband2}, or equivalently,
\begin{equation}
    \frac{\partial}{\partial \bar{\xi}}\begin{bmatrix}\ket{u_{1\bm{k}}},\dots,\ket{u_{r\bm{k}}}\end{bmatrix} = 0,
\label{eq: holomorphic gauge}
\end{equation}
where $\xi$, noted above Eq.~(\ref{eq: Kaehler band condition})
denotes a local complex coordinate in the Brillouin zone for which the band defines a holomorphic map to the Grassmannian $\Gr_r(\mathbb{C}^N)$. One can then write the Gram matrix of scalar products
\begin{align}
    S(\bm{k}) = \begin{bmatrix}
        \bra{u_{i\bm{k}}}u_{j\bm{k}}\rangle
    \end{bmatrix}_{1\leq i,j\leq r},
\end{align}
and we have the Abelian Berry curvature and the quantum metric given by
\begin{eqnarray}
    \frac{1}{2}\Omega &=& P^*\omega_{FS}=\frac{i}{2} h d\xi\wedge d\bar{\xi},\\
    g &=& P^*g_{FS}=h|d\xi|^2,
\end{eqnarray}
where $h$ is built out of second derivatives of the K\"ahler potential as
\begin{align}
    h = \frac{\partial^2 \log \det S(\bm{k})}{\partial \xi \partial \bar{\xi}}.
\end{align}

Because it will be useful for later sections, we now discuss some aspects of Riemannian curvature appropriate to K\"ahler bands. In the K\"ahler setting, one has the Ricci form, expressed in terms of $h$ as
\begin{align}
    \Ric{\omega} = \frac{i}{2}\frac{\partial^2\log h}{\partial\xi \partial \bar{\xi}}d\xi\wedge d\bar{\xi}.
\end{align}
The latter determines the Gaussian/scalar curvature of the K\"ahler metric, given by
\begin{align}
    K=-2\frac{\Ric{\omega}}{\omega}=-\frac{1}{h}\frac{\partial^2\log h}{\partial \xi \partial \bar{\xi}}.
\end{align}
The Euler characteristic of the Brillouin zone, being a genus one surface, is equal to zero and as a consequence
\begin{align}
    \chi(\BZ^2)=\frac{1}{2\pi}\int_{\BZ^2}K\;\omega=-\int_{\BZ^2}\frac{\Ric{\omega}}{\pi}=0,
    \label{eq: Euler characteristic of BZ}
\end{align}
independently of the K\"ahler band.

\subsubsection{Ideal bands}
\label{subsubsec: ideal bands}
Ideal bands form a subset of the set of all K\"ahler bands satisfying the additional constraint: at each momentum $\bm k$, the trace of the quantum metric, with respect to some $\bm{k}-$independent unimodular metric $\widetilde g$, equals to the Berry curvature which is positive; in other words, $\Tr_{\widetilde g} g(\bm k) = \Omega(\bm k) > 0$ for every $\bm k$~\cite{JieWang_exactlldescription}. In this case, the determinant bound is automatically saturated. There are also other equivalent definitions of ideal K\"ahler bands. For instance, a variational definition requires   
\begin{equation}
    \min_{\widetilde g} \left(\int \Tr_{\widetilde g} g(\bm k)\right) = \mC,\label{def_IB_integratedtrace}
\end{equation}
which means the integrated trace minimized over the space of unimodular metrics $\widetilde g$ must equal to the Chern number for an ideal band. Alternatively, a canonical way of defining ideal band is by requiring the quantum geometric tensor to have a $\bm k-$independent null vector satisfying $\sum_b \chi^{ab}(\bm k)w_{b} = 0$ for all $\bm k$~\cite{JieWang_exactlldescription}. Such null vector is the complex vector $w_a$ that determines the anisotropy of ideal bands $\widetilde g_{ab} = \bar{w}_aw_b + w_a\bar{w}_b$ and anti-symmetric tensor $\varepsilon_{ab} = -i\left(\bar w_aw_b - w_a \bar w_b\right)$. The complex vector $w_a$ is uniquely determined by these two conditions up to a $\mathrm{U}(1)$ phase. The trace condition Eq.~(\ref{def_IB_integratedtrace}) is also referred to as the ``ideal quantum geometry condition''. From now on we will simplify notation by using ``$\Tr$'' for the generalized trace defined with respect to the optimal unimodular metric $\widetilde g$ whose integrated trace value is minimized.

Therefore, an ideal band is a K\"ahler band whose complex structure is momentum independent over the Brillouin zone and is the one determined by the unimodular metric for which the trace bound is saturated. The momentum-space holomorphicity further restricts the cell-periodic part of the Bloch wavefunction to be a holomorphic function of complex momentum $k$, or $z$, up to a normalization factor,
\begin{equation}
    |u_{0\bm k}\rangle = \widetilde{\mathcal{ N}}_{0\bm k}|\widetilde u_{0\bm{k}}\rangle,\quad \text{ for all } \bm k\in\BZ^2, \label{def_holomorphic_function_idealband}
\end{equation}
where $|\widetilde{u}_{0\bm{k}}\rangle$ is holomorphic: $|\overline{\partial}\widetilde u_{0\bm{k}}\rangle=0$. The reason why we add a subscript ``$0$'' is that later, we will generalize the notion of LL so that the ideal band fits as the $n=0$ generalized LL in an infinite tower analogous to LLs. Higher LLs will not be holomorphic in momentum, just as happened in the standard LL case, but they will obey a quantized integrated trace formula.

The momentum-space holomorphicity strongly restricts the possible form of wavefunction, once the Chern number is given~\cite{JieWang_exactlldescription,Jie_Origin22,Mera_uniqueness_23}. To see this, we fix a coordinate $\bm r$, then the holomorphic wavefunction $\widetilde u_{0\bm k}(\bm r)$, if regarded as a function of $\bm k$, must satisfy appropriate boundary conditions consistent with the Chern number of the band --- as such they determine holomorphic sections of a holomorphic line bundle over the Brillouin zone. By the Riemann-Roch theorem, $\widetilde u_{0\bm k}(\bm r)$ may then be expressed in terms of $\mC$ linearly independent functions, which are given by elliptic functions such as the Jacobi theta functions or the Weierstrass Sigma functions --- in physical terms, they are essentially the standard lowest LL wavefunctions $u^{\rm LL}_{0\bm k}(\bm r)$. The type of functions, either theta functions or the sigma function, depends essentially on the gauge choice, but they are ultimately equivalent. The most general form of all ideal (a single band, {\it i.e.} rank $r = 1$) bands of Chern number $\mC > 0$, including $\mC = 1$, is given by~\cite{JieWang_exactlldescription,Jie_Origin22}:
\begin{equation}
    u_{0\bm k}(\bm r) = \mathcal N_{0\bm k} \sum_{\sigma = 0}^{\mC - 1}\mathcal B(\bm r+\sigma\bm a_1)u^{\rm LL}_{0\bm k}(\bm r+\sigma\bm a_1),\label{def_IB_wavefunction}
\end{equation}
where one magnetic unit cell defining $u^{\rm LL}_{0\bm k}$ corresponds to $\mC$ lattice unit cells. Technically, this means we have chosen the gauge such that the Weierstrass sigma function in $u^{\rm LL}_{0\bm k}$ defined in Eq.~(\ref{def_uLL}) has its quasi-periodicity in $\mC\bm a_1$ and $\bm a_2$ where $\bm a_{1,2}$ are the primitive lattice vectors. Holomorphicity also allows us to describe the fluctuations of the quantum geometry in terms of the normalization factors~\cite{JieWang_exactlldescription}: for a single ideal band, the Berry curvature and trace of quantum metric are~\footnote{For ideal band constructed as determinant states from multiple bands, normalization determines the $\bm k-$dependent part of quantum geometry in the same way as in Eq.~(\ref{idealband_Omega_N}), but the constant part is different (because the Chern number of the determinant state is the total Chern number of the band complex); see Eq.~(\ref{OmegaPsi1}).},
\begin{equation}
    \Omega_{0}(\bm k) = \Tr g_{0}(\bm k) = \mC\left(1 + \partial\bar\partial \log \mathcal N^{-2}_{0\bm k}\right),\label{idealband_Omega_N}
\end{equation}
where $\partial$ and $\bar\partial$ are defined in Eq.~(\ref{def_notation}).

Ideal bands can be realized exactly or approximately in many different systems. The chiral magic-angle twisted bilayer graphene (TBG) is one important example for ideal bands~\cite{Grisha_TBG,Grisha_TBG2,JieWang_NodalStructure}. Besides, the $\mC = 1$ ideal bands are realized in an exact way as Dirac zero modes in a periodic magnetic field~\cite{Liang_DiracNonuniformB}, the lowest LL on curved manifolds~\cite{Crepel_LLCurvedSpace23}, as well as the exactly flat band of the Kapit-Mueller model~\cite{Kapit_Mueller,ModelFCI_Zhao,Emil_constantBerry,Dong_Mueller_20}. Besides, higher Chern number ideal bands are exact in a couple of moir\'e graphene models including the chiral twisted multilayer graphene~\cite{Jie_hierarchyidealband,Eslam_highC_idealband} and twisted trilayer graphene models~\cite{GuerciMoraTTG1,GuerciMoraTTG2,GuerciMoraTTG3,Grisha_eTTG23,Grisha_TTG23}. In addition, ideal bands are also approximated across many realistic materials, including, for example, the strained graphene~\cite{Eslam_StrainedGraphene23} and the twisted bilayer TMD material~\cite{NicolasMacDonald23,NicolasMacDonald24,CrepelRegnaultRaquel23,CrepelDirac24}.

An important implication of ideal bands is the existence of exact model FCIs realized as short-ranged interactions regardless of the quantum geometric fluctuations ({\it i.e.} non-uniformity of Berry curvature and quantum metric)~\cite{Grisha_TBG2,JieWang_exactlldescription,Jie_hierarchyidealband,Eslam_highC_idealband,Jie_Origin22,junkaidonghighC22}. For instance, Laughlin $\nu=1/3$ states can be stabilized by the shortest ranged Trugman-Kivelson interaction $\sum_{i<j}\delta''(\bm r_i - \bm r_j)$ in ideal bands as exact zero modes which necessarily must be the ground state because of the non-negativity of the many-body spectrum of repulsive interactions; generalization to all other model wavefunctions are straightforward~\cite{TrugmanKivelson85,Haldane_hierarchy,GreiterWenWilczekPRL91,GreiterWenWilczekNuclearPhyB92}. Since the ideal quantum geometry Eq.~(\ref{def_IB_integratedtrace}) directly implies exact FCI ground states (if one ignores band dispersion and considers projected short-ranged interactions), engineering models or materials to approach the ideal-band limit is supposed to enhance the stability of fractionalized phase~\cite{Dan_parker21,Nicolas_WSe2_23,Dong_CFL23,Goldman_CFL_23}. This provides theoretical insights of the experimental occurrence of FCIs in twisted MoTe$_2$~\cite{NicolasMacDonald23,NicolasMacDonald24}. It is also worth to emphasize ideal quantum geometry and short-ranged interaction are sufficient but not necessary conditions for FCIs; there are examples of stable FCIs in models violating the ideal conditions~\cite{Titus_FCI_ExceedGap_14,Read_FCI_15,Oreg_GeometryFCI24}.

There are many equivalent but complementary perspectives to understand the exact model FCIs in ideal bands. For instance, the exact mapping between ideal band wavefunctions to the lowest LL at the single-particle level implies a direct construction of many-body model wavefunctions. From this construction, the short-ranged clustering property, which are essential for model states to be zero modes, is preserved~\cite{Grisha_TBG2}. Another perspective which can equally prove the existence of zero modes involves mapping the quantum geometric fluctuation effect into center of mass dependence of interactions~\cite{JieWang_exactlldescription}. Moreover, the ideal K\"ahler condition can be physically viewed as the ability to attach ``a vortex'' $\zeta(\bm r) \equiv \sum_a w_a r^a$ to the ideal band Hilbert space, thereby allowing one to explicitly construct model wavefunctions in the same way as in the lowest LLs --- such vortexability perspective will be reviewed with more details in the following section. Last but not least, the exact many-body zero modes can be understood as a consequence of the hidden exact Girvin-MacDonald-Platzman algebra~\cite{gmpb,gmpl} present in ideal bands~\cite{Jie_Origin22}.

Finally, let us discuss the physical meaning of the generalized trace and possible experimental ways of its detection. In the context of standard LLs, a special case of ideal band, as explored in Sec.~\ref{subsec: flat Landau level states}, the unimodular metric $\widetilde g$, which defines the generalized trace, characterizes the anisotropy of LLs. This anisotropy is influenced by the material properties and can be adjusted by tilting magnetic fields. Moreover, for general ideal bands, since vortexability implies perfect circular dichroism~\cite{Dong_CFL23}, the optimal metric for these bands could potentially be extracted through optical experiments by measuring excitation rates~\cite{Dong_CFL23,OzawaGoldman19,LF_ProbingQG_24}.

\subsubsection{Vortexable bands}
\label{subsubsec: vortexable band}
A related notion to K\"ahler bands is that of vortexability, which is a real-space property and was introduced in Ref.~\cite{LedwithVishwanathParker22}. A vortexable band is defined to be any band complex (either a single band or multiple bands) that obeys the following condition, called the vortexability criterion:
\begin{equation}
   \left(1-\mathcal{P}\right)\mathsf{z}(\bm r)\mathcal{P}=0,\quad \bm r = \left(x, y\right), \label{eq: vortexability criterion}
\end{equation}
where $\mathcal{P}$ is the orthogonal projector onto the band. The vortex function $\mathsf z(\bm r)$ defines a diffeomorphism $\mathsf z: \mathbb R^2 \rightarrow \mathbb C$. When $\mathsf{z}$ is a linear function of $\bm r$, the vortex function $\mathsf{z}(\bm r)$ reduces to $\zeta(\bm r)$ up to a constant shift, and the vortexable band reduces to ideal band. If the vortex function is instead a non-linear function of $\bm r$, the momentum-space ideal quantum geometric condition Eq.~(\ref{def_IB_integratedtrace}) is violated and such general vortexable band is outside the ideal band class. One concrete example of the general non-ideal vortexable band is the zero mode of Dirac fermion with periodically rotated velocity field in magnetic field~\cite{LedwithVishwanathParker22}.

The diffeomorphism $\mathsf z$ induces, by pullback, in $\mathbb{R}^2$ a K\"ahler structure induced from the flat one in $\mathbb{C}$, determined by the standard flat metric and symplectic forms $\left(|d\mathsf z|^2,\frac{i}{2}d\mathsf z\wedge d\bar{\mathsf z}\right)$,
\begin{eqnarray}
    g_{\rm vortex} &\equiv& \mathsf z^*(|d\mathsf z|^2) = \sum_{a,b=x,y}\mathrm{Re}\left(\partial_a \bar{\mathsf z}\partial_b \mathsf z\right)dx^a dx^b,\\
    \Omega_{\rm vortex} &\equiv& \mathsf z^*\left(\frac{i}{2}d\mathsf z\wedge d\bar{\mathsf z}\right) = \mathrm{Im}\left(\partial_1 \bar{\mathsf z}\partial_2 \mathsf z\right)dx^1\wedge dx^2,\nonumber
\end{eqnarray}
where $g_{\rm vortex}$ and $\Omega_{\rm vortex}$ are called ``vortex metric'' and ``vortex chirality'', respectively, following Ref.~\cite{LedwithVishwanathParker22}. Observe that by construction $\sqrt{\det(g_{\rm vortex})}dx^1\wedge dx^2 = \Omega_{\rm vortex}$, hence a real-space version of the determinant bound (K\"ahler condition) is satisfied.

Vortexable bands, defined as fulfilling vortexability condition Eq.~(\ref{eq: vortexability criterion}), allow exact FCI model states constructed from the vortex functions, either for the ideal vortexable bands whose vortexable functions are linear functions of $\bm r$ or the general vortexable bands with nonlinear vortex functions. The family of model FCIs includes model Laughlin, Halperin, Moore-Read states realized in a single band~\cite{Grisha_TBG2,junkaidonghighC22,JieWang_exactlldescription,Jie_hierarchyidealband,Jie_MonteCarlo}, as well as unprojected Jain states realized in multiple bands~\cite{Fujimoto24}. If real-space short-ranged Trugman-Kivelson type interactions~\cite{TrugmanKivelson85} can be realized, model states are also exact zero modes there.

Recently the notion of ``vortexability'' has been generalized to higher-LL-like states via ``higher vortexability'' whose definition requires a couple of projectors~\cite{Fujimoto24}. We will later see that the ``generalized Landau levels'' introduced in this work fulfill the higher vortexability criteria. Therefore, the generalized LLs introduced here belong to the set of higher vortexable bands. However, in comparison with the higher vortexability criteria, whose definition requires the notion of other vortexable bands, which is initially unknown; in contrast, in this work we discuss a systematic construction of generalized LLs (in Sec.~\ref{sec: QUANTIZED TRACE FORMULA AND CANONICAL BASIS}) and a practical criterion to search for them (in Sec.~\ref{sec:finite-N-models}). In addition, in this work we study the quantum geometries of generalized LLs, point out the existence of a series of geometric invariants, and systematically explore many-body physics associated to generalized LLs.

\subsubsection{Relations and implications to lowest Landau level physics}
\label{subsubsection: relations amongst the different notions}
To summarize, K\"ahler and vortexable bands exhibit, respectively, momentum and real-space holomorphicity. Not all K\"ahler bands are vortexable and similarly not all vortexable bands have momentum-space K\"ahler structure. Vortexability allows an explicit construction of first quantized fractional quantum Hall and FCI model wavefunctions within the Hilbert space of vortexable bands.

\begin{figure}
    \includegraphics[width=1.0\linewidth]{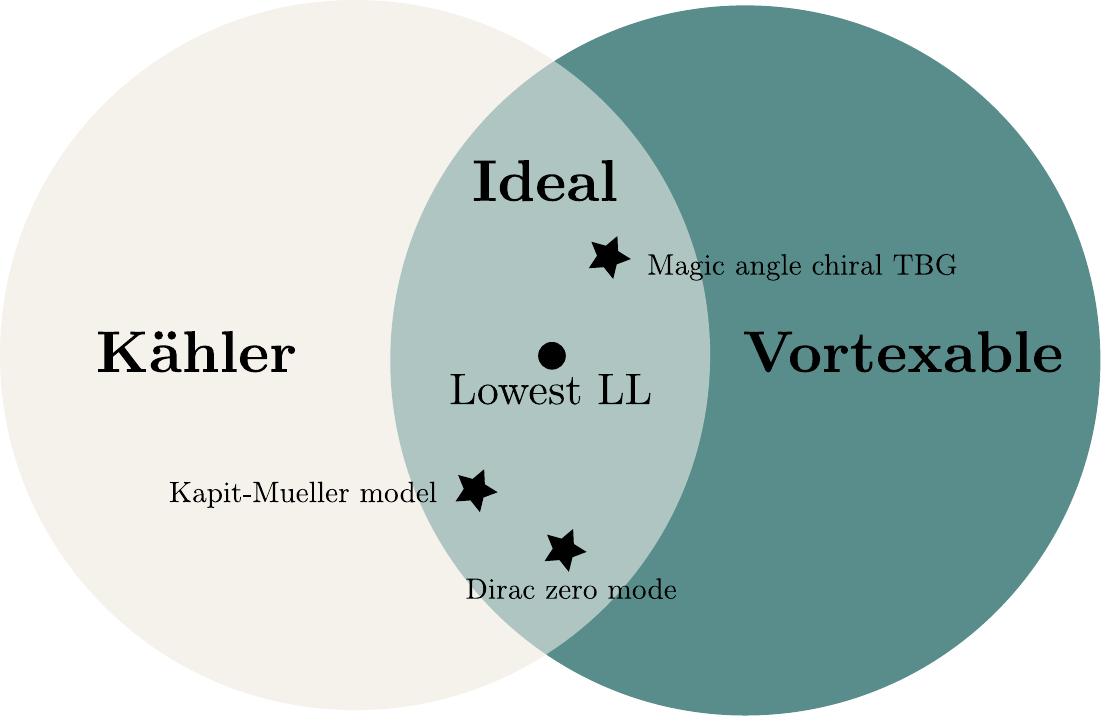} \caption{The figure summarizes the interrelationship of K\"ahler, ideal and vortexable bands. K\"ahler and vortexable bands have a notion of momentum and real space holomorphicity, respectively. Ideal bands sit exactly at the intersection of K\"ahler and vortexable bands, thus enjoy both momentum and real space holomorphicity notions. All vortexable permit exact construction of first-quantized fractional quantum Hall type model wavefunctions which becomes exact zero energy ground states when short-ranged repulsive interactions are considered. Ideal bands are exactly realized in the lowest LL, filled lowest multiple LLs, flatbands of magic angle chiral TBG, flatband of Kapit-Mueller model, zero modes of Dirac fermions in periodic magnetic fields, and are approximately realized in twisted MoTe$_2$. Ideal bands can be regarded as ``generalized lowest LLs''. This work is based on the ideal band setting and systematically generalizes this notion to higher LL analogs.} \label{fig:notion_of_bands}
\end{figure}

The set of ideal bands is precisely at the intersection of K\"ahler and vortexable bands. They hence exhibit both momentum and real-space holomorphicity. We summarize the relationships amongst K\"ahler, ideal and vortexable bands in Fig.~\ref{fig:notion_of_bands}. Because of this discussion, we will use the terms ``ideal band'', ``ideal K\"ahler band'' or ``ideal vortexable band'' interchangeably in what follows. Ideal bands can be regarded as generalized lowest LL states. In this work, we will focus on the ideal band setting and consider its higher LL generalizations. The generalized higher LLs can also be termed as higher ideal K\"ahler/vortexable bands as well.

\section{GENERALIZED LANDAU LEVELS AND QUANTIZED INTEGRATED TRACE FORMULA}
\label{sec: QUANTIZED TRACE FORMULA AND CANONICAL BASIS}
In this work, we will focus on the ideal band setting which, as reviewed above, is at the intersection between K\"ahler and vortexable bands. Also we will mainly focus on the single non-degenerate band case, meaning $P(\bm{k})$ is a rank-one projector, though multiple bands will be important for the intermediate discussions. The novelty in this work is that we will not be focusing on the holomorphic wavefunctions themselves, but rather on their higher-LL partners which themselves are not holomorphic and do not saturate any of the local geometric bounds relating the quantum metric and Berry curvature considered above [\emph{cf.} Eq.~\eqref{eq: quantum geometric bounds}]. We will point out the existence of a canonical basis constructed from higher LL analogs of ideal bands, provide invariants which characterize them and show their precise quantization in analogy to what happens with LLs.

We begin by taking Chern number $\mC = 1$ and consider adding modulations to the LL wavefunctions. We define an unnormalized ``modulated Landau level'' basis,
\begin{equation}
    e_{n\bm k}(\bm r) = \mathcal B(\bm r)u^{\rm LL}_{n\bm k}(\bm r), \quad n=0,1,\dots \label{def_modulated_LLs}
\end{equation}
which is obtained by multiplying the standard periodic part of the LL states $u^{\rm LL}_{n\bm k}$ by a same coordinate $\bm r-$dependent modulating factor $\mathcal{B}(\bm{r})$. If such factor is periodic in real space, the resulting $e_{n\bm k}$'s still obey lattice magnetic translation symmetry as the standard LLs and hence represents a LL type state; if $\mathcal B(\bm r)$ is instead quasi-periodic with the opposite quasi-periodicity as $u^{\rm LL}_{n\bm k}(\bm r)$, the modulated LL state $e_{n\bm k}(\bm r)$ can be real-space periodic and represent the cell-periodic part of a Bloch wavefunction. A concrete realization of $e_{0\bm k}$ with periodic $\mathcal B(\bm r)$ includes Dirac zero modes in a periodic magnetic field~\cite{Liang_DiracNonuniformB}; $e_{0\bm k}$ with quasi-periodic $\mathcal B(\bm r)$ can be realized as the zero mode of magic-angle chiral TBG~\cite{Grisha_TBG,Grisha_TBG2,JieWang_NodalStructure}. 

For a general $\mathcal B(\bm r)$, the modulated basis is no longer orthonormal $\langle e_{m\bm k}|e_{n\bm k}\rangle \neq \delta_{mn}$, nevertheless they transform in an identical way as LL states under the momentum-space ladder operators, because $\hat a$ and $\hat a^\dag$ do not act on $\mathcal B(\bm r)$:
\begin{eqnarray}
    \sqrt{n}|e_{n\bm k}\rangle &=& \hat a^\dag |e_{n-1,\bm k}\rangle,\label{rec1_e}\\
    \sqrt{n}|e_{n-1,\bm k}\rangle &=& \hat a|e_{n\bm k}\rangle,\label{rec2_e}\\
    0 &=& \hat a|e_{0\bm k}\rangle.\label{rec3_e}
\end{eqnarray}

The generalization to Chern number $\mC>1$ is straightforward. The $|e_{n\bm k}\rangle$ will be a modulated color-entangled $n$th LL states:
\begin{equation}
    e_{n\bm k}(\bm r) = \sum_{\sigma = 0}^{\mC - 1} \mathcal B(\bm r + \sigma\bm a_1)u^{\rm LL}_{n\bm k}(\bm r + \sigma\bm a_1), \label{def_en}
\end{equation}
where the magnetic unit cell of $u^{\rm LL}_{n\bm k}$ is defined on $\mC\bm a_1 \times \bm a_2$, the same as in Eq.~(\ref{def_IB_wavefunction}). The indices $n$ are still raised up and down in an identical way as in Eqs.~(\ref{rec1_e}) to~(\ref{rec3_e}). From now on, unless specified, when we write modulated LLs or generalized LLs, they mean states of general Chern number $\mC \geq 1$.

What will be relevant for the quantized trace formula below is the orthonormal basis, $\{\ket{u_{m \bm{k}}}\}$, with $\langle u_{m\bm k}|u_{n\bm k}\rangle = \delta_{mn}$, obtained by applying the Gram-Schmidt orthogonalization process to the basis $\{\ket{e_{n\bm{k}}}\}$ and setting $|u_{0\bm k}\rangle\equiv \mathcal{N}_{0{\bm k}}|e_{0\bm k}\rangle$. This orthogonalization process yields
\begin{equation}
    |u_{m\bm k}\rangle = \sum_{n}|e_{n\bm k}\rangle U_{nm}(\bm k),
\end{equation}
where the $U_{nm}(\bm k)$'s determine an upper-triangular matrix, {\it i.e.} $U_{nm}(\bm{k})=0$ for $n > m$. The $|u_{0\bm k}\rangle$ is an ideal band Bloch wavefunction, the $|u_{1\bm k}\rangle$ is a linear superposition of $|e_{0\bm k}\rangle$ and $|e_{1\bm k}\rangle$ with a $\bm k-$dependent coefficient, and in general $|u_{n\bm k}\rangle$ is a superposition of $|e_{m\bm k}\rangle$ with $m=0,...,n$. We will be adding a subscript $n$ to quantum geometric quantities to indicate that they are derived from $|u_{n\bm k}\rangle$. 

We now state one of the main results of this work:
\begin{theorem}\label{theorem:quantizedtrace}
    The integrated trace of the quantum metric of the orthonormal state $|u_n\rangle$ obeys the quantization formula,
    \begin{equation}
        \int\Tr g_{n}(\bm k) = (2n+1)\int\Omega_{n}(\bm k) = (2n+1)\mC,\label{def_quantized_trace}
    \end{equation}
    for all $n\geq 0$. Here $\int$ stands for momentum-space integration, conveniently normalized as defined in Eq.~(\ref{def_notation}).
\end{theorem}

This equality generalizes the standard LL trace formulas and remarks how the LL index gives an invariant regardless of quantum geometric fluctuations. Therefore, we term the orthonormal basis $\{\ket{u_{n\bm k}}\}$ the ``generalized Landau levels''. The zeroth generalized LL is the standard ideal band. We emphasize Eq.~(\ref{def_quantized_trace}) does not mean $2n+1$ is a ``topological invariant'' which is defined to be robust to any type of local perturbation (such as disorder). In contrast, the quantized trace formula in Eq.~(\ref{def_quantized_trace}) is invariant under smoothly changing the modulation function $\mathcal B(\bm r)$ and requires momentum-space ``holomorphicity'' of $\ket{u_{0\bm{k}}}$ as a prerequisite, which will become clear in the next sections.

In the following, we first discuss a proof of the integrated trace formula motivated from the physical point of view by comparing and contrasting the orthonormal basis $\ket{u_n}$ to standard LL states. We then give a geometric interpretation and proof from a mathematical point of view, namely in light of the theory of holomorphic curves and associated moving frames~\cite{lawson:71,griffiths:74, griffiths:harris:14}.

\subsection{Generalized Landau Levels, Recursions and Invariants}\label{rec:recursions}
\subsubsection{Connection coefficients} \label{sec:proof:LL:connection}
We begin by studying the structure of the non-Abelian interband Berry connection coefficients in the orthonormal basis $\{\ket{u_{n\bm k}}\}$ obtained from applying the Gram-Schmidt algorithm to $\{\ket{e_{n\bm k}}\}$. We will show that, due to holomorphicity and orthogonality of the basis, the non-Abelian Berry connection coefficients admit a particular simple form in such basis. We define the holomorphic and anti-holomorphic non-Abelian Berry connection coefficients as
\begin{eqnarray}
    A_{mn}(\bm k) &\equiv& -i\langle u_{m\bm k}|\partial u_{n\bm k}\rangle,\label{def_berry_connection}\\
    \bar A_{mn}(\bm k) &\equiv& -i\langle u_{m\bm k}|\bar\partial u_{n\bm k}\rangle,\nonumber
\end{eqnarray}
where the momentum-space derivatives $\partial$ and $\bar\partial$ are defined in Eq.~(\ref{def_notation}). The holomorphic and anti-holomorphic connection are related by Hermitian conjugation in the orthonormal basis.

Notice from Eq.~(\ref{def_ladder_k}) that the anti-holomorphic partial derivative is closely related to the LL lowering operator, therefore $|\bar\partial u_{n\bm k}\rangle$ is at most a linear superposition of $\ket{u_{m\bm{k}}}$ with $0\leq m\leq n$. As a result, $\bar A_{mn}(\bm k)$ must vanish identically for any $m > n$. Since $\bar A$ is Hermitian to $A$, we arrive at,
\begin{equation}
    A_{mn}(\bm k) = 0,\quad \forall m < n.
\end{equation}
Similarly $|\partial u_{n\bm k}\rangle$ is at most a linear superposition of $\ket{u_{m\bm{k}}}$ of $0\leq m\leq n+1$. Hence the holomorphic coefficients $A_{mn}$ must vanish identically for $m > n+1$:
\begin{equation}
    A_{mn}(\bm k) = 0,\quad \forall m > n+1.
\end{equation}

We therefore conclude that the holomorphic connection coefficient $A_{mn}$ can only take non-zero values when $m=n$ or $m=n+1$. We have chosen a gauge such that the off-diagonal component $A_{n,n-1}(\bm k)$ is purely imaginary. Therefore we parameterize the non-Abelian holomorphic connection coefficients as follows,
\begin{eqnarray}
    A_{nn}(\bm k) &=& -i\frac{\bar z_k}{2} + i\alpha_{n\bm k};\quad n\geq 0, \nonumber \\
    A_{n,n-1}(\bm k) &=& -i\mathcal N_{n\bm k}^{-1};\quad n\geq 1,\label{connections}
\end{eqnarray}
where $\mathcal N_{n\bm k} \in \mathbb{R}$ is real and $\alpha_{n\bm k} \in \mathbb{C}$. We note that for the standard LLs, $\mathcal N_{n\bm k} = 1/\sqrt{n}$ and $\alpha_{n\bm k} = 0$.

The generalized LL states are canonical in the sense that, as bands, they are completely determined by the ideal K\"ahler band $|u_{0{\bm k}}\rangle$. In particular, if we multiply the wavefunction of the ideal K\"ahler band by a momentum dependent nonvanishing smooth function, which does not change the band, then the other bands' wavefunctions just change by a common smooth momentum dependent phase factor. This last property also dictates that only the diagonal connection coefficients change, and they all change by the gradient of the common phase factor. See Appendix~\ref{sec:FSeqn_and_holomorphic_gauge_transform} for the proof.

For notational simplicity, in what follows, we will occasionally omit the momentum subscript and simply denote $\left(|u_{n\bm k}\rangle, \mathcal N_{n\bm k}, \alpha_{n\bm k}\right)$ as $\left(|u_n\rangle, \mathcal N_n, \alpha_n\right)$.

\subsubsection{Recursion of states} \label{sec:proof:LL:recursionstate}
It follows from Eq.~(\ref{connections}), that the momentum-space raising operator $\hat a^\dag$ will map $|u_n\rangle$ to a superposition of $|u_n\rangle$ and $|u_{n+1}\rangle$ only. Similarly, the lowering operator $\hat a$ will map state $|u_n\rangle$ to a superposition of $|u_n\rangle$ and $|u_{n-1}\rangle$. Using the ladder operators and Eq.~(\ref{connections}), we arrive at the following recursion relation for the $\ket{u_n}$'s: for all $n > 1$,
\begin{eqnarray}
    \mathcal N_{n}^{-1}|u_n\rangle &=& \hat a^\dag|u_{n-1}\rangle + \alpha_{n-1}|u_{n-1}\rangle,\label{rec1}\\
    \mathcal N_n^{-1}|u_{n-1}\rangle &=& \hat a|u_n\rangle + \bar{\alpha}_n|u_n\rangle,\label{rec2}
\end{eqnarray}
and for $n=0$,
\begin{equation}
    0 = \hat a|u_0\rangle + \bar{\alpha}_0|u_0\rangle.\label{rec3}
\end{equation}

The state recursion relation above is one of the key results derived in this section. They show that also for the Gram-Schmidt orthogonalized states, a ladder structure resembling that of the LL ladder operator representation is satisfied. Comparing with Eqs.~(\ref{rec1_e}) to~(\ref{rec3_e}), we see that the nontrivial information of the modulations $\mathcal B(\bm r)$ and the quantum geometric fluctuations are encoded in the coefficients $\mathcal N_n$ and $\alpha_n$.

The coefficients cannot be all independent. The ladder operators $\hat a,\hat a^\dagger$ appearing in Eqs.~(\ref{rec1}) to~(\ref{rec3}) satisfy the algebra $[\hat a, \hat a^\dag] = 1$. This algebra imposes the following constrains on the coefficients: when $n \geq 1$,
\begin{eqnarray}
    \mathcal N_{n+1}^{-2} - \mathcal N_{n}^{-2} &=& 1 - \left(\partial\bar\alpha_n + \bar\partial\alpha_n\right),\label{rec_Nn}\\
    \alpha_n - \alpha_{n-1} &=& \partial\log\mathcal N_n,\label{rec_alpha}
\end{eqnarray}
and when $n=0$,
\begin{eqnarray}
    \mathcal N_1^{-2} &=& 1 - \left(\partial\bar\alpha_0 + \bar\partial\alpha_0\right),\label{rec_Nn2}\\
    \alpha_0 &=& \partial\log\mathcal N_0.\label{rec_alpha2}
\end{eqnarray}
From Eq.~(\ref{rec_Nn}) to Eq.~(\ref{rec_alpha2}), we see that the coefficients $(\mathcal N_n, \alpha_n)$ are all uniquely determined by $\mathcal N_0$ in our problem. Explicitly, ($\mathcal N_n, \alpha_n$) of all $n \geq 1$ are recursively determined from $\mathcal N_0$ as follows:
\begin{eqnarray}
    \alpha_n &=& \sum_{m=0}^{n}\partial\log\mathcal N_m,\\
    \mathcal N_{n+1}^{-2} &=& 1+\mathcal N_n^{-2}+\sum_{m=0}^{n}\partial\bar\partial\log\mathcal N_m^{-2}.\label{recursionNn}
\end{eqnarray}
Recall that $\mathcal N_0$ is the normalization of the standard ideal band wavefunction, thereby it is uniquely determined by $\mathcal B(\bm r)$. The concrete expression of $\mathcal N_0$ in terms of $\mathcal B(\bm r)$ can be found in Refs.~\cite{JieWang_exactlldescription,Jie_Origin22}. We leave derivation details of Eq.~(\ref{rec_Nn}) to Eq.~(\ref{recursionNn}) to Appendix.~\ref{sec:constrain_coefficients}.

In the next section, we will motivate and illustrate the physical meaning of the above recursion equations. In fact, they are rooted in the intricate relations between the geometries of individual band and fully filled multiple bands of generalized LLs, as will be explained shortly. Afterwards, we will present a geometric framework confirming and motivating these recursion relations and establish uniqueness of the multi-band system $\{\ket{u_n}\}$ determined from the ideal band $\ket{u_0}$.

\subsubsection{Geometries of individual generalized Landau levels} \label{subsubsec: Quantum geometry of the Gram-Schmidt orthogonalized bands}

In this section, we discuss quantum geometric quantities derived from the $|u_{n}\rangle$ band and illustrate the physical meaning of the recursion relations presented in the previous section. Although $|u_{n}\rangle$ of $n\geq1$ itself is not holomorphic in momentum, it turns out to be beneficial to express quantum geometric quantities in terms of complex variables. For a general band $|u\rangle$, the Berry curvature and the trace of the quantum metric take the following form in terms of holomorphic and anti-holomorphic derivatives whose derivation details can be found in Appendix.~\ref{geometry_holomorphic_frame}:
\begin{eqnarray}
	\Omega &=& \mC\left[\langle\partial u|\partial u\rangle - \langle\bar\partial u|\bar\partial u\rangle\right],\label{Omega_holomorphic}\\
	\Tr g &=& \mC\left[\langle\partial u|\partial u\rangle + \langle\bar\partial u|\bar\partial u\rangle - 2|A|^2\right],\label{trg_holomorphic}
\end{eqnarray}
where $A \equiv -i\langle u|\partial u\rangle$ is the holomorphic Abelian Berry connection.

Replacing $|u\rangle$ with $|u_{n}\rangle$, and using the connection elements in Eq.~(\ref{connections}) and the state recursions Eqs.~(\ref{rec1}) to~(\ref{rec3}), one arrives at the explicit relation between geometric quantities of the $|u_{n}\rangle$ band and the normalization factors $\mathcal N_n$, which is one of the key results of this section. We have
\begin{eqnarray}
    \Omega_{n} &=& \mC\left(\mathcal N^{-2}_{n+1} - \mathcal N^{-2}_{n}\right)\label{rec_omega},\\
    \Tr g_{n} &=& \mC\left(\mathcal N^{-2}_{n+1} + \mathcal N^{-2}_{n}\right),\label{rec_trg}
\end{eqnarray}
for all $n \geq 1$, and for the zeroth state,
\begin{eqnarray}
    \Omega_{0} = \Tr g_{0} &=& \mC\mathcal N_{1}^{-2}\label{Omega0_1}\\
    &=& \mC\left(1 + \partial\bar\partial\log\mathcal N_{0}^{-2}\right).\label{Omega0_2}
\end{eqnarray}

We comment that the above illustrates the physical meaning of Eq.~(\ref{rec_Nn}) and Eq.~(\ref{rec_Nn2}) --- they are Berry curvature formula for the $n$th generalized LL $|u_n\rangle$: their right-hand sides are nothing but the standard definition of Berry curvature after noticing that $\alpha_n$ determines the diagonal components of the Berry connection; their left-hand sides are alternative expressions of Berry curvature in terms of normalization factors, consistent with Eq.~(\ref{rec_omega}).

For the zeroth state $|u_0\rangle$, holomorphicity directly yields its Berry curvature as discussed around Eq.~(\ref{idealband_Omega_N}) and Ref.~\cite{JieWang_exactlldescription}. In addition, Eq.~(\ref{Omega0_1}) provides a new expression of Berry curvature of the ideal band, not in terms of its own normalization factor, but instead in terms of the first normalization factor $\mathcal N_{1}$. Later we will see such relation can be extended to a more general way: the $\mathcal N_{N+1}^{-2}$ determines the Berry curvature of the fully filled lowest $N$ generalized LLs, and products of normalizations, more precisely the logarithm of Eq.~(\ref{def_N_PsiN}), determines the K\"ahler potential of the ideal band for the many-body determinant state.

We conclude this subsection by proving the quantized integrated trace formula, which is a consequence of the results derived above. Integrating the Berry curvature yields Chern number, and hence, Eq.~(\ref{rec_omega}) and Eq.~(\ref{Omega0_1}) give the following quantized formula for normalization factors:
\begin{equation}
    \int\mathcal N^{-2}_{n} = n,\quad \forall n\geq 1,\label{int_Nn}
\end{equation}
whose physical meaning is the quantization of Chern number of the determinant states constructed from filling the lowest $n$ generalized LL states. Eq.~(\ref{int_Nn}) determines the integrated value of the trace of quantum metric through Eq.~(\ref{rec_trg}) and Eq.~(\ref{Omega0_1}). Using these we prove the quantized integrated trace formula in Theorem.~\ref{theorem:quantizedtrace} for the generalized LL $\{|u_n\rangle\}$ constructed by us.

\subsubsection{Geometries of filled lowest $N$ generalized Landau levels}
\label{subsubsec: quantum geometry of the fully filled first N bands}
In this section, we further derive relations among $\mathcal N_n$ and $\alpha_n$ from geometric properties of the fully filled lowest generalized LLs. We will illustrate the meaning of Eq.~(\ref{rec_alpha}) and derive a recursion relation for $\mathcal N_n$. The recursion relation shows that all normalization factors descend from the first normalization factor $\mathcal N_{0}$ and its derivatives, and is in align with the Calabi rigidity theorem which will be discussed in more detail in the next section.

We will consider fully occupied lowest $N$ generalized LLs by fermions. The many-body wavefunction $\Psi_{N}$ at a fixed momentum is a Slater determinant, also known as a wedge product of $\ket{u_{n}}$, written as
\begin{eqnarray}
    |\Psi_{N}\rangle &=& \mathrm{Slaterdet}\Big[\ket{u_{0}},\dots, \ket{u_{N-1}}\Big]\nonumber\\
    &=& |u_{0}\rangle\wedge |u_{1}\rangle\wedge ... \wedge |u_{N-1}\rangle,
\end{eqnarray}
which is normalized to one. Since $\ket{u_{n}}$ is a linear superposition of $\ket{e_{n}}$ with $n=0,...,N-1$, the same state can be rewritten as wedge products of $\ket{e_{0}}$ to $\ket{e_{N-1}}$ up to a normalization factor:
\begin{equation}
    |\Psi_{N}\rangle = \mathcal N_{\Psi_N}\times |e_{0}\rangle\wedge |e_{1}\rangle\wedge ...\wedge |e_{N-1}\rangle.
\end{equation}
The state recursion relations, Eqs.~(\ref{rec1}) to~(\ref{rec3}), give a concrete form for the normalization factor,
\begin{equation}
    \mathcal N_{\Psi_N} = \prod_{n=0}^{N-1}(\sqrt{n!} \mathcal N_{n})^{N-n}. \label{def_N_PsiN}
\end{equation}

The many-body state $|\Psi_{N}\rangle$ is an ideal band. Due to holomorphicity, similar to Eq.~(\ref{idealband_Omega_N}), the Berry curvature of $|\Psi_{N}\rangle$, denoted as $\Omega_{\Psi_N}$, is given by
\begin{eqnarray}
    \Omega_{\Psi_N}/\mC &=& N + \partial\bar\partial\log\mathcal N^{-2}_\Psi,\nonumber\\
    &=& N + \sum_{n=0}^{N-1}(N-n)\partial\bar\partial\log\mathcal N^{-2}_{n}.\label{OmegaPsi1}
\end{eqnarray}
where the total Chern number of $|\Psi_{N}\rangle$ is $N\mC$. On the other side, the Berry curvature of the band complex must be the sum of that of individual bands:
\begin{equation}
    \Omega_{\Psi_N}/\mC = \sum_{n=0}^{N-1}\Omega_{n}/\mC = \mathcal N^{-2}_{N}.\label{OmegaPsi2}
\end{equation}
Equating Eq.~(\ref{OmegaPsi1}) and Eq.~(\ref{OmegaPsi2}) gives the following recursion relation that must be satisfied by the normalization factors:
\begin{equation}
    \mathcal N^{-2}_{N} = N + \sum_{n=0}^{N-1}(N-n)\partial\bar\partial\log\mathcal N^{-2}_{n},\label{rec_normalizations}
\end{equation}
valid for all $N \geq 1$. This extends the relations Eq.~(\ref{Omega0_1}) and Eq.~(\ref{Omega0_2}) to general $N \geq 1$, and points out all normalization factors $\mathcal N_n$ are recursively and uniquely determined from $\mathcal N_{0}$ and its derivatives. Eq.~(\ref{rec_normalizations}) is consistent with Eq.~(\ref{recursionNn}). Moreover, Eq.~(\ref{rec_normalizations}) also deduces the following formula:
\begin{eqnarray}
    \mathcal N^{-2}_{N+1} - 2\mathcal N^{-2}_{N} + \mathcal N^{-2}_{N-1} = \partial\bar\partial\log\mathcal N_{N}^{-2},
\end{eqnarray}
which, as we will see shortly in the next section, has the geometric meaning as being the Ricci scalar of $\Psi_N$.
\subsection{Geometric Formulation: Holomorphic Curves and Moving Frames}
\label{subsec: Geometric Formulation: Holomorphic Curves and Moving Frames}
The geometric formulation in this section offers an alternative way, using the theory of holomorphic curves and associated moving frames, to derive the results of the previous section. Here, we emphasize the underlying K\"ahler geometry of the problem and formulate in a way that does not rely on specific gauge choices. The gauge choices here are two-fold. On the one hand, by multiplying all the wavefunctions by a smooth $\bm{r}$-dependent phase factor, which alters the form of the wavefunctions and ladder operators introduced before, we can, for instance, change to a Landau gauge or to a more general gauge. Moreover, we can also multiply the Bloch wavefunction of the ideal K\"ahler band by a smooth momentum dependent nonvanishing complex number and this does not change the band, since the latter only cares about the associated orthogonal projector. Indeed, by making use of the latter choice, is how we achieve a holomorphic representative of the Bloch wavefunction, see the discussion around Eq.~\eqref{eq: holomorphic gauge}. Hereafter, by \emph{holomorphic gauge}, we mean that we take the Bloch wavefunction of the ideal K\"ahler band to be such that $\frac{\partial}{\partial \bar{z}}\ket{u_{\bm{k}}}=0$. Note that the holomorphic gauge is not unique, because we can always multiply $\ket{u_{\bm{k}}}$ by a nonvanishing holomorphic function and this does not change the band nor the holomorphicity of the resulting Bloch wavefunction. A holomorphic gauge was used before in Sec.~\ref{subsubsec: ideal bands} and denoted by $\widetilde{u}_{0\bm{k}}(\bf{r})$. Below, since there are no other types of gauges involved, we will drop the tilde to make the notation lighter.

Before discussing the holomorphic curves, we recall the classical Frenet-Serret theory for curves in Euclidean space.

\subsubsection{Classical Frenet-Serret theory for curves}
The Frenet-Serret theory describes the kinetics of a particle moving along a curve in a $n$-dimensional Euclidean space. Without loss of generality and for illustration purpose we take $n=3$. Denoting the particle's position at time $t$ as $\bm r(t)$, its velocity, acceleration, and jerk are respectively,
\begin{equation}
S(t)=\left[\frac{d\bm{r}}{dt}, \frac{d^2\bm{r}}{dt^2},\frac{d^3\bm{r}}{dt^3}\right],   
\end{equation}
where for simplicity we will take the velocity to have unit norm $|d\bm{r}/dt|^2 = 1$. We denote the velocity by $\bm v=d\bm{r}/dt$. Moreover, we will denote the vector normal to the curve as $\bm n$, and the third vector orthogonal to $\bm v$ and $\bm n$, known as the binormal, as $\bm b$. In general, the velocity $d\bm r/dt$, the acceleration $d^2\bm r/dt^2$ and the jerk $d^3\bm r/dt^3$ are not orthogonal; nevertheless they define the local frame. One can take a Gram-Schmidt orthogonalization to get an orthogonal frame by the rotation matrix,
\begin{equation}
    R(t) = \textrm{Gram-Schmidt}\left[S(t)\right] = \left[\bm{v},\bm{n},\bm{b}\right],
\end{equation}
where
\begin{equation}
    \bm{n} =\frac{ \frac{d\bm{v}}{dt}-\left(\bm{v}\cdot\frac{d\bm{v}}{dt}\right)\bm{v}}{\left|\frac{d\bm{v}}{dt}-\left(\bm{v}\cdot\frac{d\bm{v}}{dt}\right)\bm{v}\right|}=\frac{\frac{d\bm{v}}{dt}}{\left|\frac{d\bm{v}}{dt}\right|},\quad \bm{b} = \pm \bm{v}\times \bm{n},
\end{equation}
where we used the derivative of the condition $\bm{v}^2 = 1$ to note $\bm{v}\cdot d\bm{v}/dt=0$ in order to simplify the expression. The vectors $\bm{v},\bm{n},\bm{b}$ are called a Frenet-Serret frame for the curve $\bm{r}(t)$. The Frenet-Serret equations give the rate of change of the frame $R(t)$,
\begin{align}
    \frac{dR}{dt}=\left[\frac{d\bm{v}}{dt},\frac{d\bm{n}}{dt},\frac{d\bm{b}}{dt} \right]=\left[\bm{v},\bm{n},\bm{b}\right]\begin{bmatrix} 
    0 & -\kappa & 0\\
    \kappa  &  0 &  -\tau \\
    0 & \tau & 0
    \end{bmatrix},
\end{align}
the functions $\kappa(t)$ and $\tau(t)$ are geometric invariants of the curve and they are known, respectively, as the ``curvature'' and ``torsion'' of the curve. Note that $\kappa=|d\bm{v}/dt|$ is just the magnitude of the acceleration of the curve. The expression for the torsion can be derived too, but it will not be important in what follows and we omit it. Importantly, they completely determine the curve up to rigid motion, {\it i.e.} a global translation and rotation in Euclidean space.

The matrix appearing in the right-hand side is $R^{-1}dR/dt$ and is the pullback under the map $t\mapsto R(t)$ of the Maurer-Cartan one-form $R^{-1}dR$ on $\mathrm{O}(3)$. Because of that it is a skew-symmetric matrix, which follows from differentiating the equation $R^tR=I$. More generally, for a curve in $\mathbb{R}^n$, by Gram-Schmidt orthogonalization of $S(t)=\left[d\bm{r}/dt,\dots, d^n\bm{r}/dt^n\right]$ we get an orthogonal frame $R(t)=[\bm{u}_1,\dots, \bm{u}_n]$. The rate of change of $R(t)$ is described by
\begin{align}
\frac{d\bm{u}_i}{dt}=\sum_{j=1}^n \bm{u}_j \theta^{j}_{\; i},
\end{align}
where 
\begin{equation}
    \theta = [\theta^{i}_{\; j}]_{1\leq i,j\leq n} \equiv R^{-1}\frac{dR}{dt} = \left[\bm{u}_i\cdot \frac{d\bm{u}_j}{dt}\right]_{1\leq i,j\leq n}.
\end{equation}
In general, from $R$ being a rotation matrix, we know that $\theta^t=-\theta$. The specific form of the frame $R$ built from Gram-Schmidt orthogonalization of $S$ yields that $\theta$ has a very sparse form, namely,
\begin{equation}
    \theta = \begin{bmatrix}
    0 & \theta^{1}_{\ \ 2} & 0 & \cdots & 0 & 0  & 0 \\
    \theta^{2}_{\ \ 1} & 0 & \theta^{2}_{\ \ 3} & \cdots & 0 & 0  & 0\\
    0 & \theta^{3}_{\ \ 2} & 0 & \cdots & 0 & 0 & 0 \\
    \vdots & \vdots & \vdots & \ddots & \vdots & \vdots & \vdots \\
    0 & 0 & 0 & \cdots & 0 & \theta^{n-2}_{\ \ n-1} &  0 \\
    0 & 0 & 0 & \cdots & \theta^{n-1}_{\ \ n-2} & 0 & \theta^{n}_{\  n-1} \\
    0 & 0 & 0 & \cdots & 0 & \theta^{n}_{\ \ n-1} & 0 
    \end{bmatrix},
\end{equation}
yielding the Frenet-Serret equations
\begin{equation}
    \frac{d\bm{u}_i}{dt}= \bm{u}_{i-1}\theta^{i-1}_{\ i} + \bm{u}_{i+1}\theta^{i+1}_{\ i}.
\end{equation}

\subsubsection{Holomorphic curve and moving frames} \label{sec:holomorphic_curve_moving_frames}
The classical Frenet-Serret theory describes a curve, parameterized by a single real variable $t$, in $\mathbb R^n$. The geometry of such curve is completely specified, up to rigid motion, by the invariants determined by the associated moving frames. For instance, in the above section, the functions $\kappa$ and $\tau$ are the invariants and $R(t)$ is an orthogonal moving frame. This picture can be generalized to ideal K\"ahler bands and provides a geometric interpretation of the quantized trace formula discussed above.

The important observation is that ideal K\"ahler bands are, up to normalization and phase, holomorphic functions of the complex momentum variable $z$, where complex momentum $z$ is defined in Eq.~(\ref{def_notation}). This means that an ideal K\"ahler band can be understood as a ``holomorphic curve''~\cite{lawson:71, griffiths:74}, {\it i.e.} a holomorphic map from the complex Brillouin zone (of complex dimension one) to the projective space. A holomorphic curve looks like a curve in the sense that it is described by a single, yet complex, parameter $z$, which in our cases describes the Brillouin zone, but more generally can be an arbitrary Riemann surface.

\begin{figure}
    \includegraphics[width=1.0\linewidth]{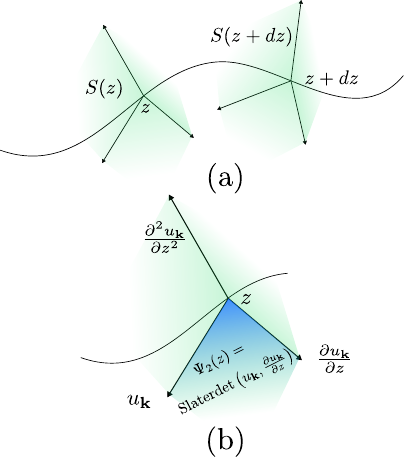} \caption{Geometric picture of an ideal K\"ahler band as a holomorphic curve, the idea of moving frames and the geometric interpretation of the associated tower of holomorphic maps. (a) Illustration of the holomorphic curve and moving frame. (b) Geometric interpretation of the holomorphic state $\Psi_{N=2}(z)$ --- it is the two-dimensional subspace spanned by the first two elements of $S(z)$.$\Psi_2$ physically corresponds to fully filling the bands determined by $\ket{u_{0\bm{k}}}$ and $\ket{u_{1\bm{k}}}$.} \label{fig:holomorphic_curve_moving_frame}
\end{figure}

We denote the periodic part of the Bloch wavefunction of the ideal K\"ahler band in a holomorphic gauge by $|u_{\bm{k}}\rangle$. We can attach, at each point of the Brillouin zone $z$, a natural holomorphic frame composed of the derivatives with respect to $z$ of $\ket{u_{\bm{k}}}$. Such moving frame is illustrated in Fig.~\ref{fig:holomorphic_curve_moving_frame} and is denoted by,
\begin{equation}
    S(z) = \begin{bmatrix}
        \ket{u_{\bm{k}}}, \frac{\partial}{\partial z}\ket{u_{\bm{k}}}, \dots, \frac{\partial^n}{\partial z^n}\ket{u_{\bm{k}}},\dots
    \end{bmatrix}, \label{def_holocurve_S}
\end{equation}
where we used the notation $S(z)$ to stress that the frame is holomorphic in $z$. The frame $S(z)$, after the Gram-Schmidt orthogonalization process, gives the unitary frame --- {\it i.e.} an orthogonal basis at each $\bm{k}$ --- of generalized LL states $\{\ket{u_{n\bm k}}\}$:
\begin{eqnarray}
    U(\bm k) &=& \text{Gram-Schmidt}\left[S(z)\right],\label{def_holocurve_U}\\
    &=& \begin{bmatrix}
\ket{u_{0\bm{k}}},\ket{u_{1\bm{k}}},\dots,\ket{u_{n\bm{k}}},\dots\end{bmatrix}.\nonumber
\end{eqnarray}

The unitary frame $U(\bm{k})$, is distinguished in the sense that if we choose a different holomorphic gauge, by multiply $\ket{u_{\bm{k}}}$ by a holomorphic nonvanishing function of $z$, only changes $U(\bm{k})$ by a smooth scalar phase, see the Appendix~\ref{sec:FSeqn_and_holomorphic_gauge_transform}. Hence, the associated bands, uniquely determined by the orthogonal projectors $\ket{u_{n}}\bra{u_n}$ with $n \geq 1$, are uniquely determined from the knowledge of $\ket{u_{0}}\bra{u_{0}}$, {\it i.e.} from the ideal K\"ahler band.

The rate of the change of the unitary frame along the holomorphic curve is described by the equation,
\begin{align} 
    d\ket{u_n} =\sum_m \ket{u_m} \theta^m_{\; n}, \label{def_change_unitary_frame}
\end{align}
where we have omitted the momentum labeling. Here $d=dz\frac{\partial}{\partial z} +d\bar{z}\frac{\partial}{\partial\bar{z}}$, and $\theta$ is the pullback of the Maurer-Cartan $1$-form on the unitary group,
\begin{eqnarray}
    \theta^{m}_{\; n} &\equiv& \left[U^{-1}dU\right]^{m}_{\; n} = \langle u_m| d|u_n\rangle,
\end{eqnarray}
which is closely related to the Berry connection in Eq.~(\ref{def_berry_connection}). The unitarity of $U(\bm k)$ implies $\theta$ is a skew-Hermitian matrix. Moreover, holomorphicity of the ideal K\"ahler band together with the fact that $U(\bm{k})$ results from the Gram-Schmidt process hence differs from $S(z)$ by an upper triangular matrix, forces $\theta$ to take a sparse form, just like the case of Frenet-Serret theory for the classical curves:
\begin{align}
    \theta= \begin{bmatrix}
    {\theta_{\ \ 0}^{0} }& {\theta_{\ \ 1}^{0} }& {0 }& {0 }& {\cdots }\\
    {\theta_{\ \ 0}^{1} }& {\theta_{\ \ 1}^{1} }& {\theta_{\ \ 2}^{1} }& {0 }& {\cdots }\\
    {0 }& {\theta_{\ \ 1}^{2} }& {\theta_{\ \ 2}^{2} }& {\theta_{\ \ 3}^{2} }& {\cdots }\\
    {0 }& {0 }& {\theta_{\ \ 2}^{3} }& {\theta_{\ \ 3}^{3} }& {\ddots }\\
    {\vdots }& {\vdots }& {\vdots }& {\ddots }& {\ddots}\\
    \end{bmatrix}.
\end{align}

Last but not least, $\theta^{n+1}_{\; n}$ is proportional to $dz$ and $\theta^{n-1}_{\; n}=-\overline{\theta^{n}_{\; n-1}}$ is proportional to $d\bar{z}$. This implies that the unitary moving frames obeys the ``Frenet-Serret'' equations along the holomorphic curve: for $n \geq 1$,
\begin{equation}
    d\ket{u_n}= \ket{u_{n-1}}\theta^{n-1}_{\; n} +\ket{u_n}\theta^{n}_{\; n} + \ket{u_{n+1}}\theta^{n+1}_{\; n}, \label{def_FS_holomorphic_curve1}
\end{equation}
and for $n=0$,
\begin{equation}
    d\ket{u_0} = \ket{u_{0}}\theta^{0}_{\;  0} + \ket{u_{1}}\theta^{1}_{\;  0}, \label{def_FS_holomorphic_curve2}
\end{equation}
where more details can be found in Appendix~\ref{sec:FSeqn_and_holomorphic_gauge_transform}.

The above Frenet-Serret equations for a holomorphic curve constitute the geometric interpretation of the state recursion relations of Eqs.~(\ref{rec1}) to~(\ref{rec3}). In the following, we will revisit the proof of the quantized integrated trace condition presented in Sec.~\ref{subsubsec: Quantum geometry of the Gram-Schmidt orthogonalized bands} from the point of view of holomorphic curves.

\subsubsection{Slater determinants, a tower of holomorphic maps and geometric recursion relations}
\label{subsubsec: Slater determinants, a tower of holomorphic maps and geometric recursion relations}
Associated to the ideal K\"ahler band determined by $|u_{\bm{k}}\rangle$ [as mentioned above Eq.~(\ref{def_holocurve_S}), we have chosen to use $|u_{\bm k}\rangle$ to denote the ideal K\"ahler band $|u_{0\bm k}\rangle$ in the holomorphic gauge], there is a tower of holomorphic maps induced by the Slater determinants
\begin{eqnarray}
    \Psi_N: z &\longmapsto& \mathrm{Slaterdet}\left( |u_{\bm{k}}\rangle, \frac{\partial}{\partial z}|u_{\bm k}\rangle, \dots, \frac{\partial^{N-1}}{\partial z^{N-1}}|u_{\bm k}\rangle \right) \nonumber \\
    &=& \ket{u_{\bm{k}}}\wedge \frac{\partial}{\partial z}\ket{u_{\bm{k}}}\wedge \dots\wedge \frac{\partial^{N-1}}{\partial z^{N-1}}\ket{u_{\bm{k}}},\nonumber
\end{eqnarray}
with $N=1,2,\dots$. The induced maps to projective space are clearly holomorphic because $\Psi_N(z)$ is holomorphic in $z$, since $\ket{u_{\bm{k}}}$ is so, and taking derivatives with respect to $z$ and performing wedge products preserves holomorphicity. At each momentum, described by the complex variable $z$, $\Psi_N(z)$ is $N$-particle fermionic state obtained by fully filling the lowest $N$ generalized LL states $\ket{u_{0}}\dots\ket{u_{N-1}}$. Geometrically, these can be interpreted as representing the $N$-dimensional plane generated by the first $N$ states, as illustrated in Fig.~\ref{fig:holomorphic_curve_moving_frame}.

Each of these holomorphic maps can be interpreted as a rank-$N$ degenerate K\"ahler band described by the orthogonal projector
\begin{equation}
    P_N=\sum_{n=0}^{N-1}\ket{u_n}\bra{u_n},
\end{equation}
carrying a multi-band quantum metric $\gamma_N$ and berry curvature $\omega_N$,
\begin{eqnarray}
    \gamma_N &=& {\rm Tr} \left(P_NdP_NdP_N\right),\\
    \omega_N &=& -\frac{i}{2} {\rm Tr} \left(P_NdP_N\wedge dP_N\right),
\end{eqnarray}
which, due to holomorphicity, saturate the quantum geometric inequalities in Eq.~\eqref{eq: quantum geometric bounds}. In particular, we can show, using the Frenet-Serret equations, that,
\begin{eqnarray}
    \gamma_N &=& |\theta^N_{\; N-1}|^2 = \left|\bra{u_N}\frac{\partial}{\partial z}\ket{u_{N-1}}\right|^2|dz|^2.\label{eq: geometry of slaterdets}\\ 
    \omega_N &=& \frac{i}{2} \theta^{N}_{\; N-1}\wedge \overline{\theta^{N}_{\; N-1}} = \frac{i}{2}\left|\bra{u_N}\frac{\partial}{\partial z}\ket{u_{N-1}}\right|^2dz\wedge d\bar{z}.\nonumber
\end{eqnarray}

For a full derivation of the above equations, we refer the reader to Appendix~\ref{sec:Derivation of the quantum metric of nth band}. The $\omega_N$'s, or, equivalently the $\gamma_N$'s, are invariants of the K\"ahler band determined by $\ket{u_{\bm{k}}}$ and they are completely specified by the interband Berry connection coefficients $\theta^{n}_{\; n-1}$. However, unlike the case of curves in Euclidean space which carry extrinsic information, holomorphic curves only care about intrinsic data, meaning everything depends on the first fundamental form, which in this setting is precisely the quantum metric~\footnote{In geometry, extrinsic data of an immersion into a Riemannian manifold is any geometric quantity which cannot be written in terms of the induced (pullback) metric. The standard example is if one takes a piece of paper and folds it into a cylinder. Both the unfold paper and the cylinder are immersions of the plane to the 3D space. The metric will not change (intrinsic geometry), but the extrinsic geometry is different.}. The intrinsic and extrinsic geometries of Bloch states is also considered in Ref.~\cite{avdoshkin:popov:23}. In particular, by Calabi's rigidity theorem~\cite{calabi:53}, if two K\"ahler bands $\ket{u_{\bm{k}}}$ and $\ket{u'_{\bm{k}}}$ give rise to the same K\"ahler metric, then they differ, up to re-scaling, by a $\bm{k}-$independent unitary transformation, so the K\"ahler metrics $\omega_N$ of $N \geq 2$ should be related to $\omega_1$. Note that if two K\"ahler bands $\ket{u_{\bm{k}}}$ and $\ket{u'_{\bm{k}}}$ differ by a $\bm{k}$-independent unitary transformation, the resulting quantum metrics are the same, but the converse is not obvious. Indeed, this is the content of the recursion relation in Eq.~\eqref{eq: geometric recursion relation} which is a local form of the so-called ``Pl\"{u}cker relations''~\cite{lawson:71,griffiths:74,griffiths:harris:14}.

The $n$th state $\ket{u_n}$ does not determine a K\"ahler band because its quantum metric and Berry curvature do not saturate the quantum geometric bounds in Eq.~\eqref{eq: quantum geometric bounds}, but its quantum metric, which we denote by $g_n$, satisfies
\begin{equation}
    g_n = \bra{du_n}\left(1-\ket{u_n}\bra{u_n}\right)\ket{du_n}=\gamma_n+\gamma_{n+1}, \label{eq: quantum metric of nth band}
\end{equation}
which can be proved using the Frenet-Serret equations and the details can be found in Appendix~\ref{sec:Derivation of the quantum metric of nth band}. Hence, when performing the integral of the trace with respect to the unimodular metric, one gets the sum of the integrals of the Berry curvatures $\omega_n$ and $\omega_{n+1}$ which are Chern numbers and, thus, quantization follows. However, to show the exact value of quantization it is convenient to proceed further in the understanding of the K\"ahler geometry of the $\Psi_N$'s and, in particular, their relation to the one of the generating K\"ahler band $\ket{u_{0}}$. For that, one can use the Frenet-Serret equation together with the Maurer-Cartan structure equation,
\begin{equation}
    d\theta +\theta\wedge \theta=0,
\end{equation}
or in components,
\begin{equation}
    d\theta^m_{\; n} +\sum_{l}\theta^{m}_{\; l} \wedge \theta^{l}_{\; n}=0,
\end{equation}
to derive the following geometric recursion relations on the K\"ahler geometry in the Brillouin zone induced by the tower of maps $\Psi_N$~\cite{griffiths:74},
\begin{equation}
    \Ric{\omega_N} = \omega_{N-1}-2\omega_N+\omega_{N+1}, \label{eq: geometric recursion relation}
\end{equation}
where $N \geq 1$ and $\Ric{\omega_N}$ is the Ricci form associated with the K\"ahler form $\omega_N$. Integrating the Ricci form yields the Euler characteristic of the Brillouin zone, see Eq.~\eqref{eq: Euler characteristic of BZ}, which is zero independently of the metric chosen. This yields the following recursion between Chern numbers,
\begin{equation}
    0 = \mC_{N-1} - 2\mC_N + \mC_{N+1}, \label{eq: Pluecker relations}
\end{equation}
with $\mC_N \equiv \int\omega_N$. Using the fact that $\mC_0 = 0$ (because $\Psi_0$ is the constant map) and $\mC_1 = \mC$ (the Chern number of the ideal K\"ahler band $|u_{0\bm k}\rangle$), one arrives at $\mathcal{C}_n=n\;\mathcal{C}$. Therefore we arrived at the quantized formula of integrated trace metric,
\begin{equation}
    \int\Tr g_n = \mC_n + \mC_{n+1} = \left(2n+1\right)\mC.
\end{equation}

We comment that the above results all require the $\omega_N$'s to be nonvanishing everywhere. LLs fulfill this condition. The existence of singularities (ramification points), which exist for finite dimensional tight-binding models, will alter the above relations Eq.~(\ref{eq: Pluecker relations}). We leave more detailed discussions about these singular points and related physics for future work.

\subsection{Summary}
In this section, we introduced the ``generalized Landau levels'' $\{\ket{u_n}\}$, which is a set of orthonormal and canonical basis. They can also be interpreted as forming a distinguished unitary moving frame over holomorphic curves. The geometric invariants and relations can be understood either from making analogies to LLs or based on holomorphic curves and moving frames.

From the viewpoint of LL analogy, the basis states $\{\ket{u_n}\}$ have a similar structure under the action of momentum-space ladder operators: the $\hat a^\dag$ and $\hat a$ respectively increase or decrease the level index at most by one. The difference of the ladder operator representation is encoded in the normalization factors $\mathcal N_{n}$ and coefficients $\alpha_{n}$ which determines the connection. Notably these coefficients completely determines the geometries of both individual and multiple generalized LLs in a concise way. All normalization factors are uniquely determined from the zeroth normalization factor $\mathcal N_{0}$. The geometries contained in normalization factor directly proves the quantized integrated trace formula.

Alternatively, we can also interpret an ideal K\"ahler band as a holomorphic curve. Associated to the holomorphic curve there is a canonical, up to an overall phase factor, unitary moving frame built by applying Gram-Schmidt orthogonalization to the derivatives of a holomorphic representative of the ideal K\"ahler band with respect to the holomorphic momentum variable $z$. Physically, this means that associated to an ideal K\"ahler band there is a tower of bands, exactly similar to the case of LLs. The ladder structure is also naturally related to the application of a holomorphic derivative, \emph{c.f.} the expression of the momentum-space raising operator in Eq.~\eqref{def_ladder_k}. Due to the geometric properties of the distinguished frame and the fact that filling $N$ of these bands, from $n=0$ up to $n=N-1$, produces a rank $N$ ideal K\"ahler band, one can relate the metric of the individual $u_n$ to the K\"ahler metric of filled bands, from which the quantization formula for the integrated trace follows.

We summarize the geometric quantities and their interrelationships, for individual and filled generalized LLs, in Table~\ref{Table:Geometries}.

\begin{table*}
\begin{tabular}{ |p{3.5cm}||p{3.5cm}|p{3.5cm}|p{5cm}| }
    \hline
    \textbf{Geometric data} & \textbf{Component} & \textbf{Expression} & \textbf{Differential form} \\
    \hline 
    \multirow{2}{3cm}{Connection coefficients} & $A_{nn}$ & $-i\bar z_k/2 + i\alpha_n$ & $\theta^{n}_{\; n} = iA_{nn} dz$ \\
    & $A_{n,n-1}$ & $-i\mathcal N^{-1}_n$ & $\theta^{n}_{\; n-1} = iA_{n,n-1}dz$ \\ \hline
    Curvature of $\ket{u_n}$ & $\Omega_n$ & $\mC\left(\mathcal N^{-2}_{n+1} - \mathcal N^{-2}_{n}\right)$ & $\omega_{n+1} - \omega_{n} = \frac{i}{2}\Omega_n dz \wedge d\bar z$ \\
    Metric of $\ket{u_n}$ & $g_n$ & $\mC\left(\mathcal N^{-2}_{n+1} + \mathcal N^{-2}_{n}\right)$ & $\gamma_{n+1} + \gamma_{n} = \left(\Tr g_n\right) |dz|^2$ \\ \hline
    Curvature of $\Psi_N$ & $\Omega_{\Psi_N}$ & $\mC \mathcal N^{-2}_N$ & $\omega_N = \frac{i}{2}\Omega_{\Psi_N}dz \wedge d\bar z$ \\
    Metric of $\Psi_N$ & $g_{\Psi_N}$ & $\mC \mathcal N^{-2}_N$ & $\gamma_N = |\theta^{N}_{\; N-1}|^2 $ \\
    Ricci curvature of $\Psi_N$ & $R_{\Psi_N}$ & $\partial\bar\partial\log\mathcal N^{-2}_N$ & ${\rm Ric}(\omega_N) = \omega_{N-1} - 2\omega_{N} + \omega_{N+1}$ \\
    \hline
\end{tabular}
\caption{Summary of geometries for individual generalized LLs $\ket{u_n}$ and filled lowest $N$ generalized LLs $\Psi_N$. The ``expression'' for metric and Ricci curvature refer to their trace part.}\label{Table:Geometries}
\end{table*}

\section{SUPERPOSED STATE GEOMETRY AND DECOMPOSITION} \label{sec:suposed_state_geometry_and_decomposition}
The previous section proved the quantized integrated trace formula and discussed other geometric properties associated to individual or filled multiple generalized LLs. In this section, we will discuss properties following from another aspect of generalized LL states: the completeness. Because generalized LLs form a complete basis, they can be used to expand general Chern band.

One of the most interesting application of the completeness is: given a Chern band, how to know the weight of each generalized LL components. Addressing this question quantifies the ``Landau level mimicry'' and is important for understanding the geometric stabilities of Abelian and non-Abelian FCIs. For instance, identifying the parameter region of the material such that the target narrow Chern band approaches the zeroth (first) generalized LL can be one of the promising guiding rules in realizing Abelian (non-Abelian) fractionalized phases.

\subsection{The $u_m - u_n$ Model} \label{sec:superposition:umunmodel}
In this section, we analyses states constructed by superposing two generalized LLs. Although the single-particle Hilbert space spanned by two generalized LLs consists of linearly superposing them with $\bm k-$dependent superposition coefficients, we start with studying a simpler yet solvable model by assuming the superposition coefficients are $\bm k$-independent. We term this model the $u_m - u_n$ model and it captures a subspace of the full Hilbert space spanned by two generalized LLs. We will discuss the quantum geometric properties of the $u_m - u_n$ model in this section. The $u_m - u_n$ model is important for later discussion on non-Abelian fractionalization in Sec.~\ref{sec:nonabelian_fractionalization}.

Concretely, the cell-period state of the $u_m - u_n$ model takes the following form,
\begin{equation}
    |u_{\bm k}\rangle = \sqrt{\lambda}|u_{m\bm k}\rangle + \sqrt{1-\lambda}e^{i\phi}|u_{n\bm k}\rangle,\label{def_um_un_model}
\end{equation}
where $\lambda \in [0,1] \in \mathbb{R}$ and $\phi \in [0,2\pi) \in \mathbb{R}$ parameterize the momentum-independent superposition coefficients. Clearly $|u_{\bm k}\rangle$ is normalized to unit.

We will denote the quantum metric and Berry curvature of $|u_{\bm k}\rangle$ as $g(\bm k)$ and $\Omega(\bm k)$, and those for individual bands as $g_{m}(\bm k)$ and $\Omega_{m}(\bm k)$. Without loss of generality, we assume $l = n-m \geq 1$. We will leave all relevant derivation details to Appendix~\ref{subsec: derivation for geometry of umun model}. The momentum dependence in Berry curvature, quantum metric and normalization factors will be omitted for notational simplicity.

\subsubsection{When $|m-n| \geq 2$}
In this case, the Berry curvature $\Omega$ can be proved to be a simple linear combination from Berry curvatures of $|u_m\rangle$ and $|u_n\rangle$,
\begin{equation}
    \Omega = \lambda\Omega_{m} + (1-\lambda)\Omega_{m+l}.\label{Int_Omega_um_un1}
\end{equation}
Clearly, the standard deviation of Berry curvature, defined as,
\begin{equation}
    \sigma_{\Omega} \equiv \left(\langle\Omega^2\rangle - \langle\Omega\rangle^2\right)^{\frac12},\label{def_delta_Omega}
\end{equation}
has no dependence on $\phi$. In the above, $\langle\dots\rangle$ means average over the first Brillouin zone.

Nevertheless, the integrated quantum metric of $|u_{\bm k}\rangle$ is slightly more complicated. Its analytical expression is given by
\begin{eqnarray}
    \int\Tr g/\mC &=& \lambda\int\Tr g_{m}/\mC + (1-\lambda)\int\Tr g_{m+l}/\mC\nonumber\\
    &+& 2\lambda(1-\lambda) \int |\sum_{p=1}^{l}\partial\log\mathcal N^{-1}_{m+p}|^2.\label{Int_trace_um_un1}
\end{eqnarray}
Same as the Berry curvature formula, the first line is a linear superposition of the integrated trace value contributed from $u_{m}$ and $u_n$, respectively. However, $\int\Tr g$ also has a nonlinear contribution. Such nonlinear term is zero in the limit of vanishing geometric fluctuation, {\it i.e.} when the generalized LLs reduces to standard LLs. More importantly the integrated trace $\int\Tr g$, as seen above, is independent on the relative superposition phase $\phi$.

\subsubsection{When $|m-n| = 1$}
In this case, even the Berry curvature $\Omega$ will have nontrivial dependence on $\lambda$,
\begin{eqnarray}
    \Omega/\mC &=& \lambda\Omega_{m}/\mC + (1-\lambda)\Omega_{m+1}/\mC\label{Int_Omega_um_un2}\\
    &+& \sqrt{\lambda(1-\lambda)} \left[e^{i\phi}\partial\mathcal N^{-1}_{m+1} + e^{-i\phi}\bar\partial \mathcal{N}^{-1}_{m+1}\right].\nonumber
\end{eqnarray}
However, since normalization factors $\mathcal N_{m+1,\bm k}$ are periodic functions under shifting momentum by reciprocal lattice vectors $\bm k \rightarrow \bm k+\bm b$, the second line vanishes identically after integrating over the first Brillouin zone. Therefore the Chern number of $|u_{\bm k}\rangle$ is still $\mC$.

The expression of integrated trace value for quantum metric for this case is found to be
\begin{eqnarray}
    \int\Tr g/\mC &=& \lambda\int\Tr g_m/\mC + (1-\lambda)\int\Tr g_{m+l}/\mC\label{Int_trace_um_un2}\\
    &+& 2\lambda(1-\lambda)\int \left[|\partial\log\mathcal N^{-1}_{m+1}|^2 - \mathcal N_{m+1}^{-2}\right],\nonumber
\end{eqnarray}
which after using Eq.~(\ref{def_quantized_trace}) and Eq.~(\ref{int_Nn}) is simplified to
\begin{eqnarray}
    \int\Tr g/\mC &=& 2(m+1)\lambda^2 - 2(m+2)\lambda + (2m+3)\nonumber\\
    &+& 2\lambda(1-\lambda)\int |\partial\log\mathcal N^{-1}_{m+1}|^2.\label{Int_trace_um_un3}
\end{eqnarray}
Notice that in this case, even in the limit of vanishing geometric fluctuation, the integrated quantum metric still has non-linear dependence on $\lambda$.

\subsubsection{Integrated trace value: phase insensitivity, non-linearity and lower bound}

We first draw the conclusion that the $\int\Tr g$ of the $u_m - u_n$ model, either $|m-n| \geq 1$ or $|m-n| \geq 2$, is independent of the relative phase $\phi$. Such feature is not only proved in Eq.~(\ref{Int_trace_um_un2}) and Eq.~(\ref{Int_trace_um_un3}) where details can be found in the Appendix~\ref{subsec: derivation for geometry of umun model}, but also can be straightforwardly numerically verified. Nevertheless, we emphasize the form factor of the $u_m - u_n$ model, hence the interacting problem of the model, do have nontrivial $\phi-$dependence, as will be discussed in Sec.~\ref{sec:nonabelian_fractionalization}.

Secondly, we notice that the quantum geometric fluctuation, {\it i.e.}, the momentum dependence of $\mathcal N_{n\bm k}$, generally enhances the integrated trace value. This means, for any fixed $\lambda$, the $\int\Tr g$ is always lower bounded by the value obtained from linearly superposing two usual LLs whose quantum geometric fluctuation is zero. It is also notable that, increasing the weight of $u_{m+l}$ in the $u_m - u_{m+1}$ model, by decreasing $\lambda$, does not necessarily always enhance $\int \Tr g$.

\begin{figure*}
    \centering
    \includegraphics[width=1.0\linewidth]{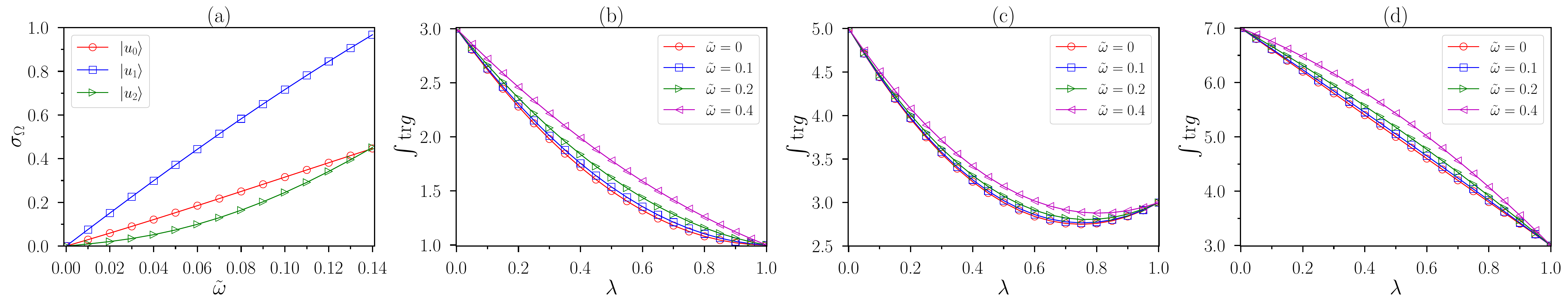} \caption{(a) The variance of Berry curvature for $|u_0\rangle$, $|u_1\rangle$ and $|u_2\rangle$ as a function of the modulation parameter $\tilde\omega$. (b)-(d) $\int\Tr g$ for the (b) $u_0 - u_1$, (c) $u_1-u_2$, and (d) $u_1-u_3$ model at different values of $\tilde\omega = 0$, 0.1, 0.2, and 0.4.} \label{fig:geometry}
\end{figure*}

The analysis presented above is consistent with the numerical examination listed in Fig.~\ref{fig:geometry}, where the enhancement of $\int \Tr g$ by quantum geometric fluctuation [Fig.~\ref{fig:geometry}(b)-(d)] and the non-monotonic behavior of $\int \Tr g$ with $\lambda$ [Fig.~\ref{fig:geometry}(c)] can be seen clearly. In the numerical calculation, we take the modulation function $\mathcal B(\bm r)$ to be periodic on triangular lattice and parameterize it as 
\begin{equation}
    \mathcal B(\bm r) = 1 + \tilde{\omega} \sum_{m=1}^6 e^{i{\bm b}_m\cdot\bm r}, \label{def_tildeB}
\end{equation}
where ${\bm b}_m$'s are the six shortest reciprocal lattice vectors of a triangular lattice and $\tilde{\omega}$ is the amplitude of the Fourier modes controlling the strength of quantum geometric fluctuation. The standard LLs correspond to the special case of $\tilde\omega = 0$. We will use the same parameterization to study interacting problem in Sec.~\ref{sec:nonabelian_fractionalization}. Details of numerical simulations are presented in Appendix~\ref{sec:numerics_details}. 

Last but not least, we discuss the lower bound of the integrated trace value of the $u_m - u_n$ model. Since at any fixed $\lambda$, the $\int \Tr g$ of the $u_m - u_n$ model is lower bounded by that value of superposed standard LLs, and since $\int \Tr g$ of the standard LL has a simple form whose global minimal value is easy to obtain, we arrive at the following exact lower bound of $\int \Tr g$ of the $u_m - u_n$ model:
\begin{equation}
    \int \Tr g/\mC \geq \begin{cases}
    2m + 1,\quad n - m \geq 2,\\
    \frac{3n}{2} - \frac{1}{2n},\quad n - m = 1,\label{intteg_lower_bound} \end{cases}
\end{equation}
which is tighter than the trace bound $\int \Tr g/\mC \geq 1$.

Recently, fundamental bounds relating the energy gap of an insulator to its ``quantum weight'' were derived~\cite{LiangFu_bounds_24_PRX,LiangFu_QuantumWeight_24}. For non-interacting insulators, the quantum weight reduces to the ground-state integrated trace of the quantum metric. Following this, the lower bound Eq.~(\ref{intteg_lower_bound}) gives a tighter upper bound of the energy gap of insulators in terms of Chern number, when the ground state is well approximated by the $u_m - u_n$ model.

\subsection{Wavefunction Anatomy} \label{sec:superposition:anotomy}
Due to the completeness of the generalized LL basis $\{\ket{u_n}\}$, any cell-periodic state of a Chern band can be decomposed as follows,
\begin{equation}
    |u_{\bm k}\rangle = \sum_{n = 0}^{N} C^{\mathcal B}_{n\bm k} |u^{\mathcal B}_{n\bm k}\rangle,\label{def_wf_decomposition}
\end{equation}
where in general $N = \infty$ and we assume $C^{\mathcal B}_{n\bm k}$ is periodic under shifting the momentum by reciprocal lattice vectors $\bm k \rightarrow \bm k+\bm b$. In the above, we have included the superscript $\mathcal B$ to emphasize the fact that both the generalized LL basis $\{\ket{u_n}\}$ and the decomposition coefficients $C_{n\bm k}$ depend on the choice of the modulation function $\mathcal B(\bm r)$. Note that the topological properties and boundary condition of two sides must be identical: Chern number of the left side $|u_{\bm k}\rangle$ is identical to those on the right hand side $|u_{n\bm k}\rangle$. Moreover if $|u_{\bm k}\rangle$ is a LL type state obeying magnetic translation symmetry then $\mathcal B(\bm r)$ is a periodic function; and if $|u_{\bm k}\rangle$ is a Bloch type state obeying standard Bloch translation symmetry then $\mathcal B(\bm r)$ must be quasi-periodic.

\subsubsection{Quantifying the Landau level mimicry}
Viewing $\mathcal B(\bm r)$ as a ``gauge choice'' and $|C^{\mathcal B}_{n\bm k}|^2$ as the distribution function defined in the parameter space $(n, \bm k)$, it is a well defined question about what is the optimal gauge $\mathcal B_*$ that minimizes the spread of the distribution function, in the same spirit of Wannier function's optimization problem~\cite{Vanderbilt_Wannier_RMP}. By definition, a single generalized state $\ket{u_n}$ constructed from $\mathcal B_*(\bm r)$ has the optimal distribution function sharply peaked at $n$ for any momentum $\bm k$, supposing the ``gauge'' choice $\mathcal B$ used for decomposition has been chosen to be identical as the modulation function $\mathcal B_*$ used for constructing the input state. Thereby we propose, for a general single-band wavefunction, the minimal spread of the distribution function $p_{n} \equiv \int~|C_{n\bm k}|^2$ can serve as a quantitative criterion of how its quantum geometry is approximated by a LL type state. In the next section, we discuss a canonical method to determine the optimal $\mathcal B_*$ if the input state is known to be constructed as linear super-positions of finite number of generalized LLs.

For a general Chern band, if there is a single peak in the distribution under the optimal decomposition gauge, the integer part of the mean value $n_* \equiv \sum_n n p_{n}$ approximates the LL index. This generalizes the $n = 0$ ideal quantum geometry criterion~\cite{Grisha_TBG2,JieWang_exactlldescription} to general $n$ in a quantitative way.

With the optimal gauge $\mathcal B_*$, the wavefunction coefficients are supposed to be concentrated near $n_*$, thereby it makes sense to give a cutoff on level index $n$ and consider wavefunction with finite $N$. Moreover although in the most general case, $N$ is required to be taken to infinity for the representation Eq.~(\ref{def_wf_decomposition}) to be exact, we anticipate that models with finite $N$ has the ability to well approximate large a class of wavefunctions of low integrated trace value. We thus define a ``finite-$N$'' model to be Eq.~(\ref{def_wf_decomposition}) with $N$ being a finite positive integer. The superposition coefficients $C^\mathcal{B}_{n\bm k}$ in the finite$-N$ model are allowed to be momentum dependent.

\subsubsection{Finite-$N$ models and the canonical decomposition algorithm} \label{sec:finite-N-models}
As mentioned at the beginning of this section, an important practical question is, given a Chern band $|u_\mathbf{k}\rangle$, such as the one for twisted MoTe$_2$ homobilayers at various twist angles, how can we decompose such Chern band into generalized LLs. Here we provide a practical algorithm to obtain the decomposition assuming a finite number of expansion basis $N$: namely assuming the finite-$N$ model.

We will denote the $N-$dimensional Hilbert space at $\bm k$ spanned by $\ket{u_{0\bm{k}}}\dots\ket{u_{N-1 \bm{k}}}$ as $\mathcal H_{N\bm k}$,
\begin{equation}
    \mathcal H_{N\bm k} \equiv \text{span}\left\{|u_{0\bm k}\rangle,...,|u_{N-1\bm k}\rangle\right\}.
\end{equation}
The recursion relations Eq.~(\ref{rec1}) to Eq.~(\ref{rec3}) yield that the space $\mathcal H_{N\bm k}$ is preserved by the action of the lowering operator $\hat a$ or any of its positive powers, {\it i.e.},
\begin{equation}
    \hat a \mathcal H_{N\bm k} \in \mathcal H_{N\bm k}.
\end{equation}
If at some momentum $\bm k$, operator $\hat a$ to some power $n$ acting on the input state $|u\rangle$ annihilates the state, it means $\hat a^{n-1}|u\rangle$ is proportional to $\ket{u_0}$ at that point, and $\ket{u_n}$ can be constructed successively from the raising ladder operators $\hat a^\dag$. A more general situation is that $\hat a^n$'s with $n = 0,...,N-1$ do not annihilate the input state $|u\rangle$. In this case, the important observation is that the generated $N$ vectors
\begin{equation}
    |v_{n\bm k}\rangle \equiv \hat a^{n}|u_{\bm k}\rangle
\end{equation}
fully span the Hilbert space $\mathcal H_{N\bm k}$, although $|v_n\rangle$'s are in general not orthonormal.

Additionally, the action of the raising operator $\hat a^\dag$ maps $|u_{\bm k}\rangle$ to a vector living in the Hilbert space $\mathcal H_{N+1,\bm k}$ which is composed of $\mathcal H_{N\bm k}$ and $|u_{N\bm k}\rangle$:
\begin{equation}
    \hat a^\dag |u_{\bm k}\rangle \in \mathcal H_{N+1,\bm k} = \mathcal H_{N\bm k} \oplus \text{span}\left\{|u_{N\bm k}\rangle\right\}.
\end{equation}
Hence projecting out $\ket{v_{0}},...,\ket{v_{N-1}}$ from $\hat a^\dag |u_{\bm k}\rangle$ yields $|u_{N\bm k}\rangle$ uniquely up to an undetermined $\bm k-$dependent $\mathrm{U}(1)$ phase. In this way, the information of $\mathcal B$ is extracted as it is contained in $N$th generalized LL state $|u_{N\bm k}\rangle$. With $|u_{N\bm k}\rangle$ in hand for every $\bm k$, according to the recursion relation Eq.~(\ref{rec2}), successively applying the lowering ladder operator followed by Gram-Schmidt orthogonalization in each step yields recursively from $|u_{N-1}\rangle$ to $|u_0\rangle$, up to undetermined $\bm k-$dependent $\mathrm{U}(1)$ phases. This procedure can be formulated as
\begin{eqnarray}
    |u''_{n-1,\bm k}\rangle &=& \hat a|u_{n,\bm k}\rangle,\\
    |u'_{n-1,\bm k}\rangle &=& |u''_{n-1,\bm k}\rangle - |u_{n,\bm k}\rangle\langle u_{n,\bm k}|u''_{n-1,\bm k}\rangle,\\
    |u_{n-1,\bm k}\rangle &=& |u'_{n-1,\bm k}\rangle/\sqrt{|\langle u'_{n-1,\bm k}|u'_{n-1,\bm k}\rangle|},
\end{eqnarray}
for $n = N, N-1, ..., 1$. In this way, the superposition weights are extracted from the wavefunction overlap $|C_{n\bm k}|^2 = |\langle u_{n\bm k}|u_{\bm k}\rangle|^2$.

In the above, we have discussed how to extract superposition weight $|C_{n\bm k}|^2$ from the finite$-N$ model in an exact manner. Since individual generalized LL is a special case of the finite-$N$ model, our exact algorithm easily reduces to the geometric criteria for single generalized LL. Comparing with Ref.~\cite{Fujimoto24}, the canonical algorithm discussed here is simpler, more practical and more general. We notice that in practice, due to the fact that numerics replaces continuous derivatives with finite difference, the numerical error is accumulated in each step using the ladder operators. The larger the $N$ is, the more times ladder operators are applied and more error is accumulated. Nevertheless, the error can be well controlled if one takes the momentum-space mesh dense enough.

In the end, we comment on extending the above discussion to a general Chern band, which involves infinite rather than finite $N$ and has the potential applications to moir\'e materials. One idea for decomposing an infinite $N$ model is to set a maximal $N_{\rm max}$ and approximate it as a finite-$N_{\rm max}$ model, which is supposed to be a good approximation if the genuine weights $|C_{n\bm k}|^2$ are indeed small when $n > N_{\rm max}$. However, whether the weight extracted from the approximated finite$-N$ model converges or not as one increases $N_{\rm max}$ and the general guiding principle behind are currently not systematically explored. We leave more systematical exploration to future work.

\section{NON-ABELIAN FRACTIONALIZATION} \label{sec:nonabelian_fractionalization}
In this section, we study the geometric stability of FCIs with a focus on non-Abelian phases. To be more concrete, ``stability'' here refers to the competition between fractionalized phases and non-fractionalized phases, such as Fermi liquids, charge density waves, or other conventional orders. In general, many factors influence this competition, including band dispersion, band mixing, and details of interactions. Therefore, theoretical studies at all levels, including first-principle calculations, model building, and many-body numerics, are crucial for comparisons with or guidance for real experiments~\cite{Dan_parker21,Kaisun_FCI21,DiXiaoFCI23,Valentin22_anomaloushallmetal,Nicolas_WSe2_23,Dong_CFL23,Goldman_CFL_23,LiangFuFCI23,YangZhangNonabelian24,DiXiaoNonabelian24,ChoNonabelian24,BAB_FCI_tTMD_1,BAB_FCI_tTMD_2,BAB_FCI_tTMD_3,DiXiao_SmallAngleTMD_2311,Jackson:2015aa,AndreasCDW,YaoWang_FCI_Semimetal_24,LiangFu_QED3_PRB23,Xueyang_Threebody_24,Xueyang_PhaseTransitions_24,Xueyang_Intertwined_24,Goldman_Criticality_24,Oreg_GeometryFCI24,Sheng_2405}. In the context of fractional quantum Hall physics, non-Abelian fractionalized states can often occur in the first $n=1$ LL with Coulomb interaction. Representative examples include the Moore-Read~\cite{MoreReadState} and Read-Rezayi state~\cite{Read_Rezayi}.

Among the various factors mentioned above, the quantum geometry is one crucial factor determining the stability of fractionalized phases. To focus on the quantum geometric effect, we consider an ``idealized'' set up with an isolated and dispersionless band. Our construction of generalized higher LL states allows us to systematically explore the non-Abelian fractionalization directly within the single-particle Hilbert space containing key features of first LL. In contrast to existing literature on the geometric stability of non-Abelian phases which focus on the twisted MoTe$_2$~\cite{YangZhangNonabelian24,ChoNonabelian24,DiXiaoNonabelian24} or related toy models~\cite{LiangFuNonabelian24}, our set up does not require a particular single-electron Hamiltonian that yields bands. We will directly employ the $u_m - u_n$ model discussed in Sec.~\ref{sec:superposition:umunmodel} as the ansatz of the wavefunction of the isolated flat band. We propose a geometric criteria in terms of integrated trace of quantum metric $\int\Tr g$ and Berry curvature standard derivation $\delta_{\Omega}$ favoring Moore-Read state. We argue the $u_m - u_n$ model captures a signature portion of the general single-particle Hilbert space of first Landau level like and thereby our geometric stability criteria is supposed to be general and can serve as a necessary condition for Moore-Read phase in general Chern band. We will compare our geometric stability criteria to existing case studies available in literature.

\subsection{Stability of the Moore-Read State} \label{Sec:MooreRead:Stability}
A representative class of non-Abelian states in the Moore-Read (MR) class includes the Pfaffian (Pf) state, its particle-hole conjugate the anti-Pfaffian (aPf) state, and the more exotic particle-hole symmetric Pfaffian state (PH-Pf), all of which can in principle occur in a half-filled $n=1$ standard LL and can be interpreted as the superconducting paired state of composite fermions on top of the composite Fermi liquid state~\cite{HalperinLeeRead,Son}. In what follows we will use ``Moore-Read'' to denote the MR class that includes Pf, aPf, and PH-Pf states.

The MR states were initially motivated by the conformal-field-theory construction of fractional quantum Hall wavefunctions~\cite{MoreReadState}. The Pf state is the densest zero-energy ground state of a three-body short-ranged repulsion in the lowest LL~\cite{GreiterWenWilczekPRL91}. Moreover, numerical studies indicate Pf and aPf are favored by the Coulomb interaction in the $n=1$ LL~\cite{PhysRevLett.80.1505,PhysRevLett.104.076803,RezayiHaldanePRL00,PhysRevLett.101.016807}. Although the MR states have been extensively studied in the context of LLs, their general stability in the presence of nonuniform quantum geometries is not explored much.

\begin{figure*}
    \centering \includegraphics[width=1.0\linewidth]{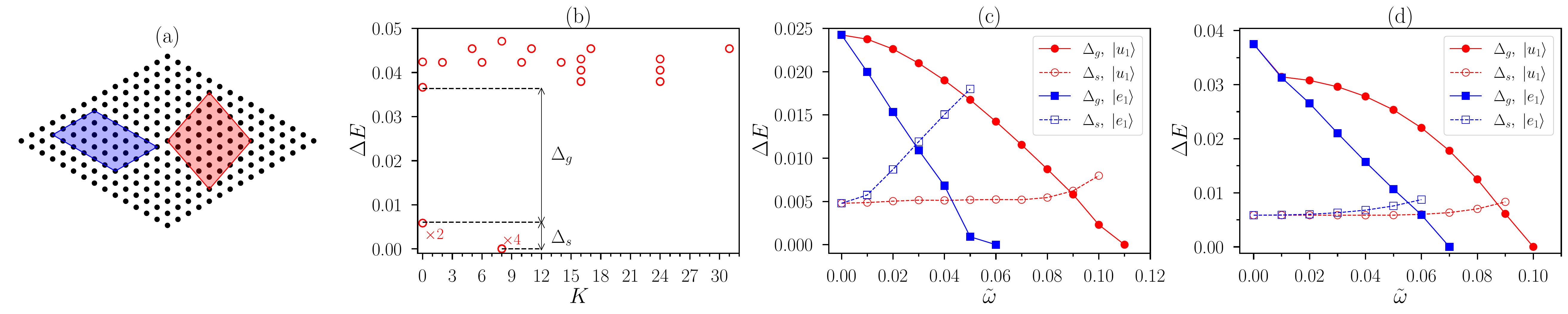} \caption{(a) Tilted lattices for $N_e=12$ (blue) and $N_e=16$ (red) electrons. (b) A representative energy spectrum of the Coulomb interaction of $N_e = 16$ electron in a half-filled first generalized LL $|u_1\rangle$ constructed by using modulation function $\mathcal B(\bm r)$ of Eq.~(\ref{def_tildeB}) with $\tilde\omega = 0.02$. $K$ labels the momentum of the many-body states. The multiplicity of levels is given when their energy difference is hardly visible. Neutral gap and ground state manifold energy splitting are denoted respectively as $\Delta_g$ and $\Delta_s$. (c) and (d): ground-state splitting $\Delta_s$ and neutral gap $\Delta_g$ for the Coulomb interaction in the first generalized LL $|u_1\rangle$ and the first modulated LL $|e_1\rangle$, as a function of the quantum geometric fluctuation strength parameter $\tilde\omega$. (c) is for $N_e=12$ electrons and (d) is for $N_e=16$ electrons.} \label{fig:moore_read}
\end{figure*}

We consider projected density-density interaction,
\begin{equation}
    H = \frac{1}{2\mathcal A}\sum_{\bm q}v_{\bm q} :\rho_{\bm q}\rho_{-\bm q}:,\label{def_H_um_un}
\end{equation}
normal ordered with respect to electron's vacuum. In the above, $\mathcal A$ is the real-space area of the 2D system. In numerics we use Coulomb interaction which has $v_{\bm q} = 2\pi/|{\bm q}|$. The projected density operator $\rho_{\bm q}$ is determined by the form factor,
\begin{eqnarray}
    \rho_{\bm q} &=& \sum_{\bm b} \sum^\prime_{\bm k,\bm k'} \delta_{\bm b,\bm k-\bm k'-\bm q} f^{\bm k\bm k'}_{\bm b}c^\dag_{\bm k}c_{\bm k'},\label{densityrho}\\
    f^{\bm k\bm k'}_{\bm b} &=& \int_{\rm u.c.} d^2\bm r~e^{-i\bm b\cdot\bm r}u^*_{\bm k}(\bm r)u_{\bm k'}(\bm r),\label{def_formfactor}
\end{eqnarray}
where $c^\dag_{\bm k}$ creates an electron at momentum $\bm k$ and $\bm b$ is a reciprocal lattice vector. In the above, $\sum'$ stands for the summation of momentum points within the first Brillouin zone and $\int_{\rm u.c.}$ denotes integrated within the real space unit cell. Substituting Eqs.~(\ref{densityrho}) and (\ref{def_formfactor}) into Eq.~(\ref{def_H_um_un}), we can derive the second quantized expression of the interaction Hamiltonian as
\begin{equation}
    H = \sum'_{\bm k_1,\dots,\bm k_4}V_{\bm k_1\bm k_2;\bm k_3\bm k_4}c^\dag_{\bm k_1}c^\dag_{\bm k_2}c_{\bm k_3}c_{\bm k_4},\label{H1234}
\end{equation}
where the matrix elements are
\begin{equation}
    V_{\bm k_1\bm k_2\bm k_3\bm k_4} =\frac{1}{2\mathcal{A}} \sum_{\bm b} v_{\bm k_1-\bm k_4-\bm b}f^{\bm k_1\bm k_4}_{\bm b}f^{\bm k_2\bm k_3}_{\delta{\bm b}-\bm b}, \label{V1234}
\end{equation}
with $\delta{\bm b}\equiv{\bm k}_1+{\bm k}_2-{\bm k}_3-{\bm k}_4$ a reciprocal lattice vector.

In this section, the wavefunction $u_{\bm k}(\bm r)$ will be chosen as either one generalized LL, or a modulated LL or states of the $u_m - u_n$ model. We set Chern number $\mC = 1$. Throughout our numerics, we will choose isotropic geometry such that complex momentum $k = (k_x + ik_y)/\sqrt{2}$ respecting triangular lattice translation symmetries, and the trace ``$\Tr$'' reduces to the standard one. Such standard trace is also often used in literature in studying twisted MoTe$_2$ and moir\'e graphene. The modulating function $\mathcal B(\bm r)$ is parameterized following Eq.~(\ref{def_tildeB}) where $\tilde\omega$ is the only parameter tuning the strength of quantum geometric fluctuations.

The numerical hallmark of the Pf and aPf at finite system sizes is the six-fold (two-fold) nearly degenerate ground-states on the torus geometry when the electron number $N_e$ is even (odd). We focus on periodic samples with even $N_e$. In the absence of quantum geometry fluctuation $\tilde\omega = 0$, any two-body translational invariant interacting Hamiltonian has exact particle-hole symmetry at half filling, leading to the equal energy of the Pf and aPf states. Therefore, we expect $12$-fold ground-state degeneracy in total in the thermodynamic limit for the MR phase stabilized by two-body interactions. However, due to finite size effect, numerical simulations often report six-fold nearly degenerate ground state manifold, corresponding to either the symmetric or the anti-symmetric superposition of Pf and aPf states in a particle-hole symmetric system where ground states spontaneously breaks particle-hole symmetry~\cite{RezayiHaldanePRL00,PhysRevB.80.241311,PhysRevLett.101.016807,PapicHaldaneRezayiPRL12}. Degeneracies and nature of ground states with and without particle-hole symmetry will be discussed in Sec.~\ref{sec::weightonMR}.

\begin{figure*}
    \centering
    \includegraphics[width=1.0\linewidth]{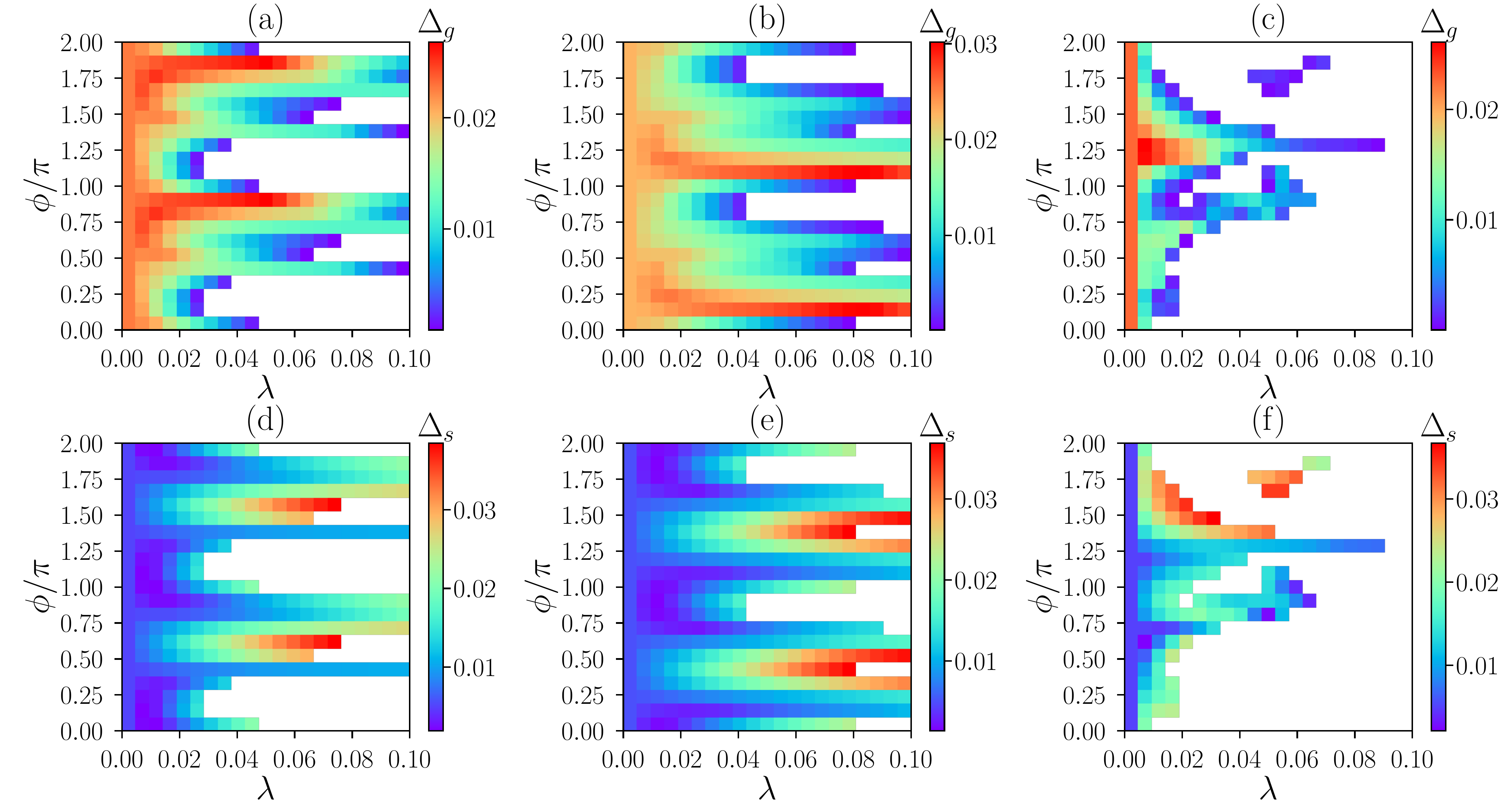} \caption{The many-body gap $\Delta_g$ (the first row) and ground-state splitting $\Delta_s$ (the second row) at $\nu=1/2$ in the $u_0 - u_1$ [(a) and (d)], $u_2 - u_1$ [(b) and (e)], and $u_3-u_1$ model [(c) and (f)] for $N_e=12$ electrons with Coulomb interaction. The modulation parameter are taken as $\tilde\omega = 0.02$ for all sub-figures. In those white regions of the $\lambda-\phi$ parameter space, there is no six-fold ground-state degeneracy found numerically at MR momenta.} \label{fig:MRgap}
\end{figure*}

To begin with, we study the MR state in the first generalized LL $|u_1\rangle$ and the first modulated LL $|e_1\rangle$. Similar to the situation in the standard $n=1$ LL, we observe six-fold ground-state degeneracy below the critical quantum geometry fluctuation in our exact diagonalization study. For various system sizes of $N_e=12$, $14$, and $16$ electrons, the momenta of the ground states match both those of Pf and aPf, which is a strong signature of the MR phase. A further confirmation from wavefunction overlap is presented in Appendix~\ref{sec:MRweight}. The sample geometry and a representative energy spectrum are displayed in Figs.~\ref{fig:moore_read}(a) and (b), respectively. We denote the ground-state energy splitting as $\Delta_s$, defined as the difference between the maximal and minimal energy of the six ground states. We also denote the neutral gap as $\Delta_g$, defined as the difference between the minimal exited energy and the maximal ground-state energy. A stable non-Abelian phase can be characterized by large $\Delta_g$ and small $\Delta_s$ [Fig.~\ref{fig:moore_read}(b)]. With increasing quantum geometry fluctuation by increasing $\tilde\omega$, the neutral gap $\Delta_g$ monotonically decays. As shown in Figs.~\ref{fig:moore_read}(c) and (d), $\Delta_g$ becomes even smaller than $\Delta_s$ at sufficiently strong $\tilde{\omega}$ and finally vanishes with the absence of six-fold ground-state degeneracy at $\tilde\omega\sim 0.1$. We also compare the evolution of $\Delta_g$ and $\Delta_s$ with $\tilde{\omega}$ for $|e_1\rangle$ and $|u_1\rangle$. For the latter, $\Delta_g$ survives up to larger $\tilde{\omega}$ [Figs.~\ref{fig:moore_read}(c) and (d)], suggesting that the putative MR phase is generally enhanced in the Schmidt orthogonalized state $|u_1\rangle$ as compared to $|e_1\rangle$.

Next we study the geometric stability of MR state in the $u_m - u_n$ model. We will focus on the $u_0 - u_1$, $u_2 - u_1$ and $u_3 - u_1$ model for the purpose of realizing the MR phase because they resembles the first LL while being enriched by geometric fluctuation and higher LL components. The wavefunction $u_{\bm k}({\bm r})$ used in Eq.~(\ref{def_formfactor}) is defined in Eq.~(\ref{def_um_un_model}) and copied in below:
\begin{equation}
    u_{\bm k}(\bm r) = \sqrt{\lambda}u_{m\bm k}(\bm r) + \sqrt{1-\lambda}e^{i\phi}u_{n\bm k}(\bm r). \label{def_umn2}
\end{equation}
This set up has a controlled limit where the MR phase is guaranteed to occur: in the limit of $\lambda = 0$ and $\tilde\omega = 0$ the above wavefunction reduces to the standard $n=1$ LL wavefunction which is known to stabilize MR phase under Coulomb interaction. We then explore the stability of the observed six-fold ground-state degeneracy against $\lambda$ and $\phi$ at fixed quantum geometric fluctuation strength $\tilde\omega$. As a representative example, we display in Fig.~\ref{fig:MRgap} the neutral gap $\Delta_g$ and the ground-state splitting $\Delta_s$ at $\tilde\omega = 0.02$  in the $(\lambda, \phi)$ parameter space for the $u_0 - u_1$, $u_2 - u_1$, and $u_3-u_1$ model of $N_e=12$ electrons, respectively. We can observe a clear tendency that the six-fold ground-state degeneracy with Pf and aPf momenta becomes worse when the weight $1-\lambda$ of $u_1$ decrease. For $\tilde\omega = 0.02$, the largest $\lambda$ which allows decent six-fold ground-state degeneracy with $\Delta_g\geq\Delta_s$ is about $0.07$, $0.18$, and $0.06$ for the $u_0 - u_1$, $u_2 - u_1$, and $u_3-u_1$ model, respectively. These results indicate that the dominant $u_1$ is essential for the stabilization of the MR phase. We also emphasize $\int\Tr g$ being close to $3$ is not a sufficient condition for MR; for instance a nearly equal weight mixing of $|u_1\rangle$ and $|u_2\rangle$ can give $\int\Tr g \approx 3$ but the wavefunction is no longer dominated by $|u_1\rangle$ as seen in Fig.~\ref{fig:geometry}(c).

\begin{figure*}
    \centering
    \includegraphics[width=1.0\linewidth]{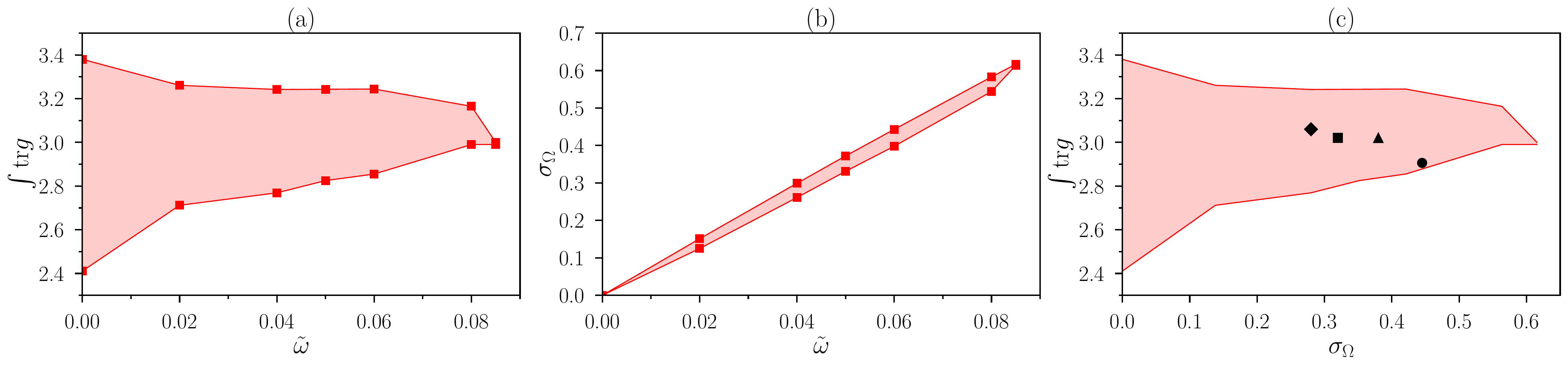} \caption{The windows of (a) $\int {\rm tr}g$ and (b) $\sigma_\Omega$ as a function of $\tilde\omega$ in which we find six-fold ground-state degeneracy with $\Delta_g\geq\Delta_s$ for $N_e=12$ with Coulomb interaction. The windows are evaluated based on the energy spectra for the half filled $u_0-u_1$, $u_2-u_1$ and $u_3-u_1$ models. (c): Combining (a) and (b), we give the region of the stable MR phase in the quantum geometric parameter space spanned by $\sigma_\Omega$ and $\int {\rm tr}g$. The markers indicate the materials and models in which the MR phase and the corresponding quantum geometric quantities were reported, including the twisted MoTe$_2$ homobilayers with twist angle $2^\circ$ (square) and $2.14^\circ$ (diamond)~\cite{DiXiaoNonabelian24}, skyrmion Chern band model (triangle)~\cite{LiangFuNonabelian24}, and the double TBG model in Sec.~\ref{Sec:TwistedMultilayerGraphene} (dot).} \label{fig:trgwindow}
\end{figure*}

In the above, we have fixed $\tilde\omega$ and studied the conditions in terms of $(\lambda, \phi)$ that allows MR. In below, we convert model parameters $(\lambda, \phi)$ into physical parameters $(\int \Tr g, \sigma_\Omega)$, and vary $\tilde\omega$ to obtain a geometric stability criteria for MR state. Here $\int \Tr g$ is the integrated trace value and $\sigma_\Omega$ is the standard deviation of Berry curvature. Concretely, for each fixed $\tilde\omega$, we first get the union of $\int\Tr g$ data or $\sigma_\Omega$ data obtained from the three $u_m - u_n$ model ($u_0 - u_1$, $u_2 - u_1$, and $u_3-u_1$), whose evolution with respect to $\tilde\omega$ can be found in Fig.~\ref{fig:trgwindow}(a) and (b). We find that the window of $\int \Tr g$ favoring the MR phase becomes narrower with increasing $\tilde\omega$ [Fig.~\ref{fig:trgwindow}(a)], while the scope of $\sigma_\Omega$ is almost located along a straight line [Fig.~\ref{fig:trgwindow}(b)]. We can thus combine Figs.~\ref{fig:trgwindow}(a) and \ref{fig:trgwindow}(b) to determine a region in terms of $(\sigma_\Omega, \int \Tr g)$ in which the MR phase is stable, as displayed in Fig.~\ref{fig:trgwindow}(c). This region provides a quantum geometric criterion for the existence of MR phase in our $u_m - u_n$ models. In what follows, we will argue such geometric criterion is a necessary condition for MR.

Compared to the interacting physics in realistic models such as twisted MoTe$_2$ homobilayers, the set up of the interacting problem in our $u_m - u_n$ model has a couple of simplifications as summarized in below. (i) First of all, we work in the exact flat band limit by ignoring single-particle dispersion which can originate from bare band dispersion as well as Hartree-Fock renormalization. (ii) Secondly, we consider a single-band projected interaction and ignores multi-band mixing effects. (iii) Thirdly, we have assumed spacial isotropy and have only retained the leading Fourier modes in $\mathcal B(\bm r)$. (iv) Last but not least, we have neglected the momentum dependence of the superposition coefficients $\lambda$ and $\phi$, therefore the $u_m - u_n$ model only captures a subspace of the Hilbert space spanned by two generalized LLs.

We will argue many of the simplifications mentioned above are conditions that favor fractionalized phases. Therefore, the geometric criteria we propose for the MR phase, arguably, sets the largest window. The argument goes as follows. (i) Weak band dispersion is generally regarded as important for fractionalized phases which are strongly correlated phases driven by interactions; suppressing single-particle dispersion helps stabilizing fractionalized phases over trivial phases such as Fermi liquids or density waves. (ii) A single isolated band is important for a flat Chern band to behave like an isolated LL, which also facilitates fractionalized phases. Band mixing often leads to competing phases. (iii) Based on the observation in LLs that, increasing the strength of anisotropy can drive phase transitions from fractional quantum Hall to classically ordered phases~\cite{Bo_BandMassAnisotropy_12}, the isotropy condition also favors fractionalized phases. Higher-order Fourier modes can be shown to have Gaussian suppressed contributions to the form factor hence arguably have negligible effects to interacting physics~\cite{JieWang_exactlldescription}. (iv) We expect allowing momentum-dependent superposition coefficients deviates flatbands from being LL like, therefore also disfavors the fractionalized phases.

Based on the argument above, we expect that the geometric criteria obtained from our $u_m - u_n$ models in Fig.~\ref{fig:trgwindow}(c) is the general geometric stability criterion for the MR phase within the consideration of Coulomb interaction. Recent first-principle calculations indicated consecutive Chern bands in the twisted MoTe$_2$ homobilayers at small twist angle ($\sim 2^\circ$)~\cite{DiXiao_SmallAngleTMD_2311}. Motivated by this, a couple of theoretical studies pointed out the possibility of realizing the MR phase at small-angle twisted MoTe$_2$ homobilayers~\cite{ChoNonabelian24,YangZhangNonabelian24,DiXiaoNonabelian24} as well as related toy models~\cite{LiangFuNonabelian24}.
We collect from these work the single-particle geometric quantities ($\int\Tr g$ and $\sigma_{\Omega}$) allowing the MR phase and demonstrate them in Fig.~\ref{fig:trgwindow}(c). Remarkably, we find they are all bounded within the window we propose based on the $u_m - u_n$ model. A practical message of our result is that: for a given model or material, if its single-particle geometric quantities, including the integrated trace value of $\int\Tr g$ and Berry curvature fluctuation $\sigma_{\Omega}$, are outside the window we propose in Fig.~\ref{fig:trgwindow}(c), the likelihood of realizing the MR phase is low, if Coulomb interaction is considered.

\subsection{Nature of the Ground States} \label{sec::weightonMR}
Numerically we have observed the six-fold ground-state degeneracy consistent with the MR phase in finite systems. A remaining question is: are ground states Pf or aPf? In the following, we will consider this issue by studying the particle-hole (PH) symmetry in our model.

Our starting point is a general two-body translational invariant interaction Hamiltonian Eq.~(\ref{H1234}). The PH transformation exchanges $c_{\bm k}$ and $c^\dag_{\bm k}$. It also exchanges the occupation number $0$ and $1$ at each ${\bm k}$ point in the Fock basis of the many-body state. PH transformation also complex conjugate scalars of the Hilbert space. Then we have the PH-transformed Hamiltonian as,
\begin{eqnarray}
	H_{\rm PH} &=& H + \sum'_{\bm k}\epsilon_{\bm k}\left(n_{\bm k} - \frac12\right), \label{def_singlebody_Pf_aPf} \\
    \epsilon_{\bm k} &=& \sum'_{\bm k'} \left(V_{\bm k\bm k'\bm k\bm k'} + V_{\bm k'\bm k\bm k'\bm k} - V_{\bm k\bm k'\bm k'\bm k} - V_{\bm k'\bm k\bm k\bm k'}\right),\nonumber
\end{eqnarray}
where $n_{\bm k} = c^\dag_{\bm k}c_{\bm k}$ and $V_{\bm k_1\bm k_2\bm k_3\bm k_4} = V^*_{\bm k_4\bm k_3\bm k_2\bm k_1}$ is used. Note that the PH transformed Hamiltonian differs with the original one by an additional dispersion $\epsilon_{\bm k}$. In standard LLs, due to the magnetic translation symmetry, $\epsilon_{\bm k} = 0$, guaranteeing any translational invariant two-body interaction in a half-filled LL is exactly PH symmetric. As discussed in Sec.~\ref{Sec:MooreRead:Stability}, such exact PH symmetry in the standard LLs gives cat state of Pf and aPf when the ground state spontaneously breaks PH symmetry; at finite system sizes, six-fold degeneracy are detected in the MR phase while one should expect twelve-fold degenerate in the thermodynamic limit.

When the quantum geometric fluctuation is turned on, both $\epsilon_{\bm k}$ and electron's occupations are nonuniform in the Brillouin zone, therefore the PH symmetry is explicitly broken. We denote the energy difference of Pf and aPf as $E_\Delta \equiv \langle H\rangle_{\rm Pf} - \langle H\rangle_{\rm aPf}$ where $\langle\mathcal O\rangle_{\Psi}$ means the expectation value of operator $\mathcal O$ evaluated with respect to a many-body state $\Psi$. Performing PH transformation, $E_\Delta$ can be rewritten as $\langle H\rangle_{\rm Pf} - \langle H_{\rm PH}\rangle_{\rm Pf}$, which following Eq.~(\ref{def_singlebody_Pf_aPf}) is given by
\begin{equation}
    E_\Delta = -\sum_{\bm k} \epsilon_{\bm k} \left(\langle n_{\bm k}\rangle_{\rm Pf} - \frac12\right). \label{eq:PfaPf}
\end{equation}
When the system size is sufficiently large, one can replace the lattice summation to continuous integration. In such limit, Eq.~(\ref{eq:PfaPf}) can be further simplified to
\begin{eqnarray}
    E_\Delta &=& -\mathcal{A} e_\Delta,\\
    e_\Delta &=& \frac{1}{4\pi^2} \int \epsilon_{\bm k} \left(\langle n_{\bm k}\rangle_{\rm Pf} - \frac12\right)d^2{\bm k}, \label{eq:PfaPfD}
\end{eqnarray}
where $e_\Delta$ is obtained by integration over the first Brillouin zone and has a converged value in the thermodynamic limit which is generally speaking non-zero. Therefore, asymptotically the energy difference $E_\Delta$ of Pf and aPf states has a linear scaling with particle number and the slope of the scaling is determined by $e_\Delta$. When PH symmetry is exact, $e_\Delta$ is forced to be zero.

This analysis allows us to construct a phenomenological classical Ising model to describe the competition between the Pf and aPf at large system sizes,
\begin{equation}
	h_{\rm Ising} = \frac{1}{2} E_\Delta \sigma_3 + \Big[\langle {\rm Pf}|H|{\rm aPf}\rangle \sigma^+ + h.c.\Big], \label{def_Pf_aPf_model}
\end{equation}
where $\sigma^+ = \sigma_1 + i\sigma_2$ and $\sigma_{1,2,3}$ are Pauli matrices in the basis of Pf and aPf states. Due to the fact that Pf and aPf are distinct topological order, they are orthogonal in the thermodynamic limit. Therefore the off-diagonal terms decay exponentially when increasing the system size. Therefore we expect that either the Pf or the aPf state is selected out by the linearly growing $E_\Delta$ for sufficiently large systems in generic flatbands outside standard LLs.

\begin{figure}
    \centering
    \includegraphics[width=0.9\linewidth]{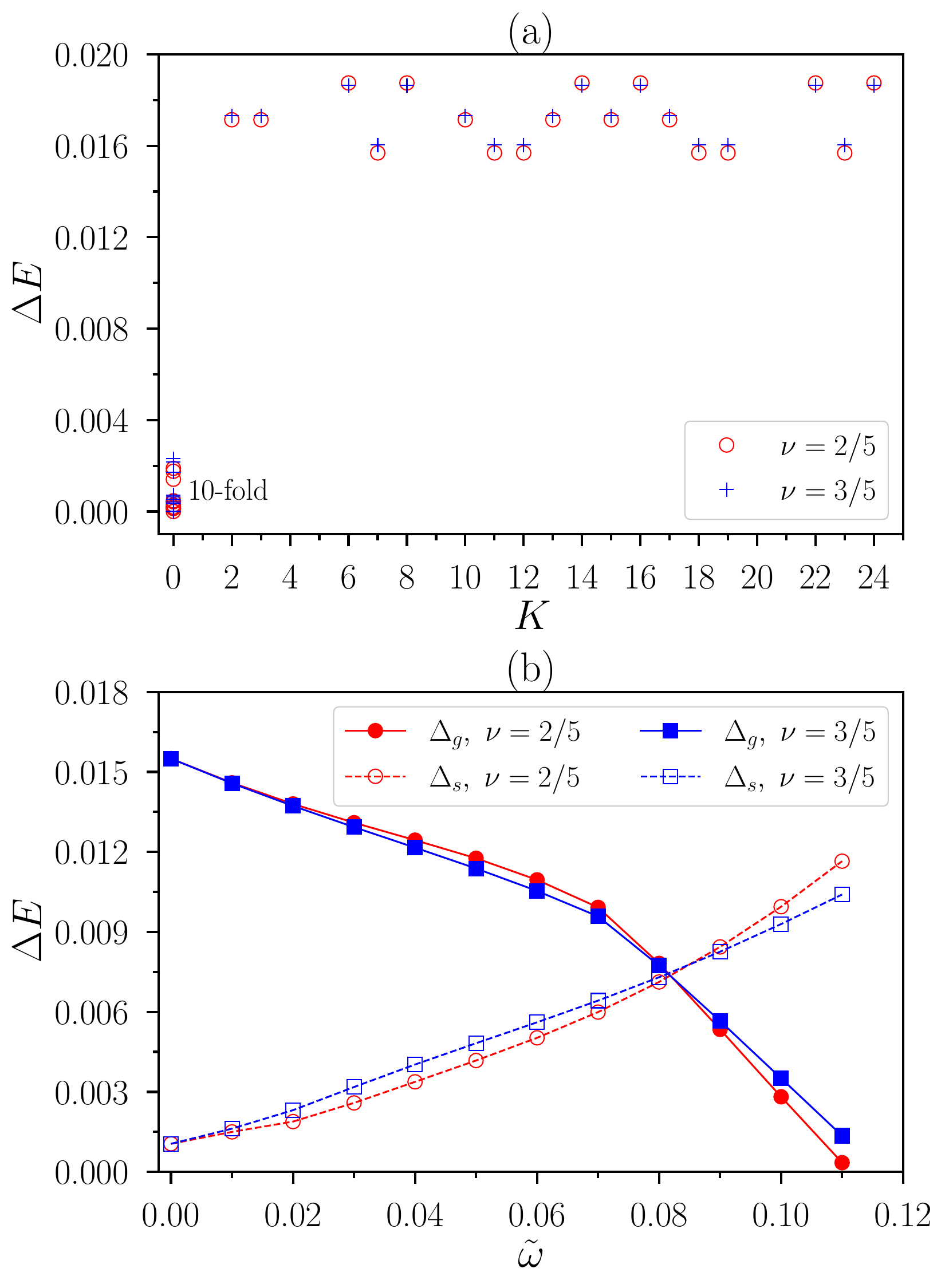} \caption{(a) The energy spectrum of Coulomb interaction in a partially filled first generalized Landau level $|u_1\rangle$ at the filling fraction for Read-Rezayi state. Screening is added to the interaction to improve the $10-$fold degeneracy. We choose $\tilde\omega = 0.02$. (b) The neutral gap $\Delta_g$ and splitting $\Delta_s$ as a function of the Berry curvature modulation parameter $\tilde\omega$. The system size is $N_e=10$ electrons at $\nu=2/5$ and $N_e=15$ electrons at $\nu=3/5$.} \label{fig:RR}
\end{figure}

\subsection{Read-Rezayi Parafermion State}
Besides the MR family, other non-Abelian fractional quantum Hall states, such as the Read-Rezayi (RR) parafermion state at $\nu=3/5$ and its particle-hole conjugate at $\nu=2/5$ have also been proposed to be realizable in the $n=1$ LL~\cite{Read_Rezayi,Slingerland_RR_12,Sheng_RR_PRL15}. One of the motivations for realizing the RR states is the topological quantum computation: the MR states support Majorana-type excitations which are not enough for universal quantum computation; instead the RR states have Fibonacci anyons which in theory is sufficient for universal topological quantum computation.

We briefly discuss the stability of the RR states against quantum geometric fluctuations. We project the Coulomb interaction to $|u_1\rangle$. Remarkably, we observe $10$-fold ground-state degeneracy at both $\nu=2/5$ and $\nu=3/5$ within a range of modulation parameter $\tilde\omega$ (Fig.~\ref{fig:RR}). This degeneracy, and the momentum sectors in which the ground states appear, are consistent with the RR states. The RR states at $\nu=2/5$ and $\nu=3/5$ are both destroyed at roughly the same strength of quantum geometric fluctuation as the MR states. As the ground-state gap $\Delta_g$ and splitting $\Delta_s$ of both the $\nu=2/5$ and $\nu=3/5$ states show similar evolution with increasing $\tilde\omega$ [Fig.~\ref{fig:RR}(b)], the PH symmetry breaking by quantum geometric fluctuation is quite weak in system sizes tractable by exact diagonalization, which is consistent with our observation for the MR states in Sec.~\ref{sec::weightonMR}. 

\subsection{Approximately First Generalized Landau Level in a Graphene-Based Moir\'e Model} \label{Sec:TwistedMultilayerGraphene}
In this section, we present a concrete graphene-based moir\'e model hosting approximately $n=1$ generalized LL at zero magnetic field. In certain parameter region, the model exhibits a nearly flat band which is separated by a finite (but small) band gap from remote bands and exhibits suitable quantum geometric properties for non-Abelian states.

\begin{figure}
    \centering
    \includegraphics[width=1.0\linewidth]{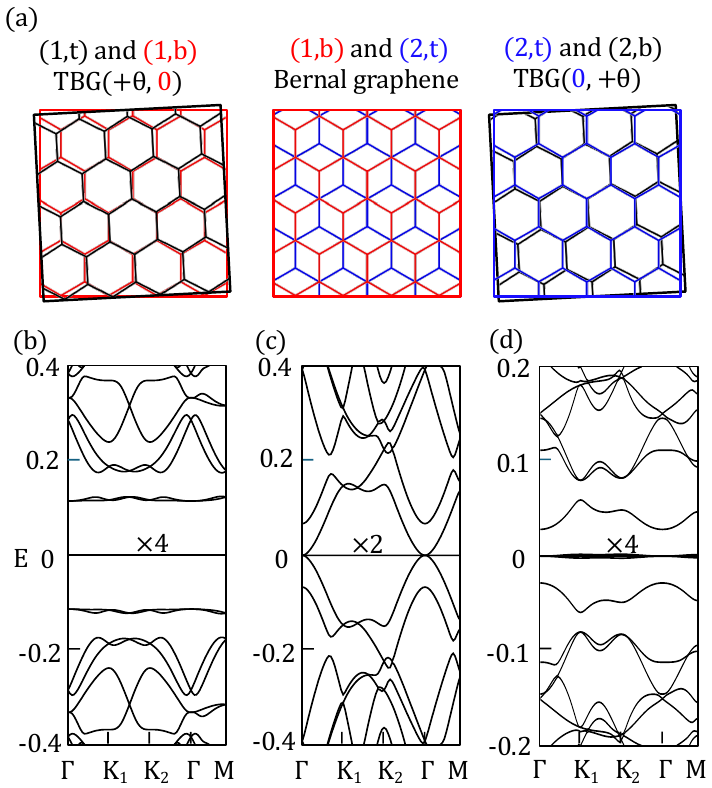} \caption{(a) The lattice structure of the double TBG model consisting of a Bernal stacked two TBG sheets. Here layers $[(1,t), (1,b)]$ and $[(2,t),(2,b)]$ are form individual TBG sheets with the relative twist angles given by $(\theta,0)$ and $(0,\theta)$, respectively. The layers $[(1,b), (2,t)]$ form Bernal bilayer graphene. (b)-(d) The band structures of Eq.~\eqref{eq:Ham_2TBGs} with parameter $(\alpha,\gamma_1,\beta)=(0.586,8,1)$, $(\alpha,\gamma_1,\beta)=(0.586,8,0)$ and $(\alpha,\gamma_1,\beta)=(2.221,8,0)$, respectively. (b) is the band structure for the unphysical limit $\beta = 1$ at the first magic angle, where four-fold exactly degenerate bands are seen. (d) and (e) are at the physical limit $\beta = 0$, at the first and second magic angles, respectively. Comparing to (c), (d) still has nearly flat bands with a finite band gap to remote bands. These bands can be split by applying displacement fields due to their different layer spinor structures.} \label{fig:2TBGs_band}
\end{figure}

We first describe the twisting and stacking geometry of this model. The model consists of four layers of graphene sheets, and we will denote their twisting angle as $(\theta,0,0,\theta)$. This corresponds to a configuration where a TBG sheet is stacked on top of another TBG sheet. Besides the twisting angles, we also need to describe their relative stacking. We denote the layers as $\left[(1,t), (1,b), (2,t), (2,b)\right]$. Here $(1,2)$ indicates the first or the second TBG sheet, and $(t,b)$ refers to the top or bottom layer within each TBG sheet. As illustrated in Fig.~\ref{fig:2TBGs_band}(a), the two TBG sheets are stacked with a lateral shift such that $[(1,b),(2,t)]$ forms Bernal stacking configuration.

The single-electron Hamiltonian in a single valley is given as
\begin{equation}
    \label{eq:Ham_2TBGs}
    H_{1}=\left(\begin{array}{cc}0 & D_{1}^{\dagger} \\ D_{1} & 0\end{array}\right), \quad D_{1}=\left(\begin{array}{cc}D_{0} & \Gamma_{1} \\ 0 & D^\prime_{0}\end{array}\right),
\end{equation}
where $H_{1}$ is written in the sublattice basis $\left[A, B\right]$, and $D_{1}$ describes the inter-sublattice coupling written in the layer basis. Since the second TBG is twisted in the opposite direction to the first TBG, $D_0^\prime=\mu_x D_0 \mu_x$ with the Pauli matrices in layer space $\mu$. Here $D_0$ is the standard chiral operator for the chiral model of TBG~\cite{Grisha_TBG},
\begin{eqnarray}
    D_{0} = \left(\begin{array}{cc} -i \bar{\partial}_{\bm r} & \alpha U(\bm{r}) \\ \alpha U(-\bm{r}) & -i \bar{\partial}_{\bm r}\end{array}\right),\quad U(\bm{r}) = \sum_{n=0}^{2} e^{i\left(\frac{2\pi n i}{3}-\boldsymbol{q}_{n} \cdot \bm{r}\right)},\nonumber
\end{eqnarray}
where $\alpha$ is proportional to the inverse of twist angle, $\bm q_j$ are the three primitive moir\'e vectors. We have followed the notation of Ref.~\cite{Grisha_TBG}. To differentiate from the momentum-space derivative notation used in this work, we have added subscript $\bm r$ for real-space holomorphic and anti-holomorphic derivatives above.  
The interlayer tunneling matrix $\Gamma_{1}$ is given by
\begin{equation}
    \Gamma_{1} = \left(\begin{array}{cc}0 & \beta\gamma_1 \\ \gamma_1 &  0 \end{array}\right),
\end{equation}
where the $\beta\gamma_1$ component corresponds to the inter-sublattice coupling between $(1, t)$ and $(2, b)$, {\it i.e.} the top layer of the first TBG and the bottom layer of the second TBG. Physically $\beta$ should be smaller than one; in the realistic situation, $\beta$ should be nearly zero.

The model is known to support two-fold exact zero modes per valley and sublattice for each momentum $\bm k$ at the magic angles at chiral limit with unphysical parameter $\beta = 1$~\cite{Fujimoto24} [see Fig.~\ref{fig:2TBGs_band}(b)]. The two sublattice polarized zero modes are
\begin{equation}
    |\mathcal{U}_{0\bm{k}}\rangle = \left(\begin{array}{c}|u_{0\bm{k}}\rangle \\ 0\end{array}\right), \quad |\mathcal{U}_{1\bm{k}}\rangle = \left(\begin{array}{c} |u_{1\bm{k}}\rangle \\ -\gamma_{1}^{-1}\mu_x|u_{0\bm{k}}\rangle\end{array}\right),\nonumber
\end{equation}
and the fact they are zero modes follows from the zero mode equations $D_0|u_{0\bm{k}}\rangle = 0$ and $D_0|u_{1\bm{k}}\rangle=\mu_x|u_{0\bm{k}}\rangle$. The $|u_{0\bm{k}}\rangle$ is the standard ideal band wavefunction of magic-angle chiral TBG with lowest LL features~\cite{Grisha_TBG,Grisha_TBG2,JieWang_NodalStructure,JieWang_exactlldescription,popov2020hidden,XiDai_PseudoLandaulevel}, and we can readily show $|u_{1\bm{k}}\rangle=\hat{a}^{\dagger}|u_{0\bm{k}}\rangle$ where $\hat a^\dag$ is the momentum-space ladder operator introduced in Eq.~(\ref{def_ladder_k}). At sufficiently large $\gamma_1$, the second zero mode $|\mathcal{U}_{1\bm{k}}\rangle$ polarizes to the first TBG sheet and becomes identical to the generalized $n=1$ LL discussed above. For finite $\gamma_1$, the $\mathcal{U}_{0}\rangle$ is still the ideal band, but $\mathcal{U}_{1}\rangle$ deviates from the generalized first LL. For example, the integrated trace of $|\mathcal{U}_{0\bm{k}}\rangle$ and $|\mathcal{U}_{1\bm{k}}\rangle$ at the first magic angle $\alpha= 0.586$ and $\gamma_1=8$ with $\beta=1$ as $\int \Tr g_{0} = 1.0$ and $\int \Tr g_{1} = 2.98$, respectively. Importantly, the exact flatness of degenerate zero modes requires not only chiral approximation, magic angle, but also $\beta = 1$ which does not hold for realistic material. Instead, for realistic materials, $\beta$ is nearly zero. Nevertheless, one can ask what are the properties of the model with more realistic parameters. We notice at the first magic angle, in the more realistic limit $\beta = 0$, the band becomes dispersive as shown in Fig.~\ref{fig:2TBGs_band}(c).

The novelty of this section is to point out that, at the second magic angle, the model allows an isolated band that is well approximated by the $n = 1$ generalized LL even in the more realistic limit with $\beta=0$. In Fig.~\ref{fig:2TBGs_band}(d), we compute the band structure at the second magic angle $\alpha = 2.221$ with parameters $(\gamma_1,\beta)=(8,0)$. The bands near charge neutrality are almost flat and there is a finite gap from the remote bands. The survival of flat bands results from the fact that the deviation from the $\beta=1$ limit is suppressed by the large moir\'e coupling $\alpha$ compared to the case at the first magic angle.

Since the $|\mathcal{U}_1\rangle$ has a nonzero second component at finite $\gamma_1$, the two sublattice polarized flatbands can in principle be separated with the help of vertical displacement fields. However, it is important to notice that there is a trade off: the displacement field will enhance both the band gap and the bandwidth. In other words, the band has already become quite dispersive when the band gap exceeds the interaction strength upon adding a sufficiently large displacement field. At small displacement field, we find the flatband remains character of the first generalized LL. Its integrated trace and the standard derivation of Berry curvature are given as follows,
\begin{equation}
    \int \Tr g_{1} = 2.91, \quad \sigma_{\Omega} = 0.45.\label{geo_double_TBG}
\end{equation}

\begin{figure}
    \centering
    \includegraphics[width=1.0\linewidth]{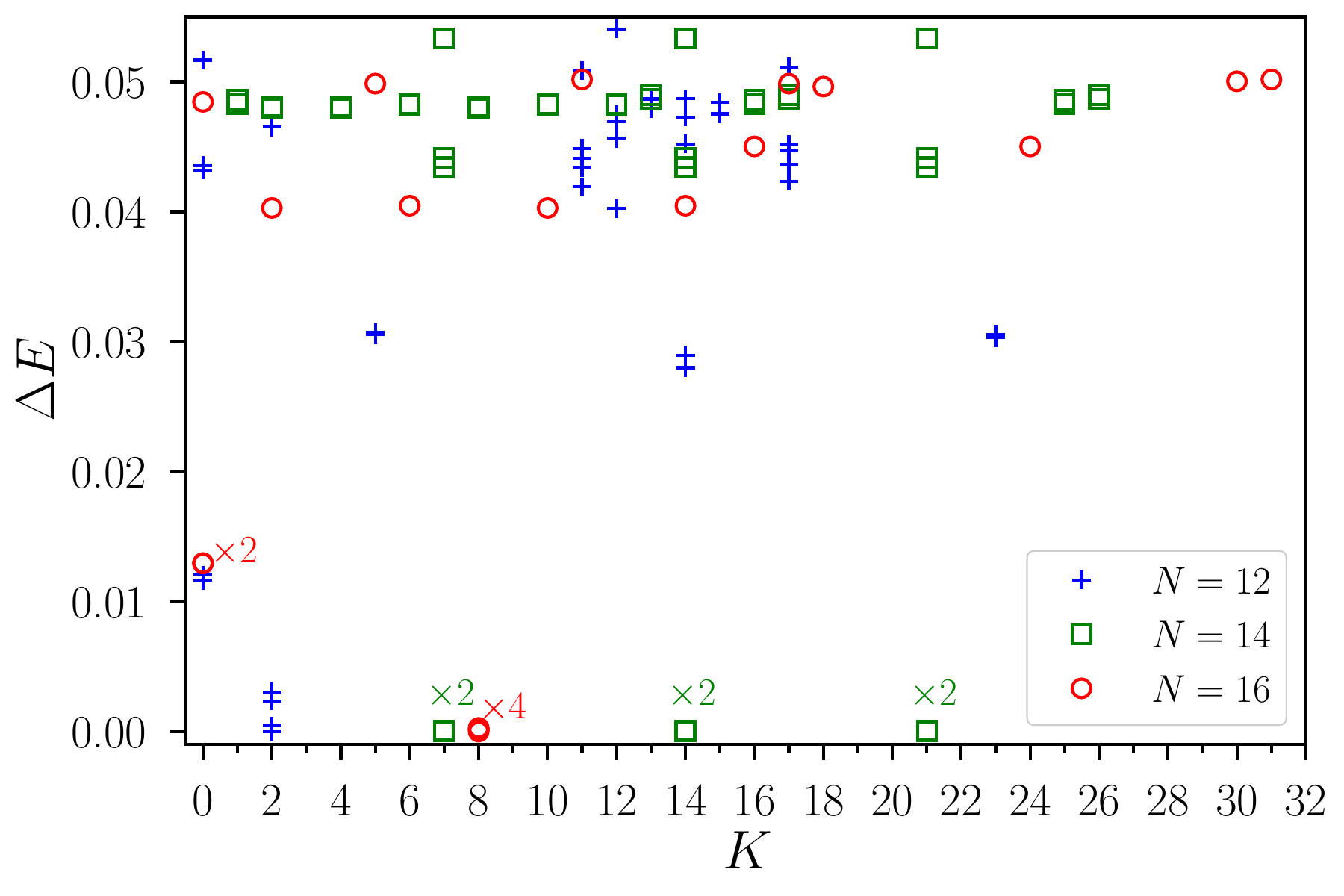} \caption{The energy spectrum of Coulomb interaction projected to the first conduction band of the double-TBG model at the second magic angle with $\beta = 0$ for $N_e = 12, 14$ and $16$ electrons. The multiplicity of levels is given when their energy difference is hardly visible. The ground-state degeneracy pattern is exactly what one would expect if the MR state is stabilized.} \label{fig:ManatoMR}
\end{figure}

On the other side, reducing the interaction strength by increasing the dielectric constant can make the band gap effectively larger. We hence proceed to neglect the band dispersion and assume valid single-band projection (which is not justified). We show numerical evidence of the non-Abelian MR state within this setting. In Fig.~\ref{fig:ManatoMR}, we provided the exact diagonalization spectrum for $N_e = 12, 14, 16$ electrons at half filling. For all particle numbers considered, a six-fold degeneracy in the right momentum sectors are observed, which is a hallmark of the MR phase.

To conclude, the double TBG model exhibits nearly flat bands of the first generalized LL type at zero magnetic field. The single-particle quantum geometric indicators fall within the range demonstrated in Fig.~\ref{fig:trgwindow}(c), and many-body calculations show evidence for the MR phase. We emphasize the purpose of studying this model is not to make connection to realistic experiments, but as a concrete example to show the consistency of the quantum geometric criterion we proposed. In below we summarize the unrealistic aspects of the model considered above. Firstly, although we have eliminated unphysical tunneling, the model still requires the chiral limit. The chiral limit requires the vanishing of the intra-sublattice moir\'e tunneling, which is unrealistic in real materials. Secondly, the model considered here requires the second magic angle which is near $0.25^\circ$. In reality, the relaxation effect is stronger at smaller angle where the strain and domain wall effect become important~\cite{Ahbay_TTG_22,Khalaf_PRB22}, which have not been taken into account in the current model. Importantly, also as mentioned above, it is difficult in this model to isolate the approximately $n=1$ generalized LL and meanwhile keep it flat. Therefore, the band projection, which is used to obtain the six-fold degeneracy, is hard to be justified although we argue reducing interaction strength can effectively enhance the band gap. Last but not least, the Hartree-Fock renormalization presumably will also strongly modify the single-particle dispersion and affect the stability of non-Abelian states. We leave a more detailed exploration of this model for future work.

\section{CONCLUSION, DISCUSSION AND OUTLOOK} \label{sec: conclusion}
In this work, we generalize the notion of ideal Chern bands, {\it i.e.} the generalized lowest LLs, to higher LL partners. We term them, including the standard ideal bands, as generalized LLs. The generalized LLs are systematically constructed, starting from the standard ideal bands, by applying ladder operators in momentum space and Gram-Schmidt orthogonalization. Generalized LLs resemble the standard LLs in many ways, while allowing for nonuniform quantum geometries. Single and multiple-band quantum geometries of generalized LLs are explicitly determined in a concise form, with the results summarized in Table~\ref{Table:Geometries}. Particularly notable is the quantized integrated trace value which determines the LL index. The generalized LLs are also understood as describing a distinguished unitary moving frame over the holomorphic curve on the space of quantum states determined by the ideal band.

As a complete orthonormal basis, the generalized LLs span the entire Hilbert space of Chern bands, including LL type and Bloch type Chern bands. We propose a variational way of quantifying the LL mimicry based on generalized LLs. For models consisting of mixing finite number of generalized LLs, a canonical algorithm exists which allows exact determination of the superposition weights. The $u_m - u_n$ model is supposed to capture a large portion of the single-particle Hilbert space of first LL type. Exact diagonalization within such model provides an arguably widest range of geometric quantities permitting non-Abelian MR phases. The geometric window is compared with properties of existing models and materials that host MR states, and the geometric window may serve as a guiding rule in future material designing in search of MR phase in moir\'e flatbands. Isolated first generalized LL can potentially exist in double moir\'e graphene at small twist angle. The model studied in this work improves upon previous graphene based examples by removing the necessity for further layer hopping, although more work is required to connect to the real materials. Moreover, geometric stability of the Read-Rezayi phases are briefly discussed.

We have shown the generalized LLs not only exhibit formal elegance but also have practical implications, for instance for the stability of non-Abelian fractionalized orders in moir\'e materials and beyond. We anticipate the generalized Landau level notion as a useful tool to decompose wavefunctions of generic Chern band for detailed analysis, in parallel with the concept of pseudo-potential as being a tool for analysing interactions. Therefore we expect the theory and implications provided in this work are potentially significant for future studies.

There are a couple of interesting future directions. First of all, finding out a practical numerical method to systematically decompose general Chern bands into the generalized LLs is an open question. This work demonstrates that for the finite-$N$ model, such algorithm exists. However, a general Chern band unavoidably involves an infinite number of generalized LLs and generalizing from finite to infinite $N$ is an important future direction. Moreover, it is interesting to systematically examine the double TBG model under realistic settings, including the competition between maximizing band gap and minimizing bandwidth upon applying an displacement field, effect of domain wall, strain and Hartree-Fock renormalization, and the possibility for MR states. Last but not least, this works focused on smooth modulation functions. Including ramification points in the description of the holomorphic curves, which physically correspond to momentum-space singularities, will enrich the recursion formulas provided in this work and alter the invariants. Their physical implications constitute promising future studies.
\section{ACKNOWLEDGEMENTS}
We acknowledge Andrei Bernevig, Junkai Dong, Liang Fu, Daniele Guerci, Guangyue Ji, Eslam Khalaf, Patrick Ledwith, Nicolas Regnault, Fengcheng Wu, Di Xiao, Fang Xie, Bo Yang and Junyi Zhang for useful discussions. We especially acknowledge Ashvin Vishwanath for illuminating discussions and feedback on the manuscript.

Z.~L. was supported by the National Natural Science Foundation of China (Grant No.~12374149 and 12350403). 
B.~M. thanks Harvard CMSA and the department of physics of Harvard University, for their support and hospitality during his visit. B.~M. acknowledges support from the Security and Quantum Information Group (SQIG) in Instituto de Telecomunica\c{c}\~{o}es, Lisbon. This work is funded by FCT (Funda\c{c}\~{a}o para a Ci\^{e}ncia e a Tecnologia) through national funds FCT I.P. and, when eligible, by COMPETE 2020 FEDER funds, under Award UIDB/50008/2020 and the Scientific Employment Stimulus --- Individual Call (CEEC Individual) --- 2022.05522.CEECIND/CP1716/CT0001, with DOI 10.54499/2022.05522.CEECIND/CP1716/CT0001. 
M.~F. acknowledges support from by JSPS KAKENHI grant no. JP23KJ0339 and JP24K16987, and the Center for the Advancement of Topological Semimetals (CATS), an Energy Frontier Research Center at the Ames National Laboratory. Work at the Ames National Laboratory is supported by the U.S. Department of Energy (DOE), Basic Energy Sciences (BES) and is operated for the U.S. DOE by Iowa State University under Contract No. DE-AC02-07CH11358. 
T.~O. acknowledges support from JSPS KAKENHI Grant No. JP24K00548, JST PRESTO Grant No. JPMJPR2353, JST CREST Grant No. JPMJCR19T1.

\bibliography{ref.bib}

\begin{thebibliography}{146}%
\makeatletter
\providecommand \@ifxundefined [1]{%
 \@ifx{#1\undefined}
}%
\providecommand \@ifnum [1]{%
 \ifnum #1\expandafter \@firstoftwo
 \else \expandafter \@secondoftwo
 \fi
}%
\providecommand \@ifx [1]{%
 \ifx #1\expandafter \@firstoftwo
 \else \expandafter \@secondoftwo
 \fi
}%
\providecommand \natexlab [1]{#1}%
\providecommand \enquote  [1]{``#1''}%
\providecommand \bibnamefont  [1]{#1}%
\providecommand \bibfnamefont [1]{#1}%
\providecommand \citenamefont [1]{#1}%
\providecommand \href@noop [0]{\@secondoftwo}%
\providecommand \href [0]{\begingroup \@sanitize@url \@href}%
\providecommand \@href[1]{\@@startlink{#1}\@@href}%
\providecommand \@@href[1]{\endgroup#1\@@endlink}%
\providecommand \@sanitize@url [0]{\catcode `\\12\catcode `\$12\catcode `\&12\catcode `\#12\catcode `\^12\catcode `\_12\catcode `\%12\relax}%
\providecommand \@@startlink[1]{}%
\providecommand \@@endlink[0]{}%
\providecommand \url  [0]{\begingroup\@sanitize@url \@url }%
\providecommand \@url [1]{\endgroup\@href {#1}{\urlprefix }}%
\providecommand \urlprefix  [0]{URL }%
\providecommand \Eprint [0]{\href }%
\providecommand \doibase [0]{https://doi.org/}%
\providecommand \selectlanguage [0]{\@gobble}%
\providecommand \bibinfo  [0]{\@secondoftwo}%
\providecommand \bibfield  [0]{\@secondoftwo}%
\providecommand \translation [1]{[#1]}%
\providecommand \BibitemOpen [0]{}%
\providecommand \bibitemStop [0]{}%
\providecommand \bibitemNoStop [0]{.\EOS\space}%
\providecommand \EOS [0]{\spacefactor3000\relax}%
\providecommand \BibitemShut  [1]{\csname bibitem#1\endcsname}%
\let\auto@bib@innerbib\@empty
\bibitem [{\citenamefont {Haldane}(2017)}]{HaldaneNobelLecture}%
  \BibitemOpen
  \bibfield  {author} {\bibinfo {author} {\bibfnamefont {F.~D.~M.}\ \bibnamefont {Haldane}},\ }\bibfield  {title} {\bibinfo {title} {Nobel lecture: Topological quantum matter},\ }\href {https://doi.org/10.1103/RevModPhys.89.040502} {\bibfield  {journal} {\bibinfo  {journal} {Rev. Mod. Phys.}\ }\textbf {\bibinfo {volume} {89}},\ \bibinfo {pages} {040502} (\bibinfo {year} {2017})}\BibitemShut {NoStop}%
\bibitem [{\citenamefont {Xiao}\ \emph {et~al.}(2010)\citenamefont {Xiao}, \citenamefont {Chang},\ and\ \citenamefont {Niu}}]{Di_Review}%
  \BibitemOpen
  \bibfield  {author} {\bibinfo {author} {\bibfnamefont {D.}~\bibnamefont {Xiao}}, \bibinfo {author} {\bibfnamefont {M.-C.}\ \bibnamefont {Chang}},\ and\ \bibinfo {author} {\bibfnamefont {Q.}~\bibnamefont {Niu}},\ }\bibfield  {title} {\bibinfo {title} {Berry phase effects on electronic properties},\ }\href {https://doi.org/10.1103/RevModPhys.82.1959} {\bibfield  {journal} {\bibinfo  {journal} {Rev. Mod. Phys.}\ }\textbf {\bibinfo {volume} {82}},\ \bibinfo {pages} {1959} (\bibinfo {year} {2010})}\BibitemShut {NoStop}%
\bibitem [{\citenamefont {T\"orm\"a}(2023)}]{Torma_Essay_23}%
  \BibitemOpen
  \bibfield  {author} {\bibinfo {author} {\bibfnamefont {P.}~\bibnamefont {T\"orm\"a}},\ }\bibfield  {title} {\bibinfo {title} {Essay: Where can quantum geometry lead us?},\ }\href {https://doi.org/10.1103/PhysRevLett.131.240001} {\bibfield  {journal} {\bibinfo  {journal} {Phys. Rev. Lett.}\ }\textbf {\bibinfo {volume} {131}},\ \bibinfo {pages} {240001} (\bibinfo {year} {2023})}\BibitemShut {NoStop}%
\bibitem [{\citenamefont {Niu}\ \emph {et~al.}(1985)\citenamefont {Niu}, \citenamefont {Thouless},\ and\ \citenamefont {Wu}}]{NiuThoulessWu}%
  \BibitemOpen
  \bibfield  {author} {\bibinfo {author} {\bibfnamefont {Q.}~\bibnamefont {Niu}}, \bibinfo {author} {\bibfnamefont {D.~J.}\ \bibnamefont {Thouless}},\ and\ \bibinfo {author} {\bibfnamefont {Y.-S.}\ \bibnamefont {Wu}},\ }\bibfield  {title} {\bibinfo {title} {Quantized hall conductance as a topological invariant},\ }\href {https://doi.org/10.1103/PhysRevB.31.3372} {\bibfield  {journal} {\bibinfo  {journal} {Phys. Rev. B}\ }\textbf {\bibinfo {volume} {31}},\ \bibinfo {pages} {3372} (\bibinfo {year} {1985})}\BibitemShut {NoStop}%
\bibitem [{\citenamefont {Sundaram}\ and\ \citenamefont {Niu}(1999)}]{Sundaram_Niu}%
  \BibitemOpen
  \bibfield  {author} {\bibinfo {author} {\bibfnamefont {G.}~\bibnamefont {Sundaram}}\ and\ \bibinfo {author} {\bibfnamefont {Q.}~\bibnamefont {Niu}},\ }\bibfield  {title} {\bibinfo {title} {Wave-packet dynamics in slowly perturbed crystals: Gradient corrections and berry-phase effects},\ }\href {https://doi.org/10.1103/PhysRevB.59.14915} {\bibfield  {journal} {\bibinfo  {journal} {Phys. Rev. B}\ }\textbf {\bibinfo {volume} {59}},\ \bibinfo {pages} {14915} (\bibinfo {year} {1999})}\BibitemShut {NoStop}%
\bibitem [{\citenamefont {Xiao}\ \emph {et~al.}(2005)\citenamefont {Xiao}, \citenamefont {Shi},\ and\ \citenamefont {Niu}}]{Di_BerryDOS}%
  \BibitemOpen
  \bibfield  {author} {\bibinfo {author} {\bibfnamefont {D.}~\bibnamefont {Xiao}}, \bibinfo {author} {\bibfnamefont {J.}~\bibnamefont {Shi}},\ and\ \bibinfo {author} {\bibfnamefont {Q.}~\bibnamefont {Niu}},\ }\bibfield  {title} {\bibinfo {title} {Berry phase correction to electron density of states in solids},\ }\href {https://doi.org/10.1103/PhysRevLett.95.137204} {\bibfield  {journal} {\bibinfo  {journal} {Phys. Rev. Lett.}\ }\textbf {\bibinfo {volume} {95}},\ \bibinfo {pages} {137204} (\bibinfo {year} {2005})}\BibitemShut {NoStop}%
\bibitem [{\citenamefont {Haldane}(2004)}]{haldaneanomaloushall}%
  \BibitemOpen
  \bibfield  {author} {\bibinfo {author} {\bibfnamefont {F.~D.~M.}\ \bibnamefont {Haldane}},\ }\bibfield  {title} {\bibinfo {title} {Berry curvature on the fermi surface: Anomalous hall effect as a topological fermi-liquid property},\ }\href {https://doi.org/10.1103/PhysRevLett.93.206602} {\bibfield  {journal} {\bibinfo  {journal} {Phys. Rev. Lett.}\ }\textbf {\bibinfo {volume} {93}},\ \bibinfo {pages} {206602} (\bibinfo {year} {2004})}\BibitemShut {NoStop}%
\bibitem [{\citenamefont {Sodemann}\ and\ \citenamefont {Fu}(2015)}]{LiangFu_Sodemann_BerryDipole15}%
  \BibitemOpen
  \bibfield  {author} {\bibinfo {author} {\bibfnamefont {I.}~\bibnamefont {Sodemann}}\ and\ \bibinfo {author} {\bibfnamefont {L.}~\bibnamefont {Fu}},\ }\bibfield  {title} {\bibinfo {title} {Quantum nonlinear hall effect induced by berry curvature dipole in time-reversal invariant materials},\ }\href {https://doi.org/10.1103/PhysRevLett.115.216806} {\bibfield  {journal} {\bibinfo  {journal} {Phys. Rev. Lett.}\ }\textbf {\bibinfo {volume} {115}},\ \bibinfo {pages} {216806} (\bibinfo {year} {2015})}\BibitemShut {NoStop}%
\bibitem [{\citenamefont {{Asteria}}\ \emph {et~al.}(2019)\citenamefont {{Asteria}}, \citenamefont {{Tran}}, \citenamefont {{Ozawa}}, \citenamefont {{Tarnowski}}, \citenamefont {{Rem}}, \citenamefont {{Fl{\"a}schner}}, \citenamefont {{Sengstock}}, \citenamefont {{Goldman}},\ and\ \citenamefont {{Weitenberg}}}]{GoldmanCrhistofNaturePhysics18}%
  \BibitemOpen
  \bibfield  {author} {\bibinfo {author} {\bibfnamefont {L.}~\bibnamefont {{Asteria}}}, \bibinfo {author} {\bibfnamefont {D.~T.}\ \bibnamefont {{Tran}}}, \bibinfo {author} {\bibfnamefont {T.}~\bibnamefont {{Ozawa}}}, \bibinfo {author} {\bibfnamefont {M.}~\bibnamefont {{Tarnowski}}}, \bibinfo {author} {\bibfnamefont {B.~S.}\ \bibnamefont {{Rem}}}, \bibinfo {author} {\bibfnamefont {N.}~\bibnamefont {{Fl{\"a}schner}}}, \bibinfo {author} {\bibfnamefont {K.}~\bibnamefont {{Sengstock}}}, \bibinfo {author} {\bibfnamefont {N.}~\bibnamefont {{Goldman}}},\ and\ \bibinfo {author} {\bibfnamefont {C.}~\bibnamefont {{Weitenberg}}},\ }\bibfield  {title} {\bibinfo {title} {{Measuring quantized circular dichroism in ultracold topological matter}},\ }\href {https://doi.org/10.1038/s41567-019-0417-8} {\bibfield  {journal} {\bibinfo  {journal} {Nature Physics}\ }\textbf {\bibinfo {volume} {15}},\ \bibinfo {pages} {449} (\bibinfo {year} {2019})},\ \Eprint {https://arxiv.org/abs/1805.11077} {arXiv:1805.11077 [cond-mat.quant-gas]}
  \BibitemShut {NoStop}%
\bibitem [{\citenamefont {Repellin}\ and\ \citenamefont {Goldman}(2019)}]{RepellinGoldman19}%
  \BibitemOpen
  \bibfield  {author} {\bibinfo {author} {\bibfnamefont {C.}~\bibnamefont {Repellin}}\ and\ \bibinfo {author} {\bibfnamefont {N.}~\bibnamefont {Goldman}},\ }\bibfield  {title} {\bibinfo {title} {Detecting fractional chern insulators through circular dichroism},\ }\href {https://doi.org/10.1103/PhysRevLett.122.166801} {\bibfield  {journal} {\bibinfo  {journal} {Phys. Rev. Lett.}\ }\textbf {\bibinfo {volume} {122}},\ \bibinfo {pages} {166801} (\bibinfo {year} {2019})}\BibitemShut {NoStop}%
\bibitem [{\citenamefont {{Goldman}}\ and\ \citenamefont {{Ozawa}}(2024)}]{GoldmanOzawa24}%
  \BibitemOpen
  \bibfield  {author} {\bibinfo {author} {\bibfnamefont {N.}~\bibnamefont {{Goldman}}}\ and\ \bibinfo {author} {\bibfnamefont {T.}~\bibnamefont {{Ozawa}}},\ }\bibfield  {title} {\bibinfo {title} {{Relating the Hall conductivity to the many-body Chern number using Fermi's Golden rule and Kramers-Kronig relations}},\ }\href {https://doi.org/10.48550/arXiv.2403.03340} {\bibfield  {journal} {\bibinfo  {journal} {arXiv e-prints}\ ,\ \bibinfo {eid} {arXiv:2403.03340}} (\bibinfo {year} {2024})},\ \Eprint {https://arxiv.org/abs/2403.03340} {arXiv:2403.03340 [cond-mat.mes-hall]} \BibitemShut {NoStop}%
\bibitem [{\citenamefont {Morimoto}\ and\ \citenamefont {Nagaosa}(2016)}]{Nagaosa_ShiftCurrent16}%
  \BibitemOpen
  \bibfield  {author} {\bibinfo {author} {\bibfnamefont {T.}~\bibnamefont {Morimoto}}\ and\ \bibinfo {author} {\bibfnamefont {N.}~\bibnamefont {Nagaosa}},\ }\bibfield  {title} {\bibinfo {title} {Topological nature of nonlinear optical effects in solids},\ }\href {https://doi.org/10.1126/sciadv.1501524} {\bibfield  {journal} {\bibinfo  {journal} {Science Advances}\ }\textbf {\bibinfo {volume} {2}},\ \bibinfo {pages} {e1501524} (\bibinfo {year} {2016})},\ \Eprint {https://arxiv.org/abs/https://www.science.org/doi/pdf/10.1126/sciadv.1501524} {https://www.science.org/doi/pdf/10.1126/sciadv.1501524} \BibitemShut {NoStop}%
\bibitem [{\citenamefont {Ozawa}\ \emph {et~al.}(2019)\citenamefont {Ozawa}, \citenamefont {Price}, \citenamefont {Amo}, \citenamefont {Goldman}, \citenamefont {Hafezi}, \citenamefont {Lu}, \citenamefont {Rechtsman}, \citenamefont {Schuster}, \citenamefont {Simon}, \citenamefont {Zilberberg},\ and\ \citenamefont {Carusotto}}]{RMP_TopologicalPhotonics19}%
  \BibitemOpen
  \bibfield  {author} {\bibinfo {author} {\bibfnamefont {T.}~\bibnamefont {Ozawa}}, \bibinfo {author} {\bibfnamefont {H.~M.}\ \bibnamefont {Price}}, \bibinfo {author} {\bibfnamefont {A.}~\bibnamefont {Amo}}, \bibinfo {author} {\bibfnamefont {N.}~\bibnamefont {Goldman}}, \bibinfo {author} {\bibfnamefont {M.}~\bibnamefont {Hafezi}}, \bibinfo {author} {\bibfnamefont {L.}~\bibnamefont {Lu}}, \bibinfo {author} {\bibfnamefont {M.~C.}\ \bibnamefont {Rechtsman}}, \bibinfo {author} {\bibfnamefont {D.}~\bibnamefont {Schuster}}, \bibinfo {author} {\bibfnamefont {J.}~\bibnamefont {Simon}}, \bibinfo {author} {\bibfnamefont {O.}~\bibnamefont {Zilberberg}},\ and\ \bibinfo {author} {\bibfnamefont {I.}~\bibnamefont {Carusotto}},\ }\bibfield  {title} {\bibinfo {title} {Topological photonics},\ }\href {https://doi.org/10.1103/RevModPhys.91.015006} {\bibfield  {journal} {\bibinfo  {journal} {Rev. Mod. Phys.}\ }\textbf {\bibinfo {volume} {91}},\ \bibinfo {pages} {015006} (\bibinfo {year} {2019})}\BibitemShut {NoStop}%
\bibitem [{\citenamefont {Ahn}\ \emph {et~al.}(2022)\citenamefont {Ahn}, \citenamefont {Guo}, \citenamefont {Nagaosa},\ and\ \citenamefont {Vishwanath}}]{Ahn:2022aa}%
  \BibitemOpen
  \bibfield  {author} {\bibinfo {author} {\bibfnamefont {J.}~\bibnamefont {Ahn}}, \bibinfo {author} {\bibfnamefont {G.-Y.}\ \bibnamefont {Guo}}, \bibinfo {author} {\bibfnamefont {N.}~\bibnamefont {Nagaosa}},\ and\ \bibinfo {author} {\bibfnamefont {A.}~\bibnamefont {Vishwanath}},\ }\bibfield  {title} {\bibinfo {title} {Riemannian geometry of resonant optical responses},\ }\href {https://doi.org/10.1038/s41567-021-01465-z} {\bibfield  {journal} {\bibinfo  {journal} {Nature Physics}\ }\textbf {\bibinfo {volume} {18}},\ \bibinfo {pages} {290} (\bibinfo {year} {2022})}\BibitemShut {NoStop}%
\bibitem [{\citenamefont {Peotta}\ and\ \citenamefont {T{\"o}rm{\"a}}(2015)}]{Peotta:2015aa}%
  \BibitemOpen
  \bibfield  {author} {\bibinfo {author} {\bibfnamefont {S.}~\bibnamefont {Peotta}}\ and\ \bibinfo {author} {\bibfnamefont {P.}~\bibnamefont {T{\"o}rm{\"a}}},\ }\bibfield  {title} {\bibinfo {title} {Superfluidity in topologically nontrivial flat bands},\ }\href {https://doi.org/10.1038/ncomms9944} {\bibfield  {journal} {\bibinfo  {journal} {Nature Communications}\ }\textbf {\bibinfo {volume} {6}},\ \bibinfo {pages} {8944} (\bibinfo {year} {2015})}\BibitemShut {NoStop}%
\bibitem [{\citenamefont {Hazra}\ \emph {et~al.}(2019)\citenamefont {Hazra}, \citenamefont {Verma},\ and\ \citenamefont {Randeria}}]{Randeria_19PRX}%
  \BibitemOpen
  \bibfield  {author} {\bibinfo {author} {\bibfnamefont {T.}~\bibnamefont {Hazra}}, \bibinfo {author} {\bibfnamefont {N.}~\bibnamefont {Verma}},\ and\ \bibinfo {author} {\bibfnamefont {M.}~\bibnamefont {Randeria}},\ }\bibfield  {title} {\bibinfo {title} {Bounds on the superconducting transition temperature: Applications to twisted bilayer graphene and cold atoms},\ }\href {https://doi.org/10.1103/PhysRevX.9.031049} {\bibfield  {journal} {\bibinfo  {journal} {Phys. Rev. X}\ }\textbf {\bibinfo {volume} {9}},\ \bibinfo {pages} {031049} (\bibinfo {year} {2019})}\BibitemShut {NoStop}%
\bibitem [{\citenamefont {Verma}\ \emph {et~al.}(2021)\citenamefont {Verma}, \citenamefont {Hazra},\ and\ \citenamefont {Randeria}}]{Randeria_PNAS_21}%
  \BibitemOpen
  \bibfield  {author} {\bibinfo {author} {\bibfnamefont {N.}~\bibnamefont {Verma}}, \bibinfo {author} {\bibfnamefont {T.}~\bibnamefont {Hazra}},\ and\ \bibinfo {author} {\bibfnamefont {M.}~\bibnamefont {Randeria}},\ }\bibfield  {title} {\bibinfo {title} {Optical spectral weight, phase stiffness, and $t_c$ bounds for trivial and topological flat band superconductors},\ }\href {https://doi.org/10.1073/pnas.2106744118} {\bibfield  {journal} {\bibinfo  {journal} {Proceedings of the National Academy of Sciences}\ }\textbf {\bibinfo {volume} {118}},\ \bibinfo {pages} {e2106744118} (\bibinfo {year} {2021})}\BibitemShut {NoStop}%
\bibitem [{\citenamefont {T{\"o}rm{\"a}}\ \emph {et~al.}(2022)\citenamefont {T{\"o}rm{\"a}}, \citenamefont {Peotta},\ and\ \citenamefont {Bernevig}}]{BAB_Torma_NatureReview22}%
  \BibitemOpen
  \bibfield  {author} {\bibinfo {author} {\bibfnamefont {P.}~\bibnamefont {T{\"o}rm{\"a}}}, \bibinfo {author} {\bibfnamefont {S.}~\bibnamefont {Peotta}},\ and\ \bibinfo {author} {\bibfnamefont {B.~A.}\ \bibnamefont {Bernevig}},\ }\bibfield  {title} {\bibinfo {title} {Superconductivity, superfluidity and quantum geometry in twisted multilayer systems},\ }\href {https://doi.org/10.1038/s42254-022-00466-y} {\bibfield  {journal} {\bibinfo  {journal} {Nature Reviews Physics}\ }\textbf {\bibinfo {volume} {4}},\ \bibinfo {pages} {528} (\bibinfo {year} {2022})}\BibitemShut {NoStop}%
\bibitem [{\citenamefont {Mao}\ and\ \citenamefont {Chowdhury}(2023)}]{Debanjan_23_PNAS}%
  \BibitemOpen
  \bibfield  {author} {\bibinfo {author} {\bibfnamefont {D.}~\bibnamefont {Mao}}\ and\ \bibinfo {author} {\bibfnamefont {D.}~\bibnamefont {Chowdhury}},\ }\bibfield  {title} {\bibinfo {title} {Diamagnetic response and phase stiffness for interacting isolated narrow bands},\ }\href {https://doi.org/10.1073/pnas.2217816120} {\bibfield  {journal} {\bibinfo  {journal} {Proceedings of the National Academy of Sciences}\ }\textbf {\bibinfo {volume} {120}},\ \bibinfo {pages} {e2217816120} (\bibinfo {year} {2023})},\ \Eprint {https://arxiv.org/abs/https://www.pnas.org/doi/pdf/10.1073/pnas.2217816120} {https://www.pnas.org/doi/pdf/10.1073/pnas.2217816120} \BibitemShut {NoStop}%
\bibitem [{\citenamefont {Onishi}\ and\ \citenamefont {Fu}(2024)}]{LiangFu_bounds_24_PRX}%
  \BibitemOpen
  \bibfield  {author} {\bibinfo {author} {\bibfnamefont {Y.}~\bibnamefont {Onishi}}\ and\ \bibinfo {author} {\bibfnamefont {L.}~\bibnamefont {Fu}},\ }\bibfield  {title} {\bibinfo {title} {Fundamental bound on topological gap},\ }\href {https://doi.org/10.1103/PhysRevX.14.011052} {\bibfield  {journal} {\bibinfo  {journal} {Phys. Rev. X}\ }\textbf {\bibinfo {volume} {14}},\ \bibinfo {pages} {011052} (\bibinfo {year} {2024})}\BibitemShut {NoStop}%
\bibitem [{\citenamefont {Haldane}(2011)}]{Haldanegeometry}%
  \BibitemOpen
  \bibfield  {author} {\bibinfo {author} {\bibfnamefont {F.~D.~M.}\ \bibnamefont {Haldane}},\ }\bibfield  {title} {\bibinfo {title} {Geometrical description of the fractional quantum hall effect},\ }\href {https://doi.org/10.1103/PhysRevLett.107.116801} {\bibfield  {journal} {\bibinfo  {journal} {Phys. Rev. Lett.}\ }\textbf {\bibinfo {volume} {107}},\ \bibinfo {pages} {116801} (\bibinfo {year} {2011})}\BibitemShut {NoStop}%
\bibitem [{\citenamefont {Nguyen}\ and\ \citenamefont {Son}(2021)}]{Son_PRR_21}%
  \BibitemOpen
  \bibfield  {author} {\bibinfo {author} {\bibfnamefont {D.~X.}\ \bibnamefont {Nguyen}}\ and\ \bibinfo {author} {\bibfnamefont {D.~T.}\ \bibnamefont {Son}},\ }\bibfield  {title} {\bibinfo {title} {Probing the spin structure of the fractional quantum hall magnetoroton with polarized raman scattering},\ }\href {https://doi.org/10.1103/PhysRevResearch.3.023040} {\bibfield  {journal} {\bibinfo  {journal} {Phys. Rev. Res.}\ }\textbf {\bibinfo {volume} {3}},\ \bibinfo {pages} {023040} (\bibinfo {year} {2021})}\BibitemShut {NoStop}%
\bibitem [{\citenamefont {Liang}\ \emph {et~al.}(2024)\citenamefont {Liang}, \citenamefont {Liu}, \citenamefont {Yang}, \citenamefont {Huang}, \citenamefont {Wurstbauer}, \citenamefont {Dean}, \citenamefont {West}, \citenamefont {Pfeiffer}, \citenamefont {Du},\ and\ \citenamefont {Pinczuk}}]{LingjieDu_Graviton_Nature_24}%
  \BibitemOpen
  \bibfield  {author} {\bibinfo {author} {\bibfnamefont {J.}~\bibnamefont {Liang}}, \bibinfo {author} {\bibfnamefont {Z.}~\bibnamefont {Liu}}, \bibinfo {author} {\bibfnamefont {Z.}~\bibnamefont {Yang}}, \bibinfo {author} {\bibfnamefont {Y.}~\bibnamefont {Huang}}, \bibinfo {author} {\bibfnamefont {U.}~\bibnamefont {Wurstbauer}}, \bibinfo {author} {\bibfnamefont {C.~R.}\ \bibnamefont {Dean}}, \bibinfo {author} {\bibfnamefont {K.~W.}\ \bibnamefont {West}}, \bibinfo {author} {\bibfnamefont {L.~N.}\ \bibnamefont {Pfeiffer}}, \bibinfo {author} {\bibfnamefont {L.}~\bibnamefont {Du}},\ and\ \bibinfo {author} {\bibfnamefont {A.}~\bibnamefont {Pinczuk}},\ }\bibfield  {title} {\bibinfo {title} {Evidence for chiral graviton modes in fractional quantum hall liquids},\ }\href {https://doi.org/10.1038/s41586-024-07201-w} {\bibfield  {journal} {\bibinfo  {journal} {Nature}\ }\textbf {\bibinfo {volume} {628}},\ \bibinfo {pages} {78} (\bibinfo {year} {2024})}\BibitemShut {NoStop}%
\bibitem [{\citenamefont {Roy}(2014)}]{RahulRoy14}%
  \BibitemOpen
  \bibfield  {author} {\bibinfo {author} {\bibfnamefont {R.}~\bibnamefont {Roy}},\ }\bibfield  {title} {\bibinfo {title} {Band geometry of fractional topological insulators},\ }\href {https://doi.org/10.1103/PhysRevB.90.165139} {\bibfield  {journal} {\bibinfo  {journal} {Phys. Rev. B}\ }\textbf {\bibinfo {volume} {90}},\ \bibinfo {pages} {165139} (\bibinfo {year} {2014})}\BibitemShut {NoStop}%
\bibitem [{\citenamefont {Jackson}\ \emph {et~al.}(2015)\citenamefont {Jackson}, \citenamefont {M{\"o}ller},\ and\ \citenamefont {Roy}}]{Jackson:2015aa}%
  \BibitemOpen
  \bibfield  {author} {\bibinfo {author} {\bibfnamefont {T.~S.}\ \bibnamefont {Jackson}}, \bibinfo {author} {\bibfnamefont {G.}~\bibnamefont {M{\"o}ller}},\ and\ \bibinfo {author} {\bibfnamefont {R.}~\bibnamefont {Roy}},\ }\bibfield  {title} {\bibinfo {title} {Geometric stability of topological lattice phases},\ }\href {https://doi.org/10.1038/ncomms9629} {\bibfield  {journal} {\bibinfo  {journal} {Nature Communications}\ }\textbf {\bibinfo {volume} {6}},\ \bibinfo {pages} {8629} (\bibinfo {year} {2015})}\BibitemShut {NoStop}%
\bibitem [{\citenamefont {Claassen}\ \emph {et~al.}(2015)\citenamefont {Claassen}, \citenamefont {Lee}, \citenamefont {Thomale}, \citenamefont {Qi},\ and\ \citenamefont {Devereaux}}]{Martin_PositionMomentumDuality}%
  \BibitemOpen
  \bibfield  {author} {\bibinfo {author} {\bibfnamefont {M.}~\bibnamefont {Claassen}}, \bibinfo {author} {\bibfnamefont {C.~H.}\ \bibnamefont {Lee}}, \bibinfo {author} {\bibfnamefont {R.}~\bibnamefont {Thomale}}, \bibinfo {author} {\bibfnamefont {X.-L.}\ \bibnamefont {Qi}},\ and\ \bibinfo {author} {\bibfnamefont {T.~P.}\ \bibnamefont {Devereaux}},\ }\bibfield  {title} {\bibinfo {title} {Position-momentum duality and fractional quantum hall effect in chern insulators},\ }\href {https://doi.org/10.1103/PhysRevLett.114.236802} {\bibfield  {journal} {\bibinfo  {journal} {Phys. Rev. Lett.}\ }\textbf {\bibinfo {volume} {114}},\ \bibinfo {pages} {236802} (\bibinfo {year} {2015})}\BibitemShut {NoStop}%
\bibitem [{\citenamefont {Lee}\ \emph {et~al.}(2015)\citenamefont {Lee}, \citenamefont {Papi\ifmmode~\acute{c}\else \'{c}\fi{}},\ and\ \citenamefont {Thomale}}]{ChinaHuaLeePRX15}%
  \BibitemOpen
  \bibfield  {author} {\bibinfo {author} {\bibfnamefont {C.~H.}\ \bibnamefont {Lee}}, \bibinfo {author} {\bibfnamefont {Z.}~\bibnamefont {Papi\ifmmode~\acute{c}\else \'{c}\fi{}}},\ and\ \bibinfo {author} {\bibfnamefont {R.}~\bibnamefont {Thomale}},\ }\bibfield  {title} {\bibinfo {title} {Geometric construction of quantum hall clustering hamiltonians},\ }\href {https://doi.org/10.1103/PhysRevX.5.041003} {\bibfield  {journal} {\bibinfo  {journal} {Phys. Rev. X}\ }\textbf {\bibinfo {volume} {5}},\ \bibinfo {pages} {041003} (\bibinfo {year} {2015})}\BibitemShut {NoStop}%
\bibitem [{\citenamefont {Ozawa}\ and\ \citenamefont {Mera}(2021)}]{kahlerband1}%
  \BibitemOpen
  \bibfield  {author} {\bibinfo {author} {\bibfnamefont {T.}~\bibnamefont {Ozawa}}\ and\ \bibinfo {author} {\bibfnamefont {B.}~\bibnamefont {Mera}},\ }\bibfield  {title} {\bibinfo {title} {Relations between topology and the quantum metric for chern insulators},\ }\href {https://doi.org/10.1103/PhysRevB.104.045103} {\bibfield  {journal} {\bibinfo  {journal} {Phys. Rev. B}\ }\textbf {\bibinfo {volume} {104}},\ \bibinfo {pages} {045103} (\bibinfo {year} {2021})}\BibitemShut {NoStop}%
\bibitem [{\citenamefont {Mera}\ and\ \citenamefont {Ozawa}(2021{\natexlab{a}})}]{kahlerband2}%
  \BibitemOpen
  \bibfield  {author} {\bibinfo {author} {\bibfnamefont {B.}~\bibnamefont {Mera}}\ and\ \bibinfo {author} {\bibfnamefont {T.}~\bibnamefont {Ozawa}},\ }\bibfield  {title} {\bibinfo {title} {K\"ahler geometry and chern insulators: Relations between topology and the quantum metric},\ }\href {https://doi.org/10.1103/PhysRevB.104.045104} {\bibfield  {journal} {\bibinfo  {journal} {Phys. Rev. B}\ }\textbf {\bibinfo {volume} {104}},\ \bibinfo {pages} {045104} (\bibinfo {year} {2021}{\natexlab{a}})}\BibitemShut {NoStop}%
\bibitem [{\citenamefont {Mera}\ and\ \citenamefont {Ozawa}(2021{\natexlab{b}})}]{kahlerband3}%
  \BibitemOpen
  \bibfield  {author} {\bibinfo {author} {\bibfnamefont {B.}~\bibnamefont {Mera}}\ and\ \bibinfo {author} {\bibfnamefont {T.}~\bibnamefont {Ozawa}},\ }\bibfield  {title} {\bibinfo {title} {Engineering geometrically flat chern bands with fubini-study k\"ahler structure},\ }\href {https://doi.org/10.1103/PhysRevB.104.115160} {\bibfield  {journal} {\bibinfo  {journal} {Phys. Rev. B}\ }\textbf {\bibinfo {volume} {104}},\ \bibinfo {pages} {115160} (\bibinfo {year} {2021}{\natexlab{b}})}\BibitemShut {NoStop}%
\bibitem [{\citenamefont {Regnault}\ and\ \citenamefont {Bernevig}(2011)}]{Regnault_Bernevig_CFI_PRX11}%
  \BibitemOpen
  \bibfield  {author} {\bibinfo {author} {\bibfnamefont {N.}~\bibnamefont {Regnault}}\ and\ \bibinfo {author} {\bibfnamefont {B.~A.}\ \bibnamefont {Bernevig}},\ }\bibfield  {title} {\bibinfo {title} {Fractional chern insulator},\ }\href {https://doi.org/10.1103/PhysRevX.1.021014} {\bibfield  {journal} {\bibinfo  {journal} {Phys. Rev. X}\ }\textbf {\bibinfo {volume} {1}},\ \bibinfo {pages} {021014} (\bibinfo {year} {2011})}\BibitemShut {NoStop}%
\bibitem [{\citenamefont {Neupert}\ \emph {et~al.}(2011)\citenamefont {Neupert}, \citenamefont {Santos}, \citenamefont {Chamon},\ and\ \citenamefont {Mudry}}]{Titus_FCI_11}%
  \BibitemOpen
  \bibfield  {author} {\bibinfo {author} {\bibfnamefont {T.}~\bibnamefont {Neupert}}, \bibinfo {author} {\bibfnamefont {L.}~\bibnamefont {Santos}}, \bibinfo {author} {\bibfnamefont {C.}~\bibnamefont {Chamon}},\ and\ \bibinfo {author} {\bibfnamefont {C.}~\bibnamefont {Mudry}},\ }\bibfield  {title} {\bibinfo {title} {Fractional quantum hall states at zero magnetic field},\ }\href {https://doi.org/10.1103/PhysRevLett.106.236804} {\bibfield  {journal} {\bibinfo  {journal} {Phys. Rev. Lett.}\ }\textbf {\bibinfo {volume} {106}},\ \bibinfo {pages} {236804} (\bibinfo {year} {2011})}\BibitemShut {NoStop}%
\bibitem [{\citenamefont {{Sheng}}\ \emph {et~al.}(2011)\citenamefont {{Sheng}}, \citenamefont {{Gu}}, \citenamefont {{Sun}},\ and\ \citenamefont {{Sheng}}}]{Sheng_FCI_11}%
  \BibitemOpen
  \bibfield  {author} {\bibinfo {author} {\bibfnamefont {D.~N.}\ \bibnamefont {{Sheng}}}, \bibinfo {author} {\bibfnamefont {Z.-C.}\ \bibnamefont {{Gu}}}, \bibinfo {author} {\bibfnamefont {K.}~\bibnamefont {{Sun}}},\ and\ \bibinfo {author} {\bibfnamefont {L.}~\bibnamefont {{Sheng}}},\ }\bibfield  {title} {\bibinfo {title} {{Fractional quantum Hall effect in the absence of Landau levels}},\ }\href {https://doi.org/10.1038/ncomms1380} {\bibfield  {journal} {\bibinfo  {journal} {Nature Communications}\ }\textbf {\bibinfo {volume} {2}},\ \bibinfo {eid} {389} (\bibinfo {year} {2011})},\ \Eprint {https://arxiv.org/abs/1102.2658} {arXiv:1102.2658 [cond-mat.str-el]} \BibitemShut {NoStop}%
\bibitem [{\citenamefont {BERGHOLTZ}\ and\ \citenamefont {LIU}(2013)}]{zhao_review}%
  \BibitemOpen
  \bibfield  {author} {\bibinfo {author} {\bibfnamefont {E.~J.}\ \bibnamefont {BERGHOLTZ}}\ and\ \bibinfo {author} {\bibfnamefont {Z.}~\bibnamefont {LIU}},\ }\bibfield  {title} {\bibinfo {title} {Topological flat band models and fractional chern insulators},\ }\href {https://doi.org/10.1142/S021797921330017X} {\bibfield  {journal} {\bibinfo  {journal} {International Journal of Modern Physics B}\ }\textbf {\bibinfo {volume} {27}},\ \bibinfo {pages} {1330017} (\bibinfo {year} {2013})},\ \Eprint {https://arxiv.org/abs/https://doi.org/10.1142/S021797921330017X} {https://doi.org/10.1142/S021797921330017X} \BibitemShut {NoStop}%
\bibitem [{\citenamefont {Ledwith}\ \emph {et~al.}(2020)\citenamefont {Ledwith}, \citenamefont {Tarnopolsky}, \citenamefont {Khalaf},\ and\ \citenamefont {Vishwanath}}]{Grisha_TBG2}%
  \BibitemOpen
  \bibfield  {author} {\bibinfo {author} {\bibfnamefont {P.~J.}\ \bibnamefont {Ledwith}}, \bibinfo {author} {\bibfnamefont {G.}~\bibnamefont {Tarnopolsky}}, \bibinfo {author} {\bibfnamefont {E.}~\bibnamefont {Khalaf}},\ and\ \bibinfo {author} {\bibfnamefont {A.}~\bibnamefont {Vishwanath}},\ }\bibfield  {title} {\bibinfo {title} {Fractional chern insulator states in twisted bilayer graphene: An analytical approach},\ }\href {https://doi.org/10.1103/PhysRevResearch.2.023237} {\bibfield  {journal} {\bibinfo  {journal} {Phys. Rev. Research}\ }\textbf {\bibinfo {volume} {2}},\ \bibinfo {pages} {023237} (\bibinfo {year} {2020})}\BibitemShut {NoStop}%
\bibitem [{\citenamefont {Wang}\ \emph {et~al.}(2021{\natexlab{a}})\citenamefont {Wang}, \citenamefont {Cano}, \citenamefont {Millis}, \citenamefont {Liu},\ and\ \citenamefont {Yang}}]{JieWang_exactlldescription}%
  \BibitemOpen
  \bibfield  {author} {\bibinfo {author} {\bibfnamefont {J.}~\bibnamefont {Wang}}, \bibinfo {author} {\bibfnamefont {J.}~\bibnamefont {Cano}}, \bibinfo {author} {\bibfnamefont {A.~J.}\ \bibnamefont {Millis}}, \bibinfo {author} {\bibfnamefont {Z.}~\bibnamefont {Liu}},\ and\ \bibinfo {author} {\bibfnamefont {B.}~\bibnamefont {Yang}},\ }\bibfield  {title} {\bibinfo {title} {Exact landau level description of geometry and interaction in a flatband},\ }\href {https://doi.org/10.1103/PhysRevLett.127.246403} {\bibfield  {journal} {\bibinfo  {journal} {Phys. Rev. Lett.}\ }\textbf {\bibinfo {volume} {127}},\ \bibinfo {pages} {246403} (\bibinfo {year} {2021}{\natexlab{a}})}\BibitemShut {NoStop}%
\bibitem [{\citenamefont {{Ledwith}}\ \emph {et~al.}(2022)\citenamefont {{Ledwith}}, \citenamefont {{Vishwanath}},\ and\ \citenamefont {{Parker}}}]{LedwithVishwanathParker22}%
  \BibitemOpen
  \bibfield  {author} {\bibinfo {author} {\bibfnamefont {P.~J.}\ \bibnamefont {{Ledwith}}}, \bibinfo {author} {\bibfnamefont {A.}~\bibnamefont {{Vishwanath}}},\ and\ \bibinfo {author} {\bibfnamefont {D.~E.}\ \bibnamefont {{Parker}}},\ }\bibfield  {title} {\bibinfo {title} {{Vortexability: A Unifying Criterion for Ideal Fractional Chern Insulators}},\ }\href@noop {} {\bibfield  {journal} {\bibinfo  {journal} {arXiv e-prints}\ ,\ \bibinfo {eid} {arXiv:2209.15023}} (\bibinfo {year} {2022})},\ \Eprint {https://arxiv.org/abs/2209.15023} {arXiv:2209.15023 [cond-mat.str-el]} \BibitemShut {NoStop}%
\bibitem [{\citenamefont {Wang}\ and\ \citenamefont {Liu}(2022)}]{Jie_hierarchyidealband}%
  \BibitemOpen
  \bibfield  {author} {\bibinfo {author} {\bibfnamefont {J.}~\bibnamefont {Wang}}\ and\ \bibinfo {author} {\bibfnamefont {Z.}~\bibnamefont {Liu}},\ }\bibfield  {title} {\bibinfo {title} {Hierarchy of ideal flatbands in chiral twisted multilayer graphene models},\ }\href {https://doi.org/10.1103/PhysRevLett.128.176403} {\bibfield  {journal} {\bibinfo  {journal} {Phys. Rev. Lett.}\ }\textbf {\bibinfo {volume} {128}},\ \bibinfo {pages} {176403} (\bibinfo {year} {2022})}\BibitemShut {NoStop}%
\bibitem [{\citenamefont {Halperin}\ \emph {et~al.}(1993)\citenamefont {Halperin}, \citenamefont {Lee},\ and\ \citenamefont {Read}}]{HalperinLeeRead}%
  \BibitemOpen
  \bibfield  {author} {\bibinfo {author} {\bibfnamefont {B.~I.}\ \bibnamefont {Halperin}}, \bibinfo {author} {\bibfnamefont {P.~A.}\ \bibnamefont {Lee}},\ and\ \bibinfo {author} {\bibfnamefont {N.}~\bibnamefont {Read}},\ }\bibfield  {title} {\bibinfo {title} {Theory of the half-filled landau level},\ }\href {https://doi.org/10.1103/PhysRevB.47.7312} {\bibfield  {journal} {\bibinfo  {journal} {Phys. Rev. B}\ }\textbf {\bibinfo {volume} {47}},\ \bibinfo {pages} {7312} (\bibinfo {year} {1993})}\BibitemShut {NoStop}%
\bibitem [{\citenamefont {Son}(2015)}]{Son}%
  \BibitemOpen
  \bibfield  {author} {\bibinfo {author} {\bibfnamefont {D.~T.}\ \bibnamefont {Son}},\ }\bibfield  {title} {\bibinfo {title} {Is the composite fermion a dirac particle?},\ }\href {https://doi.org/10.1103/PhysRevX.5.031027} {\bibfield  {journal} {\bibinfo  {journal} {Phys. Rev. X}\ }\textbf {\bibinfo {volume} {5}},\ \bibinfo {pages} {031027} (\bibinfo {year} {2015})}\BibitemShut {NoStop}%
\bibitem [{\citenamefont {Cai}\ \emph {et~al.}(2023)\citenamefont {Cai}, \citenamefont {Anderson}, \citenamefont {Wang}, \citenamefont {Zhang}, \citenamefont {Liu}, \citenamefont {Holtzmann}, \citenamefont {Zhang}, \citenamefont {Fan}, \citenamefont {Taniguchi}, \citenamefont {Watanabe}, \citenamefont {Ran}, \citenamefont {Cao}, \citenamefont {Fu}, \citenamefont {Xiao}, \citenamefont {Yao},\ and\ \citenamefont {Xu}}]{XiaodongXu_Signature23}%
  \BibitemOpen
  \bibfield  {author} {\bibinfo {author} {\bibfnamefont {J.}~\bibnamefont {Cai}}, \bibinfo {author} {\bibfnamefont {E.}~\bibnamefont {Anderson}}, \bibinfo {author} {\bibfnamefont {C.}~\bibnamefont {Wang}}, \bibinfo {author} {\bibfnamefont {X.}~\bibnamefont {Zhang}}, \bibinfo {author} {\bibfnamefont {X.}~\bibnamefont {Liu}}, \bibinfo {author} {\bibfnamefont {W.}~\bibnamefont {Holtzmann}}, \bibinfo {author} {\bibfnamefont {Y.}~\bibnamefont {Zhang}}, \bibinfo {author} {\bibfnamefont {F.}~\bibnamefont {Fan}}, \bibinfo {author} {\bibfnamefont {T.}~\bibnamefont {Taniguchi}}, \bibinfo {author} {\bibfnamefont {K.}~\bibnamefont {Watanabe}}, \bibinfo {author} {\bibfnamefont {Y.}~\bibnamefont {Ran}}, \bibinfo {author} {\bibfnamefont {T.}~\bibnamefont {Cao}}, \bibinfo {author} {\bibfnamefont {L.}~\bibnamefont {Fu}}, \bibinfo {author} {\bibfnamefont {D.}~\bibnamefont {Xiao}}, \bibinfo {author} {\bibfnamefont {W.}~\bibnamefont {Yao}},\ and\ \bibinfo {author} {\bibfnamefont {X.}~\bibnamefont {Xu}},\ }\bibfield  {title}
  {\bibinfo {title} {Signatures of fractional quantum anomalous hall states in twisted mote2},\ }\href {https://doi.org/10.1038/s41586-023-06289-w} {\bibfield  {journal} {\bibinfo  {journal} {Nature}\ }\textbf {\bibinfo {volume} {622}},\ \bibinfo {pages} {63} (\bibinfo {year} {2023})}\BibitemShut {NoStop}%
\bibitem [{\citenamefont {Zeng}\ \emph {et~al.}(2023)\citenamefont {Zeng}, \citenamefont {Xia}, \citenamefont {Kang}, \citenamefont {Zhu}, \citenamefont {Kn{\"u}ppel}, \citenamefont {Vaswani}, \citenamefont {Watanabe}, \citenamefont {Taniguchi}, \citenamefont {Mak},\ and\ \citenamefont {Shan}}]{KinfaiMak_Thermodynamic23}%
  \BibitemOpen
  \bibfield  {author} {\bibinfo {author} {\bibfnamefont {Y.}~\bibnamefont {Zeng}}, \bibinfo {author} {\bibfnamefont {Z.}~\bibnamefont {Xia}}, \bibinfo {author} {\bibfnamefont {K.}~\bibnamefont {Kang}}, \bibinfo {author} {\bibfnamefont {J.}~\bibnamefont {Zhu}}, \bibinfo {author} {\bibfnamefont {P.}~\bibnamefont {Kn{\"u}ppel}}, \bibinfo {author} {\bibfnamefont {C.}~\bibnamefont {Vaswani}}, \bibinfo {author} {\bibfnamefont {K.}~\bibnamefont {Watanabe}}, \bibinfo {author} {\bibfnamefont {T.}~\bibnamefont {Taniguchi}}, \bibinfo {author} {\bibfnamefont {K.~F.}\ \bibnamefont {Mak}},\ and\ \bibinfo {author} {\bibfnamefont {J.}~\bibnamefont {Shan}},\ }\bibfield  {title} {\bibinfo {title} {Thermodynamic evidence of fractional chern insulator in moir{\'e}mote2},\ }\href {https://doi.org/10.1038/s41586-023-06452-3} {\bibfield  {journal} {\bibinfo  {journal} {Nature}\ }\textbf {\bibinfo {volume} {622}},\ \bibinfo {pages} {69} (\bibinfo {year} {2023})}\BibitemShut {NoStop}%
\bibitem [{\citenamefont {Park}\ \emph {et~al.}(2023)\citenamefont {Park}, \citenamefont {Cai}, \citenamefont {Anderson}, \citenamefont {Zhang}, \citenamefont {Zhu}, \citenamefont {Liu}, \citenamefont {Wang}, \citenamefont {Holtzmann}, \citenamefont {Hu}, \citenamefont {Liu}, \citenamefont {Taniguchi}, \citenamefont {Watanabe}, \citenamefont {Chu}, \citenamefont {Cao}, \citenamefont {Fu}, \citenamefont {Yao}, \citenamefont {Chang}, \citenamefont {Cobden}, \citenamefont {Xiao},\ and\ \citenamefont {Xu}}]{XiaoDongXu_Observation23}%
  \BibitemOpen
  \bibfield  {author} {\bibinfo {author} {\bibfnamefont {H.}~\bibnamefont {Park}}, \bibinfo {author} {\bibfnamefont {J.}~\bibnamefont {Cai}}, \bibinfo {author} {\bibfnamefont {E.}~\bibnamefont {Anderson}}, \bibinfo {author} {\bibfnamefont {Y.}~\bibnamefont {Zhang}}, \bibinfo {author} {\bibfnamefont {J.}~\bibnamefont {Zhu}}, \bibinfo {author} {\bibfnamefont {X.}~\bibnamefont {Liu}}, \bibinfo {author} {\bibfnamefont {C.}~\bibnamefont {Wang}}, \bibinfo {author} {\bibfnamefont {W.}~\bibnamefont {Holtzmann}}, \bibinfo {author} {\bibfnamefont {C.}~\bibnamefont {Hu}}, \bibinfo {author} {\bibfnamefont {Z.}~\bibnamefont {Liu}}, \bibinfo {author} {\bibfnamefont {T.}~\bibnamefont {Taniguchi}}, \bibinfo {author} {\bibfnamefont {K.}~\bibnamefont {Watanabe}}, \bibinfo {author} {\bibfnamefont {J.-H.}\ \bibnamefont {Chu}}, \bibinfo {author} {\bibfnamefont {T.}~\bibnamefont {Cao}}, \bibinfo {author} {\bibfnamefont {L.}~\bibnamefont {Fu}}, \bibinfo {author} {\bibfnamefont {W.}~\bibnamefont {Yao}}, \bibinfo {author}
  {\bibfnamefont {C.-Z.}\ \bibnamefont {Chang}}, \bibinfo {author} {\bibfnamefont {D.}~\bibnamefont {Cobden}}, \bibinfo {author} {\bibfnamefont {D.}~\bibnamefont {Xiao}},\ and\ \bibinfo {author} {\bibfnamefont {X.}~\bibnamefont {Xu}},\ }\bibfield  {title} {\bibinfo {title} {Observation of fractionally quantized anomalous hall effect},\ }\href {https://doi.org/10.1038/s41586-023-06536-0} {\bibfield  {journal} {\bibinfo  {journal} {Nature}\ }\textbf {\bibinfo {volume} {622}},\ \bibinfo {pages} {74} (\bibinfo {year} {2023})}\BibitemShut {NoStop}%
\bibitem [{\citenamefont {Xu}\ \emph {et~al.}(2023)\citenamefont {Xu}, \citenamefont {Sun}, \citenamefont {Jia}, \citenamefont {Liu}, \citenamefont {Xu}, \citenamefont {Li}, \citenamefont {Gu}, \citenamefont {Watanabe}, \citenamefont {Taniguchi}, \citenamefont {Tong}, \citenamefont {Jia}, \citenamefont {Shi}, \citenamefont {Jiang}, \citenamefont {Zhang}, \citenamefont {Liu},\ and\ \citenamefont {Li}}]{TingxinLiPRX23}%
  \BibitemOpen
  \bibfield  {author} {\bibinfo {author} {\bibfnamefont {F.}~\bibnamefont {Xu}}, \bibinfo {author} {\bibfnamefont {Z.}~\bibnamefont {Sun}}, \bibinfo {author} {\bibfnamefont {T.}~\bibnamefont {Jia}}, \bibinfo {author} {\bibfnamefont {C.}~\bibnamefont {Liu}}, \bibinfo {author} {\bibfnamefont {C.}~\bibnamefont {Xu}}, \bibinfo {author} {\bibfnamefont {C.}~\bibnamefont {Li}}, \bibinfo {author} {\bibfnamefont {Y.}~\bibnamefont {Gu}}, \bibinfo {author} {\bibfnamefont {K.}~\bibnamefont {Watanabe}}, \bibinfo {author} {\bibfnamefont {T.}~\bibnamefont {Taniguchi}}, \bibinfo {author} {\bibfnamefont {B.}~\bibnamefont {Tong}}, \bibinfo {author} {\bibfnamefont {J.}~\bibnamefont {Jia}}, \bibinfo {author} {\bibfnamefont {Z.}~\bibnamefont {Shi}}, \bibinfo {author} {\bibfnamefont {S.}~\bibnamefont {Jiang}}, \bibinfo {author} {\bibfnamefont {Y.}~\bibnamefont {Zhang}}, \bibinfo {author} {\bibfnamefont {X.}~\bibnamefont {Liu}},\ and\ \bibinfo {author} {\bibfnamefont {T.}~\bibnamefont {Li}},\ }\bibfield  {title} {\bibinfo {title}
  {Observation of integer and fractional quantum anomalous hall effects in twisted bilayer ${\mathrm{mote}}_{2}$},\ }\href {https://doi.org/10.1103/PhysRevX.13.031037} {\bibfield  {journal} {\bibinfo  {journal} {Phys. Rev. X}\ }\textbf {\bibinfo {volume} {13}},\ \bibinfo {pages} {031037} (\bibinfo {year} {2023})}\BibitemShut {NoStop}%
\bibitem [{\citenamefont {Lu}\ \emph {et~al.}(2024)\citenamefont {Lu}, \citenamefont {Han}, \citenamefont {Yao}, \citenamefont {Reddy}, \citenamefont {Yang}, \citenamefont {Seo}, \citenamefont {Watanabe}, \citenamefont {Taniguchi}, \citenamefont {Fu},\ and\ \citenamefont {Ju}}]{LongJuFCI23}%
  \BibitemOpen
  \bibfield  {author} {\bibinfo {author} {\bibfnamefont {Z.}~\bibnamefont {Lu}}, \bibinfo {author} {\bibfnamefont {T.}~\bibnamefont {Han}}, \bibinfo {author} {\bibfnamefont {Y.}~\bibnamefont {Yao}}, \bibinfo {author} {\bibfnamefont {A.~P.}\ \bibnamefont {Reddy}}, \bibinfo {author} {\bibfnamefont {J.}~\bibnamefont {Yang}}, \bibinfo {author} {\bibfnamefont {J.}~\bibnamefont {Seo}}, \bibinfo {author} {\bibfnamefont {K.}~\bibnamefont {Watanabe}}, \bibinfo {author} {\bibfnamefont {T.}~\bibnamefont {Taniguchi}}, \bibinfo {author} {\bibfnamefont {L.}~\bibnamefont {Fu}},\ and\ \bibinfo {author} {\bibfnamefont {L.}~\bibnamefont {Ju}},\ }\bibfield  {title} {\bibinfo {title} {Fractional quantum anomalous hall effect in multilayer graphene},\ }\href {https://doi.org/10.1038/s41586-023-07010-7} {\bibfield  {journal} {\bibinfo  {journal} {Nature}\ }\textbf {\bibinfo {volume} {626}},\ \bibinfo {pages} {759} (\bibinfo {year} {2024})}\BibitemShut {NoStop}%
\bibitem [{\citenamefont {Moore}\ and\ \citenamefont {Read}(1991)}]{MoreReadState}%
  \BibitemOpen
  \bibfield  {author} {\bibinfo {author} {\bibfnamefont {G.}~\bibnamefont {Moore}}\ and\ \bibinfo {author} {\bibfnamefont {N.}~\bibnamefont {Read}},\ }\bibfield  {title} {\bibinfo {title} {Nonabelions in the fractional quantum hall effect},\ }\href {https://doi.org/https://doi.org/10.1016/0550-3213(91)90407-O} {\bibfield  {journal} {\bibinfo  {journal} {Nuclear Physics B}\ }\textbf {\bibinfo {volume} {360}},\ \bibinfo {pages} {362} (\bibinfo {year} {1991})}\BibitemShut {NoStop}%
\bibitem [{\citenamefont {Read}\ and\ \citenamefont {Rezayi}(1999)}]{Read_Rezayi}%
  \BibitemOpen
  \bibfield  {author} {\bibinfo {author} {\bibfnamefont {N.}~\bibnamefont {Read}}\ and\ \bibinfo {author} {\bibfnamefont {E.}~\bibnamefont {Rezayi}},\ }\bibfield  {title} {\bibinfo {title} {Beyond paired quantum hall states: Parafermions and incompressible states in the first excited landau level},\ }\href {https://doi.org/10.1103/PhysRevB.59.8084} {\bibfield  {journal} {\bibinfo  {journal} {Phys. Rev. B}\ }\textbf {\bibinfo {volume} {59}},\ \bibinfo {pages} {8084} (\bibinfo {year} {1999})}\BibitemShut {NoStop}%
\bibitem [{\citenamefont {{Reddy}}\ \emph {et~al.}(2024)\citenamefont {{Reddy}}, \citenamefont {{Paul}}, \citenamefont {{Abouelkomsan}},\ and\ \citenamefont {{Fu}}}]{LiangFuNonabelian24}%
  \BibitemOpen
  \bibfield  {author} {\bibinfo {author} {\bibfnamefont {A.~P.}\ \bibnamefont {{Reddy}}}, \bibinfo {author} {\bibfnamefont {N.}~\bibnamefont {{Paul}}}, \bibinfo {author} {\bibfnamefont {A.}~\bibnamefont {{Abouelkomsan}}},\ and\ \bibinfo {author} {\bibfnamefont {L.}~\bibnamefont {{Fu}}},\ }\bibfield  {title} {\bibinfo {title} {{Non-Abelian fractionalization in topological minibands}},\ }\href {https://doi.org/10.48550/arXiv.2403.00059} {\bibfield  {journal} {\bibinfo  {journal} {arXiv e-prints}\ ,\ \bibinfo {eid} {arXiv:2403.00059}} (\bibinfo {year} {2024})},\ \Eprint {https://arxiv.org/abs/2403.00059} {arXiv:2403.00059 [cond-mat.mes-hall]} \BibitemShut {NoStop}%
\bibitem [{\citenamefont {{Xu}}\ \emph {et~al.}(2024)\citenamefont {{Xu}}, \citenamefont {{Mao}}, \citenamefont {{Zeng}},\ and\ \citenamefont {{Zhang}}}]{YangZhangNonabelian24}%
  \BibitemOpen
  \bibfield  {author} {\bibinfo {author} {\bibfnamefont {C.}~\bibnamefont {{Xu}}}, \bibinfo {author} {\bibfnamefont {N.}~\bibnamefont {{Mao}}}, \bibinfo {author} {\bibfnamefont {T.}~\bibnamefont {{Zeng}}},\ and\ \bibinfo {author} {\bibfnamefont {Y.}~\bibnamefont {{Zhang}}},\ }\bibfield  {title} {\bibinfo {title} {{Multiple Chern bands in twisted MoTe$_2$ and possible non-Abelian states}},\ }\href {https://doi.org/10.48550/arXiv.2403.17003} {\bibfield  {journal} {\bibinfo  {journal} {arXiv e-prints}\ ,\ \bibinfo {eid} {arXiv:2403.17003}} (\bibinfo {year} {2024})},\ \Eprint {https://arxiv.org/abs/2403.17003} {arXiv:2403.17003 [cond-mat.str-el]} \BibitemShut {NoStop}%
\bibitem [{\citenamefont {{Ahn}}\ \emph {et~al.}(2024)\citenamefont {{Ahn}}, \citenamefont {{Lee}}, \citenamefont {{Yananose}}, \citenamefont {{Kim}},\ and\ \citenamefont {{Cho}}}]{ChoNonabelian24}%
  \BibitemOpen
  \bibfield  {author} {\bibinfo {author} {\bibfnamefont {C.-E.}\ \bibnamefont {{Ahn}}}, \bibinfo {author} {\bibfnamefont {W.}~\bibnamefont {{Lee}}}, \bibinfo {author} {\bibfnamefont {K.}~\bibnamefont {{Yananose}}}, \bibinfo {author} {\bibfnamefont {Y.}~\bibnamefont {{Kim}}},\ and\ \bibinfo {author} {\bibfnamefont {G.~Y.}\ \bibnamefont {{Cho}}},\ }\bibfield  {title} {\bibinfo {title} {{First Landau Level Physics in Second Moir{\'e} Band of $2.1^\circ$ Twisted Bilayer MoTe${}_2$}},\ }\href {https://doi.org/10.48550/arXiv.2403.19155} {\bibfield  {journal} {\bibinfo  {journal} {arXiv e-prints}\ ,\ \bibinfo {eid} {arXiv:2403.19155}} (\bibinfo {year} {2024})},\ \Eprint {https://arxiv.org/abs/2403.19155} {arXiv:2403.19155 [cond-mat.str-el]} \BibitemShut {NoStop}%
\bibitem [{\citenamefont {{Wang}}\ \emph {et~al.}(2024)\citenamefont {{Wang}}, \citenamefont {{Zhang}}, \citenamefont {{Liu}}, \citenamefont {{Wang}}, \citenamefont {{Cao}},\ and\ \citenamefont {{Xiao}}}]{DiXiaoNonabelian24}%
  \BibitemOpen
  \bibfield  {author} {\bibinfo {author} {\bibfnamefont {C.}~\bibnamefont {{Wang}}}, \bibinfo {author} {\bibfnamefont {X.-W.}\ \bibnamefont {{Zhang}}}, \bibinfo {author} {\bibfnamefont {X.}~\bibnamefont {{Liu}}}, \bibinfo {author} {\bibfnamefont {J.}~\bibnamefont {{Wang}}}, \bibinfo {author} {\bibfnamefont {T.}~\bibnamefont {{Cao}}},\ and\ \bibinfo {author} {\bibfnamefont {D.}~\bibnamefont {{Xiao}}},\ }\bibfield  {title} {\bibinfo {title} {{Higher Landau-Level Analogues and Signatures of Non-Abelian States in Twisted Bilayer MoTe$_2$}},\ }\href {https://doi.org/10.48550/arXiv.2404.05697} {\bibfield  {journal} {\bibinfo  {journal} {arXiv e-prints}\ ,\ \bibinfo {eid} {arXiv:2404.05697}} (\bibinfo {year} {2024})},\ \Eprint {https://arxiv.org/abs/2404.05697} {arXiv:2404.05697 [cond-mat.str-el]} \BibitemShut {NoStop}%
\bibitem [{\citenamefont {{Fujimoto}}\ \emph {et~al.}(2024)\citenamefont {{Fujimoto}}, \citenamefont {{Parker}}, \citenamefont {{Dong}}, \citenamefont {{Khalaf}}, \citenamefont {{Vishwanath}},\ and\ \citenamefont {{Ledwith}}}]{Fujimoto24}%
  \BibitemOpen
  \bibfield  {author} {\bibinfo {author} {\bibfnamefont {M.}~\bibnamefont {{Fujimoto}}}, \bibinfo {author} {\bibfnamefont {D.~E.}\ \bibnamefont {{Parker}}}, \bibinfo {author} {\bibfnamefont {J.}~\bibnamefont {{Dong}}}, \bibinfo {author} {\bibfnamefont {E.}~\bibnamefont {{Khalaf}}}, \bibinfo {author} {\bibfnamefont {A.}~\bibnamefont {{Vishwanath}}},\ and\ \bibinfo {author} {\bibfnamefont {P.}~\bibnamefont {{Ledwith}}},\ }\bibfield  {title} {\bibinfo {title} {{Higher vortexability: zero field realization of higher Landau levels}},\ }\href {https://doi.org/10.48550/arXiv.2403.00856} {\bibfield  {journal} {\bibinfo  {journal} {arXiv e-prints}\ ,\ \bibinfo {eid} {arXiv:2403.00856}} (\bibinfo {year} {2024})},\ \Eprint {https://arxiv.org/abs/2403.00856} {arXiv:2403.00856 [cond-mat.mes-hall]} \BibitemShut {NoStop}%
\bibitem [{\citenamefont {{Liu}}\ \emph {et~al.}(2024)\citenamefont {{Liu}}, \citenamefont {{Liu}},\ and\ \citenamefont {{Bergholtz}}}]{Emil_2405}%
  \BibitemOpen
  \bibfield  {author} {\bibinfo {author} {\bibfnamefont {H.}~\bibnamefont {{Liu}}}, \bibinfo {author} {\bibfnamefont {Z.}~\bibnamefont {{Liu}}},\ and\ \bibinfo {author} {\bibfnamefont {E.~J.}\ \bibnamefont {{Bergholtz}}},\ }\bibfield  {title} {\bibinfo {title} {{Non-Abelian Fractional Chern Insulators and Competing States in Flat Moir{\'e} Bands}},\ }\href {https://doi.org/10.48550/arXiv.2405.08887} {\bibfield  {journal} {\bibinfo  {journal} {arXiv e-prints}\ ,\ \bibinfo {eid} {arXiv:2405.08887}} (\bibinfo {year} {2024})},\ \Eprint {https://arxiv.org/abs/2405.08887} {arXiv:2405.08887 [cond-mat.str-el]} \BibitemShut {NoStop}%
\bibitem [{\citenamefont {Griffiths}(1974)}]{griffiths:74}%
  \BibitemOpen
  \bibfield  {author} {\bibinfo {author} {\bibfnamefont {P.}~\bibnamefont {Griffiths}},\ }\bibfield  {title} {\bibinfo {title} {{On Cartan's method of Lie groups and moving frames as applied to uniqueness and existence questions in differential geometry}},\ }\href {https://doi.org/10.1215/S0012-7094-74-04180-5} {\bibfield  {journal} {\bibinfo  {journal} {Duke Mathematical Journal}\ }\textbf {\bibinfo {volume} {41}},\ \bibinfo {pages} {775 } (\bibinfo {year} {1974})}\BibitemShut {NoStop}%
\bibitem [{\citenamefont {Griffiths}\ and\ \citenamefont {Harris}(2014)}]{griffiths:harris:14}%
  \BibitemOpen
  \bibfield  {author} {\bibinfo {author} {\bibfnamefont {P.}~\bibnamefont {Griffiths}}\ and\ \bibinfo {author} {\bibfnamefont {J.}~\bibnamefont {Harris}},\ }\href@noop {} {\emph {\bibinfo {title} {{Principles of algebraic geometry}}}}\ (\bibinfo  {publisher} {John Wiley \& Sons},\ \bibinfo {year} {2014})\BibitemShut {NoStop}%
\bibitem [{\citenamefont {Wang}\ \emph {et~al.}(2023)\citenamefont {Wang}, \citenamefont {Klevtsov},\ and\ \citenamefont {Liu}}]{Jie_Origin22}%
  \BibitemOpen
  \bibfield  {author} {\bibinfo {author} {\bibfnamefont {J.}~\bibnamefont {Wang}}, \bibinfo {author} {\bibfnamefont {S.}~\bibnamefont {Klevtsov}},\ and\ \bibinfo {author} {\bibfnamefont {Z.}~\bibnamefont {Liu}},\ }\bibfield  {title} {\bibinfo {title} {Origin of model fractional chern insulators in all topological ideal flatbands: Explicit color-entangled wave function and exact density algebra},\ }\href {https://doi.org/10.1103/PhysRevResearch.5.023167} {\bibfield  {journal} {\bibinfo  {journal} {Phys. Rev. Res.}\ }\textbf {\bibinfo {volume} {5}},\ \bibinfo {pages} {023167} (\bibinfo {year} {2023})}\BibitemShut {NoStop}%
\bibitem [{\citenamefont {Haldane}(2018{\natexlab{a}})}]{haldanemodularinv}%
  \BibitemOpen
  \bibfield  {author} {\bibinfo {author} {\bibfnamefont {F.~D.~M.}\ \bibnamefont {Haldane}},\ }\bibfield  {title} {\bibinfo {title} {A modular-invariant modified weierstrass sigma-function as a building block for lowest-landau-level wavefunctions on the torus},\ }\href {https://doi.org/10.1063/1.5042618} {\bibfield  {journal} {\bibinfo  {journal} {Journal of Mathematical Physics}\ }\textbf {\bibinfo {volume} {59}},\ \bibinfo {pages} {071901} (\bibinfo {year} {2018}{\natexlab{a}})},\ \Eprint {https://arxiv.org/abs/https://doi.org/10.1063/1.5042618} {https://doi.org/10.1063/1.5042618} \BibitemShut {NoStop}%
\bibitem [{\citenamefont {Haldane}(2018{\natexlab{b}})}]{haldaneholomorphic}%
  \BibitemOpen
  \bibfield  {author} {\bibinfo {author} {\bibfnamefont {F.~D.~M.}\ \bibnamefont {Haldane}},\ }\bibfield  {title} {\bibinfo {title} {The origin of holomorphic states in landau levels from non-commutative geometry and a new formula for their overlaps on the torus},\ }\href {https://doi.org/10.1063/1.5046122} {\bibfield  {journal} {\bibinfo  {journal} {Journal of Mathematical Physics}\ }\textbf {\bibinfo {volume} {59}},\ \bibinfo {pages} {081901} (\bibinfo {year} {2018}{\natexlab{b}})},\ \Eprint {https://arxiv.org/abs/https://doi.org/10.1063/1.5046122} {https://doi.org/10.1063/1.5046122} \BibitemShut {NoStop}%
\bibitem [{\citenamefont {Wang}\ \emph {et~al.}(2019)\citenamefont {Wang}, \citenamefont {Geraedts}, \citenamefont {Rezayi},\ and\ \citenamefont {Haldane}}]{Jie_MonteCarlo}%
  \BibitemOpen
  \bibfield  {author} {\bibinfo {author} {\bibfnamefont {J.}~\bibnamefont {Wang}}, \bibinfo {author} {\bibfnamefont {S.~D.}\ \bibnamefont {Geraedts}}, \bibinfo {author} {\bibfnamefont {E.~H.}\ \bibnamefont {Rezayi}},\ and\ \bibinfo {author} {\bibfnamefont {F.~D.~M.}\ \bibnamefont {Haldane}},\ }\bibfield  {title} {\bibinfo {title} {Lattice monte carlo for quantum hall states on a torus},\ }\href {https://doi.org/10.1103/PhysRevB.99.125123} {\bibfield  {journal} {\bibinfo  {journal} {Phys. Rev. B}\ }\textbf {\bibinfo {volume} {99}},\ \bibinfo {pages} {125123} (\bibinfo {year} {2019})}\BibitemShut {NoStop}%
\bibitem [{\citenamefont {Wang}\ \emph {et~al.}(2021{\natexlab{b}})\citenamefont {Wang}, \citenamefont {Zheng}, \citenamefont {Millis},\ and\ \citenamefont {Cano}}]{JieWang_NodalStructure}%
  \BibitemOpen
  \bibfield  {author} {\bibinfo {author} {\bibfnamefont {J.}~\bibnamefont {Wang}}, \bibinfo {author} {\bibfnamefont {Y.}~\bibnamefont {Zheng}}, \bibinfo {author} {\bibfnamefont {A.~J.}\ \bibnamefont {Millis}},\ and\ \bibinfo {author} {\bibfnamefont {J.}~\bibnamefont {Cano}},\ }\bibfield  {title} {\bibinfo {title} {Chiral approximation to twisted bilayer graphene: Exact intravalley inversion symmetry, nodal structure, and implications for higher magic angles},\ }\href {https://doi.org/10.1103/PhysRevResearch.3.023155} {\bibfield  {journal} {\bibinfo  {journal} {Phys. Rev. Research}\ }\textbf {\bibinfo {volume} {3}},\ \bibinfo {pages} {023155} (\bibinfo {year} {2021}{\natexlab{b}})}\BibitemShut {NoStop}%
\bibitem [{\citenamefont {Wang}(2019)}]{Jie_Dirac}%
  \BibitemOpen
  \bibfield  {author} {\bibinfo {author} {\bibfnamefont {J.}~\bibnamefont {Wang}},\ }\bibfield  {title} {\bibinfo {title} {Dirac fermion hierarchy of composite fermi liquids},\ }\href {https://doi.org/10.1103/PhysRevLett.122.257203} {\bibfield  {journal} {\bibinfo  {journal} {Phys. Rev. Lett.}\ }\textbf {\bibinfo {volume} {122}},\ \bibinfo {pages} {257203} (\bibinfo {year} {2019})}\BibitemShut {NoStop}%
\bibitem [{\citenamefont {Wu}\ \emph {et~al.}(2013)\citenamefont {Wu}, \citenamefont {Regnault},\ and\ \citenamefont {Bernevig}}]{YangleWu_ColorEntanglement13}%
  \BibitemOpen
  \bibfield  {author} {\bibinfo {author} {\bibfnamefont {Y.-L.}\ \bibnamefont {Wu}}, \bibinfo {author} {\bibfnamefont {N.}~\bibnamefont {Regnault}},\ and\ \bibinfo {author} {\bibfnamefont {B.~A.}\ \bibnamefont {Bernevig}},\ }\bibfield  {title} {\bibinfo {title} {Bloch model wave functions and pseudopotentials for all fractional chern insulators},\ }\href {https://doi.org/10.1103/PhysRevLett.110.106802} {\bibfield  {journal} {\bibinfo  {journal} {Phys. Rev. Lett.}\ }\textbf {\bibinfo {volume} {110}},\ \bibinfo {pages} {106802} (\bibinfo {year} {2013})}\BibitemShut {NoStop}%
\bibitem [{\citenamefont {Wu}\ \emph {et~al.}(2014)\citenamefont {Wu}, \citenamefont {Regnault},\ and\ \citenamefont {Bernevig}}]{Yangle_haldanestatistics}%
  \BibitemOpen
  \bibfield  {author} {\bibinfo {author} {\bibfnamefont {Y.-L.}\ \bibnamefont {Wu}}, \bibinfo {author} {\bibfnamefont {N.}~\bibnamefont {Regnault}},\ and\ \bibinfo {author} {\bibfnamefont {B.~A.}\ \bibnamefont {Bernevig}},\ }\bibfield  {title} {\bibinfo {title} {Haldane statistics for fractional chern insulators with an arbitrary chern number},\ }\href {https://doi.org/10.1103/PhysRevB.89.155113} {\bibfield  {journal} {\bibinfo  {journal} {Phys. Rev. B}\ }\textbf {\bibinfo {volume} {89}},\ \bibinfo {pages} {155113} (\bibinfo {year} {2014})}\BibitemShut {NoStop}%
\bibitem [{\citenamefont {{Mera}}\ and\ \citenamefont {{Ozawa}}(2023)}]{Mera_uniqueness_23}%
  \BibitemOpen
  \bibfield  {author} {\bibinfo {author} {\bibfnamefont {B.}~\bibnamefont {{Mera}}}\ and\ \bibinfo {author} {\bibfnamefont {T.}~\bibnamefont {{Ozawa}}},\ }\bibfield  {title} {\bibinfo {title} {{Uniqueness of Landau levels and their analogs with higher Chern numbers}},\ }\href {https://doi.org/10.48550/arXiv.2304.00866} {\bibfield  {journal} {\bibinfo  {journal} {arXiv e-prints}\ ,\ \bibinfo {eid} {arXiv:2304.00866}} (\bibinfo {year} {2023})},\ \Eprint {https://arxiv.org/abs/2304.00866} {arXiv:2304.00866 [cond-mat.mes-hall]} \BibitemShut {NoStop}%
\bibitem [{\citenamefont {Yang}\ \emph {et~al.}(2012)\citenamefont {Yang}, \citenamefont {Papi\ifmmode~\acute{c}\else \'{c}\fi{}}, \citenamefont {Rezayi}, \citenamefont {Bhatt},\ and\ \citenamefont {Haldane}}]{Bo_BandMassAnisotropy_12}%
  \BibitemOpen
  \bibfield  {author} {\bibinfo {author} {\bibfnamefont {B.}~\bibnamefont {Yang}}, \bibinfo {author} {\bibfnamefont {Z.}~\bibnamefont {Papi\ifmmode~\acute{c}\else \'{c}\fi{}}}, \bibinfo {author} {\bibfnamefont {E.~H.}\ \bibnamefont {Rezayi}}, \bibinfo {author} {\bibfnamefont {R.~N.}\ \bibnamefont {Bhatt}},\ and\ \bibinfo {author} {\bibfnamefont {F.~D.~M.}\ \bibnamefont {Haldane}},\ }\bibfield  {title} {\bibinfo {title} {Band mass anisotropy and the intrinsic metric of fractional quantum hall systems},\ }\href {https://doi.org/10.1103/PhysRevB.85.165318} {\bibfield  {journal} {\bibinfo  {journal} {Phys. Rev. B}\ }\textbf {\bibinfo {volume} {85}},\ \bibinfo {pages} {165318} (\bibinfo {year} {2012})}\BibitemShut {NoStop}%
\bibitem [{\citenamefont {Hasdemir}\ \emph {et~al.}(2015)\citenamefont {Hasdemir}, \citenamefont {Liu}, \citenamefont {Deng}, \citenamefont {Shayegan}, \citenamefont {Pfeiffer}, \citenamefont {West}, \citenamefont {Baldwin},\ and\ \citenamefont {Winkler}}]{Shayegen_PRB_15}%
  \BibitemOpen
  \bibfield  {author} {\bibinfo {author} {\bibfnamefont {S.}~\bibnamefont {Hasdemir}}, \bibinfo {author} {\bibfnamefont {Y.}~\bibnamefont {Liu}}, \bibinfo {author} {\bibfnamefont {H.}~\bibnamefont {Deng}}, \bibinfo {author} {\bibfnamefont {M.}~\bibnamefont {Shayegan}}, \bibinfo {author} {\bibfnamefont {L.~N.}\ \bibnamefont {Pfeiffer}}, \bibinfo {author} {\bibfnamefont {K.~W.}\ \bibnamefont {West}}, \bibinfo {author} {\bibfnamefont {K.~W.}\ \bibnamefont {Baldwin}},\ and\ \bibinfo {author} {\bibfnamefont {R.}~\bibnamefont {Winkler}},\ }\bibfield  {title} {\bibinfo {title} {$\ensuremath{\nu}=1/2$ fractional quantum hall effect in tilted magnetic fields},\ }\href {https://doi.org/10.1103/PhysRevB.91.045113} {\bibfield  {journal} {\bibinfo  {journal} {Phys. Rev. B}\ }\textbf {\bibinfo {volume} {91}},\ \bibinfo {pages} {045113} (\bibinfo {year} {2015})}\BibitemShut {NoStop}%
\bibitem [{\citenamefont {Hossain}\ \emph {et~al.}(2018)\citenamefont {Hossain}, \citenamefont {Ma}, \citenamefont {Chung}, \citenamefont {Pfeiffer}, \citenamefont {West}, \citenamefont {Baldwin},\ and\ \citenamefont {Shayegan}}]{Shayegen_PRL_18}%
  \BibitemOpen
  \bibfield  {author} {\bibinfo {author} {\bibfnamefont {M.~S.}\ \bibnamefont {Hossain}}, \bibinfo {author} {\bibfnamefont {M.~K.}\ \bibnamefont {Ma}}, \bibinfo {author} {\bibfnamefont {Y.~J.}\ \bibnamefont {Chung}}, \bibinfo {author} {\bibfnamefont {L.~N.}\ \bibnamefont {Pfeiffer}}, \bibinfo {author} {\bibfnamefont {K.~W.}\ \bibnamefont {West}}, \bibinfo {author} {\bibfnamefont {K.~W.}\ \bibnamefont {Baldwin}},\ and\ \bibinfo {author} {\bibfnamefont {M.}~\bibnamefont {Shayegan}},\ }\bibfield  {title} {\bibinfo {title} {Unconventional anisotropic even-denominator fractional quantum hall state in a system with mass anisotropy},\ }\href {https://doi.org/10.1103/PhysRevLett.121.256601} {\bibfield  {journal} {\bibinfo  {journal} {Phys. Rev. Lett.}\ }\textbf {\bibinfo {volume} {121}},\ \bibinfo {pages} {256601} (\bibinfo {year} {2018})}\BibitemShut {NoStop}%
\bibitem [{\citenamefont {Hossain}\ \emph {et~al.}(2022)\citenamefont {Hossain}, \citenamefont {Ma}, \citenamefont {Villegas-Rosales}, \citenamefont {Chung}, \citenamefont {Pfeiffer}, \citenamefont {West}, \citenamefont {Baldwin},\ and\ \citenamefont {Shayegan}}]{Shayegen_PRL_22}%
  \BibitemOpen
  \bibfield  {author} {\bibinfo {author} {\bibfnamefont {M.~S.}\ \bibnamefont {Hossain}}, \bibinfo {author} {\bibfnamefont {M.~K.}\ \bibnamefont {Ma}}, \bibinfo {author} {\bibfnamefont {K.~A.}\ \bibnamefont {Villegas-Rosales}}, \bibinfo {author} {\bibfnamefont {Y.~J.}\ \bibnamefont {Chung}}, \bibinfo {author} {\bibfnamefont {L.~N.}\ \bibnamefont {Pfeiffer}}, \bibinfo {author} {\bibfnamefont {K.~W.}\ \bibnamefont {West}}, \bibinfo {author} {\bibfnamefont {K.~W.}\ \bibnamefont {Baldwin}},\ and\ \bibinfo {author} {\bibfnamefont {M.}~\bibnamefont {Shayegan}},\ }\bibfield  {title} {\bibinfo {title} {Anisotropic two-dimensional disordered wigner solid},\ }\href {https://doi.org/10.1103/PhysRevLett.129.036601} {\bibfield  {journal} {\bibinfo  {journal} {Phys. Rev. Lett.}\ }\textbf {\bibinfo {volume} {129}},\ \bibinfo {pages} {036601} (\bibinfo {year} {2022})}\BibitemShut {NoStop}%
\bibitem [{\citenamefont {Lee}\ \emph {et~al.}(2013)\citenamefont {Lee}, \citenamefont {Thomale},\ and\ \citenamefont {Qi}}]{ChingHuaLeePRB13}%
  \BibitemOpen
  \bibfield  {author} {\bibinfo {author} {\bibfnamefont {C.~H.}\ \bibnamefont {Lee}}, \bibinfo {author} {\bibfnamefont {R.}~\bibnamefont {Thomale}},\ and\ \bibinfo {author} {\bibfnamefont {X.-L.}\ \bibnamefont {Qi}},\ }\bibfield  {title} {\bibinfo {title} {Pseudopotential formalism for fractional chern insulators},\ }\href {https://doi.org/10.1103/PhysRevB.88.035101} {\bibfield  {journal} {\bibinfo  {journal} {Phys. Rev. B}\ }\textbf {\bibinfo {volume} {88}},\ \bibinfo {pages} {035101} (\bibinfo {year} {2013})}\BibitemShut {NoStop}%
\bibitem [{\citenamefont {Lee}\ \emph {et~al.}(2017)\citenamefont {Lee}, \citenamefont {Claassen},\ and\ \citenamefont {Thomale}}]{Lee_engineering_PRB17}%
  \BibitemOpen
  \bibfield  {author} {\bibinfo {author} {\bibfnamefont {C.~H.}\ \bibnamefont {Lee}}, \bibinfo {author} {\bibfnamefont {M.}~\bibnamefont {Claassen}},\ and\ \bibinfo {author} {\bibfnamefont {R.}~\bibnamefont {Thomale}},\ }\bibfield  {title} {\bibinfo {title} {Band structure engineering of ideal fractional chern insulators},\ }\href {https://doi.org/10.1103/PhysRevB.96.165150} {\bibfield  {journal} {\bibinfo  {journal} {Phys. Rev. B}\ }\textbf {\bibinfo {volume} {96}},\ \bibinfo {pages} {165150} (\bibinfo {year} {2017})}\BibitemShut {NoStop}%
\bibitem [{Note1()}]{Note1}%
  \BibitemOpen
  \bibinfo {note} {We added a quotation mark to ``periodic part'' because for Landau level type states, $u_{\protect \bm {k}}(\protect \bm {r})$ rigorously speaking is not a periodic function upon translations; only periodic wavefunctions of Bloch states are. Nevertheless, the translation phase are independent on $\protect \bm {k}$ hence $u_{\protect \bm {k}}(\protect \bm {r})$ of different $\protect \bm {k}$ are still within the same Hilbert space.}\BibitemShut {Stop}%
\bibitem [{\citenamefont {Ledwith}\ \emph {et~al.}(2021)\citenamefont {Ledwith}, \citenamefont {Khalaf},\ and\ \citenamefont {Vishwanath}}]{LEDWITH2021168646}%
  \BibitemOpen
  \bibfield  {author} {\bibinfo {author} {\bibfnamefont {P.~J.}\ \bibnamefont {Ledwith}}, \bibinfo {author} {\bibfnamefont {E.}~\bibnamefont {Khalaf}},\ and\ \bibinfo {author} {\bibfnamefont {A.}~\bibnamefont {Vishwanath}},\ }\bibfield  {title} {\bibinfo {title} {Strong coupling theory of magic-angle graphene: A pedagogical introduction},\ }\href {https://doi.org/https://doi.org/10.1016/j.aop.2021.168646} {\bibfield  {journal} {\bibinfo  {journal} {Annals of Physics}\ }\textbf {\bibinfo {volume} {435}},\ \bibinfo {pages} {168646} (\bibinfo {year} {2021})},\ \bibinfo {note} {special issue on Philip W. Anderson}\BibitemShut {NoStop}%
\bibitem [{\citenamefont {Koulakov}\ \emph {et~al.}(1996)\citenamefont {Koulakov}, \citenamefont {Fogler},\ and\ \citenamefont {Shklovskii}}]{FoglerPRL96}%
  \BibitemOpen
  \bibfield  {author} {\bibinfo {author} {\bibfnamefont {A.~A.}\ \bibnamefont {Koulakov}}, \bibinfo {author} {\bibfnamefont {M.~M.}\ \bibnamefont {Fogler}},\ and\ \bibinfo {author} {\bibfnamefont {B.~I.}\ \bibnamefont {Shklovskii}},\ }\bibfield  {title} {\bibinfo {title} {Charge density wave in two-dimensional electron liquid in weak magnetic field},\ }\href {https://doi.org/10.1103/PhysRevLett.76.499} {\bibfield  {journal} {\bibinfo  {journal} {Phys. Rev. Lett.}\ }\textbf {\bibinfo {volume} {76}},\ \bibinfo {pages} {499} (\bibinfo {year} {1996})}\BibitemShut {NoStop}%
\bibitem [{\citenamefont {Fogler}\ \emph {et~al.}(1996)\citenamefont {Fogler}, \citenamefont {Koulakov},\ and\ \citenamefont {Shklovskii}}]{FoglerPRB96}%
  \BibitemOpen
  \bibfield  {author} {\bibinfo {author} {\bibfnamefont {M.~M.}\ \bibnamefont {Fogler}}, \bibinfo {author} {\bibfnamefont {A.~A.}\ \bibnamefont {Koulakov}},\ and\ \bibinfo {author} {\bibfnamefont {B.~I.}\ \bibnamefont {Shklovskii}},\ }\bibfield  {title} {\bibinfo {title} {Ground state of a two-dimensional electron liquid in a weak magnetic field},\ }\href {https://doi.org/10.1103/PhysRevB.54.1853} {\bibfield  {journal} {\bibinfo  {journal} {Phys. Rev. B}\ }\textbf {\bibinfo {volume} {54}},\ \bibinfo {pages} {1853} (\bibinfo {year} {1996})}\BibitemShut {NoStop}%
\bibitem [{\citenamefont {Rezayi}\ and\ \citenamefont {Haldane}(2000)}]{RezayiHaldanePRL00}%
  \BibitemOpen
  \bibfield  {author} {\bibinfo {author} {\bibfnamefont {E.~H.}\ \bibnamefont {Rezayi}}\ and\ \bibinfo {author} {\bibfnamefont {F.~D.~M.}\ \bibnamefont {Haldane}},\ }\bibfield  {title} {\bibinfo {title} {Incompressible paired hall state, stripe order, and the composite fermion liquid phase in half-filled landau levels},\ }\href {https://doi.org/10.1103/PhysRevLett.84.4685} {\bibfield  {journal} {\bibinfo  {journal} {Phys. Rev. Lett.}\ }\textbf {\bibinfo {volume} {84}},\ \bibinfo {pages} {4685} (\bibinfo {year} {2000})}\BibitemShut {NoStop}%
\bibitem [{Note2()}]{Note2}%
  \BibitemOpen
  \bibinfo {note} {For ideal band constructed as determinant states from multiple bands, normalization determines the $\protect \bm {k}-$dependent part of quantum geometry in the same way as in Eq.~(\ref {idealband_Omega_N}), but the constant part is different (because the Chern number of the determinant state is the total Chern number of the band complex); see Eq.~(\ref {OmegaPsi1}).}\BibitemShut {Stop}%
\bibitem [{\citenamefont {Tarnopolsky}\ \emph {et~al.}(2019)\citenamefont {Tarnopolsky}, \citenamefont {Kruchkov},\ and\ \citenamefont {Vishwanath}}]{Grisha_TBG}%
  \BibitemOpen
  \bibfield  {author} {\bibinfo {author} {\bibfnamefont {G.}~\bibnamefont {Tarnopolsky}}, \bibinfo {author} {\bibfnamefont {A.~J.}\ \bibnamefont {Kruchkov}},\ and\ \bibinfo {author} {\bibfnamefont {A.}~\bibnamefont {Vishwanath}},\ }\bibfield  {title} {\bibinfo {title} {Origin of magic angles in twisted bilayer graphene},\ }\href {https://doi.org/10.1103/PhysRevLett.122.106405} {\bibfield  {journal} {\bibinfo  {journal} {Phys. Rev. Lett.}\ }\textbf {\bibinfo {volume} {122}},\ \bibinfo {pages} {106405} (\bibinfo {year} {2019})}\BibitemShut {NoStop}%
\bibitem [{\citenamefont {{Dong}}\ \emph {et~al.}(2022{\natexlab{a}})\citenamefont {{Dong}}, \citenamefont {{Wang}},\ and\ \citenamefont {{Fu}}}]{Liang_DiracNonuniformB}%
  \BibitemOpen
  \bibfield  {author} {\bibinfo {author} {\bibfnamefont {J.}~\bibnamefont {{Dong}}}, \bibinfo {author} {\bibfnamefont {J.}~\bibnamefont {{Wang}}},\ and\ \bibinfo {author} {\bibfnamefont {L.}~\bibnamefont {{Fu}}},\ }\bibfield  {title} {\bibinfo {title} {{Dirac electron under periodic magnetic field: Platform for fractional Chern insulator and generalized Wigner crystal}},\ }\href@noop {} {\bibfield  {journal} {\bibinfo  {journal} {arXiv e-prints}\ ,\ \bibinfo {eid} {arXiv:2208.10516}} (\bibinfo {year} {2022}{\natexlab{a}})},\ \Eprint {https://arxiv.org/abs/2208.10516} {arXiv:2208.10516 [cond-mat.mes-hall]} \BibitemShut {NoStop}%
\bibitem [{\citenamefont {{Estienne}}\ \emph {et~al.}(2023)\citenamefont {{Estienne}}, \citenamefont {{Regnault}},\ and\ \citenamefont {{Cr{\'e}pel}}}]{Crepel_LLCurvedSpace23}%
  \BibitemOpen
  \bibfield  {author} {\bibinfo {author} {\bibfnamefont {B.}~\bibnamefont {{Estienne}}}, \bibinfo {author} {\bibfnamefont {N.}~\bibnamefont {{Regnault}}},\ and\ \bibinfo {author} {\bibfnamefont {V.}~\bibnamefont {{Cr{\'e}pel}}},\ }\bibfield  {title} {\bibinfo {title} {{Ideal Chern bands as Landau levels in curved space}},\ }\href {https://doi.org/10.1103/PhysRevResearch.5.L032048} {\bibfield  {journal} {\bibinfo  {journal} {Physical Review Research}\ }\textbf {\bibinfo {volume} {5}},\ \bibinfo {eid} {L032048} (\bibinfo {year} {2023})},\ \Eprint {https://arxiv.org/abs/2304.01251} {arXiv:2304.01251 [cond-mat.mes-hall]} \BibitemShut {NoStop}%
\bibitem [{\citenamefont {Kapit}\ and\ \citenamefont {Mueller}(2010)}]{Kapit_Mueller}%
  \BibitemOpen
  \bibfield  {author} {\bibinfo {author} {\bibfnamefont {E.}~\bibnamefont {Kapit}}\ and\ \bibinfo {author} {\bibfnamefont {E.}~\bibnamefont {Mueller}},\ }\bibfield  {title} {\bibinfo {title} {Exact parent hamiltonian for the quantum hall states in a lattice},\ }\href {https://doi.org/10.1103/PhysRevLett.105.215303} {\bibfield  {journal} {\bibinfo  {journal} {Phys. Rev. Lett.}\ }\textbf {\bibinfo {volume} {105}},\ \bibinfo {pages} {215303} (\bibinfo {year} {2010})}\BibitemShut {NoStop}%
\bibitem [{\citenamefont {Behrmann}\ \emph {et~al.}(2016)\citenamefont {Behrmann}, \citenamefont {Liu},\ and\ \citenamefont {Bergholtz}}]{ModelFCI_Zhao}%
  \BibitemOpen
  \bibfield  {author} {\bibinfo {author} {\bibfnamefont {J.}~\bibnamefont {Behrmann}}, \bibinfo {author} {\bibfnamefont {Z.}~\bibnamefont {Liu}},\ and\ \bibinfo {author} {\bibfnamefont {E.~J.}\ \bibnamefont {Bergholtz}},\ }\bibfield  {title} {\bibinfo {title} {Model fractional chern insulators},\ }\href {https://doi.org/10.1103/PhysRevLett.116.216802} {\bibfield  {journal} {\bibinfo  {journal} {Phys. Rev. Lett.}\ }\textbf {\bibinfo {volume} {116}},\ \bibinfo {pages} {216802} (\bibinfo {year} {2016})}\BibitemShut {NoStop}%
\bibitem [{\citenamefont {Varjas}\ \emph {et~al.}(2022)\citenamefont {Varjas}, \citenamefont {Abouelkomsan}, \citenamefont {Yang},\ and\ \citenamefont {Bergholtz}}]{Emil_constantBerry}%
  \BibitemOpen
  \bibfield  {author} {\bibinfo {author} {\bibfnamefont {D.}~\bibnamefont {Varjas}}, \bibinfo {author} {\bibfnamefont {A.}~\bibnamefont {Abouelkomsan}}, \bibinfo {author} {\bibfnamefont {K.}~\bibnamefont {Yang}},\ and\ \bibinfo {author} {\bibfnamefont {E.~J.}\ \bibnamefont {Bergholtz}},\ }\bibfield  {title} {\bibinfo {title} {{Topological lattice models with constant Berry curvature}},\ }\href {https://doi.org/10.21468/SciPostPhys.12.4.118} {\bibfield  {journal} {\bibinfo  {journal} {SciPost Phys.}\ }\textbf {\bibinfo {volume} {12}},\ \bibinfo {pages} {118} (\bibinfo {year} {2022})}\BibitemShut {NoStop}%
\bibitem [{\citenamefont {Dong}\ and\ \citenamefont {Mueller}(2020)}]{Dong_Mueller_20}%
  \BibitemOpen
  \bibfield  {author} {\bibinfo {author} {\bibfnamefont {J.}~\bibnamefont {Dong}}\ and\ \bibinfo {author} {\bibfnamefont {E.~J.}\ \bibnamefont {Mueller}},\ }\bibfield  {title} {\bibinfo {title} {Exact topological flat bands from continuum landau levels},\ }\href {https://doi.org/10.1103/PhysRevA.101.013629} {\bibfield  {journal} {\bibinfo  {journal} {Phys. Rev. A}\ }\textbf {\bibinfo {volume} {101}},\ \bibinfo {pages} {013629} (\bibinfo {year} {2020})}\BibitemShut {NoStop}%
\bibitem [{\citenamefont {Ledwith}\ \emph {et~al.}(2022)\citenamefont {Ledwith}, \citenamefont {Vishwanath},\ and\ \citenamefont {Khalaf}}]{Eslam_highC_idealband}%
  \BibitemOpen
  \bibfield  {author} {\bibinfo {author} {\bibfnamefont {P.~J.}\ \bibnamefont {Ledwith}}, \bibinfo {author} {\bibfnamefont {A.}~\bibnamefont {Vishwanath}},\ and\ \bibinfo {author} {\bibfnamefont {E.}~\bibnamefont {Khalaf}},\ }\bibfield  {title} {\bibinfo {title} {Family of ideal chern flatbands with arbitrary chern number in chiral twisted graphene multilayers},\ }\href {https://doi.org/10.1103/PhysRevLett.128.176404} {\bibfield  {journal} {\bibinfo  {journal} {Phys. Rev. Lett.}\ }\textbf {\bibinfo {volume} {128}},\ \bibinfo {pages} {176404} (\bibinfo {year} {2022})}\BibitemShut {NoStop}%
\bibitem [{\citenamefont {{Guerci}}\ \emph {et~al.}(2023{\natexlab{a}})\citenamefont {{Guerci}}, \citenamefont {{Mao}},\ and\ \citenamefont {{Mora}}}]{GuerciMoraTTG1}%
  \BibitemOpen
  \bibfield  {author} {\bibinfo {author} {\bibfnamefont {D.}~\bibnamefont {{Guerci}}}, \bibinfo {author} {\bibfnamefont {Y.}~\bibnamefont {{Mao}}},\ and\ \bibinfo {author} {\bibfnamefont {C.}~\bibnamefont {{Mora}}},\ }\bibfield  {title} {\bibinfo {title} {{Chern mosaic and ideal flat bands in equal-twist trilayer graphene}},\ }\href {https://doi.org/10.48550/arXiv.2305.03702} {\bibfield  {journal} {\bibinfo  {journal} {arXiv e-prints}\ ,\ \bibinfo {eid} {arXiv:2305.03702}} (\bibinfo {year} {2023}{\natexlab{a}})},\ \Eprint {https://arxiv.org/abs/2305.03702} {arXiv:2305.03702 [cond-mat.mes-hall]} \BibitemShut {NoStop}%
\bibitem [{\citenamefont {{Guerci}}\ \emph {et~al.}(2023{\natexlab{b}})\citenamefont {{Guerci}}, \citenamefont {{Mao}},\ and\ \citenamefont {{Mora}}}]{GuerciMoraTTG2}%
  \BibitemOpen
  \bibfield  {author} {\bibinfo {author} {\bibfnamefont {D.}~\bibnamefont {{Guerci}}}, \bibinfo {author} {\bibfnamefont {Y.}~\bibnamefont {{Mao}}},\ and\ \bibinfo {author} {\bibfnamefont {C.}~\bibnamefont {{Mora}}},\ }\bibfield  {title} {\bibinfo {title} {{Nature of even and odd magic angles in helical twisted trilayer graphene}},\ }\href {https://doi.org/10.48550/arXiv.2308.02638} {\bibfield  {journal} {\bibinfo  {journal} {arXiv e-prints}\ ,\ \bibinfo {eid} {arXiv:2308.02638}} (\bibinfo {year} {2023}{\natexlab{b}})},\ \Eprint {https://arxiv.org/abs/2308.02638} {arXiv:2308.02638 [cond-mat.mes-hall]} \BibitemShut {NoStop}%
\bibitem [{\citenamefont {{Datta}}\ \emph {et~al.}(2024)\citenamefont {{Datta}}, \citenamefont {{Guerci}}, \citenamefont {{Goerbig}},\ and\ \citenamefont {{Mora}}}]{GuerciMoraTTG3}%
  \BibitemOpen
  \bibfield  {author} {\bibinfo {author} {\bibfnamefont {A.}~\bibnamefont {{Datta}}}, \bibinfo {author} {\bibfnamefont {D.}~\bibnamefont {{Guerci}}}, \bibinfo {author} {\bibfnamefont {M.~O.}\ \bibnamefont {{Goerbig}}},\ and\ \bibinfo {author} {\bibfnamefont {C.}~\bibnamefont {{Mora}}},\ }\bibfield  {title} {\bibinfo {title} {{Helical trilayer graphene in magnetic field: Chern mosaic and higher Chern number ideal flat bands}},\ }\href {https://doi.org/10.48550/arXiv.2404.15452} {\bibfield  {journal} {\bibinfo  {journal} {arXiv e-prints}\ ,\ \bibinfo {eid} {arXiv:2404.15452}} (\bibinfo {year} {2024})},\ \Eprint {https://arxiv.org/abs/2404.15452} {arXiv:2404.15452 [cond-mat.mes-hall]} \BibitemShut {NoStop}%
\bibitem [{\citenamefont {Popov}\ and\ \citenamefont {Tarnopolsky}(2023{\natexlab{a}})}]{Grisha_eTTG23}%
  \BibitemOpen
  \bibfield  {author} {\bibinfo {author} {\bibfnamefont {F.~K.}\ \bibnamefont {Popov}}\ and\ \bibinfo {author} {\bibfnamefont {G.}~\bibnamefont {Tarnopolsky}},\ }\bibfield  {title} {\bibinfo {title} {Magic angles in equal-twist trilayer graphene},\ }\href {https://doi.org/10.1103/PhysRevB.108.L081124} {\bibfield  {journal} {\bibinfo  {journal} {Phys. Rev. B}\ }\textbf {\bibinfo {volume} {108}},\ \bibinfo {pages} {L081124} (\bibinfo {year} {2023}{\natexlab{a}})}\BibitemShut {NoStop}%
\bibitem [{\citenamefont {Popov}\ and\ \citenamefont {Tarnopolsky}(2023{\natexlab{b}})}]{Grisha_TTG23}%
  \BibitemOpen
  \bibfield  {author} {\bibinfo {author} {\bibfnamefont {F.~K.}\ \bibnamefont {Popov}}\ and\ \bibinfo {author} {\bibfnamefont {G.}~\bibnamefont {Tarnopolsky}},\ }\bibfield  {title} {\bibinfo {title} {Magic angle butterfly in twisted trilayer graphene},\ }\href {https://doi.org/10.1103/PhysRevResearch.5.043079} {\bibfield  {journal} {\bibinfo  {journal} {Phys. Rev. Res.}\ }\textbf {\bibinfo {volume} {5}},\ \bibinfo {pages} {043079} (\bibinfo {year} {2023}{\natexlab{b}})}\BibitemShut {NoStop}%
\bibitem [{\citenamefont {Gao}\ \emph {et~al.}(2023)\citenamefont {Gao}, \citenamefont {Dong}, \citenamefont {Ledwith}, \citenamefont {Parker},\ and\ \citenamefont {Khalaf}}]{Eslam_StrainedGraphene23}%
  \BibitemOpen
  \bibfield  {author} {\bibinfo {author} {\bibfnamefont {Q.}~\bibnamefont {Gao}}, \bibinfo {author} {\bibfnamefont {J.}~\bibnamefont {Dong}}, \bibinfo {author} {\bibfnamefont {P.}~\bibnamefont {Ledwith}}, \bibinfo {author} {\bibfnamefont {D.}~\bibnamefont {Parker}},\ and\ \bibinfo {author} {\bibfnamefont {E.}~\bibnamefont {Khalaf}},\ }\bibfield  {title} {\bibinfo {title} {Untwisting moir\'e physics: Almost ideal bands and fractional chern insulators in periodically strained monolayer graphene},\ }\href {https://doi.org/10.1103/PhysRevLett.131.096401} {\bibfield  {journal} {\bibinfo  {journal} {Phys. Rev. Lett.}\ }\textbf {\bibinfo {volume} {131}},\ \bibinfo {pages} {096401} (\bibinfo {year} {2023})}\BibitemShut {NoStop}%
\bibitem [{\citenamefont {Morales-Dur\'an}\ \emph {et~al.}(2024)\citenamefont {Morales-Dur\'an}, \citenamefont {Wei}, \citenamefont {Shi},\ and\ \citenamefont {MacDonald}}]{NicolasMacDonald23}%
  \BibitemOpen
  \bibfield  {author} {\bibinfo {author} {\bibfnamefont {N.}~\bibnamefont {Morales-Dur\'an}}, \bibinfo {author} {\bibfnamefont {N.}~\bibnamefont {Wei}}, \bibinfo {author} {\bibfnamefont {J.}~\bibnamefont {Shi}},\ and\ \bibinfo {author} {\bibfnamefont {A.~H.}\ \bibnamefont {MacDonald}},\ }\bibfield  {title} {\bibinfo {title} {Magic angles and fractional chern insulators in twisted homobilayer transition metal dichalcogenides},\ }\href {https://doi.org/10.1103/PhysRevLett.132.096602} {\bibfield  {journal} {\bibinfo  {journal} {Phys. Rev. Lett.}\ }\textbf {\bibinfo {volume} {132}},\ \bibinfo {pages} {096602} (\bibinfo {year} {2024})}\BibitemShut {NoStop}%
\bibitem [{\citenamefont {{Shi}}\ \emph {et~al.}(2024)\citenamefont {{Shi}}, \citenamefont {{Morales-Dur{\'a}n}}, \citenamefont {{Khalaf}},\ and\ \citenamefont {{MacDonald}}}]{NicolasMacDonald24}%
  \BibitemOpen
  \bibfield  {author} {\bibinfo {author} {\bibfnamefont {J.}~\bibnamefont {{Shi}}}, \bibinfo {author} {\bibfnamefont {N.}~\bibnamefont {{Morales-Dur{\'a}n}}}, \bibinfo {author} {\bibfnamefont {E.}~\bibnamefont {{Khalaf}}},\ and\ \bibinfo {author} {\bibfnamefont {A.~H.}\ \bibnamefont {{MacDonald}}},\ }\bibfield  {title} {\bibinfo {title} {{Adiabatic Approximation and Aharonov-Casher Bands in Twisted Homobilayer TMDs}},\ }\href {https://doi.org/10.48550/arXiv.2404.13455} {\bibfield  {journal} {\bibinfo  {journal} {arXiv e-prints}\ ,\ \bibinfo {eid} {arXiv:2404.13455}} (\bibinfo {year} {2024})},\ \Eprint {https://arxiv.org/abs/2404.13455} {arXiv:2404.13455 [cond-mat.mes-hall]} \BibitemShut {NoStop}%
\bibitem [{\citenamefont {{Cr{\'e}pel}}\ \emph {et~al.}(2023)\citenamefont {{Cr{\'e}pel}}, \citenamefont {{Regnault}},\ and\ \citenamefont {{Queiroz}}}]{CrepelRegnaultRaquel23}%
  \BibitemOpen
  \bibfield  {author} {\bibinfo {author} {\bibfnamefont {V.}~\bibnamefont {{Cr{\'e}pel}}}, \bibinfo {author} {\bibfnamefont {N.}~\bibnamefont {{Regnault}}},\ and\ \bibinfo {author} {\bibfnamefont {R.}~\bibnamefont {{Queiroz}}},\ }\bibfield  {title} {\bibinfo {title} {{The chiral limits of moir{\'e} semiconductors: origin of flat bands and topology in twisted transition metal dichalcogenides homobilayers}},\ }\href {https://doi.org/10.48550/arXiv.2305.10477} {\bibfield  {journal} {\bibinfo  {journal} {arXiv e-prints}\ ,\ \bibinfo {eid} {arXiv:2305.10477}} (\bibinfo {year} {2023})},\ \Eprint {https://arxiv.org/abs/2305.10477} {arXiv:2305.10477 [cond-mat.mes-hall]} \BibitemShut {NoStop}%
\bibitem [{\citenamefont {{Cr{\'e}pel}}\ \emph {et~al.}(2024)\citenamefont {{Cr{\'e}pel}}, \citenamefont {{Ding}}, \citenamefont {{Verma}}, \citenamefont {{Regnault}},\ and\ \citenamefont {{Queiroz}}}]{CrepelDirac24}%
  \BibitemOpen
  \bibfield  {author} {\bibinfo {author} {\bibfnamefont {V.}~\bibnamefont {{Cr{\'e}pel}}}, \bibinfo {author} {\bibfnamefont {P.}~\bibnamefont {{Ding}}}, \bibinfo {author} {\bibfnamefont {N.}~\bibnamefont {{Verma}}}, \bibinfo {author} {\bibfnamefont {N.}~\bibnamefont {{Regnault}}},\ and\ \bibinfo {author} {\bibfnamefont {R.}~\bibnamefont {{Queiroz}}},\ }\bibfield  {title} {\bibinfo {title} {{Topologically protected flatness in chiral moir{\'e} heterostructures}},\ }\href {https://doi.org/10.48550/arXiv.2403.19656} {\bibfield  {journal} {\bibinfo  {journal} {arXiv e-prints}\ ,\ \bibinfo {eid} {arXiv:2403.19656}} (\bibinfo {year} {2024})},\ \Eprint {https://arxiv.org/abs/2403.19656} {arXiv:2403.19656 [cond-mat.mes-hall]} \BibitemShut {NoStop}%
\bibitem [{\citenamefont {{Dong}}\ \emph {et~al.}(2022{\natexlab{b}})\citenamefont {{Dong}}, \citenamefont {{Ledwith}}, \citenamefont {{Khalaf}}, \citenamefont {{Lee}},\ and\ \citenamefont {{Vishwanath}}}]{junkaidonghighC22}%
  \BibitemOpen
  \bibfield  {author} {\bibinfo {author} {\bibfnamefont {J.}~\bibnamefont {{Dong}}}, \bibinfo {author} {\bibfnamefont {P.~J.}\ \bibnamefont {{Ledwith}}}, \bibinfo {author} {\bibfnamefont {E.}~\bibnamefont {{Khalaf}}}, \bibinfo {author} {\bibfnamefont {J.~Y.}\ \bibnamefont {{Lee}}},\ and\ \bibinfo {author} {\bibfnamefont {A.}~\bibnamefont {{Vishwanath}}},\ }\bibfield  {title} {\bibinfo {title} {{Exact Many-Body Ground States from Decomposition of Ideal Higher Chern Bands: Applications to Chirally Twisted Graphene Multilayers}},\ }\href@noop {} {\bibfield  {journal} {\bibinfo  {journal} {arXiv e-prints}\ ,\ \bibinfo {eid} {arXiv:2210.13477}} (\bibinfo {year} {2022}{\natexlab{b}})},\ \Eprint {https://arxiv.org/abs/2210.13477} {arXiv:2210.13477 [cond-mat.mes-hall]} \BibitemShut {NoStop}%
\bibitem [{\citenamefont {Trugman}\ and\ \citenamefont {Kivelson}(1985)}]{TrugmanKivelson85}%
  \BibitemOpen
  \bibfield  {author} {\bibinfo {author} {\bibfnamefont {S.~A.}\ \bibnamefont {Trugman}}\ and\ \bibinfo {author} {\bibfnamefont {S.}~\bibnamefont {Kivelson}},\ }\bibfield  {title} {\bibinfo {title} {Exact results for the fractional quantum hall effect with general interactions},\ }\href {https://doi.org/10.1103/PhysRevB.31.5280} {\bibfield  {journal} {\bibinfo  {journal} {Phys. Rev. B}\ }\textbf {\bibinfo {volume} {31}},\ \bibinfo {pages} {5280} (\bibinfo {year} {1985})}\BibitemShut {NoStop}%
\bibitem [{\citenamefont {Haldane}(1983)}]{Haldane_hierarchy}%
  \BibitemOpen
  \bibfield  {author} {\bibinfo {author} {\bibfnamefont {F.~D.~M.}\ \bibnamefont {Haldane}},\ }\bibfield  {title} {\bibinfo {title} {Fractional quantization of the hall effect: A hierarchy of incompressible quantum fluid states},\ }\href {https://doi.org/10.1103/PhysRevLett.51.605} {\bibfield  {journal} {\bibinfo  {journal} {Phys. Rev. Lett.}\ }\textbf {\bibinfo {volume} {51}},\ \bibinfo {pages} {605} (\bibinfo {year} {1983})}\BibitemShut {NoStop}%
\bibitem [{\citenamefont {Greiter}\ \emph {et~al.}(1991)\citenamefont {Greiter}, \citenamefont {Wen},\ and\ \citenamefont {Wilczek}}]{GreiterWenWilczekPRL91}%
  \BibitemOpen
  \bibfield  {author} {\bibinfo {author} {\bibfnamefont {M.}~\bibnamefont {Greiter}}, \bibinfo {author} {\bibfnamefont {X.-G.}\ \bibnamefont {Wen}},\ and\ \bibinfo {author} {\bibfnamefont {F.}~\bibnamefont {Wilczek}},\ }\bibfield  {title} {\bibinfo {title} {Paired hall state at half filling},\ }\href {https://doi.org/10.1103/PhysRevLett.66.3205} {\bibfield  {journal} {\bibinfo  {journal} {Phys. Rev. Lett.}\ }\textbf {\bibinfo {volume} {66}},\ \bibinfo {pages} {3205} (\bibinfo {year} {1991})}\BibitemShut {NoStop}%
\bibitem [{\citenamefont {Greiter}\ \emph {et~al.}(1992)\citenamefont {Greiter}, \citenamefont {Wen},\ and\ \citenamefont {Wilczek}}]{GreiterWenWilczekNuclearPhyB92}%
  \BibitemOpen
  \bibfield  {author} {\bibinfo {author} {\bibfnamefont {M.}~\bibnamefont {Greiter}}, \bibinfo {author} {\bibfnamefont {X.}~\bibnamefont {Wen}},\ and\ \bibinfo {author} {\bibfnamefont {F.}~\bibnamefont {Wilczek}},\ }\bibfield  {title} {\bibinfo {title} {Paired hall states},\ }\href {https://doi.org/https://doi.org/10.1016/0550-3213(92)90401-V} {\bibfield  {journal} {\bibinfo  {journal} {Nuclear Physics B}\ }\textbf {\bibinfo {volume} {374}},\ \bibinfo {pages} {567} (\bibinfo {year} {1992})}\BibitemShut {NoStop}%
\bibitem [{\citenamefont {{Parker}}\ \emph {et~al.}(2021)\citenamefont {{Parker}}, \citenamefont {{Ledwith}}, \citenamefont {{Khalaf}}, \citenamefont {{Soejima}}, \citenamefont {{Hauschild}}, \citenamefont {{Xie}}, \citenamefont {{Pierce}}, \citenamefont {{Zaletel}}, \citenamefont {{Yacoby}},\ and\ \citenamefont {{Vishwanath}}}]{Dan_parker21}%
  \BibitemOpen
  \bibfield  {author} {\bibinfo {author} {\bibfnamefont {D.}~\bibnamefont {{Parker}}}, \bibinfo {author} {\bibfnamefont {P.}~\bibnamefont {{Ledwith}}}, \bibinfo {author} {\bibfnamefont {E.}~\bibnamefont {{Khalaf}}}, \bibinfo {author} {\bibfnamefont {T.}~\bibnamefont {{Soejima}}}, \bibinfo {author} {\bibfnamefont {J.}~\bibnamefont {{Hauschild}}}, \bibinfo {author} {\bibfnamefont {Y.}~\bibnamefont {{Xie}}}, \bibinfo {author} {\bibfnamefont {A.}~\bibnamefont {{Pierce}}}, \bibinfo {author} {\bibfnamefont {M.~P.}\ \bibnamefont {{Zaletel}}}, \bibinfo {author} {\bibfnamefont {A.}~\bibnamefont {{Yacoby}}},\ and\ \bibinfo {author} {\bibfnamefont {A.}~\bibnamefont {{Vishwanath}}},\ }\bibfield  {title} {\bibinfo {title} {{Field-tuned and zero-field fractional Chern insulators in magic angle graphene}},\ }\href@noop {} {\bibfield  {journal} {\bibinfo  {journal} {arXiv e-prints}\ ,\ \bibinfo {eid} {arXiv:2112.13837}} (\bibinfo {year} {2021})},\ \Eprint {https://arxiv.org/abs/2112.13837} {arXiv:2112.13837
  [cond-mat.str-el]} \BibitemShut {NoStop}%
\bibitem [{\citenamefont {Morales-Dur\'an}\ \emph {et~al.}(2023)\citenamefont {Morales-Dur\'an}, \citenamefont {Wang}, \citenamefont {Schleder}, \citenamefont {Angeli}, \citenamefont {Zhu}, \citenamefont {Kaxiras}, \citenamefont {Repellin},\ and\ \citenamefont {Cano}}]{Nicolas_WSe2_23}%
  \BibitemOpen
  \bibfield  {author} {\bibinfo {author} {\bibfnamefont {N.}~\bibnamefont {Morales-Dur\'an}}, \bibinfo {author} {\bibfnamefont {J.}~\bibnamefont {Wang}}, \bibinfo {author} {\bibfnamefont {G.~R.}\ \bibnamefont {Schleder}}, \bibinfo {author} {\bibfnamefont {M.}~\bibnamefont {Angeli}}, \bibinfo {author} {\bibfnamefont {Z.}~\bibnamefont {Zhu}}, \bibinfo {author} {\bibfnamefont {E.}~\bibnamefont {Kaxiras}}, \bibinfo {author} {\bibfnamefont {C.}~\bibnamefont {Repellin}},\ and\ \bibinfo {author} {\bibfnamefont {J.}~\bibnamefont {Cano}},\ }\bibfield  {title} {\bibinfo {title} {Pressure-enhanced fractional chern insulators along a magic line in moir\'e transition metal dichalcogenides},\ }\href {https://doi.org/10.1103/PhysRevResearch.5.L032022} {\bibfield  {journal} {\bibinfo  {journal} {Phys. Rev. Res.}\ }\textbf {\bibinfo {volume} {5}},\ \bibinfo {pages} {L032022} (\bibinfo {year} {2023})}\BibitemShut {NoStop}%
\bibitem [{\citenamefont {Dong}\ \emph {et~al.}(2023)\citenamefont {Dong}, \citenamefont {Wang}, \citenamefont {Ledwith}, \citenamefont {Vishwanath},\ and\ \citenamefont {Parker}}]{Dong_CFL23}%
  \BibitemOpen
  \bibfield  {author} {\bibinfo {author} {\bibfnamefont {J.}~\bibnamefont {Dong}}, \bibinfo {author} {\bibfnamefont {J.}~\bibnamefont {Wang}}, \bibinfo {author} {\bibfnamefont {P.~J.}\ \bibnamefont {Ledwith}}, \bibinfo {author} {\bibfnamefont {A.}~\bibnamefont {Vishwanath}},\ and\ \bibinfo {author} {\bibfnamefont {D.~E.}\ \bibnamefont {Parker}},\ }\bibfield  {title} {\bibinfo {title} {Composite fermi liquid at zero magnetic field in twisted ${\mathrm{mote}}_{2}$},\ }\href {https://doi.org/10.1103/PhysRevLett.131.136502} {\bibfield  {journal} {\bibinfo  {journal} {Phys. Rev. Lett.}\ }\textbf {\bibinfo {volume} {131}},\ \bibinfo {pages} {136502} (\bibinfo {year} {2023})}\BibitemShut {NoStop}%
\bibitem [{\citenamefont {Goldman}\ \emph {et~al.}(2023)\citenamefont {Goldman}, \citenamefont {Reddy}, \citenamefont {Paul},\ and\ \citenamefont {Fu}}]{Goldman_CFL_23}%
  \BibitemOpen
  \bibfield  {author} {\bibinfo {author} {\bibfnamefont {H.}~\bibnamefont {Goldman}}, \bibinfo {author} {\bibfnamefont {A.~P.}\ \bibnamefont {Reddy}}, \bibinfo {author} {\bibfnamefont {N.}~\bibnamefont {Paul}},\ and\ \bibinfo {author} {\bibfnamefont {L.}~\bibnamefont {Fu}},\ }\bibfield  {title} {\bibinfo {title} {Zero-field composite fermi liquid in twisted semiconductor bilayers},\ }\href {https://doi.org/10.1103/PhysRevLett.131.136501} {\bibfield  {journal} {\bibinfo  {journal} {Phys. Rev. Lett.}\ }\textbf {\bibinfo {volume} {131}},\ \bibinfo {pages} {136501} (\bibinfo {year} {2023})}\BibitemShut {NoStop}%
\bibitem [{\citenamefont {Kourtis}\ \emph {et~al.}(2014)\citenamefont {Kourtis}, \citenamefont {Neupert}, \citenamefont {Chamon},\ and\ \citenamefont {Mudry}}]{Titus_FCI_ExceedGap_14}%
  \BibitemOpen
  \bibfield  {author} {\bibinfo {author} {\bibfnamefont {S.}~\bibnamefont {Kourtis}}, \bibinfo {author} {\bibfnamefont {T.}~\bibnamefont {Neupert}}, \bibinfo {author} {\bibfnamefont {C.}~\bibnamefont {Chamon}},\ and\ \bibinfo {author} {\bibfnamefont {C.}~\bibnamefont {Mudry}},\ }\bibfield  {title} {\bibinfo {title} {Fractional chern insulators with strong interactions that far exceed band gaps},\ }\href {https://doi.org/10.1103/PhysRevLett.112.126806} {\bibfield  {journal} {\bibinfo  {journal} {Phys. Rev. Lett.}\ }\textbf {\bibinfo {volume} {112}},\ \bibinfo {pages} {126806} (\bibinfo {year} {2014})}\BibitemShut {NoStop}%
\bibitem [{\citenamefont {Simon}\ \emph {et~al.}(2015)\citenamefont {Simon}, \citenamefont {Harper},\ and\ \citenamefont {Read}}]{Read_FCI_15}%
  \BibitemOpen
  \bibfield  {author} {\bibinfo {author} {\bibfnamefont {S.~H.}\ \bibnamefont {Simon}}, \bibinfo {author} {\bibfnamefont {F.}~\bibnamefont {Harper}},\ and\ \bibinfo {author} {\bibfnamefont {N.}~\bibnamefont {Read}},\ }\bibfield  {title} {\bibinfo {title} {Fractional chern insulators in bands with zero berry curvature},\ }\href {https://doi.org/10.1103/PhysRevB.92.195104} {\bibfield  {journal} {\bibinfo  {journal} {Phys. Rev. B}\ }\textbf {\bibinfo {volume} {92}},\ \bibinfo {pages} {195104} (\bibinfo {year} {2015})}\BibitemShut {NoStop}%
\bibitem [{\citenamefont {{Shavit}}\ and\ \citenamefont {{Oreg}}(2024)}]{Oreg_GeometryFCI24}%
  \BibitemOpen
  \bibfield  {author} {\bibinfo {author} {\bibfnamefont {G.}~\bibnamefont {{Shavit}}}\ and\ \bibinfo {author} {\bibfnamefont {Y.}~\bibnamefont {{Oreg}}},\ }\bibfield  {title} {\bibinfo {title} {{Quantum Geometry and Stabilization of Fractional Chern Insulators Far from the Ideal Limit}},\ }\href {https://doi.org/10.48550/arXiv.2405.09627} {\bibfield  {journal} {\bibinfo  {journal} {arXiv e-prints}\ ,\ \bibinfo {eid} {arXiv:2405.09627}} (\bibinfo {year} {2024})},\ \Eprint {https://arxiv.org/abs/2405.09627} {arXiv:2405.09627 [cond-mat.str-el]} \BibitemShut {NoStop}%
\bibitem [{\citenamefont {Girvin}\ \emph {et~al.}(1986)\citenamefont {Girvin}, \citenamefont {MacDonald},\ and\ \citenamefont {Platzman}}]{gmpb}%
  \BibitemOpen
  \bibfield  {author} {\bibinfo {author} {\bibfnamefont {S.~M.}\ \bibnamefont {Girvin}}, \bibinfo {author} {\bibfnamefont {A.~H.}\ \bibnamefont {MacDonald}},\ and\ \bibinfo {author} {\bibfnamefont {P.~M.}\ \bibnamefont {Platzman}},\ }\bibfield  {title} {\bibinfo {title} {Magneto-roton theory of collective excitations in the fractional quantum hall effect},\ }\href {https://doi.org/10.1103/PhysRevB.33.2481} {\bibfield  {journal} {\bibinfo  {journal} {Phys. Rev. B}\ }\textbf {\bibinfo {volume} {33}},\ \bibinfo {pages} {2481} (\bibinfo {year} {1986})}\BibitemShut {NoStop}%
\bibitem [{\citenamefont {Girvin}\ \emph {et~al.}(1985)\citenamefont {Girvin}, \citenamefont {MacDonald},\ and\ \citenamefont {Platzman}}]{gmpl}%
  \BibitemOpen
  \bibfield  {author} {\bibinfo {author} {\bibfnamefont {S.~M.}\ \bibnamefont {Girvin}}, \bibinfo {author} {\bibfnamefont {A.~H.}\ \bibnamefont {MacDonald}},\ and\ \bibinfo {author} {\bibfnamefont {P.~M.}\ \bibnamefont {Platzman}},\ }\bibfield  {title} {\bibinfo {title} {Collective-excitation gap in the fractional quantum hall effect},\ }\href {https://doi.org/10.1103/PhysRevLett.54.581} {\bibfield  {journal} {\bibinfo  {journal} {Phys. Rev. Lett.}\ }\textbf {\bibinfo {volume} {54}},\ \bibinfo {pages} {581} (\bibinfo {year} {1985})}\BibitemShut {NoStop}%
\bibitem [{\citenamefont {Ozawa}\ and\ \citenamefont {Goldman}(2019)}]{OzawaGoldman19}%
  \BibitemOpen
  \bibfield  {author} {\bibinfo {author} {\bibfnamefont {T.}~\bibnamefont {Ozawa}}\ and\ \bibinfo {author} {\bibfnamefont {N.}~\bibnamefont {Goldman}},\ }\bibfield  {title} {\bibinfo {title} {Probing localization and quantum geometry by spectroscopy},\ }\href {https://doi.org/10.1103/PhysRevResearch.1.032019} {\bibfield  {journal} {\bibinfo  {journal} {Phys. Rev. Res.}\ }\textbf {\bibinfo {volume} {1}},\ \bibinfo {pages} {032019} (\bibinfo {year} {2019})}\BibitemShut {NoStop}%
\bibitem [{\citenamefont {{Ghosh}}\ \emph {et~al.}(2024)\citenamefont {{Ghosh}}, \citenamefont {{Onishi}}, \citenamefont {{Xu}}, \citenamefont {{Lin}}, \citenamefont {{Fu}},\ and\ \citenamefont {{Bansil}}}]{LF_ProbingQG_24}%
  \BibitemOpen
  \bibfield  {author} {\bibinfo {author} {\bibfnamefont {B.}~\bibnamefont {{Ghosh}}}, \bibinfo {author} {\bibfnamefont {Y.}~\bibnamefont {{Onishi}}}, \bibinfo {author} {\bibfnamefont {S.-Y.}\ \bibnamefont {{Xu}}}, \bibinfo {author} {\bibfnamefont {H.}~\bibnamefont {{Lin}}}, \bibinfo {author} {\bibfnamefont {L.}~\bibnamefont {{Fu}}},\ and\ \bibinfo {author} {\bibfnamefont {A.}~\bibnamefont {{Bansil}}},\ }\bibfield  {title} {\bibinfo {title} {{Probing quantum geometry through optical conductivity and magnetic circular dichroism}},\ }\href {https://doi.org/10.48550/arXiv.2401.09689} {\bibfield  {journal} {\bibinfo  {journal} {arXiv e-prints}\ ,\ \bibinfo {eid} {arXiv:2401.09689}} (\bibinfo {year} {2024})},\ \Eprint {https://arxiv.org/abs/2401.09689} {arXiv:2401.09689 [cond-mat.mtrl-sci]} \BibitemShut {NoStop}%
\bibitem [{\citenamefont {Lawson~Jr}(1971)}]{lawson:71}%
  \BibitemOpen
  \bibfield  {author} {\bibinfo {author} {\bibfnamefont {H.~B.}\ \bibnamefont {Lawson~Jr}},\ }\bibfield  {title} {\bibinfo {title} {{The Riemannian geometry of holomorphic curves}},\ }\href {https://doi.org/10.1007/BF02584806} {\bibfield  {journal} {\bibinfo  {journal} {Boletim da Sociedade Brasileira de Matem{\'a}tica}\ }\textbf {\bibinfo {volume} {2}},\ \bibinfo {pages} {45} (\bibinfo {year} {1971})}\BibitemShut {NoStop}%
\bibitem [{Note3()}]{Note3}%
  \BibitemOpen
  \bibinfo {note} {In geometry, extrinsic data of an immersion into a Riemannian manifold is any geometric quantity which cannot be written in terms of the induced (pullback) metric. The standard example is if one takes a piece of paper and folds it into a cylinder. Both the unfold paper and the cylinder are immersions of the plane to the 3D space. The metric will not change (intrinsic geometry), but the extrinsic geometry is different.}\BibitemShut {Stop}%
\bibitem [{\citenamefont {Avdoshkin}\ and\ \citenamefont {Popov}(2023)}]{avdoshkin:popov:23}%
  \BibitemOpen
  \bibfield  {author} {\bibinfo {author} {\bibfnamefont {A.}~\bibnamefont {Avdoshkin}}\ and\ \bibinfo {author} {\bibfnamefont {F.~K.}\ \bibnamefont {Popov}},\ }\bibfield  {title} {\bibinfo {title} {{Extrinsic geometry of quantum states}},\ }\href {https://doi.org/10.1103/PhysRevB.107.245136} {\bibfield  {journal} {\bibinfo  {journal} {Phys. Rev. B}\ }\textbf {\bibinfo {volume} {107}},\ \bibinfo {pages} {245136} (\bibinfo {year} {2023})}\BibitemShut {NoStop}%
\bibitem [{\citenamefont {Calabi}(1953)}]{calabi:53}%
  \BibitemOpen
  \bibfield  {author} {\bibinfo {author} {\bibfnamefont {E.}~\bibnamefont {Calabi}},\ }\bibfield  {title} {\bibinfo {title} {{Isometric Imbedding of Complex Manifolds}},\ }\href {http://www.jstor.org/stable/1969817} {\bibfield  {journal} {\bibinfo  {journal} {Annals of Mathematics}\ }\textbf {\bibinfo {volume} {58}},\ \bibinfo {pages} {1} (\bibinfo {year} {1953})}\BibitemShut {NoStop}%
\bibitem [{\citenamefont {{Onishi}}\ and\ \citenamefont {{Fu}}(2024)}]{LiangFu_QuantumWeight_24}%
  \BibitemOpen
  \bibfield  {author} {\bibinfo {author} {\bibfnamefont {Y.}~\bibnamefont {{Onishi}}}\ and\ \bibinfo {author} {\bibfnamefont {L.}~\bibnamefont {{Fu}}},\ }\bibfield  {title} {\bibinfo {title} {{Quantum weight}},\ }\href {https://doi.org/10.48550/arXiv.2401.13847} {\bibfield  {journal} {\bibinfo  {journal} {arXiv e-prints}\ ,\ \bibinfo {eid} {arXiv:2401.13847}} (\bibinfo {year} {2024})},\ \Eprint {https://arxiv.org/abs/2401.13847} {arXiv:2401.13847 [cond-mat.str-el]} \BibitemShut {NoStop}%
\bibitem [{\citenamefont {Marzari}\ \emph {et~al.}(2012)\citenamefont {Marzari}, \citenamefont {Mostofi}, \citenamefont {Yates}, \citenamefont {Souza},\ and\ \citenamefont {Vanderbilt}}]{Vanderbilt_Wannier_RMP}%
  \BibitemOpen
  \bibfield  {author} {\bibinfo {author} {\bibfnamefont {N.}~\bibnamefont {Marzari}}, \bibinfo {author} {\bibfnamefont {A.~A.}\ \bibnamefont {Mostofi}}, \bibinfo {author} {\bibfnamefont {J.~R.}\ \bibnamefont {Yates}}, \bibinfo {author} {\bibfnamefont {I.}~\bibnamefont {Souza}},\ and\ \bibinfo {author} {\bibfnamefont {D.}~\bibnamefont {Vanderbilt}},\ }\bibfield  {title} {\bibinfo {title} {Maximally localized wannier functions: Theory and applications},\ }\href {https://doi.org/10.1103/RevModPhys.84.1419} {\bibfield  {journal} {\bibinfo  {journal} {Rev. Mod. Phys.}\ }\textbf {\bibinfo {volume} {84}},\ \bibinfo {pages} {1419} (\bibinfo {year} {2012})}\BibitemShut {NoStop}%
\bibitem [{\citenamefont {Li}\ \emph {et~al.}(2021)\citenamefont {Li}, \citenamefont {Kumar}, \citenamefont {Sun},\ and\ \citenamefont {Lin}}]{Kaisun_FCI21}%
  \BibitemOpen
  \bibfield  {author} {\bibinfo {author} {\bibfnamefont {H.}~\bibnamefont {Li}}, \bibinfo {author} {\bibfnamefont {U.}~\bibnamefont {Kumar}}, \bibinfo {author} {\bibfnamefont {K.}~\bibnamefont {Sun}},\ and\ \bibinfo {author} {\bibfnamefont {S.-Z.}\ \bibnamefont {Lin}},\ }\bibfield  {title} {\bibinfo {title} {Spontaneous fractional chern insulators in transition metal dichalcogenide moir\'e superlattices},\ }\href {https://doi.org/10.1103/PhysRevResearch.3.L032070} {\bibfield  {journal} {\bibinfo  {journal} {Phys. Rev. Research}\ }\textbf {\bibinfo {volume} {3}},\ \bibinfo {pages} {L032070} (\bibinfo {year} {2021})}\BibitemShut {NoStop}%
\bibitem [{\citenamefont {Wang}\ \emph {et~al.}(2024)\citenamefont {Wang}, \citenamefont {Zhang}, \citenamefont {Liu}, \citenamefont {He}, \citenamefont {Xu}, \citenamefont {Ran}, \citenamefont {Cao},\ and\ \citenamefont {Xiao}}]{DiXiaoFCI23}%
  \BibitemOpen
  \bibfield  {author} {\bibinfo {author} {\bibfnamefont {C.}~\bibnamefont {Wang}}, \bibinfo {author} {\bibfnamefont {X.-W.}\ \bibnamefont {Zhang}}, \bibinfo {author} {\bibfnamefont {X.}~\bibnamefont {Liu}}, \bibinfo {author} {\bibfnamefont {Y.}~\bibnamefont {He}}, \bibinfo {author} {\bibfnamefont {X.}~\bibnamefont {Xu}}, \bibinfo {author} {\bibfnamefont {Y.}~\bibnamefont {Ran}}, \bibinfo {author} {\bibfnamefont {T.}~\bibnamefont {Cao}},\ and\ \bibinfo {author} {\bibfnamefont {D.}~\bibnamefont {Xiao}},\ }\bibfield  {title} {\bibinfo {title} {Fractional chern insulator in twisted bilayer ${\mathrm{mote}}_{2}$},\ }\href {https://doi.org/10.1103/PhysRevLett.132.036501} {\bibfield  {journal} {\bibinfo  {journal} {Phys. Rev. Lett.}\ }\textbf {\bibinfo {volume} {132}},\ \bibinfo {pages} {036501} (\bibinfo {year} {2024})}\BibitemShut {NoStop}%
\bibitem [{\citenamefont {{Cr{\'e}pel}}\ and\ \citenamefont {{Fu}}(2022)}]{Valentin22_anomaloushallmetal}%
  \BibitemOpen
  \bibfield  {author} {\bibinfo {author} {\bibfnamefont {V.}~\bibnamefont {{Cr{\'e}pel}}}\ and\ \bibinfo {author} {\bibfnamefont {L.}~\bibnamefont {{Fu}}},\ }\bibfield  {title} {\bibinfo {title} {{Anomalous Hall metal and fractional Chern insulator in twisted transition metal dichalcogenides}},\ }\href@noop {} {\bibfield  {journal} {\bibinfo  {journal} {arXiv e-prints}\ ,\ \bibinfo {eid} {arXiv:2207.08895}} (\bibinfo {year} {2022})},\ \Eprint {https://arxiv.org/abs/2207.08895} {arXiv:2207.08895 [cond-mat.str-el]} \BibitemShut {NoStop}%
\bibitem [{\citenamefont {Reddy}\ \emph {et~al.}(2023)\citenamefont {Reddy}, \citenamefont {Alsallom}, \citenamefont {Zhang}, \citenamefont {Devakul},\ and\ \citenamefont {Fu}}]{LiangFuFCI23}%
  \BibitemOpen
  \bibfield  {author} {\bibinfo {author} {\bibfnamefont {A.~P.}\ \bibnamefont {Reddy}}, \bibinfo {author} {\bibfnamefont {F.}~\bibnamefont {Alsallom}}, \bibinfo {author} {\bibfnamefont {Y.}~\bibnamefont {Zhang}}, \bibinfo {author} {\bibfnamefont {T.}~\bibnamefont {Devakul}},\ and\ \bibinfo {author} {\bibfnamefont {L.}~\bibnamefont {Fu}},\ }\bibfield  {title} {\bibinfo {title} {Fractional quantum anomalous hall states in twisted bilayer ${\mathrm{mote}}_{2}$ and ${\mathrm{wse}}_{2}$},\ }\href {https://doi.org/10.1103/PhysRevB.108.085117} {\bibfield  {journal} {\bibinfo  {journal} {Phys. Rev. B}\ }\textbf {\bibinfo {volume} {108}},\ \bibinfo {pages} {085117} (\bibinfo {year} {2023})}\BibitemShut {NoStop}%
\bibitem [{\citenamefont {Jia}\ \emph {et~al.}(2024)\citenamefont {Jia}, \citenamefont {Yu}, \citenamefont {Liu}, \citenamefont {Herzog-Arbeitman}, \citenamefont {Qi}, \citenamefont {Pi}, \citenamefont {Regnault}, \citenamefont {Weng}, \citenamefont {Bernevig},\ and\ \citenamefont {Wu}}]{BAB_FCI_tTMD_1}%
  \BibitemOpen
  \bibfield  {author} {\bibinfo {author} {\bibfnamefont {Y.}~\bibnamefont {Jia}}, \bibinfo {author} {\bibfnamefont {J.}~\bibnamefont {Yu}}, \bibinfo {author} {\bibfnamefont {J.}~\bibnamefont {Liu}}, \bibinfo {author} {\bibfnamefont {J.}~\bibnamefont {Herzog-Arbeitman}}, \bibinfo {author} {\bibfnamefont {Z.}~\bibnamefont {Qi}}, \bibinfo {author} {\bibfnamefont {H.}~\bibnamefont {Pi}}, \bibinfo {author} {\bibfnamefont {N.}~\bibnamefont {Regnault}}, \bibinfo {author} {\bibfnamefont {H.}~\bibnamefont {Weng}}, \bibinfo {author} {\bibfnamefont {B.~A.}\ \bibnamefont {Bernevig}},\ and\ \bibinfo {author} {\bibfnamefont {Q.}~\bibnamefont {Wu}},\ }\bibfield  {title} {\bibinfo {title} {Moir\'e fractional chern insulators. i. first-principles calculations and continuum models of twisted bilayer ${\mathrm{mote}}_{2}$},\ }\href {https://doi.org/10.1103/PhysRevB.109.205121} {\bibfield  {journal} {\bibinfo  {journal} {Phys. Rev. B}\ }\textbf {\bibinfo {volume} {109}},\ \bibinfo {pages} {205121} (\bibinfo {year}
  {2024})}\BibitemShut {NoStop}%
\bibitem [{\citenamefont {Herzog-Arbeitman}\ \emph {et~al.}(2024)\citenamefont {Herzog-Arbeitman}, \citenamefont {Wang}, \citenamefont {Liu}, \citenamefont {Tam}, \citenamefont {Qi}, \citenamefont {Jia}, \citenamefont {Efetov}, \citenamefont {Vafek}, \citenamefont {Regnault}, \citenamefont {Weng}, \citenamefont {Wu}, \citenamefont {Bernevig},\ and\ \citenamefont {Yu}}]{BAB_FCI_tTMD_2}%
  \BibitemOpen
  \bibfield  {author} {\bibinfo {author} {\bibfnamefont {J.}~\bibnamefont {Herzog-Arbeitman}}, \bibinfo {author} {\bibfnamefont {Y.}~\bibnamefont {Wang}}, \bibinfo {author} {\bibfnamefont {J.}~\bibnamefont {Liu}}, \bibinfo {author} {\bibfnamefont {P.~M.}\ \bibnamefont {Tam}}, \bibinfo {author} {\bibfnamefont {Z.}~\bibnamefont {Qi}}, \bibinfo {author} {\bibfnamefont {Y.}~\bibnamefont {Jia}}, \bibinfo {author} {\bibfnamefont {D.~K.}\ \bibnamefont {Efetov}}, \bibinfo {author} {\bibfnamefont {O.}~\bibnamefont {Vafek}}, \bibinfo {author} {\bibfnamefont {N.}~\bibnamefont {Regnault}}, \bibinfo {author} {\bibfnamefont {H.}~\bibnamefont {Weng}}, \bibinfo {author} {\bibfnamefont {Q.}~\bibnamefont {Wu}}, \bibinfo {author} {\bibfnamefont {B.~A.}\ \bibnamefont {Bernevig}},\ and\ \bibinfo {author} {\bibfnamefont {J.}~\bibnamefont {Yu}},\ }\bibfield  {title} {\bibinfo {title} {Moir\'e fractional chern insulators. ii. first-principles calculations and continuum models of rhombohedral graphene superlattices},\ }\href
  {https://doi.org/10.1103/PhysRevB.109.205122} {\bibfield  {journal} {\bibinfo  {journal} {Phys. Rev. B}\ }\textbf {\bibinfo {volume} {109}},\ \bibinfo {pages} {205122} (\bibinfo {year} {2024})}\BibitemShut {NoStop}%
\bibitem [{\citenamefont {{Kwan}}\ \emph {et~al.}(2023)\citenamefont {{Kwan}}, \citenamefont {{Yu}}, \citenamefont {{Herzog-Arbeitman}}, \citenamefont {{Efetov}}, \citenamefont {{Regnault}},\ and\ \citenamefont {{Bernevig}}}]{BAB_FCI_tTMD_3}%
  \BibitemOpen
  \bibfield  {author} {\bibinfo {author} {\bibfnamefont {Y.~H.}\ \bibnamefont {{Kwan}}}, \bibinfo {author} {\bibfnamefont {J.}~\bibnamefont {{Yu}}}, \bibinfo {author} {\bibfnamefont {J.}~\bibnamefont {{Herzog-Arbeitman}}}, \bibinfo {author} {\bibfnamefont {D.~K.}\ \bibnamefont {{Efetov}}}, \bibinfo {author} {\bibfnamefont {N.}~\bibnamefont {{Regnault}}},\ and\ \bibinfo {author} {\bibfnamefont {B.~A.}\ \bibnamefont {{Bernevig}}},\ }\bibfield  {title} {\bibinfo {title} {{Moir{\'e} Fractional Chern Insulators III: Hartree-Fock Phase Diagram, Magic Angle Regime for Chern Insulator States, the Role of the Moir{\'e} Potential and Goldstone Gaps in Rhombohedral Graphene Superlattices}},\ }\href {https://doi.org/10.48550/arXiv.2312.11617} {\bibfield  {journal} {\bibinfo  {journal} {arXiv e-prints}\ ,\ \bibinfo {eid} {arXiv:2312.11617}} (\bibinfo {year} {2023})},\ \Eprint {https://arxiv.org/abs/2312.11617} {arXiv:2312.11617 [cond-mat.str-el]} \BibitemShut {NoStop}%
\bibitem [{\citenamefont {{Zhang}}\ \emph {et~al.}(2023)\citenamefont {{Zhang}}, \citenamefont {{Wang}}, \citenamefont {{Liu}}, \citenamefont {{Fan}}, \citenamefont {{Cao}},\ and\ \citenamefont {{Xiao}}}]{DiXiao_SmallAngleTMD_2311}%
  \BibitemOpen
  \bibfield  {author} {\bibinfo {author} {\bibfnamefont {X.-W.}\ \bibnamefont {{Zhang}}}, \bibinfo {author} {\bibfnamefont {C.}~\bibnamefont {{Wang}}}, \bibinfo {author} {\bibfnamefont {X.}~\bibnamefont {{Liu}}}, \bibinfo {author} {\bibfnamefont {Y.}~\bibnamefont {{Fan}}}, \bibinfo {author} {\bibfnamefont {T.}~\bibnamefont {{Cao}}},\ and\ \bibinfo {author} {\bibfnamefont {D.}~\bibnamefont {{Xiao}}},\ }\bibfield  {title} {\bibinfo {title} {{Polarization-driven band topology evolution in twisted MoTe$_2$ and WSe$_2$}},\ }\href {https://doi.org/10.48550/arXiv.2311.12776} {\bibfield  {journal} {\bibinfo  {journal} {arXiv e-prints}\ ,\ \bibinfo {eid} {arXiv:2311.12776}} (\bibinfo {year} {2023})},\ \Eprint {https://arxiv.org/abs/2311.12776} {arXiv:2311.12776 [cond-mat.mtrl-sci]} \BibitemShut {NoStop}%
\bibitem [{\citenamefont {Wilhelm}\ \emph {et~al.}(2021)\citenamefont {Wilhelm}, \citenamefont {Lang},\ and\ \citenamefont {L\"auchli}}]{AndreasCDW}%
  \BibitemOpen
  \bibfield  {author} {\bibinfo {author} {\bibfnamefont {P.}~\bibnamefont {Wilhelm}}, \bibinfo {author} {\bibfnamefont {T.~C.}\ \bibnamefont {Lang}},\ and\ \bibinfo {author} {\bibfnamefont {A.~M.}\ \bibnamefont {L\"auchli}},\ }\bibfield  {title} {\bibinfo {title} {Interplay of fractional chern insulator and charge density wave phases in twisted bilayer graphene},\ }\href {https://doi.org/10.1103/PhysRevB.103.125406} {\bibfield  {journal} {\bibinfo  {journal} {Phys. Rev. B}\ }\textbf {\bibinfo {volume} {103}},\ \bibinfo {pages} {125406} (\bibinfo {year} {2021})}\BibitemShut {NoStop}%
\bibitem [{\citenamefont {{Yang}}\ \emph {et~al.}(2024)\citenamefont {{Yang}}, \citenamefont {{Zhai}}, \citenamefont {{Fan}},\ and\ \citenamefont {{Yao}}}]{YaoWang_FCI_Semimetal_24}%
  \BibitemOpen
  \bibfield  {author} {\bibinfo {author} {\bibfnamefont {W.}~\bibnamefont {{Yang}}}, \bibinfo {author} {\bibfnamefont {D.}~\bibnamefont {{Zhai}}}, \bibinfo {author} {\bibfnamefont {F.-R.}\ \bibnamefont {{Fan}}},\ and\ \bibinfo {author} {\bibfnamefont {W.}~\bibnamefont {{Yao}}},\ }\bibfield  {title} {\bibinfo {title} {{Fractional quantum anomalous Hall effect in a semimetal}},\ }\href {https://doi.org/10.48550/arXiv.2405.01829} {\bibfield  {journal} {\bibinfo  {journal} {arXiv e-prints}\ ,\ \bibinfo {eid} {arXiv:2405.01829}} (\bibinfo {year} {2024})},\ \Eprint {https://arxiv.org/abs/2405.01829} {arXiv:2405.01829 [cond-mat.mes-hall]} \BibitemShut {NoStop}%
\bibitem [{\citenamefont {Song}\ \emph {et~al.}(2023)\citenamefont {Song}, \citenamefont {Goldman},\ and\ \citenamefont {Fu}}]{LiangFu_QED3_PRB23}%
  \BibitemOpen
  \bibfield  {author} {\bibinfo {author} {\bibfnamefont {X.-Y.}\ \bibnamefont {Song}}, \bibinfo {author} {\bibfnamefont {H.}~\bibnamefont {Goldman}},\ and\ \bibinfo {author} {\bibfnamefont {L.}~\bibnamefont {Fu}},\ }\bibfield  {title} {\bibinfo {title} {Emergent ${\mathrm{qed}}_{3}$ from half-filled flat chern bands},\ }\href {https://doi.org/10.1103/PhysRevB.108.205123} {\bibfield  {journal} {\bibinfo  {journal} {Phys. Rev. B}\ }\textbf {\bibinfo {volume} {108}},\ \bibinfo {pages} {205123} (\bibinfo {year} {2023})}\BibitemShut {NoStop}%
\bibitem [{\citenamefont {{Zhang}}\ and\ \citenamefont {{Song}}(2024)}]{Xueyang_Threebody_24}%
  \BibitemOpen
  \bibfield  {author} {\bibinfo {author} {\bibfnamefont {L.}~\bibnamefont {{Zhang}}}\ and\ \bibinfo {author} {\bibfnamefont {X.-Y.}\ \bibnamefont {{Song}}},\ }\bibfield  {title} {\bibinfo {title} {{Moore-Read state in Half-filled Moir{\'e} Chern band from three-body Pseudo-potential}},\ }\href {https://doi.org/10.48550/arXiv.2403.11478} {\bibfield  {journal} {\bibinfo  {journal} {arXiv e-prints}\ ,\ \bibinfo {eid} {arXiv:2403.11478}} (\bibinfo {year} {2024})},\ \Eprint {https://arxiv.org/abs/2403.11478} {arXiv:2403.11478 [cond-mat.str-el]} \BibitemShut {NoStop}%
\bibitem [{\citenamefont {Song}\ \emph {et~al.}(2024{\natexlab{a}})\citenamefont {Song}, \citenamefont {Zhang},\ and\ \citenamefont {Senthil}}]{Xueyang_PhaseTransitions_24}%
  \BibitemOpen
  \bibfield  {author} {\bibinfo {author} {\bibfnamefont {X.-Y.}\ \bibnamefont {Song}}, \bibinfo {author} {\bibfnamefont {Y.-H.}\ \bibnamefont {Zhang}},\ and\ \bibinfo {author} {\bibfnamefont {T.}~\bibnamefont {Senthil}},\ }\bibfield  {title} {\bibinfo {title} {Phase transitions out of quantum hall states in moir\'e materials},\ }\href {https://doi.org/10.1103/PhysRevB.109.085143} {\bibfield  {journal} {\bibinfo  {journal} {Phys. Rev. B}\ }\textbf {\bibinfo {volume} {109}},\ \bibinfo {pages} {085143} (\bibinfo {year} {2024}{\natexlab{a}})}\BibitemShut {NoStop}%
\bibitem [{\citenamefont {Song}\ \emph {et~al.}(2024{\natexlab{b}})\citenamefont {Song}, \citenamefont {Jian}, \citenamefont {Fu},\ and\ \citenamefont {Xu}}]{Xueyang_Intertwined_24}%
  \BibitemOpen
  \bibfield  {author} {\bibinfo {author} {\bibfnamefont {X.-Y.}\ \bibnamefont {Song}}, \bibinfo {author} {\bibfnamefont {C.-M.}\ \bibnamefont {Jian}}, \bibinfo {author} {\bibfnamefont {L.}~\bibnamefont {Fu}},\ and\ \bibinfo {author} {\bibfnamefont {C.}~\bibnamefont {Xu}},\ }\bibfield  {title} {\bibinfo {title} {Intertwined fractional quantum anomalous hall states and charge density waves},\ }\href {https://doi.org/10.1103/PhysRevB.109.115116} {\bibfield  {journal} {\bibinfo  {journal} {Phys. Rev. B}\ }\textbf {\bibinfo {volume} {109}},\ \bibinfo {pages} {115116} (\bibinfo {year} {2024}{\natexlab{b}})}\BibitemShut {NoStop}%
\bibitem [{\citenamefont {{Darius Shi}}\ \emph {et~al.}(2024)\citenamefont {{Darius Shi}}, \citenamefont {{Goldman}}, \citenamefont {{Dong}},\ and\ \citenamefont {{Senthil}}}]{Goldman_Criticality_24}%
  \BibitemOpen
  \bibfield  {author} {\bibinfo {author} {\bibfnamefont {Z.}~\bibnamefont {{Darius Shi}}}, \bibinfo {author} {\bibfnamefont {H.}~\bibnamefont {{Goldman}}}, \bibinfo {author} {\bibfnamefont {Z.}~\bibnamefont {{Dong}}},\ and\ \bibinfo {author} {\bibfnamefont {T.}~\bibnamefont {{Senthil}}},\ }\bibfield  {title} {\bibinfo {title} {{Excitonic quantum criticality: from bilayer graphene to narrow Chern bands}},\ }\href {https://doi.org/10.48550/arXiv.2402.12436} {\bibfield  {journal} {\bibinfo  {journal} {arXiv e-prints}\ ,\ \bibinfo {eid} {arXiv:2402.12436}} (\bibinfo {year} {2024})},\ \Eprint {https://arxiv.org/abs/2402.12436} {arXiv:2402.12436 [cond-mat.str-el]} \BibitemShut {NoStop}%
\bibitem [{\citenamefont {{Sharma}}\ \emph {et~al.}(2024)\citenamefont {{Sharma}}, \citenamefont {{Peng}},\ and\ \citenamefont {{Sheng}}}]{Sheng_2405}%
  \BibitemOpen
  \bibfield  {author} {\bibinfo {author} {\bibfnamefont {P.}~\bibnamefont {{Sharma}}}, \bibinfo {author} {\bibfnamefont {Y.}~\bibnamefont {{Peng}}},\ and\ \bibinfo {author} {\bibfnamefont {D.~N.}\ \bibnamefont {{Sheng}}},\ }\bibfield  {title} {\bibinfo {title} {{Topological quantum phase transitions driven by displacement fields in the twisted MoTe2 bilayers}},\ }\href {https://doi.org/10.48550/arXiv.2405.08181} {\bibfield  {journal} {\bibinfo  {journal} {arXiv e-prints}\ ,\ \bibinfo {eid} {arXiv:2405.08181}} (\bibinfo {year} {2024})},\ \Eprint {https://arxiv.org/abs/2405.08181} {arXiv:2405.08181 [cond-mat.mes-hall]} \BibitemShut {NoStop}%
\bibitem [{\citenamefont {Morf}(1998)}]{PhysRevLett.80.1505}%
  \BibitemOpen
  \bibfield  {author} {\bibinfo {author} {\bibfnamefont {R.~H.}\ \bibnamefont {Morf}},\ }\bibfield  {title} {\bibinfo {title} {Transition from quantum hall to compressible states in the second landau level: New light on the $\ensuremath{\nu}\phantom{\rule{0ex}{0ex}}=\phantom{\rule{0ex}{0ex}}5/2$ enigma},\ }\href {https://doi.org/10.1103/PhysRevLett.80.1505} {\bibfield  {journal} {\bibinfo  {journal} {Phys. Rev. Lett.}\ }\textbf {\bibinfo {volume} {80}},\ \bibinfo {pages} {1505} (\bibinfo {year} {1998})}\BibitemShut {NoStop}%
\bibitem [{\citenamefont {Storni}\ \emph {et~al.}(2010)\citenamefont {Storni}, \citenamefont {Morf},\ and\ \citenamefont {Das~Sarma}}]{PhysRevLett.104.076803}%
  \BibitemOpen
  \bibfield  {author} {\bibinfo {author} {\bibfnamefont {M.}~\bibnamefont {Storni}}, \bibinfo {author} {\bibfnamefont {R.~H.}\ \bibnamefont {Morf}},\ and\ \bibinfo {author} {\bibfnamefont {S.}~\bibnamefont {Das~Sarma}},\ }\bibfield  {title} {\bibinfo {title} {Fractional quantum hall state at $\ensuremath{\nu}=\frac{5}{2}$ and the moore-read pfaffian},\ }\href {https://doi.org/10.1103/PhysRevLett.104.076803} {\bibfield  {journal} {\bibinfo  {journal} {Phys. Rev. Lett.}\ }\textbf {\bibinfo {volume} {104}},\ \bibinfo {pages} {076803} (\bibinfo {year} {2010})}\BibitemShut {NoStop}%
\bibitem [{\citenamefont {Peterson}\ \emph {et~al.}(2008)\citenamefont {Peterson}, \citenamefont {Jolicoeur},\ and\ \citenamefont {Das~Sarma}}]{PhysRevLett.101.016807}%
  \BibitemOpen
  \bibfield  {author} {\bibinfo {author} {\bibfnamefont {M.~R.}\ \bibnamefont {Peterson}}, \bibinfo {author} {\bibfnamefont {T.}~\bibnamefont {Jolicoeur}},\ and\ \bibinfo {author} {\bibfnamefont {S.}~\bibnamefont {Das~Sarma}},\ }\bibfield  {title} {\bibinfo {title} {Finite-layer thickness stabilizes the pfaffian state for the 5/2 fractional quantum hall effect: Wave function overlap and topological degeneracy},\ }\href {https://doi.org/10.1103/PhysRevLett.101.016807} {\bibfield  {journal} {\bibinfo  {journal} {Phys. Rev. Lett.}\ }\textbf {\bibinfo {volume} {101}},\ \bibinfo {pages} {016807} (\bibinfo {year} {2008})}\BibitemShut {NoStop}%
\bibitem [{\citenamefont {Wang}\ \emph {et~al.}(2009)\citenamefont {Wang}, \citenamefont {Sheng},\ and\ \citenamefont {Haldane}}]{PhysRevB.80.241311}%
  \BibitemOpen
  \bibfield  {author} {\bibinfo {author} {\bibfnamefont {H.}~\bibnamefont {Wang}}, \bibinfo {author} {\bibfnamefont {D.~N.}\ \bibnamefont {Sheng}},\ and\ \bibinfo {author} {\bibfnamefont {F.~D.~M.}\ \bibnamefont {Haldane}},\ }\bibfield  {title} {\bibinfo {title} {Particle-hole symmetry breaking and the $\ensuremath{\nu}=\frac{5}{2}$ fractional quantum hall effect},\ }\href {https://doi.org/10.1103/PhysRevB.80.241311} {\bibfield  {journal} {\bibinfo  {journal} {Phys. Rev. B}\ }\textbf {\bibinfo {volume} {80}},\ \bibinfo {pages} {241311} (\bibinfo {year} {2009})}\BibitemShut {NoStop}%
\bibitem [{\citenamefont {Papi\ifmmode~\acute{c}\else \'{c}\fi{}}\ \emph {et~al.}(2012)\citenamefont {Papi\ifmmode~\acute{c}\else \'{c}\fi{}}, \citenamefont {Haldane},\ and\ \citenamefont {Rezayi}}]{PapicHaldaneRezayiPRL12}%
  \BibitemOpen
  \bibfield  {author} {\bibinfo {author} {\bibfnamefont {Z.}~\bibnamefont {Papi\ifmmode~\acute{c}\else \'{c}\fi{}}}, \bibinfo {author} {\bibfnamefont {F.~D.~M.}\ \bibnamefont {Haldane}},\ and\ \bibinfo {author} {\bibfnamefont {E.~H.}\ \bibnamefont {Rezayi}},\ }\bibfield  {title} {\bibinfo {title} {Quantum phase transitions and the $\ensuremath{\nu}\mathbf{=}5/2$ fractional hall state in wide quantum wells},\ }\href {https://doi.org/10.1103/PhysRevLett.109.266806} {\bibfield  {journal} {\bibinfo  {journal} {Phys. Rev. Lett.}\ }\textbf {\bibinfo {volume} {109}},\ \bibinfo {pages} {266806} (\bibinfo {year} {2012})}\BibitemShut {NoStop}%
\bibitem [{\citenamefont {Bonderson}\ \emph {et~al.}(2012)\citenamefont {Bonderson}, \citenamefont {Feiguin}, \citenamefont {M\"oller},\ and\ \citenamefont {Slingerland}}]{Slingerland_RR_12}%
  \BibitemOpen
  \bibfield  {author} {\bibinfo {author} {\bibfnamefont {P.}~\bibnamefont {Bonderson}}, \bibinfo {author} {\bibfnamefont {A.~E.}\ \bibnamefont {Feiguin}}, \bibinfo {author} {\bibfnamefont {G.}~\bibnamefont {M\"oller}},\ and\ \bibinfo {author} {\bibfnamefont {J.~K.}\ \bibnamefont {Slingerland}},\ }\bibfield  {title} {\bibinfo {title} {Competing topological orders in the $\ensuremath{\nu}=12/5$ quantum hall state},\ }\href {https://doi.org/10.1103/PhysRevLett.108.036806} {\bibfield  {journal} {\bibinfo  {journal} {Phys. Rev. Lett.}\ }\textbf {\bibinfo {volume} {108}},\ \bibinfo {pages} {036806} (\bibinfo {year} {2012})}\BibitemShut {NoStop}%
\bibitem [{\citenamefont {Zhu}\ \emph {et~al.}(2015)\citenamefont {Zhu}, \citenamefont {Gong}, \citenamefont {Haldane},\ and\ \citenamefont {Sheng}}]{Sheng_RR_PRL15}%
  \BibitemOpen
  \bibfield  {author} {\bibinfo {author} {\bibfnamefont {W.}~\bibnamefont {Zhu}}, \bibinfo {author} {\bibfnamefont {S.~S.}\ \bibnamefont {Gong}}, \bibinfo {author} {\bibfnamefont {F.~D.~M.}\ \bibnamefont {Haldane}},\ and\ \bibinfo {author} {\bibfnamefont {D.~N.}\ \bibnamefont {Sheng}},\ }\bibfield  {title} {\bibinfo {title} {Fractional quantum hall states at $\ensuremath{\nu}=13/5$ and $12/5$ and their non-abelian nature},\ }\href {https://doi.org/10.1103/PhysRevLett.115.126805} {\bibfield  {journal} {\bibinfo  {journal} {Phys. Rev. Lett.}\ }\textbf {\bibinfo {volume} {115}},\ \bibinfo {pages} {126805} (\bibinfo {year} {2015})}\BibitemShut {NoStop}%
\bibitem [{\citenamefont {Popov}\ and\ \citenamefont {Milekhin}(2021)}]{popov2020hidden}%
  \BibitemOpen
  \bibfield  {author} {\bibinfo {author} {\bibfnamefont {F.~K.}\ \bibnamefont {Popov}}\ and\ \bibinfo {author} {\bibfnamefont {A.}~\bibnamefont {Milekhin}},\ }\bibfield  {title} {\bibinfo {title} {Hidden wave function of twisted bilayer graphene: The flat band as a landau level},\ }\href {https://doi.org/10.1103/PhysRevB.103.155150} {\bibfield  {journal} {\bibinfo  {journal} {Phys. Rev. B}\ }\textbf {\bibinfo {volume} {103}},\ \bibinfo {pages} {155150} (\bibinfo {year} {2021})}\BibitemShut {NoStop}%
\bibitem [{\citenamefont {Liu}\ \emph {et~al.}(2019)\citenamefont {Liu}, \citenamefont {Liu},\ and\ \citenamefont {Dai}}]{XiDai_PseudoLandaulevel}%
  \BibitemOpen
  \bibfield  {author} {\bibinfo {author} {\bibfnamefont {J.}~\bibnamefont {Liu}}, \bibinfo {author} {\bibfnamefont {J.}~\bibnamefont {Liu}},\ and\ \bibinfo {author} {\bibfnamefont {X.}~\bibnamefont {Dai}},\ }\bibfield  {title} {\bibinfo {title} {Pseudo landau level representation of twisted bilayer graphene: Band topology and implications on the correlated insulating phase},\ }\href {https://doi.org/10.1103/PhysRevB.99.155415} {\bibfield  {journal} {\bibinfo  {journal} {Phys. Rev. B}\ }\textbf {\bibinfo {volume} {99}},\ \bibinfo {pages} {155415} (\bibinfo {year} {2019})}\BibitemShut {NoStop}%
\bibitem [{\citenamefont {Turkel}\ \emph {et~al.}(2022)\citenamefont {Turkel}, \citenamefont {Swann}, \citenamefont {Zhu}, \citenamefont {Christos}, \citenamefont {Watanabe}, \citenamefont {Taniguchi}, \citenamefont {Sachdev}, \citenamefont {Scheurer}, \citenamefont {Kaxiras}, \citenamefont {Dean},\ and\ \citenamefont {Pasupathy}}]{Ahbay_TTG_22}%
  \BibitemOpen
  \bibfield  {author} {\bibinfo {author} {\bibfnamefont {S.}~\bibnamefont {Turkel}}, \bibinfo {author} {\bibfnamefont {J.}~\bibnamefont {Swann}}, \bibinfo {author} {\bibfnamefont {Z.}~\bibnamefont {Zhu}}, \bibinfo {author} {\bibfnamefont {M.}~\bibnamefont {Christos}}, \bibinfo {author} {\bibfnamefont {K.}~\bibnamefont {Watanabe}}, \bibinfo {author} {\bibfnamefont {T.}~\bibnamefont {Taniguchi}}, \bibinfo {author} {\bibfnamefont {S.}~\bibnamefont {Sachdev}}, \bibinfo {author} {\bibfnamefont {M.~S.}\ \bibnamefont {Scheurer}}, \bibinfo {author} {\bibfnamefont {E.}~\bibnamefont {Kaxiras}}, \bibinfo {author} {\bibfnamefont {C.~R.}\ \bibnamefont {Dean}},\ and\ \bibinfo {author} {\bibfnamefont {A.~N.}\ \bibnamefont {Pasupathy}},\ }\bibfield  {title} {\bibinfo {title} {Orderly disorder in magic-angle twisted trilayer graphene},\ }\href {https://doi.org/10.1126/science.abk1895} {\bibfield  {journal} {\bibinfo  {journal} {Science}\ }\textbf {\bibinfo {volume} {376}},\ \bibinfo {pages} {193} (\bibinfo {year} {2022})},\
  \Eprint {https://arxiv.org/abs/https://www.science.org/doi/pdf/10.1126/science.abk1895} {https://www.science.org/doi/pdf/10.1126/science.abk1895} \BibitemShut {NoStop}%
\bibitem [{\citenamefont {Gao}\ and\ \citenamefont {Khalaf}(2022)}]{Khalaf_PRB22}%
  \BibitemOpen
  \bibfield  {author} {\bibinfo {author} {\bibfnamefont {Q.}~\bibnamefont {Gao}}\ and\ \bibinfo {author} {\bibfnamefont {E.}~\bibnamefont {Khalaf}},\ }\bibfield  {title} {\bibinfo {title} {Symmetry origin of lattice vibration modes in twisted multilayer graphene: Phasons versus moir\'e phonons},\ }\href {https://doi.org/10.1103/PhysRevB.106.075420} {\bibfield  {journal} {\bibinfo  {journal} {Phys. Rev. B}\ }\textbf {\bibinfo {volume} {106}},\ \bibinfo {pages} {075420} (\bibinfo {year} {2022})}\BibitemShut {NoStop}%
\bibitem [{\citenamefont {Papi\ifmmode~\acute{c}\else \'{c}\fi{}}\ and\ \citenamefont {Abanin}(2014)}]{PhysRevLett.112.046602}%
  \BibitemOpen
  \bibfield  {author} {\bibinfo {author} {\bibfnamefont {Z.}~\bibnamefont {Papi\ifmmode~\acute{c}\else \'{c}\fi{}}}\ and\ \bibinfo {author} {\bibfnamefont {D.~A.}\ \bibnamefont {Abanin}},\ }\bibfield  {title} {\bibinfo {title} {Topological phases in the zeroth landau level of bilayer graphene},\ }\href {https://doi.org/10.1103/PhysRevLett.112.046602} {\bibfield  {journal} {\bibinfo  {journal} {Phys. Rev. Lett.}\ }\textbf {\bibinfo {volume} {112}},\ \bibinfo {pages} {046602} (\bibinfo {year} {2014})}\BibitemShut {NoStop}%
\bibitem [{\citenamefont {L\"auchli}\ \emph {et~al.}(2013)\citenamefont {L\"auchli}, \citenamefont {Liu}, \citenamefont {Bergholtz},\ and\ \citenamefont {Moessner}}]{hierarchy_FCI}%
  \BibitemOpen
  \bibfield  {author} {\bibinfo {author} {\bibfnamefont {A.~M.}\ \bibnamefont {L\"auchli}}, \bibinfo {author} {\bibfnamefont {Z.}~\bibnamefont {Liu}}, \bibinfo {author} {\bibfnamefont {E.~J.}\ \bibnamefont {Bergholtz}},\ and\ \bibinfo {author} {\bibfnamefont {R.}~\bibnamefont {Moessner}},\ }\bibfield  {title} {\bibinfo {title} {Hierarchy of fractional chern insulators and competing compressible states},\ }\href {https://doi.org/10.1103/PhysRevLett.111.126802} {\bibfield  {journal} {\bibinfo  {journal} {Phys. Rev. Lett.}\ }\textbf {\bibinfo {volume} {111}},\ \bibinfo {pages} {126802} (\bibinfo {year} {2013})}\BibitemShut {NoStop}%
\bibitem [{\citenamefont {Repellin}\ \emph {et~al.}(2014)\citenamefont {Repellin}, \citenamefont {Bernevig},\ and\ \citenamefont {Regnault}}]{PhysRevB.90.245401}%
  \BibitemOpen
  \bibfield  {author} {\bibinfo {author} {\bibfnamefont {C.}~\bibnamefont {Repellin}}, \bibinfo {author} {\bibfnamefont {B.~A.}\ \bibnamefont {Bernevig}},\ and\ \bibinfo {author} {\bibfnamefont {N.}~\bibnamefont {Regnault}},\ }\bibfield  {title} {\bibinfo {title} {Z2 fractional topological insulators in two dimensions},\ }\href {https://doi.org/10.1103/PhysRevB.90.245401} {\bibfield  {journal} {\bibinfo  {journal} {Phys. Rev. B}\ }\textbf {\bibinfo {volume} {90}},\ \bibinfo {pages} {245401} (\bibinfo {year} {2014})}\BibitemShut {NoStop}%
\end{thebibliography}%

\appendix
\section*{--- APPENDIX ---}

\section{Review of Useful Concepts} \label{sec:reviews}
In this section, we provide more detailed reviews on relevant concepts, such as Grassmannian and Pl\"cker embedding.

\subsection{Grassmannian} \label{sec:review:grassmannian}
For a multi-band system, say the $r$ lowest bands below the chemical potential at zero temperature, the band complex is locally described by Eq.~(\ref{eq: local frame field for a rank r band}). In this case, arbitrary linear combination amongst the $\ket{u_{i\bm{k}}}$'s does not change the band complex and hence the $N\times r$ matrix $Z(\bf{k})=\left[Z_{ij}(\bm{k})\right]_{1\leq i\leq N, 1\leq j\leq r}$ determines the bands up to multiplication on the right by an $r\times r$ invertible matrix with entries smooth functions in the Brillouin zone. The projector,
\begin{equation}
    P(\bm{k}) = \sum_{i=1}^{r}|u_{i\bm k}\rangle\langle u_{i\bm k}|,
\end{equation}
where the right-hand side assume an orthonormal basis choice, uniquely determines the band complex and determines a map from the Brillouin zone torus to the Grassmannian $\Gr_r(\mathbb{C}^N)$, a manifold which consists of all $r$-dimensional subspaces in $\mathbb{C}^N$ denoted.

When $r=1$, the Grassmannian $\Gr_r(\mathbb{C}^N)$ reduces to $\mathbb{C}P^{N-1}$. Similarly to what happened in the $r=1$ case, here the Abelian Berry curvature and the (Abelian) quantum metric, given by the traces over band indices of their non-Abelian counterparts, are respectively, the pullback of (twice) the Fubini-Study symplectic form and metric by the mapping $P:\BZ^2\to\Gr_r(\mathbb{C}^N)$.

The Grassmannian $\Gr_r(\mathbb{C}^N)$ are K\"ahler manifolds with the Fubini-Study symplectic form taking the role of the K\"ahler (symplectic) form. What this means is that these spaces carry the structure of complex manifolds in such a way that the Fubini-Study metric and symplectic forms are compatible, locally determined by derivatives of a local K\"ahler potential. Assuming that the first $r\times r$ block of $Z$ is invertible (this defines a chart in the Grassmannian), we can by multiplication by the inverse of this block on the right bring $Z$ to the form,
\begin{equation}
    Z = \begin{bmatrix} I_{r} \\ W \end{bmatrix},\label{append_def_grassmannian}
\end{equation}
where $W$ is an $(N-r)\times r$ matrix analog to the homogeneous coordinates for the case $r=1$. The local K\"ahler potential is given by
\begin{align}
\varphi=\log \det (I_r +W^\dagger W),
\label{eq: Kaehler potential Grassmannian}
\end{align}
with $I_r$ denoting the $r\times r$ identity matrix, and we have that the Fubini-Study symplectic form and metric are determined by
\begin{eqnarray}
    \omega_{FS} &=& \frac{i}{2} \sum_{i,j,k,l} \frac{\partial^2 \varphi}{\partial W_{ij}\partial \bar{W}_{kl}} dW_{ij}\wedge d\bar{W}_{kl}, \nonumber\\
    g_{FS} &=& \sum_{i,j,k,l} \frac{\partial^2 \varphi}{\partial W_{ij}\partial \bar{W}_{kl}} dW_{ij}d\bar{W}_{kl}. \label{eq: FS Kaehler stucture}
\end{eqnarray}

\subsection{Pl\"ucker embedding} \label{sec:review:plucker}
The Pl\"ucker embedding is an isometric holomorphic embedding,
\begin{equation}
    \Gr_r(\mathbb{C}^N)\hookrightarrow \mathbb{P}\Lambda^r \mathbb{C}^N\cong \mathbb{C}P^{\binom{N}{r}-1}.
\end{equation}

Here $\Lambda^r\mathbb{C}^N$ means the $r$th exterior power of $\mathbb{C}^N$, which physically is the Hilbert space of $r$-particle fermionic states built of $\mathbb{C}^N$ interpreted as the single particle Hilbert space. The Pl\"ucker embedding assigns to a subspace of dimension $r$,
\begin{equation}
    E = \mathrm{span}\{\ket{u_1},\dots, \ket{u_{r}}\}\subset \mathbb{C}^N,\nonumber
\end{equation}
the associated $r$-particle state corresponding to filling the $r$-states spanning $E$, namely the state determined by,
\begin{equation}
    \mathrm{Slaterdet}\left(\ket{u_1}, \dots, \ket{u_{r}}\right)=\ket{u_1}\wedge \dots \wedge \ket{u_r}.\nonumber
\end{equation}

In the local coordinates defined above for the Grassmannian $\Gr_r(\mathbb{C}^N)$ and for $\mathbb{C}P^{\binom{N}{r}-1}$, this map simply takes the $(N-r)\times r$ matrix $Z$, defined in Eq.~(\ref{append_def_grassmannian}), and maps it to $\binom{N}{r}-1$ ratios of size $r\times r$ minors $M_{k_1,\dots, k_r}/M_{1,\dots, r}$, with
\begin{equation}
    M_{k_1,\dots, k_r} = \sum_{j_1,\dots,j_r}\sum_{\sigma\in S_r}\mathrm{sgn}(\sigma) Z_{k_{\sigma(1)} 1}\dots Z_{k_{\sigma(r)} r},\nonumber
\end{equation}
which is obviously holomorphic in $w$ which only appears in the numerator in products of its matrix elements. Being a isometric embedding means in particular that the metric obtained by restriction of the Fubini-Study metric in $\mathbb{C}P^{\binom{N}{r}-1}$ yields the Fubini-Study metric in $\Gr_{r}(\mathbb{C}^N)$. This is readily understood from the fact that,
\begin{equation}
    ||\mathrm{Slaterdet}(\ket{u_1},\dots,\ket{u_r})||^2 = \det(\langle u_i|u_j\rangle )_{1\leq i,j\leq r},\nonumber
\end{equation}
which yields that the restriction of the K\"ahler potential in $\mathbb{C}P^{\binom{N}{r}-1}$ to the image is the K\"ahler potential of the Grassmannian --- hence implying that the map is an isometry {\it i.e.} it preserves the K\"ahler structure. In particular this implies the useful result that if we have a complex of $r$ bands determined by a projector,
\begin{equation}
    P(\bm{k}) = \sum_{n=1}^r\ket{u_{n\bm{k}}}\bra{u_{n\bm{k}}},\nonumber
\end{equation}
we can form the $r$-particle state,
\begin{equation}
    \ket{\Psi(\bm{k})} = \mathrm{Slaterdet}(\ket{u_{1\bm{k}}},\dots,\ket{u_{r\bm{k}}}),\nonumber
\end{equation}
and we have the equality between the quantum metric determined by $P(\bm{k})$ and that determined by $\ket{\Psi_{\bm{k}}}$:
\begin{equation}
    \Tr\left(PdPdP\right) = \bra{d\Psi}\left(1-\ket{\Psi}\bra{\Psi}\right)\ket{d\Psi}, \label{metric_filled_bands}
\end{equation}
where we assumed states $\ket{u_{n\bm{k}}}$ are orthonormal and $d=dk_a \partial_{\bm{k}}^a$. The imaginary part, relation between Berry curvatures in Eq.~(\ref{trace_nonabelian_det_abelian}), follows from the same reasoning.

\section{Proof Details}
In this section, we provide proofs for certain results appearing in the main text. We will work with holomorphic functions which are not normalized. For notational simplicity, we will omit the momentum labeling of state.

\subsection{Quantum geometries of filled bands} \label{sec:quantummetricfilledbands}
In the main text, we discussed the relation between quantum geometries of filled bands and the non-Abelian geometric quantities associated to the band complex, in Eq.~(\ref{trace_nonabelian_det_abelian}). It can be understood as a consequence of the fact that Pl\"ucker embedding is an isometric embedding~\cite{kahlerband1,kahlerband2,kahlerband3}. In below, we provide a direct proof which perhaps is more accessible to physicists.

The basic idea used in this direct computation is that the connection and quantum metric can be identified by expanding the overlap functions $f_{mn}(\bm k,\delta\bm k) \equiv \langle u_{m\bm k}|u_{n,\bm k+\delta\bm k}\rangle$ to first and second order,
\begin{eqnarray}
    f_{mn}(\bm k, \delta\bm k) &=& 1 + \delta k \Tr A_{mn}(\bm k) + O(|\delta\bm k|^2),\\
    |f_{mn}(\bm k, \delta\bm k)|^2 &=& 1 - \delta k^2 g_{mn}(\bm k) + O(|\delta\bm k|^3),
\end{eqnarray}
where we have tensor contraction notation $\delta k \Tr A = \sum_{a}\delta k_a\Tr A^a$ and $\delta k^2 g_{mn} = \sum_{ab}\delta k_a\delta k_b g_{mn}^{ab}$. We will denote the determinant state as,
\begin{equation}
    \mathcal U_{\bm k} = \det_{i,j\in [1,r]} u_{i\bm k}(\bm r_j),
\end{equation}
whose form factor $\mathcal F$ and the Abelian quantum metric $g_{\rm det}$ are derived to be,
\begin{eqnarray}
    \mathcal F(\bm k, \delta\bm k) &\equiv& \langle\mathcal U_{\bm k}|\mathcal U_{\bm k+\delta\bm k}\rangle = \det_{mn}f_{mn}(\bm k, \delta\bm k),\\
    |\mathcal F(\bm k, \delta\bm k)|^2 &=& 1 - \delta k^2 g_{\rm det}(\bm k) + (|\delta\bm k|^3).\label{app_def_det_g}
\end{eqnarray}

We will now derive Eq.~(\ref{trace_nonabelian_det_abelian}) from direct expansion. First of all, one can obtain the expansion for $f_{mn}$ and $\mathcal F$ as,
\begin{equation}
    f_{mn}(\bm k, \delta\bm k) = \delta_{mn} + i\delta k A_{mn} + \frac{i\delta k^2}{2}\left(\nabla A_{mn} + i\langle\nabla u_m|\nabla u_n\rangle\right),\nonumber
\end{equation}
and
\begin{eqnarray}
    \mathcal F(\bm k, \delta\bm k) &=& 1 + i\delta k\Tr A,\nonumber\\
    &+& \frac{\delta k^2}{2}\left[-(\Tr A)^2 + \Tr A^2 + i\nabla\Tr A - \Tr\langle\nabla u|\nabla u\rangle\right],\nonumber
\end{eqnarray}
where $\nabla_a \equiv \partial_{k_a}$ and identity $\det(I + tA) = 1 + t\Tr(A) + \frac{t^2}{2}\left[(\Tr A)^2 - \Tr A^2\right] + O(t^3)$ is used. Since the leading order expansion of overlap function is defined to be the connection, the above shows the trace of the non-Abelian connection, which proves the imaginary part (for Berry curvature) of Eq.~(\ref{trace_nonabelian_det_abelian}) {\it i.e.} $\Omega_{\rm det} = {\rm Tr}\Omega$ traced over band indices. To get the metric relation, one compute $|\mathcal F|^2$,
\begin{equation}
    |\mathcal F(\bm k, \delta\bm k)|^2 = 1 - \delta k^2\left[{\rm Tr}\langle\nabla u|\nabla u\rangle - {\rm Tr}A^2\right],
\end{equation}
which, following Eq.~(\ref{app_def_det_g}), proves the real part of Eq.~(\ref{trace_nonabelian_det_abelian}), {\it i.e.} $g^{ab}_{\rm det} = {\rm Tr} g^{ab}$ where the trace acts on the band indices. Hence we finished the proof of Eq.~(\ref{trace_nonabelian_det_abelian}) from a more direct way compared to the reasoning from Pl\"ucker embedding.

\subsection{Constrains from the ladder algebra} \label{sec:constrain_coefficients}
In this section we provide derivation details for Eq.~(\ref{rec_Nn}) and Eq.~(\ref{rec_alpha}). Using the state recursion Eq.~(\ref{rec1}) to Eq.~(\ref{rec3}), one can derive the action of $\hat a\hat a^\dag|u_n\rangle$ and $\hat a^\dag\hat a|u_n\rangle$ on a given generalized Landau level state $|u_n\rangle$, with $n\geq 1$, as follows,
\begin{eqnarray}
    \hat a \hat a^\dag |u_n\rangle &=& -\mathcal N_n^{-1}\alpha_n|u_{n-1}\rangle,\nonumber\\
    &+& \left(\mathcal N_{n+1}^{-2} + |\alpha_n|^2 + \bar\partial\alpha_n\right)|u_n\rangle,\nonumber\\
    &-& \left(\mathcal N_{n+1}^{-1}\bar\alpha_{n+1} + \bar\partial N_{n+1}^{-1}\right) |u_{n+1}\rangle,\nonumber
\end{eqnarray}
and
\begin{eqnarray}
    \hat a^\dag \hat a |u_n\rangle &=& \left(\partial\mathcal N_n^{-1} - \mathcal N_n^{-1}\alpha_{n-1}\right)|u_{n-1}\rangle,\nonumber\\
    &+& \left(\mathcal N_n^{-2} + |\alpha_n|^2 - \partial\bar\alpha_n\right)|u_n\rangle,\nonumber\\
    &-& \mathcal N_{n+1}^{-1}\bar\alpha_n |u_{n+1}\rangle.\nonumber
\end{eqnarray}

Since their difference must be $|u_n\rangle$ because of the algebra $[\hat a, \hat a^\dag] = 1$, the above two equations yield the two recursions Eq.~(\ref{rec_Nn}) and Eq.~(\ref{rec_alpha}).

When $n=0$, we instead have,
\begin{eqnarray}
    \hat a\hat a^\dag |u_0\rangle &=& \left(\mathcal N_1^{-2} + |\alpha_0|^2 + \bar\partial\alpha_0\right)|u_0\rangle,\nonumber\\
    &-& \left(\mathcal N_1^{-1}\bar\alpha_1 + \bar\partial\mathcal N_1^{-1}\right)|u_1\rangle,
\end{eqnarray}
and
\begin{eqnarray}
    \hat a^\dag\hat a |u_0\rangle &=& \left(|\alpha_0|^2 - \partial\bar\alpha_0\right) |u_0\rangle - \mathcal N_1^{-1}\bar\alpha_0|u_1\rangle,
\end{eqnarray}
which gives Eq.~(\ref{rec_Nn2}) after using $[\hat a,\hat a^\dag]=1$. In order to derive Eq.~(\ref{rec_alpha2}), we first of all rewrite Eq.~(\ref{rec3}) as follows,
\begin{equation}
    \hat a\left(\mathcal N_0e^{-\frac12z\bar z}|\tilde u_0\rangle\right) = -\bar\alpha_0\left(\mathcal N_0e^{-\frac12z\bar z}|\tilde u_0\rangle\right),
\end{equation}
where $z = -i\sqrt{\mC} k$ and we have replaced $|u_0\rangle$ by $\mathcal N_0\exp(-\frac12z\bar z)|\tilde u_0\rangle$ where $|\tilde u_0\rangle$ is a holomorphic function annihilated by $\hat a$ or $\bar\partial$. Plug into the expression of $\hat a$ one arrives at Eq.~(\ref{rec_alpha2}).

\subsection{Frenet-Serret equation and the effect of holomorphic gauge transformations}
\label{sec:FSeqn_and_holomorphic_gauge_transform}
In this section, we derive the Frenet-Serret equation. Moreover we also prove that the orthonormal basis $\{\ket{u_n}\}$ is canonical in the sense that they do not mix upon gauge transformations of the ideal K\"ahler band Bloch wavefunction. The derivation of the Frenet-Serret formula is parallel to the discussion of connections and the derivation of the state recursion presented in Section.~\ref{sec:proof:LL:connection} and Section.~\ref{sec:proof:LL:recursionstate}. Here we follow the notation used in Sec.~\ref{sec:holomorphic_curve_moving_frames}.

The Gram-Schmidt process is equivalent to multiplication on the right by an upper triangular matrix $U(\bm{k})= S(z) B(\bm{k})$, where $U(\bm{k})$ and $S(z)$ are defined in Eq.~(\ref{def_holocurve_U}) and Eq.~(\ref{def_holocurve_S}). From now on, we omit the momentum variable. The standard ideal K\"ahler band in a holomorphic gauge is denoted as $\ket{u_{\bm{k}}}$, the holomorphic gauge condition beling $\frac{\partial}{\partial \bar{z}}\ket{u_{\bm{k}}}=0$. Explicitly the $n$th orthonormal state $\ket{u_n}$ is given by,
\begin{equation}
    \ket{u_{n}} = \sum_{m=0}^{\infty}\frac{\partial ^m \ket{u}}{\partial z^m} B^{m}_{\; n} =  \sum_{m=0}^{n}\frac{\partial ^m \ket{u}}{\partial z^m} B^{m}_{\; n}.\nonumber
\end{equation}
Using the fact that $\ket{u}$ is holomorphic in $z$, and the fact that the inverse of an upper triangular matrix is also an upper triangular matrix, we get the following,
\begin{widetext}
\begin{equation}
    d\ket{u_{n}} = dz\sum_{m=0}^{n}\frac{\partial ^{m+1} \ket{u}}{\partial z^{m+1}} B^{m}_{\; n} + \sum_{m=0}^{n}\frac{\partial ^m \ket{u}}{\partial z^m} dB^{m}_{\; n} = dz\sum_{m=0}^{n}\sum_{l=0}^{m+1} \ket{u_{l}}\left(B^{-1}\right)^{l}_{\; m+1} B^{m}_{\; n} + \sum_{m=0}^{n}\sum_{l=0}^{m}\ket{u_{l}}\left(B^{-1}\right)^{l}_{\; m}dB^{m}_{\; n},\nonumber
\end{equation}
\end{widetext}
where in the derivation we have used the fact that $\ket{u}$ is holomorphic. Since $B$ is an upper-triangular matrix, its inverse is also upper-triangular. From this we obtain two important pieces of information,
\begin{itemize}
    \item [(i)] $\theta^{m}_{\;n}=0$, for any $m> n+1$;
    \item [(ii)] $\theta^{m+1}_{\; m}$ is a $(1,0)$-form, {\it i.e.} it only contains $dz$.
\end{itemize}

This discussion allows one to derive the Frenet type formulas, in Eq.~(\ref{def_FS_holomorphic_curve1}) and Eq.~(\ref{def_FS_holomorphic_curve2}), which are in parallel to the state recursion relations shown in Eq.~(\ref{rec1}) to Eq.~(\ref{rec3}).

Next, we want to understand the effect of gauge transformations of $\ket{u}$ induces on the $\ket{u_n}$'s. Suppose we choose a different holomorphic gauge by performing the transformation
\begin{equation}
    \ket{u_{\bf{k}}}\rightarrow G(z)\ket{u_{\bf{k}}}, \label{append_def_holo_trans}
\end{equation}
where $G(z)$ is holomorphic and non-vanishing. Then, we have 
\begin{equation}
    \frac{\partial^{n-1}}{\partial z^{n-1}}\ket{u} \rightarrow \sum_{m=0}^{n-1}\binom{n-1}{m} G^{(m)}(z) \frac{\partial^{n-1-m}}{\partial z^{n-1-m}}\ket{u}, \label{eq: change of holomorphic derivatives}
\end{equation}
hence the orthogonal projector $P_N$ determined by $\begin{bmatrix} \ket{u},\dots, \frac{\partial^{N-1}}{\partial z^{N-1}}\ket{u} \end{bmatrix}$ is the same as the one determined after gauge transformation $\begin{bmatrix} G(z)\ket{u},\dots, \frac{\partial^{N-1}}{\partial z^{N-1}}\left(G(z)\ket{u}\right) \end{bmatrix}$. This is consistent with the fact that the map to projective space induced by $\ket{u}\wedge \dots\wedge\frac{\partial^{N-1}}{\partial z^{N-1}}\ket{u}$ is invariant under this choice and with the fact that Pl\"{u}cker map which sends an orthogonal projector of rank $N$---the set of which is the Grassmannian of $N-$planes---to a Slater determinant of a basis determining it is an embedding. In terms of orthogonal projectors, we have that,
\begin{equation}
    \ket{u_{n}} = \frac{\left(1-P_n\right)\frac{\partial^{n}}{\partial z^{n}}\ket{u}}{||\left(1-P_n\right)\frac{\partial^{n}}{\partial z^{n}}\ket{u}||}.
\end{equation}
Since in the above transformation $\left(1-P_n\right)$ is left invariant, because $P_n$ is so, and since this orthogonal projector will kill all the terms in Eq.~\eqref{eq: change of holomorphic derivatives} but the one containing $\frac{\partial^{n}}{\partial z^{n}}\ket{u}$, it is clear that all generalized Landau level states acquire the same phase under holomorphic gauge transformation Eq.~(\ref{append_def_holo_trans}),
\begin{equation}
    \ket{u_{n\bm{k}}} \longrightarrow \frac{G(z)}{|G(z)|} \ket{u_{n\bm{k}}},\quad \text{for all } n \geq 0,
\end{equation}
hence proving that the associated bands, determined by $\ket{u_n}\bra{u_n}$ are uniquely determined from $\ket{u_0}\bra{u_0}$.
\subsection{Geometry of filled bands} \label{subsec: derivation of geometry of slater dets}
Here we provide a derivation of Eq.~\eqref{eq: geometry of slaterdets}. We consider the map induced by $\Psi_N$ to projective space. The pullback of the Fubini-Study form with respect to this map, which we will denote by $\omega_N$, is equal to the pullback of the Fubini-Study form in the Grassmannian under the orthogonal projector $P_N(\bm{k})=\sum_{n=0}^{N-1}\ket{u_{n\bm{k}}}\bra{u_{n\bm{k}}}$ (because the Pl\"ucker embedding, induced by $\Psi_N$, is an isometric embedding of the Grassmannian onto its image in projective space in the main text). Hence,
\begin{align}
    \omega_N &= -\frac{i}{2} {\rm Tr} \left(P_NdP_N\wedge dP_N\right) \nonumber \\
    &=-\frac{i}{2}\sum_{n=0}^{N-1}\sum_{m=N}^{\infty}\bra{du_{n}}u_{m}\rangle\wedge  \langle u_{m}\ket{d u_{n}} \nonumber\\
    &=\frac{i}{2}\sum_{n=0}^{N-1}\sum_{m=N}^{\infty}\theta^{n}_{\;  m}\wedge  \theta^{m}_{\; n} \nonumber \\
    &=\frac{i}{2}\theta^{N-1}_{\; N}\wedge \theta^{N}_{\;N-1}=\frac{i}{2}\theta^{N}_{\;N-1}\wedge \overline{\theta}^{N}_{\;N-1},
\end{align}
where we used the Frenet-Serret formulas. The quantum metric, denoted by $\gamma_N$ is, due to holomorphicity, just
\begin{align}
\gamma_N=|\theta^{N}_{\; N-1}|^2 =\left|\langle u_{N}|\frac{\partial}{\partial z}|u_{N-1}\rangle \right|^2|dz|^2.
\end{align}
\subsection{Geometry of the generalized Landau levels and relation to the geometry of filled bands} \label{sec:Derivation of the quantum metric of nth band}
Here we derive Eq.~(\ref{eq: quantum metric of nth band}). The quantum metric $g_n$, associated with $u_{n\bm{k}}$ is given by,
\begin{align}
    g_{n}&=\bra{du_n}\left(1-\ket{u_n}\bra{u_n}\right)\ket{du_n}\nonumber\\
    & =\sum_{n\neq m}\bra{d u_{n}}u_{m}\rangle \langle u_{m}|d\ket{u_{n}} \nonumber \\
    &= |\theta^{n+1}_{\; n}|^2 + |\theta^{n}_{\; n-1}|^2= \gamma_{n+1} + \gamma_{n}.
\end{align}

Although it is not needed for the proof of our main result, we will write the Berry curvature associated to $\ket{u_{n\bm{k}}}$,
\begin{align}
    F_{n} &=\sum_{m\neq n}\bra{d u_{n}}u_{m}\rangle \wedge \langle u_{m}|d\ket{u_{n}}\\
    &= -\theta^{n}_{\; n+1}\wedge \theta^{n+1}_{\; n} -\theta^{n}_{\; n-1}\wedge \theta^{n-1}_{\; n} \nonumber \\
    &=-\theta^{n+1}_{\; n}\wedge \overline{\theta}^{n+1}_{\;n} +\theta^{n}_{\; n-1}\wedge \overline{\theta}^{n}_{\; n-1}= 2i \left(\omega_{n+1}-\omega_{n}\right).\nonumber
\end{align}
\subsection{Geometries in holomorphic frame} \label{geometry_holomorphic_frame}
In this section, we provide details for Eq.~(\ref{Omega_holomorphic}) and Eq.~(\ref{trg_holomorphic}). The definition of the Berry curvature and quantum metric components for a given Bloch state $|u_{\bm k}\rangle$ are,
\begin{eqnarray}
	\Omega(\bm k) &=& \varepsilon_{ab}\langle\partial_{\bm k}^au|\partial_{\bm k}^bu\rangle,\\
	g^{ab}(\bm k) &=& \frac{1}{2}\langle\partial_{\bm k}^au|\partial_{\bm k}^bu\rangle - \frac{1}{2}A^a_{\bm k}A^b_{\bm k} + \left(a\leftrightarrow b\right),\nonumber
\end{eqnarray}
where $A^a_{\bm k} = -i\langle u_{\bm k}|\partial_{\bm k}^au_{\bm k}\rangle$ is the connection. Here we will derive equivalent expressions in terms of holomorphic derivatives. Following the convention of the main text, introducing momentum space complex coordinates,
\begin{eqnarray}
	z &\equiv& -i\sqrt{\mC}w^a k_a,\quad\partial\equiv\partial_{z}=i\mC^{-\frac12}w^*_a\partial_{\bm k}^a,
\end{eqnarray}
such that $[\partial, z] = 1$ where repeated indices are summed implicitly. Inversely, in terms of holomorphic coordinates, the momentum space derivative is given by,
\begin{equation}
	\partial_{\bm k}^a = -i\sqrt{\mC}(w^a\partial  - w^{a*}\bar\partial).
\end{equation}

Plugging this into the expression of curvature and metric and using $w_aw^a = 0$, $w_aw^{a*}=1$, one arrives at Eq.~(\ref{Omega_holomorphic}) and Eq.~(\ref{trg_holomorphic}) where $\langle\partial u|$ denotes the Hermitian conjugate of $|\partial u\rangle$, and the holomorphic connection is given by,
\begin{equation}
	A \equiv iw_a^*A^a/\sqrt{\mC} = -i\langle u|\partial u\rangle.
\end{equation}
\subsection{Ricci scalar of filled states}
\label{subsec: derivation of the geometric recursion relation}
Here we will prove the result in Eq.~(\ref{eq: geometric recursion relation}). The Ricci form for $\omega_N=(i/2) \theta^{N}_{N-1}\wedge \overline{\theta}^{N}_{N-1}=:(i/2)hdz\wedge d\bar{z}$ (of course $h=\left|\langle u_{N}|\frac{\partial}{\partial z}|u_{N-1}\rangle \right|^2$), denoted $\Ric{\omega_N}$ is defined as $1/2$ of the curvature of the Chern connection on the Hermitian holomorphic line bundle $\left(T^*\BZ^2\right)^{(1,0)}$, with the metric described by
\begin{align}
    \langle dz,dz\rangle = h^{-1}.
\end{align}
Accordingly,
\begin{align}
    \Ric{\omega_N}=\frac{i}{2}\partial \overline{\partial} \log h.
\end{align}
Because the Chern connection coincides with the Levi-Civita connection (the unique torsion free connection compatible with the metric) and $\theta^{N}_{\; N-1}$ is a unitary frame in $\left(T^*\BZ^2\right)^{(1,0)}$, obtaining the connection coefficient is equivalent to determining the one-form $\varphi$ such that
\begin{align}
    d\theta^{N}_{\; N-1}=\varphi\wedge \theta^{N}_{\; N-1},
\end{align}
and the Ricci form is just $\Ric{\omega_N}=id\varphi/2$. Now the pullback of the Maurer-Cartan one-form satisfies 
\begin{align}
    d\theta=-\theta\wedge \theta,
\end{align}
or, in components,
\begin{align}
    d\theta^{m}_{\; n}=-\sum_{k}\theta^{m}_{\; k}\wedge \theta^{k}_{\; n}.
\end{align}
The above equations imply, setting $m=N+1$ and $n=N$,
\begin{align}
    d\theta^{N}_{N-1} &=-\sum_{n=0}^{\infty} \theta^{N}_{\; n}\wedge \theta^n_{\; N-1} \nonumber \\
    &=-\sum_{n=0}^{\infty} \theta^{N}_{\;N-1}\wedge \theta^{N-1}_{\; N-2} -\theta^{N}_{\; N}\wedge \theta^{N}_{\; N-1} \nonumber \\
    &= \left(\theta^{N-1}_{\; N-1}-\theta^{N}_{\; N}\right)\wedge \theta^{N}_{\; N-1},
\end{align}
from which $\varphi= \theta^{N-1}_{\; N-1}-\theta^{N}_{\; N}$. Using the Maurer-Cartan equation once again,
\begin{align}
    d\theta^{N-1}_{\; N-1}  &=-\theta^{N-1}_{\;N-2}\wedge  \theta^{N-2}_{\;N-1} -\theta^{N-1}_{\;N}\wedge\theta^{N}_{\;N-1} \text{ and }\nonumber \\
    d\theta^{N}_{\; N}  &=-\theta^{N}_{\;N-1}\wedge  \theta^{N-1}_{\;N} -\theta^{N}_{\;N+1}\wedge\theta^{N+1}_{\;N}, 
\end{align}
from which we get the recursion relation:
\begin{align}
    \Ric{\omega_N}=\frac{id\varphi}{2}=  \omega_{N-1} -2\omega_{N} +\omega_{N+1}.
\end{align}

\subsection{Geometries of the $u_m - u_n$ model} \label{subsec: derivation for geometry of umun model}
In this section, we use Eq.~(\ref{Omega_holomorphic}) and Eq.~(\ref{trg_holomorphic}) to derive the Berry curvature and trace of quantum metric for the $u_m - u_n$ model, whose wavefunction is given as a linear superposition of two generalized Landau level basis $u_m$ and $u_n$ with a momentum independent coefficients $\alpha, \beta$ that are normalized to one $|\alpha|^2 + |\beta|^2 = 1$,
\begin{equation}
    |u_{\bm k}\rangle \equiv \alpha|u_{m\bm k}\rangle + \beta|u_{n\bm k}\rangle.
\end{equation}

We first discuss $n = m + l$ with $l \geq 2$. In this case, the following band-off-diagonal overlaps are all zero, seen from the state recursion relations: $\langle\partial u_m|\partial u_n\rangle$, $\langle\partial u_m|\bar\partial u_n\rangle$, $\langle\bar\partial u_m|\partial u_n\rangle$ and $\langle\bar\partial u_m|\bar\partial u_n\rangle$. Moreover, connection is a decoupled sum $A = |\alpha|^2A_m + |\beta|^2A_n$ where $A_m$ is the diagonal element $A_{mm}$. Consequently, the Berry curvature is a simply a decoupled sum $\Omega = |\alpha|^2\Omega_m + |\beta|^2\Omega_n$ and the trace of metric is found to be,
\begin{equation}
    \Tr g = |\alpha|^2 \Tr g_m + |\beta|^2 \Tr g_n + 2\mC |\alpha\beta|^2 |\left(A_m-A_n\right)|^2. \nonumber
\end{equation}

The final formula for integrated trace, shown in Eq.~(\ref{Int_trace_um_un1}), is derived by noticing $|A_m - A_n| = |\alpha_m - \alpha_n|$ and Eq.~(\ref{rec_alpha}) copied below,
\begin{equation}
    \alpha_n - \alpha_{n-1} = -\partial\log\mathcal N^{-1}_n.\label{app_ricci}
\end{equation}

We proceed to discuss the case with $n = m+1$. For simplicity, we introduce notations $X_{mn} \equiv \langle\partial u_m|\partial u_n\rangle$ and $Y_{mn} \equiv \langle\bar\partial u_m|\bar\partial u_n\rangle$ and we denote their diagonal elements as $X_m \equiv X_{mm}$. After some algebra, we arrive at the following useful formula,
\begin{eqnarray}
    \langle\partial u|\partial u\rangle &=& |\alpha|^2 X_{m} + |\beta|^2 X_{m+1} + \left(\bar\alpha\beta X_{m,m+1} + c.c. \right),\nonumber\\
    \langle\bar\partial u|\bar\partial u\rangle &=& |\alpha|^2 Y_{m} + |\beta|^2 Y_{m+1} + \left(\bar\alpha\beta Y_{m,m+1} + c.c.\right),\nonumber
\end{eqnarray}
which gives the analytical expression of the Berry curvature, following Eq.~(\ref{Omega_holomorphic}),
\begin{eqnarray}
    \Omega/\mC &=& |\alpha|^2\Omega_m/\mC + |\beta|^2\Omega_{m+1}/\mC,\\
    &+& \mathcal N_{m+1}^{-1} \left[\bar\alpha\beta(\alpha_m-\alpha_{m+1}) + c.c.\right],\nonumber
\end{eqnarray}
which yields Eq.~(\ref{Int_Omega_um_un2}) presented in the main text. The second line can be proved to be vanish identically after the integration, following Eq.~(\ref{app_ricci}), due to the periodicity of normalization factors. Therefore, the Chern number of the model is preserved to be $\mC$.

Next we derive the trace of the quantum metric by using Eq.~(\ref{trg_holomorphic}). We first of all prove the integrated trace of metric is independent on the relative phase of $\alpha, \beta$. This is equivalent as proving terms linear to $\alpha\bar\beta$ or $\bar\alpha\beta$ vanishes after integration. After some algebra, we obtain the linear coefficient of $\bar\alpha\beta$, given as follows,
\begin{eqnarray}
    & & i\mathcal N_{m+1}^{-1} \left[\left(A_{m} + A_{m+1}\right) - 2\left(|\alpha|^2 A_{m} + |\beta|^2 A_{m+1}\right)\right],\nonumber\\
    &=& i\mathcal N_{m+1}^{-1}(|\alpha|^2 - |\beta|^2)(A_{m+1} - A_{m}),\nonumber\\
    &=& \mathcal N_{m+1}^{-1}(|\alpha|^2 - |\beta|^2)\partial\log\mathcal N^{-1}_{m+1},\nonumber\\
    &=& (|\alpha|^2 - |\beta|^2)\partial\mathcal N^{-1}_{m+1},
\end{eqnarray}
which clearly vanishes after integration because of the periodicity of normalizations. Similarly terms linear to $\alpha\bar\beta$ also vanishes after integration. Eq.~(\ref{app_ricci}) is used in the above derivation.

Having proved integrated trace is independent on the relative phase, we proceed to complete the derivation of the remaining terms. After some algebra, we arrive at the expression for the integrated trace, given as follows,
\begin{widetext}
    \begin{eqnarray}
        \int\Tr g/\mC &=& |\alpha|^2 \int\Tr g_m/\mC + |\beta|^2 \int\Tr g_{m+1}/\mC + 2|\alpha\beta|^2 \int \left[|A_m - A_{m+1}|^2 - \mathcal N_{m+1}^{-2}\right],\nonumber\\
        &=& |\alpha|^2 \int\Tr g_m/\mC + |\beta|^2 \int\Tr g_{m+1}/\mC + 2|\alpha\beta|^2 \int \left[|\partial\log\mathcal N^{-1}_{m+1}|^2 - (m+1)\right],
    \end{eqnarray}
\end{widetext}
where we used again Eq.~(\ref{app_ricci}) and the quantized invariant $\int \mathcal N^{-2}_{m+1} = m+1$ derived in the main text. Therefore we have finished the derivation details for Eq.~(\ref{Int_trace_um_un2}) and Eq.~(\ref{Int_trace_um_un3}) listed in the main text.

\section{Numerics Details} \label{sec:numerics_details}
\subsection{Construction of generalized Landau levels}
Here we present numerical details of constructing the modulated LL and generalized LL states.

We denote the magic unit cell basis vector as $\bm a_1$ and $\bm a_2$, which encloses an area of $|\bm a_1\times\bm a_2|=2\pi l_B^2$ with $l_B$ being the magnetic length. In this work we focus on the triangular lattice with,
\begin{equation}
    {\bm a}_{1,2} = a\left(\pm\frac{\sqrt{3}}{2},\frac{1}{2}\right),\quad a = l_B\sqrt{\frac{4\pi}{\sqrt{3}}}.\label{def_a12}
\end{equation}
The corresponding reciprocal vectors ${\bm b}_1$ and ${\bm b}_2$ are defined to satisfy $\bm a_i\cdot\bm b_j = 2\pi\delta_{ij}$.

We denote the wavefunction of the $n$th conventional Landau level as $|\Phi^{\rm LL}_{n\bm k}\rangle$ and the cell periodic part as $|u^{\rm LL}_{n\bm k}\rangle$. They are related by,
\begin{equation}
    \langle\bm r|\Phi^{\rm LL}_{n\bm k}\rangle = e^{i\bm k\cdot\bm r}\langle\bm r|u^{\rm LL}_{n\bm k}\rangle.
\end{equation}

We make a special gauge choice, following Ref.~\cite{JieWang_exactlldescription}, such that the boundary conditions of $|\Phi^{\rm LL}_{n\bm k}\rangle$ are similar to those of lattice Bloch bands. Denoting the guiding center operator as $\bm R$ which obeys $[R^a, R^b] = -i\varepsilon^{ab}l_B^2$, the states transform as follows:
\begin{eqnarray}
	e^{i\bm q\cdot\bm R}|\Phi^{\rm LL}_{n\bm k}\rangle &=& e^{\frac{i}{2}\bm q\times\bm k}|\Phi^{\rm LL}_{n,\bm k+\bm q}\rangle,\\
	|\Phi^{\rm LL}_{n,\bm k+\bm b}\rangle &=& \eta_{\bm b}e^{\frac{i}{2}\bm b\times\bm k}|\Phi^{\rm LL}_{n\bm k}\rangle,
\end{eqnarray}
where $\eta_{\bm b} = (-1)^{m+n+mn}$ for $\bm b = m\bm b_1 + n\bm b_2$. Under this gauge, the form factors of conventional Landau levels are
\begin{eqnarray}
    f^{\bm k\bm k'}_{mn}(\bm b) &=& \langle\Phi^{\rm LL}_{m\bm k}|e^{i(\bm k-\bm k'-\bm b)\cdot\bm r}|\Phi^{\rm LL}_{n\bm k'}\rangle\\
    &=&\langle u^{\rm LL}_{m\bm k}|e^{-i{\bm b}\cdot\bm r}|u^{\rm LL}_{n\bm k'}\rangle\nonumber\\
    &=& \eta_{\bm b}e^{\frac{i}{2}(\bm k+\bm k')\times\bm b}e^{\frac{i}{2}\bm k\times\bm k'}g_{mn}(\bm k-\bm k'-\bm b),\nonumber
\end{eqnarray}
where 
\begin{eqnarray}
    g_{n+m,n}(\bm q) &=& g^*_{n,n+m}(-\bm q) \nonumber\\
    &=& \sqrt\frac{n!}{(n+m)!}L_n^m(qq^*)(iq)^m\exp\left(-\frac12qq^*\right),\nonumber
\end{eqnarray}
$q=(q_x+iq_y)/\sqrt{2}$, and $L_n^m(x)$ is the generalized Laguerre polynomial. In the following, for notational simplicity we will denote $f^{\bm k\bm k'}_{nn'}(\bm 0)$ as $f^{\bm k\bm k'}_{nn'}$. 

The modulated Landau level Bloch states and their periodic parts are,
\begin{eqnarray}
    \langle\bm r|\Phi_{n\bm k}\rangle &=& \mathcal B(\bm r)\langle\bm r|\Phi^{\rm LL}_{n\bm k}\rangle,\\
    \langle\bm r|e_{n\bm k}\rangle &=& \mathcal B(\bm r)\langle\bm r|u^{\rm LL}_{n\bm k}\rangle,
\end{eqnarray}
where $\mathcal B(\bm r) = \sum_{\bm b}w_{\bm b}\exp(i\bm b\cdot\bm r)$ is the modulation function with the Fourier modes $w_{\bm b}$. We can expand $|\Phi_{n\bm k}\rangle$ under the orthonormal and complete basis $\{|\Phi^{\rm LL}_{n\bm k}\rangle\}$ as,
\begin{eqnarray}
	|\Phi_{n\bm k}\rangle &=& \mathcal{B}(\bm r) |\Phi^{\rm LL}_{n,\bm k}\rangle=\sum_m \langle \Phi^{\rm LL}_{m\bm k}|\mathcal{B}(\bm r)|\Phi^{\rm LL}_{n\bm k}\rangle |\Phi^{\rm LL}_{m\bm k}\rangle\nonumber\\
    &=&\sum_m\left(\sum_{\bm b}w_{\bm b}f^{\bm k\bm k}_{mn}(-\bm b)\right)|\Phi^{\rm LL}_{m\bm k}\rangle.\label{app_Lam}
\end{eqnarray}
Then we have
\begin{eqnarray}
	|e_{n\bm k}\rangle =\sum_m\left(\sum_{\bm b}w_{\bm b}f^{\bm k\bm k}_{mn}(-\bm b)\right)e^{-i{\bm k}\cdot{\bm r}}|\Phi^{\rm LL}_{m\bm k}\rangle.\nonumber
\end{eqnarray}
Because 
\begin{eqnarray}
    e^{-i{\bm k}\cdot{\bm r}}|\Phi^{\rm LL}_{m\bm k}\rangle&=&\sum_{m',\bm k'\in\rm{1BZ}}|\Phi^{\rm LL}_{m'\bm k'}\rangle\langle\Phi^{\rm LL}_{m'\bm k'}|e^{-i\bm k\cdot\bm r}|\Phi^{\rm LL}_{m\bm k}\rangle\nonumber\\
    &=&\sum_{m'}|\Phi^{\rm LL}_{m'\bm 0}\rangle f^{\bm 0\bm k}_{m'm},
\end{eqnarray}
we arrive at the final expression of the modulated LL states,
\begin{eqnarray}
    |e_{n\bm k}\rangle &=&\sum_m \left(\sum_{\bm b}\sum_{m'}w_{\bm b}f^{\bm k,\bm k}_{m',n}(-\bm b)f^{\bm 0\bm k}_{mm'}\right)|\Phi^{\rm LL}_{m\bm 0}\rangle\nonumber\\
    &=&\sum_m \left(\sum_{\bm b}w_{\bm b}f^{\bm 0\bm k}_{mn}(-\bm b)\right)|\Phi^{\rm LL}_{m\bm 0}\rangle. \label{app_uQH_1BZ}
\end{eqnarray}

The generalized LL basis states can be obtained straightforwardly by applying Schmidt orthogonalization to Eq.~(\ref{app_uQH_1BZ}). Numerically we keep the leading terms of $w_{\bm b}$: $w_{\bm 0}=1$ and $w_{\bm b}=\tilde w$ for the shortest ${\bm b}$. We also choose a cutoff for the sum over Landau level index $m$ in Eq.~(\ref{app_uQH_1BZ}). The required cutoff increases with $w$, which we vary between $10$ and $80$.

Note that $|u_{n\bm k}\rangle$ is expanded under the $\bm k-$independent basis $\{|\Phi^{\rm LL}_{m\bm 0}\rangle\}$, so it is convenient for computing the derivatives with respect to ${\bm k}$. Numerically the momentum-space derivatives, including the holomorphic and anti-holomorphic derivatives (equivalently the implication of the momentum-space ladder operators), are achieved by the covariant derivatives with $\bm q_a\rightarrow0$,
\begin{equation}
	|D^au_{\bm k}\rangle \approx \frac{1}{2|\bm q_a|}\left(\frac{|u_{\bm k+\bm q_a}\rangle}{\langle u_{\bm k}|u_{\bm k+\bm q_a}\rangle} - \frac{|u_{\bm k-\bm q_a}\rangle}{\langle u_{\bm k}|u_{\bm k-\bm q_a}\rangle}\right).
\end{equation}

\subsection{Interacting Hamiltonian}
A generic two-body interaction can be expressed as 
\begin{eqnarray}
    H &=&\frac{1}{2\mathcal{A}} \sum_{\bm q }v_{\bm q}:\rho_{\bm q}\rho_{-\bm q}:,
\end{eqnarray}
where $\mathcal{A}$ is the area of the 2D system and $v({\bm q})$ is the Fourier transform of interaction potential. For the bare Coulomb interaction we have $v_{\bm q}\propto 2\pi/|{\bm q}|$. In the presence of screening, $v({\bm q})$ decays faster with $|{\bm q}|$. We adopt the screening description in Ref.~\cite{PhysRevLett.112.046602} to stabilize the RR states (Fig.~\ref{fig:RR} in the main text). The energies of both bare and screened Coulomb interactions are given in unit of $\frac{e^2}{4\pi\epsilon a}$, where $-e$ is the electron's charge and $\epsilon$ is the dielectric constant.

Suppose a band Bloch state can be expanded by the conventional Landau level Bloch-like states as 
\begin{eqnarray}
   |\Phi_{\bm k}\rangle=\sum_n \alpha_{n{\bm k}}|\Phi^{\rm LL}_{n{\bm k}}\rangle,
   \label{app_cn}
\end{eqnarray}
whose periodic part is $|u_{\bm k}\rangle=\sum_n \alpha_{n{\bm k}}|u^{\rm LL}_{n{\bm k}}\rangle$. Then the matrix element of the band projected density operator $\rho_{\bm q}$ is,
\begin{eqnarray}
   \langle\Phi({\bm k})|\rho_{\bm q}|\Phi({\bm k}')\rangle&=&\sum_{m,n}\alpha_{m,{\bm k}}^*\alpha_{n,{\bm k}}\langle\Phi^{\rm LL}_{m{\bm k}}|e^{i{\bm q}\cdot{\bm r}}|\Phi^{\rm LL}_{n{\bm k}'}\rangle\nonumber\\
   &=&\delta_{{\bm k}-{\bm k}'-{\bm b},{\bm q}}\sum_{m,n}\alpha_{m,{\bm k}}^*\alpha_{n,{\bm k}}f^{{\bm k}{\bm k}'}_{mn}({\bm b})\nonumber\\
   &=&\delta_{{\bm k}-{\bm k}'-{\bm b},{\bm q}}\langle u_{\bm k}|e^{-i{\bm b}\cdot{\bm r}}|u_{{\bm k}'}\rangle,\nonumber
\end{eqnarray}
where $f_{\bm b}^{{\bm k},{\bm k}'}\equiv\langle u_{\bm k}|e^{-i{\bm b}\cdot{\bm r}}|u_{{\bm k}'}\rangle$ is the band form factor. Then we can find the second quantization form of the two-body interaction as Eqs.~(\ref{H1234}) and (\ref{V1234}) in the main text. For the $u_m-u_n$ model, the coefficients $\alpha_{n,{\bm k}}$ in Eq.~(\ref{app_cn}) can be obtained from Schmidt orthogonalization of Eq.~(\ref{app_Lam}) at point ${\bm k}$.

Similarly, the band projected three-body interaction can be written as
\begin{eqnarray}
    H_{\rm 3b} &=&\frac{1}{6\mathcal{A}^2} \sum_{{\bm q }_1,{\bm q}_2}v_{{\bm q}_1,{\bm q}_2}:\rho_{{\bm q}_1}\rho_{{\bm q}_2}\rho_{{-\bm q}_1-{\bm q}_2}:,
\end{eqnarray}
which, under second quantization, takes the form of
\begin{equation}
	H_{\rm 3b} = \sum_{{\bm k}_1,\cdots,{\bm k}_6\in{\rm 1BZ}} V_{{\bm k}_1,\cdots,{\bm k}_6}c^\dag_{\bm k_1}c^\dag_{\bm k_2}c^\dag_{\bm k_3}c_{\bm k_4}c_{\bm k_5}c_{\bm k_6},
 \label{H123456}
\end{equation}
where $c^\dag_{\bm k}$ creates an electron in $|\Phi_{\bf k}\rangle$ and the momentum conservation is implicitly imposed. The matrix element
\begin{equation}
    V_{\bm k_1,\cdots,\bm k_6} =\frac{1}{6\mathcal{A}^2} \sum_{{\bm b},{\bm b}'} v_{\bm k_1-\bm k_6-\bm b,{\bm k}_2-{\bm k}_5-{\bm b}'}f^{\bm k_1\bm k_6}_{\bm b}f^{\bm k_2\bm k_5}_{{\bm b}'}f^{\bm k_3\bm k_4}_{\delta{\bm b}-\bm b-{\bm b}'},\nonumber
\end{equation}
where $\delta{\bm b}={\bm k}_1+{\bm k}_2+{\bm k}_3-{\bm k}_4-{\bm k}_5-{\bm k}_6$. For the parent repulsion of the Pf state, we have $v_{{\bm q}_1,{\bm q}_2}\propto-|{\bm q}_1|^2|{\bm q}_1-{\bm q}_2|^4$.

\subsection{Tilted sample geometry}
In the system sizes tractable by exact diagonalization, robust fractionalized states, especially the non-Abelian ones, often require the geometry of the sample to be isotropic, {\it i.e.}, close to the 2D limit. Otherwise some charge ordered phases could be favored. In our numerical simulations, we carefully choose tilted samples whose aspect ratio is close to one. The details of tilted samples can be found in Refs.~\cite{hierarchy_FCI,PhysRevB.90.245401}. In Table~\ref{Table:tilted}, we summarize the sample geometries used in this work. 

\begin{table}
    \begin{ruledtabular}
        \begin{tabular}{ccc}
        Number of ${\bm k}$ points $N_s$ in the 1BZ & ${\bm T}_1$ & ${\bm T}_2$ \\ \hline 
        16 & (4,0) & (0,4) \\ \hline
        24 & (4,0) & (0,6) \\ \hline
        25 & (5,0) & (0,5) \\ \hline
        28 & (6,4) & (2,6) \\ \hline
        32 & (6,2) & (2,6) \\ 
        \end{tabular}
    \end{ruledtabular} \caption{Sample geometries used in our numerical simulations. The vectors ${\bm T}_1$ and ${\bm T}_2$ give the periodic boundaries of the sample. The two integers $(m,n)$ in the second and the third columns mean ${\bm T}_i=m{\bm a}_1+n{\bm a}_2$ where ${\bm a}_{1,2}$ are given in Eq.~(\ref{def_a12}).}\label{Table:tilted}
\end{table}

\subsection{Moore-Read stability for $N_e=14$ electrons}
\begin{figure*}
    \centering
    \includegraphics[width=1.0\linewidth]{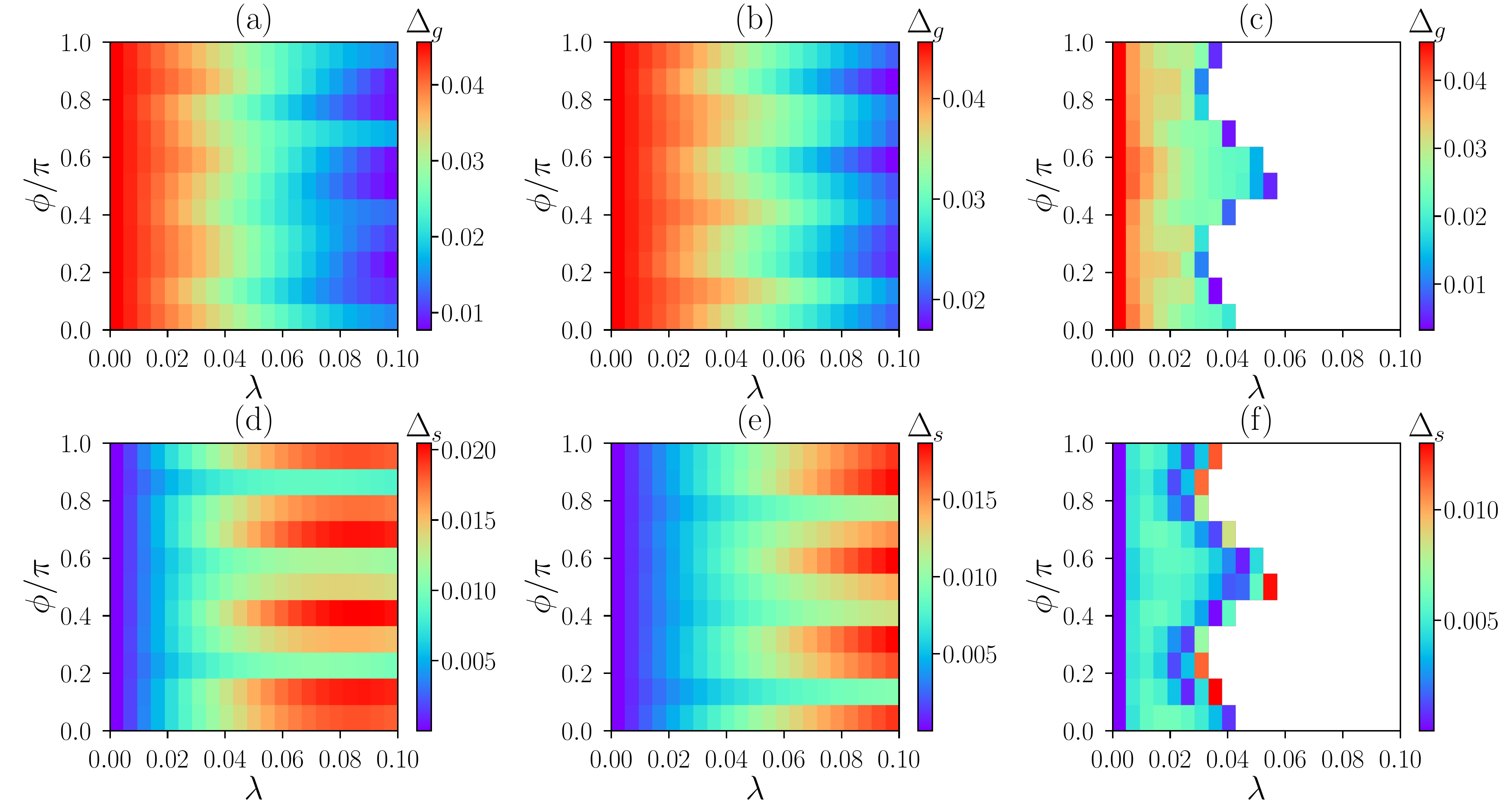} \caption{The many-body gap $\Delta_g$ (the first row) and ground-state splitting $\Delta_s$ (the second row) at $\nu=1/2$ in the $u_0 - u_1$ [(a) and (d)], $u_2 - u_1$ [(b) and (e)], and $u_3-u_1$ model [(c) and (f)] for $N_e=14$ electrons with Coulomb interaction. The modulation parameter are taken as $\tilde\omega = 0.02$ for all sub-figures. In those white regions of the $(\lambda, \phi)$ parameter space, there is no six-fold ground-state degeneracy with MR momenta.} \label{fig:MRgap_14}
\end{figure*}

In Fig.~\ref{fig:MRgap} of the main text, we have shown the neutral gap $\Delta_g$ and the ground-state splitting $\Delta_s$ at $\tilde\omega = 0.02$ in the $(\lambda, \phi)$ parameter space for the $u_0 - u_1$, $u_2 - u_1$, and $u_3-u_1$ models. The system size in that case is $N_e=12$ electrons. Here we show some data for $N_e=14$ electrons also at $\tilde\omega = 0.02$. As displayed in Fig.~\ref{fig:MRgap_14}, we find similar behavior of the six-fold degeneracy with increasing $\lambda$ --- it becomes worse when the weight of $u_1$ decays. It would be interesting to investigate how the quantum geometric region of the stable MR phase vary with growing system size.

\subsection{Ground-state weight for Moore-Read phase} \label{sec:MRweight}
\begin{figure}
    \centering
    \includegraphics[width=1.0\linewidth]{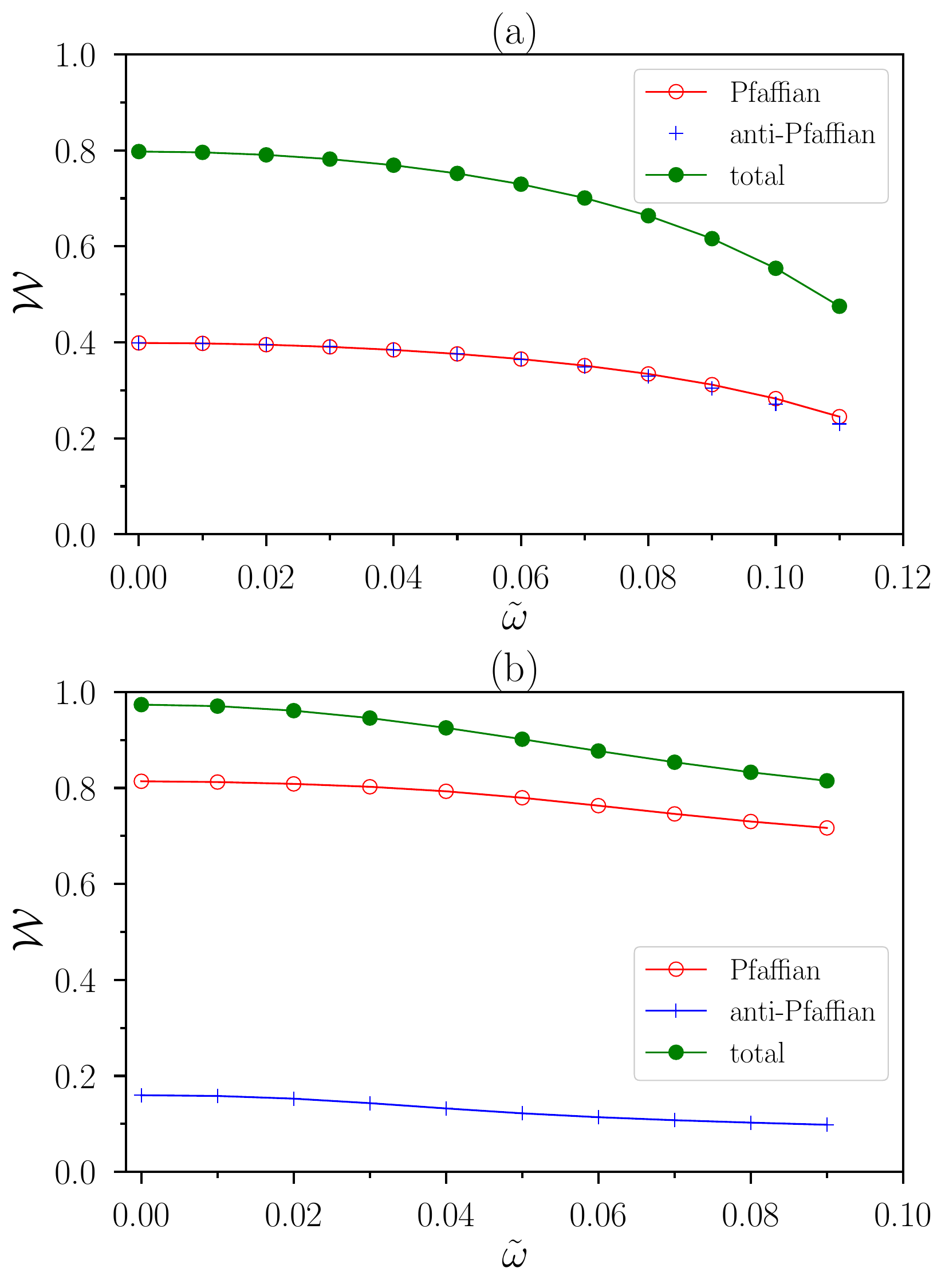} \caption{(a) The weight of the Coulomb ground state in $|u_1\rangle$ on Pf and aPf model states as well as the total weight for $N_e=14$ electrons. (b) The weight of the ground state on Pf and aPf model states as well as the total weight when a small amount of the three-body Pfaffian parent repulsion (projected to $|u_0\rangle$) is added to the Coulomb interaction (projected to $|u_1\rangle$). Here the system size is $N_e=8$ electrons. We show data up to the $\tilde\omega$ where the six-fold degeneracy is broken by level crossing with other states.} \label{fig:MRweight}
\end{figure}

In this section, we probe the quantum-geometry-induced selection of the Pf or aPf state discussed in Sec.~\ref{sec::weightonMR}. We compute the weight of the Coulomb ground state projected to $|u_1\rangle$ on the Pf and aPf states, as displayed in Fig.~\ref{fig:MRweight}(a). Here the Pf state at $\tilde\omega > 0$ Eq.~(\ref{def_tildeB})] is defined as the exact zero-energy mode of the three-body Pfaffian parent repulsion projected to $|u_0\rangle$, which can be obtained from numerical diagonalization. Notice $|u_0\rangle$ is the generalized lowest LL, or the standard ideal band, therefore it has non-trivial quantum geometric background and the three-body interaction zero mode is a generalized Pfaffian wavefunction. The corresponding aPf state is obtained by applying the PH transformation to the Pf state. All weights have been averaged over the six ground states. The total weight of the Coulomb ground state on the Pf and aPf states is decent until the critical value of $\tilde\omega \sim 0.1$ where the six-fold ground-state degeneracy disappears. This is another strong evidence of the MR phase beside the six-fold ground-state degeneracy. However, we always find almost identical weights on the Pf and aPf states even when the quantum geometric fluctuation is present. The absence of the expected quantum-geometry-induced selection of the Pf or aPf state in the system sizes studied by us is probably because the MR phase only exists for weak quantum geometric fluctuation, {\it i.e.} at relatively small $\tilde\omega$. In this case, $\epsilon_{\bm k}$ and $\langle n_{\bm k}\rangle_{\rm Pf}$ in Eq.~(\ref{eq:PfaPf}) only weakly vary with ${\bm k}$, leading to small $\Delta$ for small system sizes, such that Eq.~(\ref{def_Pf_aPf_model}) is dominated by the off-diagonal elements and the ground state is still approximately a cat state of Pf and aPf. We hence expect that much larger system sizes beyond the limit of exact diagonalization are necessary to suppress $\Delta_{1,2}$ and enhance the effects of $\Delta$. 

Another mechanism to break the PH symmetry in a single LL is the three-body interaction (being included either in purpose of model study or originating from LL mixing)~\cite{GreiterWenWilczekPRL91,GreiterWenWilczekNuclearPhyB92}. When this three-body interaction is the Pfaffian parent repulsion, we expect that the Pf state is favored and can be selected out even in small systems. We test this in the presence of quantum geometric fluctuation, as shown in Fig.~\ref{fig:MRweight}(b). Once we add a small amount of Pfaffian parent repulsion in the Hamiltonian, the ground-state weight on the Pf state dominates over that on the aPf state.

\end{document}